\documentclass[12pt]{article}
\usepackage{amsmath,amssymb,amsfonts,color,graphicx,cite,color,feynarts}
\usepackage{epsfig,psfrag,rotating,soul}
\usepackage{rotfloat}
\input paperdef

\evensidemargin \oddsidemargin
\allowdisplaybreaks

\begin{document}
\thispagestyle{empty}

\def\thefootnote{\fnsymbol{footnote}}

\begin{flushright}
IFT-UAM/CSIC-11-57\\
arXiv:1109.6232 [hep-ph]
\end{flushright}

\vspace{0.5cm}

\begin{center}

{\large\sc {\bf 
Higgs Boson masses and \boldmath{$B$}-Physics Constraints\\[.5em]
in Non-Minimal Flavor Violating SUSY scenarios}}

\vspace{1cm}

{\sc
M.~Arana-Catania$^{1}$%
\footnote{email: miguel.arana@uam.es}%
, S.~Heinemeyer$^{2}$%
\footnote{email: Sven.Heinemeyer@cern.ch}%
, M.J.~Herrero$^{1}$%
\footnote{email: Maria.Herrero@uam.es}%
~and S.~Pe\~naranda$^{3}$%
\footnote{email: siannah@unizar.es}%
}

\vspace*{.7cm}

{\sl
$^1$Departamento de F\'isica Te\'orica and Instituto de F\'isica Te\'orica,
IFT-UAM/CSIC\\
Universidad Aut\'onoma de Madrid, Cantoblanco, Madrid, Spain

\vspace*{0.1cm}

$^2$Instituto de F\'isica de Cantabria (CSIC-UC), Santander, Spain

\vspace*{0.1cm}

$^3$Departamento de F\'isica Te\'orica, Universidad de Zaragoza, 
Zaragoza, Spain

}

\end{center}

\vspace*{0.1cm}

\begin{abstract}
\noindent
We present one-loop corrections to the Higgs boson masses in the MSSM
with Non-Minimal Flavor Violation. 
%whose origen is beyond the CKM matrix. 
The flavor violation is generated from the hypothesis of general
flavor mixing in the squark mass matrices, and these are parametrized
by a complete set of $\deXYij$ ($X,Y=L,R$; $i,j=t,c,u$ or $b,s,d$).
We calculate the corrections to the Higgs masses in terms of these 
$\deXYij$ taking into account all relevant restrictions from $B$-physics
data. This includes constraints from \bsg, \bmm\ and \dmbs.
After taking into account these constraints we find sizable corrections
to the Higgs boson masses, in the case of the lightest MSSM Higgs boson
mass exceeding tens of GeV. These corrections are found mainly for the low  $\tan
\beta$ case.  
In the case of a Higgs boson mass measurement these corrections might be
used to set further constraints on $\deXYij$. 

\end{abstract}

\def\thefootnote{\arabic{footnote}}
\setcounter{page}{0}
\setcounter{footnote}{0}

\newpage
%%%%%%%%%%%%%%%%%%%%%%%%%%%%%%%%%%%%%%%%%%%%%%%%%%%%%%%%%%%%%%%%%%%%%%%%%%%%%%%
%%%%%%%%%%%%%%%%%%%%%%%%%%%%%%%%%%%%%%%%%%%%%%%%%%%%%%%%%%%%%%%%%%%%%%%%%%%%%%%

\section{Introduction}

The Higgs sector of the Minimal Supersymmetric
Standard Model (MSSM)~\cite{mssm} with two scalar doublets
accommodates five physical Higgs bosons. In
lowest order these are the light and heavy $\cp$-even $h$
and $H$, the $\cp$-odd $A$, and the charged Higgs bosons $H^\pm$.
The Higgs sector of the MSSM can be expressed at lowest
order in terms of the gauge couplings, $\MA$ and $\tb \equiv v_2/v_1$, 
the ratio of the two vacuum expectation values. All other masses and
mixing angles can therefore be predicted.
Higher-order contributions can give 
large corrections to the tree-level relations (see
e.g.\ \citere{reviews} for reviews). 

The MSSM predicts scalar partners for each fermionic degree of freedom
in the Standard Model (SM), and fermionic partners for Higgs and gauge
bosons. So far, the direct search for SUSY
particles has not been successful. One can only set lower bounds of
several hundreds of~GeV, depending on the particle and the model
specifications~\cite{ATLASsusy11,CMSsusy11}. To lift the masses of the
SUSY partners from the corresponding SM values, soft SUSY-breaking terms
are introduced~\cite{mssm}. 
The most general flavor structure of the soft SUSY-breaking
sector with flavor non-diagonal terms would
induce large flavor-changing neutral-currents, contradicting
the experimental results~\cite{Nakamura:2010zzi}. Attempts to avoid this kind of
problem include flavor-blind SUSY-breaking scenarios, like minimal
Supergravity or gauge-mediated SUSY-breaking. In these
scenarios, the sfermion-mass matrices are flavor diagonal in the same
basis as the quark matrices at the SUSY-breaking scale. However, a certain
amount of flavor mixing is generated due to the renormalization-group
evolution from the SUSY-breaking scale down to the electroweak
scale. 
In a more agnostic approach all flavor-violating terms are introduced
as free parameters, and each model point, i.e.\ each combination of
flavor-violating parameters, is tested against experimental data. 

Similarly strong bounds exist for the MSSM Higgs sector from the
non-observation of Higgs bosons at LEP~\cite{LEPHiggsSM,LEPHiggsMSSM}, 
the Tevatron~\cite{Tevbounds} and most recently from LHC 
searches~\cite{LHCHiggsbounds}. 
The LHC has good prospects to discover at least one neutral Higgs boson over
the full MSSM parameter space. A precision on the mass of a SM-like Higgs boson of $\sim 200 \mev$ is
expected~\cite{lhctdrs,jakobs}.
At the ILC a determination of the Higgs boson properties (within the
kinematic reach) will be possible, and an accuracy on the mass could
reach the $50 \mev$~level~\cite{tesla,Snowmass05Higgs}. 
The interplay of the LHC and the ILC in the neutral MSSM Higgs sector is
discussed in \citere{lhcilc}.

For the MSSM%
\footnote{We concentrate here on the case with real parameters. For
  complex parameters see \citeres{mhcMSSMlong,mhcMSSM2L} and references
  therein.}
the status of higher-order corrections to the masses and mixing angles
in the neutral Higgs sector is quite advanced. 
The full one-loop and potentially all leading two-loop corrections are
known, see~\citere{mhiggsAEC} for a review. Most recently leading
three-loop corrections became available~\cite{mhiggsEP3l,mhiggsFD3l}. 

However, the impact of non-minimal flavor violation (NMFV) on the
MSSM Higgs-boson masses and mixing angles, entering already at the 
one-loop level, has not been explored very deeply so far.
A one-loop calculation taking into account the $LL$-mixing between the
third and second generation of scalar up-type quarks has been performed
in \citere{mhNMFVearly}. A full one-loop calculation of the Higgs-boson
self-energies including all NMFV mixing terms had been implemented into
the Fortran code \fh~\cite{feynhiggs,mhiggslong,mhiggsAEC,mhcMSSMlong}, 
however no cross checks or numerical evaluations analyzing the
effects of the new mixing terms were performed.
Possible effects from NMFV on Higgs boson decays
were investigated in \cite{HdecNMFV,HdecNMFV2,HdecNMFV3}. Within a similar context of NMFV
there have been also studied some effects of scharm-stop flavor mixing in top-quark
FCNC processes \cite{Cao1} and charged Higgs processes \cite{Dittmaier:2007uw} as well as the implications for LHC
\cite{Cao2,Dittmaier:2007uw}. 
Some previous studies on the induced radiative corrections on 
the Higgs mass from scharm-stop flavor mixing have also been performed in
\cite{Cao1,Cao2}, but any effects from mixing involving the first generation of scalar quarks have been
neglected. The numerical estimates in \cite{Cao1,Cao2} also neglect all the flavor mixings in the scalar down-type sector, except for those of $LL$-type that are induced from the scalar up-type sector by SU(2) invariance. In \cite{Cao2} they also consider one example with a particular numerical value of non-vanishing ${\tilde s}_L-{\tilde b}_R$ mixing.
 
We study in this paper the consequences from NMFV 
for the MSSM Higgs-boson spectrum, where our results are obtained
  in full generality, i.e.\ all generations in the scalar up- and
  down-type quark sector are included in our analytical results.
In the numerical analysis we focus particularly on the
flavor mixing between the second and third generations which is the 
relevant one in $B$ physics. Our estimates include all type of flavor mixings, $LL$, $LR$, $RL$, and $RR$. We devote special attention to the $LR/RL$
 sector. These kind of mixing effects are expected
to be sizable, since they enter the off-diagonal $A$~parameters, which
appear directly in the coupling of the Higgs bosons to scalar quarks.

Concerning the constraints from flavor observables we take into account the
most up-to-date evaluations in the NMFV MSSM for \bsg\footnote{Subleading NLO 
MSSM corrections were evaluated in ~\cite{Degrassi:2007kj,Degrassi:2006eh}
. However, their
 effect on our evaluations would be minor.}, \bmm\ and \dmbs,
based on the BPHYSICS subroutine included in the SuFla code~\cite{sufla}. 
For the evaluation of \dmbs\
we have incorporated into this subroutine additional contributions from the
one-loop gluino boxes~\cite{Baek:2001kh} which are known to be very relevant
in the context of NMFV 
scenarios~\cite{Foster:2005wb,Gabbiani:1996hi,Becirevic:2001jj}. 

In the first step of the analysis we scan over the NMFV parameters, and
contrast them with the experimental bounds on \bsg, \bmm\ and \dmbs.
In the second step we analyze the one-loop contributions of NMFV to the
MSSM Higgs boson masses, focusing on the parameter space still allowed
by the experimental flavor constraints. In this way the full possible
impact of NMFV in the MSSM on the Higgs sector can be explored.
The paper is organized as follows: in \refse{sec:nmfv} we introduce our
notation for the NMFV MSSM and define certain benchmark scenarios that
are used for the subsequent analysis. 
Our implementation and new results on $B$-physics observables is given
in \refse{sec:Bphysics}, where we also analyze in detail which 
combination of NMFV
parameters are still allowed by current experimental constraints. 
The calculation of the corrections to Higgs boson masses in the NMFV
MSSM is presented in \refse{sec:mhiggs}. The numerical analysis of the
impact of the one-loop Higgs mass corrections is given in \refse{sec:numanal}. Our
conclusions can be found in \refse{sec:conclusions}.

%%%%%%%%%%%%%%%%%%%%%%%%%%%%%%%%%%%%%%%%%%%%%%%%%%%%%%%%%%%%%%%%%%%%%%%%%%%%%%
%%%%%%%%%%%%%%%%%%%%%%%%%%%%%%%%%%%%%%%%%%%%%%%%%%%%%%%%%%%%%%%%%%%%%%%%%%%%%%%

\section{SUSY scenarios with Non-Minimal Flavor Violation}
\label{sec:nmfv}

We work in SUSY scenarios with the same particle content as the
MSSM, but with 
general flavor mixing hypothesis in the squark sector. Within these
SUSY-NMFV scenarios, besides the usual flavor violation originated by
the CKM matrix of the quark sector, the  general flavor mixing in the
squark mass matrices additionally generates flavor violation from the
squark sector. 
These squark flavor mixings are usually described in terms of a set of
dimensionless parameters $\deXYij$ ($X,Y=L,R$; $i,j=t,c,u$ or $b,s,d$),
which for simplicity in the computations are frequently  considered
within the Mass Insertion Approximation (MIA)~\cite{Hall:1985dx}. We
will not use here this approximation, but instead we will solve exactly
the diagonalization of the squark mass matrices.   

In this section we summarize the main features of the squark flavor mixing within the SUSY-NMFV scenarios and set the notation.

The relevant MSSM superpotential terms are: 
\begin{equation}
\label{W:Hl:def}
W\,=\,\epsilon_{\alpha \beta}\left[  \,\hat H_2^\alpha \hat Q^\beta Y^u \hat U \,-\, 
 \hat H_1^\alpha\,\hat Q^\beta Y^d \hat D   \,+\,\mu \hat H_1^\alpha \hat H_2^\beta \right] \,,
\end{equation}
where the involved superfields are: $\hat Q$, containing the  
quark $(u_L\,\,\, d_L)^T$ and squark $(\tilde u_L\,\,\, \tilde d_L)^T$ 
$SU(2)$ doublets; $\hat U$, containing the quark $(u_R)^c$ and
squark $\tilde u_R^*$ $SU(2)$ singlets; $\hat D$, containing the quark $(d_R)^c$ and squark $\tilde d_R^*$ $SU(2)$ singlets; and $\hat H _{1,2}$ containing
the Higgs bosons $SU(2)$ doublets, ${\cal H}_1= ({\cal H}^0_1\,\,\, {\cal H}^-_1)^T$ and ${\cal H}_2= ({\cal H}^+_2 \,\,\,{\cal H}^0_2)^T$, 
 and their SUSY partners. 
$f^c$ 
denotes here the particle-antiparticle conjugate  of  fermion $f$, 
and $\tilde f^*$ denotes the complex conjugate of sfermion $\tilde f$. We follow the notation of ~\cite{Gunion:1986yn}, but with the the convention 
$\epsilon_{12}=-1$.
The generation indices in the superfields, $\hat Q_i$,$\hat U_i$, $\hat D_i$, quarks $q_i$, squarks  $\tilde q_i$, $(i=1,2,3)$, and Yukawa coupling $3\times 3$ matrices, $Y^u_{ij}$, $Y^d_{ij}$, ($i,j=1,2,3$), have been omitted above for brevity.

Usually one starts rotating the quark fields from the $SU(2)$ (interaction) eigenstate basis, $q^{\rm int}_{L,R}$, to the mass (physical), 
$q_{L,R}^{\rm phys}$ eigenstate basis by unitary transformations, $V^{u,d}_{L,R}$:

\begin{equation}
\VL u^{\rm phys}_{L,R} \\ c^{\rm phys}_{L,R} \\ t^{\rm phys}_{L,R} \VR =
V^u_{L,R} \VL u^{\rm int}_{L,R} \\ c^{\rm int}_{L,R} \\ t^{\rm int}_{L,R} \VR~,~~~~
\VL d^{\rm phys}_{L,R} \\ s^{\rm phys}_{L,R} \\ b^{\rm phys}_{L,R} \VR =
V^d_{L,R} \VL d^{\rm int}_{L,R} \\ s^{\rm int}_{L,R} \\ b^{\rm int}_{L,R} \VR~,
\end{equation}
such that the fermion mass matrices in the physical basis are diagonal:
\begin{eqnarray}
V^u_L Y^{u*}V^{u\dagger}_R&=&{\rm diag}(y_u,y_c,y_t)=
{\rm diag}\left(\frac{m_u}{v_2},\frac{m_c}{v_2},\frac{m_t}{v_2}\right),
\\
V^d_L Y^{d*}V^{d\dagger}_R&=&{\rm diag}(y_d,y_s,y_b)=
{\rm diag}\left(\frac{m_d}{v_1},\frac{m_s}{v_1},\frac{m_b}{v_1}\right),
\end{eqnarray} 
where $v_1=\left< {\cal H}_1^0\right>$ and $v_2=\left< {\cal H}_2^0\right>$ are the vacuum expectation values of the neutral Higgs fields.
The CKM matrix, which is responsible for the flavor violation in
the quark sector, is given as usual as, 
\begin{equation}
\VCKM= V^u_L V^{d\dagger}_L.
\end{equation}
A simultaneous (parallel) rotation of the squarks with the same above unitary matrices as their corresponding quark partners  leads to the so-called Super-CKM basis. In the NMFV scenarios, contrary to the MFV ones, the Super-CKM basis does not coincide with  the physical squark basis, i.e, their corresponding squark mass matrices are not yet diagonal. 
More concretely, rotating the squarks from the interaction basis, ${\tilde q}^{\rm int}_{L,R}$ to the Super-CKM basis, ${\tilde q}_{L,R}$, by
\begin{equation}
\VL \tilde u_{L,R} \\ \tilde c_{L,R} \\ \tilde t_{L,R} \VR =
V^u_{L,R} \VL \tilde u^{\rm int}_{L,R} \\ \tilde c^{\rm int}_{L,R} \\ \tilde t^{\rm int}_{L,R} \VR~,~~~
\VL \tilde d_{L,R} \\ \tilde s_{L,R} \\ \tilde b_{L,R} \VR =
V^d_{L,R} \VL \tilde d^{\rm int}_{L,R} \\ \tilde s^{\rm int}_{L,R} \\ \tilde b^{\rm int}_{L,R} \VR~,
\end{equation}
one gets the soft-SUSY-breaking Lagrangian transformed from:
\begin{align}
    \mathcal{L}_{\rm soft} 
 &= -  \tilde {\cal U}^{{\rm int}*}_i m_{\tilde U ij}^2 \, \tilde {\cal U}^{\rm int}_j
   - \tilde {\cal D}^{{\rm int}*}_i m_{\tilde D ij}^2 \tilde {\cal D}^{\rm int}_j
   - \tilde {\cal Q}^{{\rm int}\dagger}_i m_{\tilde Q ij}^2 \tilde {\cal
 Q}^{\rm int}_j  \non \\
  &-\left[ \tilde {\cal Q}^{\rm int}_i {\bar {\cal  A}}^{u}_{ij} \tilde {\cal U}^{{\rm int}*}_j {\cal H}_2 
         - \tilde {\cal Q}^{\rm int}_i {\bar   {\cal A}}^{d}_{ij} \tilde {\cal D}^{{\rm int}*}_j {\cal H}_1 \; + \; \text{h.c.} 
    \right]  
\label{eq:lsoft}
\end{align}
to:
\begin{align}
\label{eq:lsoft-superCKM}
   &\mathcal{L}_{\rm soft}
 = -  \tilde {\cal U}_{Ri}^* m_{\tilde U_R ij }^2 \tilde {\cal U}_{Rj} 
   - \tilde {\cal D}_{Ri}^* m_{\tilde D_R ij}^2 \tilde {\cal D}_{Rj}
-  \tilde {\cal U}_{Li}^* m_{\tilde U_L ij}^2 \tilde {\cal U}_{Lj} 
   - \tilde {\cal D}_{Li}^* m_{\tilde D_L ij}^2 \tilde {\cal D}_{Lj} \\
 &\quad
-  \left[ \tilde {\cal U}_{Li} {\cal A}^u_{ij} \tilde {\cal U}^*_{Rj} {\cal H}_2^0
- \tilde {\cal D}_{Li} (\VCKM)_{ki} {\cal A}^u_{kj} \tilde {\cal U}^*_{Rj} 
{\cal H}_2^+  
 -  \tilde {\cal U}_{Li} (\VCKM^*)_{ik} {\cal A}^d_{kj} \tilde {\cal D}^*_{Rj} 
{\cal H}_1^-
+ \tilde {\cal D}_{Li} {\cal A}^d_{ij} \tilde {\cal D}^*_{Rj} {\cal H}_1^0 +  \text{h.c.}
    \right], \nonumber 
\end{align}
where we have used calligraphic capital letters for the
squark fields with generation indexes,
\begin{equation}
\tilde {\cal U}^{{\rm int}}_{1,2,3}=\tilde u^{{\rm int}}_R,\tilde c^{{\rm int}}_R,\tilde t^{{\rm int}}_R; 
\tilde {\cal D}^{{\rm int}}_{1,2,3}=\tilde d^{{\rm int}}_R,\tilde s^{{\rm int}}_R,\tilde b^{{\rm int}}_R;
\tilde {\cal Q}^{{\rm int}}_{1,2,3}=(\tilde u^{{\rm int}}_L \, \tilde d^{{\rm int}}_L)^T, (\tilde c^{{\rm int}}_L\, \tilde s^{{\rm int}}_L)^T, (\tilde t^{{\rm int}}_L \, \tilde b^{{\rm int}}_L)^T  ;
\end{equation}
\begin{equation}
\tilde {\cal U}_{L1,2,3}=\tilde u_L,\tilde c_L,\tilde t_L; 
\tilde {\cal D}_{L1,2,3}=\tilde d_L,\tilde s_L,\tilde b_L;
\tilde {\cal U}_{R1,2,3}=\tilde u_R,\tilde c_R,\tilde t_R; 
\tilde {\cal D}_{R1,2,3}=\tilde d_R,\tilde s_R,\tilde b_R;
\end{equation}
and ($q=u, d$)
\begin{equation}
{\cal A}^{q} = V^{q}_L {\bar {\cal A}}^q V^{q \dagger}_R, 
 m_{\tilde U_R}^2 = V_R^u  m_{\tilde U}^2 V_R^{u \dagger}, 
 m_{\tilde D_R}^2 = V_R^d  m_{\tilde D}^2 V_R^{d \dagger}, 
 m_{\tilde U_L}^2 = V_L^u  m_{\tilde Q}^2 V_L^{u \dagger}, 
 m_{\tilde D_L}^2 = V_L^d  m_{\tilde Q}^2 V_L^{d \dagger}. 
\label{eq:su2}
\end{equation}
The usual procedure to introduce general flavor mixing in the squark sector is to include the non-diagonality in flavor space at this stage, namely, in the Super-CKM basis. Thus, one usually writes the $6\times 6$ non-diagonal mass matrices, ${\mathcal M}_{\tilde u}^2$ and ${\mathcal M}_{\tilde d}^2$, referred to the Super-CKM basis, being ordered respectively as $(\SupL, \SchaL, \StopL, \SupR, \SchaR, \StopR)$ and  $(\SdownL, \SstrL, \SbotL, \SdownR, \SstrR, \SbotR)$, and write them in terms of left- and right-handed blocks $M^2_{\tilde q \, AB}$ ($q=u,d$, $A,B=L,R$), which are non-diagonal $3\times 3$ matrices,
\begin{equation}
{\mathcal M}_{\tilde q}^2 =\left( \begin{array}{cc}
M^2_{\tilde q \, LL} & M^2_{\tilde q \, LR} \\ 
M_{\tilde q \, LR}^{2 \, \dagger} & M^2_{\tilde q \,RR}
\end{array} \right), \qquad \tilde q= \tilde u, \tilde d.
\label{eq:blocks-matrix}
\end{equation} 
 where:
 \begin{alignat}{5}
 M_{\tilde u \, LL \, ij}^2 
  = &  m_{\tilde U_L \, ij}^2 + \left( m_{u_i}^2
     + (T_3^u-Q_u\sin^2 \theta_W ) M_Z^2 \cos 2\beta \right) \delta_{ij},  \notag\\
 M^2_{\tilde u \, RR \, ij}
  = &  m_{\tilde U_R \, ij}^2 + \left( m_{u_i}^2
     + Q_u\sin^2 \theta_W M_Z^2 \cos 2\beta \right) \delta_{ij} \notag, \\
  M^2_{\tilde u \, LR \, ij}
  = &  \left< {\cal H}_2^0 \right> {\cal A}_{ij}^u- m_{u_{i}} \mu \cot \beta \, \delta_{ij},
 \notag, \\
 M_{\tilde d \, LL \, ij}^2 
  = &  m_{\tilde D_L \, ij}^2 + \left( m_{d_i}^2
     + (T_3^d-Q_d \sin^2 \theta_W ) M_Z^2 \cos 2\beta \right) \delta_{ij},  \notag\\
 M^2_{\tilde d \, RR \, ij}
  = &  m_{\tilde D_R \, ij}^2 + \left( m_{d_i}^2
     + Q_d\sin^2 \theta_W M_Z^2 \cos 2\beta \right) \delta_{ij} \notag, \\
  M^2_{\tilde d \, LR \, ij}
  = &  \left< {\cal H}_1^0 \right> {\cal A}_{ij}^d- m_{d_{i}} \mu \tb \, \delta_{ij}.
\label{eq:SCKM-entries}
\end{alignat}
with, $i,j=1,2,3$, $Q_u=2/3$, $Q_d=-1/3$, $T_3^u=1/2$ and $T_3^d=-1/2$. $\theta_W$ is the weak angle, $M_Z$ is the $Z$ gauge boson mass, and $(m_{u_1},m_{u_2}, m_{u_3})=(m_u,m_c,m_t)$, $(m_{d_1},m_{d_2}, m_{d_3})=(m_d,m_s,m_b)$. It should be noted that the non-diagonality in flavor comes from the values of $m_{\tilde U_L \, ij}^2$, $m_{\tilde U_R \, ij}^2$, $m_{\tilde D_L \, ij}^2$, $m_{\tilde D_R \, ij}^2$, ${\cal A}_{ij}^u$ and ${\cal A}_{ij}^d$ for $i \neq j$.

The next step is to rotate the squark states from the Super-CKM basis, 
${\tilde q}_{L,R}$, to the
physical basis, ${\tilde q}^{\rm phys}$. If we set the order in the Super-CKM basis as above, $(\SupL, \SchaL, \StopL, \SupR, \SchaR, \StopR)$ and  $(\SdownL, \SstrL, \SbotL, \SdownR, \SstrR, \SbotR)$, and in the physical basis as
${\tilde u}_{1,..6}$ and ${\tilde d}_{1,..6}$, respectively, these last rotations are given by two $6 \times 6$ matrices, $R^{\tilde u}$ and $R^{\tilde d}$, 
\BE
\VL  \tiu_{1} \\ \tiu_{2}  \\ \tiu_{3} \\
                                    \tiu_{4}   \\ \tiu_{5}  \\\tiu_{6}   \VR
  \; = \; R^{\tiu}  \VL \SupL \\ \SchaL \\\StopL \\ 
  \SupR \\ \SchaR \\ \StopR \VR ~,~~~~
\VL  \tid_{1} \\ \tid_{2}  \\  \tid_{3} \\
                                   \tid_{4}     \\ \tid_{5} \\ \tid_{6}  \VR             \; = \; R^{\tid}  \VL \SdownL \\ \SstrL \\ \SbotL \\
                                      \SdownR \\ \SstrR \\ \SbotR \VR ~,
\label{newsquarks}
\end{equation} 
yielding the diagonal mass-squared matrices as follows,
\BEA
{\rm diag}\{m_{\tiu_1}^2, m_{\tiu_2}^2, 
          m_{\tiu_3}^2, m_{\tiu_4}^2, m_{\tiu_5}^2, m_{\tiu_6}^2 
           \}  & = &
R^{\tiu}  \;  {\cal M}_{\tiu}^2   \; 
 R^{\tiu \dagger}    ~,\\
{\rm diag}\{m_{\tid_1}^2, m_{\tid_2}^2, 
          m_{\tid_3}^2, m_{\tid_4}^2, m_{\tid_5}^2, m_{\tid_6}^2 
          \}  & = &
R^{\tid}  \;   {\cal M}_{\tid}^2   \; 
 R^{\tid \dagger}    ~.
\EEA 

The corresponding Feynman rules in the physical basis for the vertices
needed  for our applications, i.e.\ the interaction of one
and two Higgs or gauge bosons with two squarks, can be found in the
Appendix A. This new set of generalized vertices had been 
implemented into the program packages
\fa/{\em FormCalc}\cite{feynarts,formcalc} extending the previous MSSM model
file~\cite{famssm}. The extended \fa\ version  
was used for the evaluation of the Feynman diagrams along this
paper to obtain the general analytical results.

\bigskip
In the numerical part of the present study we will restrict ourselves
  to the case where there are flavor mixings exclusively between the second and third squark generation. These mixings are known to produce the largest flavor violation effects in $B$ meson physics since their size are usually governed by the third generation quark masses. On the other hand, and in order to reduce further the number of independent parameters, we will focus in the following analysis on 
 constrained  SUSY scenarios, where the soft mass parameters fulfill universality hypothesis at the gauge unification (GUT) scale. Concretely, we will work with the so-called Constrained MSSM (CMSSM) and Non Universal Higgs Mass (NUHM) scenarios, which are defined by(see~\cite{AbdusSalam:2011fc} and references therein),  
 \begin{eqnarray}
 {\rm CMSSM}:& ~~~~~m_0, m_{1/2}, A_0, {\rm sign}(\mu), \tb 
 \nonumber \\
 {\rm NUHM}:& ~~~~~m_0,  m_{1/2}, A_0, {\rm sign}(\mu), \tb, 
 m_{H_1}, m_{H_2},
 \end{eqnarray} 
where, $A_0$ is the universal trilinear coupling, $m_0$, $m_{1/2}$, $m_{H_1}$, $m_{H_2}$, are the universal scalar mass,
gaugino mass, and $H_1$ and $H_2$ Higgs masses, respectively, at the GUT scale,   ${\rm sign}(\mu)$ is the sign of the $\mu$ parameter and again $\tb =v_2/v_1$.  The soft Higgs masses in the NUHM are usually parametrized as 
$m^2_{H_{1,2}}=(1+\delta_{1,2})m_0^2$, such that by taking $\delta_{1,2}=0$ one 
moves from the NUHM to the CMSSM. 

It should be noted that the condition of universal squark soft masses, $m^2_{\tilde U_Lij}= m^2_{\tilde U_Rij}=m^2_{\tilde D_Lij}= 
m^2_{\tilde D_Rij}=m^2_0 \delta_{ij}$, is fulfilled just at the GUT scale. When 
running these soft mass matrices from the GUT scale down to the relevant low energy, they will generically turn non-diagonal in flavor. However, in MFV scenarios the non-diagonal terms are exclusively generated in this running by off-diagonal terms in the $\VCKM$, and therefore they can be safely neglected at low energy. Contrary, in NMFV scenarios, the universal hypothesis in these 
squark mass matrices is by definition not fulfilled at low energies.        
 
Our final settings for the numerical evaluation of the squark flavor mixings in NMFV scenarios are  fixed (after RGE running) at low energy as follows,
\noindent \begin{equation}  
m^2_{\tilde U_L}= \left(\begin{array}{ccc}
 m^2_{\tilde U_{L11}} & 0 & 0\\
0 & m^2_{\tilde U_{L22}}  & \delta_{23}^{LL} m_{\tilde U_{L22}}m_{\tilde U_{L33}}\\
0 & \delta_{23}^{LL} m_{\tilde U_{L22}}m_{\tilde U_{L33}}& m^2_{\tilde U_{L33}} \end{array}\right)\end{equation}

\noindent \begin{equation}
v_2 {\cal A}^u  =\left(\begin{array}{ccc}
0 & 0 & 0\\
0 & 0 & \delta_{ct}^{LR} m_{\tilde U_{L22}}m_{\tilde U_{R33}}\\
0 & \delta_{ct}^{RL} m_{\tilde U_{R22}}m_{\tilde U_{L33}} & m_{t}A_{t}\end{array}\right)\end{equation}

\noindent \begin{equation}  
m^2_{\tilde U_R}= \left(\begin{array}{ccc}
 m^2_{\tilde U_{R11}} & 0 & 0\\
0 & m^2_{\tilde U_{R22}}  & \delta_{ct}^{RR} m_{\tilde U_{R22}}m_{\tilde U_{R33}}\\
0 & \delta_{ct}^{RR} m_{\tilde U_{R22}}m_{\tilde U_{R33}}& m^2_{\tilde U_{R33}} \end{array}\right)\end{equation}

\noindent \begin{equation}
m^2_{\tilde D_L} =\VCKM^{\dagger}m^2_{\tilde U_L}\VCKM
\label{eq:relac-mu2ll-md2ll-su2}\end{equation}

\noindent \begin{equation}
v_1 {\cal A}^d   =\left(\begin{array}{ccc}
0 & 0 & 0\\
0 & 0 & \delta_{sb}^{LR} m_{\tilde D_{L22}}m_{\tilde D_{R33}} \\
0 & \delta_{sb}^{RL} m_{\tilde D_{R22}}m_{\tilde D_{L33}} & m_{b}A_{b}\end{array}\right)\end{equation}

\noindent \begin{equation}  
m^2_{\tilde D_R}= \left(\begin{array}{ccc}
 m^2_{\tilde D_{R11}} & 0 & 0\\
0 & m^2_{\tilde D_{R22}}  & \delta_{sb}^{RR} m_{\tilde D_{R22}}m_{\tilde D_{R33}}\\
0 & \delta_{sb}^{RR} m_{\tilde D_{R22}}m_{\tilde D_{R33}}& m^2_{\tilde D_{R33}} 
\end{array}\right)\end{equation}
It is worth mentioning that the relation between the two soft squark mass matrices in the 'Left' sector (\ref{eq:relac-mu2ll-md2ll-su2}) is due to SU(2) gauge
invariance. Eq. \ref{eq:relac-mu2ll-md2ll-su2} can be derived from the two last relations of eq. \ref{eq:su2}. This dependence between the non-diagonal terms of these squark mass matrices is the reason why is introduced $\delta_{23}^{LL}$ instead of two independent deltas $\delta_{ct}^{LL}$ and $\delta_{sb}^{LL}$.  
To get the needed running of the soft parameters from the GUT scale down to low energy, that we set here 1 TeV,  we solve numerically the one-loop RGEs with the code {\it SPHENO}~\cite{Porod:2011nf}. 
The diagonalization of all the mass matrices is
performed with the program
\fh~\cite{feynhiggs,mhiggslong,mhiggsAEC,mhcMSSMlong}.  

In CMSSM and other SUSY-GUT scenarios the flavor changing deltas go (in the leading logarithmic approximation) as 
$\de^{LL}_{23} \simeq   -\frac{1}{8\pi^2}\frac{(3m_0^2+A_0^2)}{\tilde
  m^2}(Y^{q\dagger}Y^q)_{23}\log(\frac{M_{{\rm GUT}}}{M_{{\rm EW}}})$ ($\tilde m^2$ is usually taken as the geometric mean of the involved flavor diagonal squared squark mass matrix entries, see eq. \ref{deltasdefs}), 
whereas the $LR$, $RL$ and $RR$ deltas are
suppressed instead by small mass ratios, $\sim \frac{(m_q A_0)}{\tilde
  m^2}$ and   
$\sim \frac{(m_q^2)}{\tilde m^2}$, respectively. Furthermore, in these scenarios the mixing involving the first generation squarks is naturally suppressed by the smallness of the corresponding Yukawa couplings.
In order to keep the number of free parameters manageable, this
motivated our above choice of neglecting in the numerical 
analysis the mixing of the first generation squarks. However, we will
not assume any explicit hierarchy in the various mixing terms between
the second and third generation.

It should be noted that in the 'Left-Right' sector we have 
normalized the trilinear couplings at low energies as ${\cal A}^q_{ij}= y_{q_i} A^q_{ij}$  (with $A^u_{33}=A_t$ and $A^d_{33}=A_b$) and we have neglected the $A_i$ couplings of the first and second generations. Finally, it should be noted that the dimensionless parameters $\deXYij$ defining the non-diagonal entries in flavor space 
$(i \neq j)$ are normalized respect the  geometric mean of the corresponding diagonal squared soft masses. For instance, 
\begin{eqnarray}
&&\delta^{LL}_{23}= m^2_{\tilde U_{L23}}/(m_{\tilde U_{L22}}m_{\tilde U_{L33}}), 
\delta^{RR}_{ct}= m^2_{\tilde U_{R23}}/(m_{\tilde U_{R22}}m_{\tilde U_{R33}}),
~~~\nonumber\\
&&\delta^{LR}_{ct}= (v_2 {\cal A}^u)_{23}/(m_{\tilde U_{L22}}m_{\tilde U_{R33}}), 
\delta^{RL}_{ct}= (v_2 {\cal A}^u)_{32}/(m_{\tilde U_{R22}}m_{\tilde U_{L33}}),
~~~{\rm etc}.
\label{deltasdefs}
\end{eqnarray}     
 
For definiteness and simplicity, in the present work we will perform our estimates in specific constrained SUSY scenarios, CMSSM and NUHM, whose input parameters $m_0$, $m_{1/2}$, $A_0$, $\tb$, sign($\mu$), $\delta_{1,2}$, are summarized in table \ref{points}\footnote{We adopt here the definition in terms of the GUT-scale input parameters, while the original definition in  \cite{Allanach:2002nj}
was based on the weak-scale parameters.}, and supplemented with $\deXYij$ as
 described above. Regarding CMSSM, we have chosen six SPS benchmark points~\cite{Allanach:2002nj}, SPS1a, SPS1b, SPS2, SPS3, SPS4, and SPS5 and one  more point with 
very heavy spectra (VHeavyS). It should be noted that several of these
SPSX points are already in conflict with recent LHC
data~\cite{ATLASsusy11, CMSsusy11}, but we have chosen them here as
reference points to study the effects of SUSY on the Higgs mass
corrections, since they have been studied at length in the
literature. At present, a heavier SUSY spectrum, as for instance
our point VHeavyS is certainly more realistic and compatible
with LHC data. In general an analysis of LHC data including NMFV effects 
in the squark sector would be very desirable.
Regarding NUHM, we have chosen a point with heavy SUSY spectra and light Higgs sector (HeavySLightH) and a point (BFP) that has been proven in   \cite{Buchmueller:2011ki} to give the best fit to the set of low energy observables. For later reference, needed in our posterior analysis of the Higgs mass corrections, we also include in the table the corresponding MSSM Higgs masses,  computed with \fh~\cite{feynhiggs,mhiggslong,mhiggsAEC,mhcMSSMlong} and with all flavor changing deltas switched off, i.e., $\deXYij=0$.     

\begin{center}
\begin{table}[h!]
\begin{tabular}{|c|c|c|c|c|c|c|c|c|c|c|}
\hline 
points & $m_{1/2}$ & $m_0$ & $A_0$ & $\tb$ &  $\delta_1$ & $\delta_2$ &$m_{h}$ &$m_{H}$ &$M_{A}$ &$m_{H^{\pm}}$ \\\hline 
SPS1\,a & 250 & 100 & -100 & 10 &  $0$ & $0$ &112 &394 &394&402 \\
 SPS1\,b & 400 & 200 & 0 & 30 &  $0$ & $0$  &116 &526 &526 &532\\
 SPS2 &  300 & 1450 & 0 & 10 &  $0$ & $0$  &115 &1443 &1443 &1445 \\
 SPS3 &  400 & 90 & 0 & 10 &   $0$ & $0$   &115 &573 &572 &578\\
 SPS4 &  300 & 400 & 0 & 50 &  $0$ & $0$   &114 &404 &404 &414\\
 SPS5 &  300 & 150 & -1000 & 5 &  $0$ & $0$   &111 &694 &694 &698\\
VHeavyS &  800 & 800 & -800 & 5 &  $0$ & $0$   &120 &1524 &1524 &1526\\
HeavySLightH &600 &600 &0 & 5 &   $-1.86$ & $+1.86 $ &114 &223&219 &233\\
BFP &530 &110 &-370 & 27 &   $-84.7$ & $-84.7$ &120 &507 &507 &514\\
\hline
\end{tabular}
\caption{Values of $m_{1/2}$, $m_0$, $A_0$, $\tb$, $\delta_{1}$, $\delta_{2}$ and Higgs boson masses $m_{h}$, $m_{H}$, $M_{A}$ and $m_{H^{\pm}}$ for the points considered in the analysis. All parameters with mass dimension are in GeV, and ${\rm sign}(\mu)>0$ for all points.\label{points}}
\end{table}
\end{center}

%%%%%%%%%%%%%%%%%%%%%%%%%%%%%%%%%%%%%%%%%%%%%%%%%%%%%%%%%%%%%%%%%%%%%%%%%%%%%%

\section{Constraints on Non-Minimal Flavor Violating SUSY 
scenarios from \boldmath{$B$}-Physics} 
\label{sec:Bphysics}
In this section we analyze the constraints on Non-Minimal Flavor
Violating SUSY scenarios from $B$-Physics. Since we are mainly interested
in the phenomenological consequences of the flavor mixing between the
third and second generations we will focus% 
\footnote{
We have checked that electroweak precision observables, where NMFV
  effects enter, for instance, via $\De\rho$~\cite{mhNMFVearly}, do not
  lead to relevant additional constraints on the allowed parameter space. Our results on this constraint are in agreement with \citere{Cao1}.}%
~on the following three B
meson observables: 1) Branching ratio of the $B$ radiative decay \bsg, 
2) Branching ratio of the $B_s$ muonic decay \bmm, 
and 3) $B_s-{\bar B_s}$ mass difference \dmbs.  Another
$B$ observable of interest in the present context is \bsll.
However, we have not included this in our study, because the
predicted rates in NMFV-SUSY scenarios for this observable are closely
correlated with those from \bsg\ due to the dipole
operators dominance in the photon-penguin diagrams mediating \bsll\ 
decays.  It implies that the restrictions on the flavor
mixing $\deXYij$ parameters from \bsll
 are also expected to be correlated with those
from the radiative decays.    

The summary of the relevant features for our analysis of these three B
meson observables is given in the following. 

\subsection{\boldmath{\bsg}}

The relevant effective
Hamiltonian for this decay is given in terms of the Wilson coefficients $C_i$ and operators $O_i$ by:

\noindent \begin{equation}
\mathcal{H}_{\rm eff}=-\frac{4G_{F}}{\sqrt{2}}\VCKM^{ts*}\VCKM^{tb}\sum_{i=1}^{8}(C_{i}O_{i}+C'_{i}O'_{i}). 
\end{equation}
Where the primed operators can be obtained from the unprimed ones by replacing 
$L \leftrightarrow R$.
The complete list of operators can be found, for instance, in \cite{Gambino:2001ew}.  
In the context of SUSY scenarios with the MSSM particle content and MFV, only two of these operators get relevant contributions, the so-called photonic dipole operator $O_{7}$ and gluonic dipole operator
$O_{8}$ given, respectively, by:

\noindent \begin{equation}
O_{7}=\frac{e}{16\pi^{2}}m_{b}\left(\bar{s}_{L}\sigma^{\mu\nu}b_{R}\right)F_{\mu\nu},\end{equation}

\noindent \begin{equation}
O_{8}=\frac{g_{3}}{16\pi^{2}}m_{b}\left(\bar{s}_{L}\sigma^{\mu\nu}T^{a}b_{R}\right)G_{\mu\nu}^{a}.\end{equation}
We have omitted the color indices here for brevity.
Within NMFV also $O'_{7,8}$ have to be taken into account:

\noindent \begin{equation}
\label{opo7p}
O'_{7}=\frac{e}{16\pi^{2}}m_{b}\left(\bar{s}_{R}\sigma^{\mu\nu}b_{L}\right)F_{\mu\nu},\end{equation}

\noindent \begin{equation}
\label{opo8p}
O'_{8}=\frac{g_{3}}{16\pi^{2}}m_{b}\left(\bar{s}_{R}\sigma^{\mu\nu}T^{a}b_{L}\right)G_{\mu\nu}^{a}.\end{equation}

The Wilson coefficients at the SUSY scale are obtained as usual by the 
matching procedure of the proper matrix element being computed from the 
previous 
effective Hamiltonian to the corresponding matrix elements being computed 
from the SUSY model operating at that SUSY scale, the NMFV-MSSM in our case. 
These Wilson coefficients encode, therefore, the contributions to 
${\rm BR}(B \to X_s \ga$) from the loops of the SUSY and Higgs sectors of the 
MSSM. The effects from squark flavor mixings that are parametrized by the 
$\deXYij$, are included in this observable via the squark physical 
masses and rotation matrices, given in the previous section, that appear in 
the computation of the matrix element at the one loop level and, therefore,
are also encoded in the Wilson coefficients. The explicit expressions
for these  coefficients in the MSSM, in terms of the physical basis, can
be found, for instance, in
refs.~\cite{Bertolini:1990if,Cho:1996we,Degrassi:2000qf}. 
We have included in our analysis the most relevant loop contributions to
the Wilson coefficients, concretely: 1) loops with Higgs bosons, 2)
loops with charginos and 3) loops with gluinos. It should be noted that, at one loop order, the gluino
loops do not contribute in MFV scenarios, but they are very relevant
(dominant in many cases) in the present NMFV scenarios.   
 
Once the Wilson coefficients are known at the SUSY scale, they
are evolved with the proper Renormalization Group Equations (RGEs) down
to the proper low-energy scale. As a consequence of this running the
previous operators mix and the corresponding Wilson coefficients,
$C_{7,8}$ get involved in the computation of the $B \to X_s \ga$
decay rate. The RGE-running is done in two steps: The first one is from the SUSY scale down to the electroweak scale, and the second one is from this electroweak scale down to the B-physics scale.  For the first step, we use the LO-RGEs for the relevant Wilson coefficients as in
\cite{Degrassi:2000qf} and fix six active quark flavors in this
running. For the second running we use the NLO-RGEs as in \cite{Hurth:2003dk} and fix, correspondingly, five active quark flavors. For the charged Higgs sector we use the NLO formulas for the
Wilson coefficients as in \cite{Ciuchini:1997xe}.  

 The resummation of scalar induced large $\tb$ effects is
 performed, as usual,  by the effective Lagrangian approach that
 parametrizes the one-loop generated effective couplings between the
  ${\cal H}_2$ Higgs doublet and the down type quarks in softly broken SUSY
 models~\cite{Hall:1993gn}. A necessary condition to take into account
 all $\tb$-enhanced terms in flavor changing amplitudes is the
 diagonalization of the down-type quark mass matrix in the presence of
 these effective couplings \cite{Carena:2000uj,Carena:1999py,Isidori:2001fv}. A summary of this effective Lagrangian formalism
 for the resummation of large $\tb$ effects in the three $B$
 observables of our interest, within the context of MFV scenarios, can be found in
 \cite{Buras:2002vd}. We follow here the treatment of
 \cite{Isidori:2002qe} where the resummation of large $\tb$
 effects via effective Lagrangians is generalized to the case where the
 effective ${\bar d}^i_R d^j_L {\cal H}^0_2$ coupling contains also non-minimal
 sources of flavor mixing. It should be noted that the most relevant scalar induced
 large $\tb$ effects for the present work are those generated by
 one-loop diagrams with gluino-sbottom and chargino-stop inside the
 loops. The large $\tb$ resummation effects and the relevance of other chirally enhanced corrections
to FCNC processes within the NMFV context have recently been
studied exhaustively also in \cite{Crivellin1,Crivellin2}
(previous studies can be found, for instance, in
\citeres{Okumura:2002wa,Okumura:2003hy,Foster:2004vp}). 

The total branching ratio for this decay is finally estimated by adding
the new contributions from the SUSY and Higgs sectors to the  
SM rate. More specifically, we use eq.(42) of \cite{Hurth:2003dk} for
the estimate of  \bsg\ in terms of the ratios of 
the Wilson coefficients $C_{7,8}$ and  $C'_{7,8}$ (including all the mentioned new contributions)
divided by the corresponding $C_{7,8}^{\rm SM}$ in the SM. 

For the numerical estimates of \bsg\ we use the FORTRAN
subroutine BPHYSICS (modified as to include the contributions from  $C'_{7,8}$ which were not included in its original version) included in the SuFla code, that
incorporates all the above mentioned ingredients\cite{sufla}.  

Finally, for completeness, we include below the experimental measurement
of this observable \cite{Nakamura:2010zzi,Asner:2010qj}\footnote{We have added the various contributions to the experimental error in quadrature.}, and its
prediction within the SM \cite{Misiak:2009nr}: 

\vspace{0.5cm}
\noindent \begin{equation}
\bsg_{\rm exp}=(3.55 \pm 0.26)\times10^{-4}
\label{bsgamma-exp}
\end{equation}

\noindent \begin{equation}
\bsg_{\rm SM}= (3.15 \pm 0.23)\times10^{-4}
\label{bsgamma-SM}
\end{equation}

%%%%%%%%%%%%%%%%%%%%%%%%%%%%%%%%%%%%%%%%%%%%%%%%%%%%%%%%%%%%%%%%%%%%%%%%%%%%%%

\subsection{\boldmath{\bmm}}

The relevant effective Hamiltonian for this process is \cite{Chankowski:2000ng} \cite{Bobeth:2002ch}: 

\noindent \begin{equation}
\mathcal{H}_{\rm eff}=-\frac{G_{F}\alpha}{\sqrt{2} \pi}\VCKM^{ts*}\VCKM^{tb}\sum_{i} (C_{i}O_{i}+C'_{i} O'_{i}), 
\end{equation}
where the operators $O_i$ are given by:
\begin{align}
{O}_{10}&=\left(\bar{s}\ga^{\nu}P_Lb\right)\left(\bar{\mu}\ga_{\nu}\ga_5\mu\right),
& {O}_{10}^{\prime}&=\left(\bar{s}\ga^{\nu}P_Rb\right)\left(\bar{\mu}\ga_{\nu}\ga_5\mu\right),\nonumber\\
 {O}_{S}&=m_b\left(\bar{s}P_Rb\right)\left(\bar{\mu}\mu\right),
& {O}_{S}^{\prime}&=m_s\left(\bar{s} P_Lb \right)\left(\bar{\mu}\mu\right),\nonumber\\
 {O}_{P}&=m_b\left(\bar{s} P_Rb \right)\left(\bar{\mu}\ga_5\mu\right),
& {O}_{P}^{\prime}&=m_s\left(\bar{s} P_Lb \right)\left(\bar{\mu}\ga_5\mu\right).\label{bsm:Ops}
\end{align}
We have again omitted the color indices here for brevity.

In this case, the RG running is straightforward since the anomalous dimensions of 
the above involved operators are zero, 
and the 
prediction for the decay rate is simply expressed by: 
\begin{align}
\bmm &= \frac{G_F^2\alpha^2 m_{B_s}^2 f_{B_s}^2\tau_{B_s}}{64 \pi^3}\lvert \VCKM^{ts*}\VCKM^{tb}\rvert^2\sqrt{1-4\hat{m}_{\mu}^2}
\nonumber\\
&\times\left[\left(1-4\hat{m}_{\mu}^2\right)\lvert F_S\rvert^2+\lvert F_P+2\hat{m}_{\mu}^2 F_{10}\rvert^2\right],
\label{bsm:br}
\end{align}
where $\hat{m}_{\mu}=m_{\mu}/m_{B_s}$ and the 
$F_i$ are given by
\begin{align}
F_{S,P}&=m_{B_s}\left[\frac{C_{S,P}m_b-C_{S,P}^{\prime}m_s}{m_b+m_s}\right],
&F_{10}=C_{10}-C_{10}^{\prime}.
\nonumber
\end{align}

Within the SM the most relevant operator is $O_{10}$ as the Higgs mediated 
contributions to $O_{S,P}$ can be safely neglected. It should be noted that the 
contribution from $O_{10}$ to the decay rate is helicity suppressed by a 
factor of  $\hat{m}_{\mu}^2$ since the $B_s$ meson has spin zero. 
In contrast, in SUSY scenarios the scalar and pseudo-scalar operators, 
$O_{S,P}$, can be very important, particularly at large $\tb
\gsim 30$ where the contributions to $C_S$ and $C_P$ from neutral Higgs
penguin diagrams can become large and dominate the branching ratio,
because in this case the branching ratio grows with $\tb$ as
${\tan}^6\beta$.  The studies in the literature of these MSSM
Higgs-penguin contributions to ${\rm BR}(B_s\to \mu ^ + \mu^-)$ have focused
on both MFV \cite{Babu:1999hn,Bobeth:2001sq,Isidori:2001fv} and NMFV  
scenarios
\cite{Chankowski:2000ng,Isidori:2002qe,Foster:2004vp,Foster:2005wb}.  
In both cases the rates for ${\rm BR}(B_{s}\to\mu^{+}\mu^{-})$ at large $\tan
\beta$ can be enhanced by a few orders of magnitude compared with the
prediction in the SM, therefore providing an optimal window for SUSY
signals.

In the present context of SUSY-NMFV, with no preference for large $\tb$ 
values, there are in general three types of one-loop diagrams that contribute 
to the previous $C_i$ Wilson coefficients for this $B_s \to \mu^+ \mu^-$
decay: 1) Box diagrams, 2) $Z$-penguin diagrams and 3) neutral Higgs
boson $H$-penguin diagrams, where $H$ denotes the three neutral MSSM Higgs bosons. In our numerical estimates
we have included what are known to be the dominant contributions to
these three types of diagrams \cite{Chankowski:2000ng}: chargino
contributions to box and Z-penguin diagrams and chargino and gluino
contributions to $H$-penguin diagrams.   

Regarding the resummation of large $\tb$ effects we have followed
the same effective Lagrangian formalism as explained in the previous
case of $B \to X_s \ga$. In the case of contributions from
$H$-penguin diagrams to  
$B_s \to \mu^+ \mu^-$ this resummation is very relevant and it has been
generalized to NMFV-SUSY scenarios in \cite{Isidori:2002qe}. 

For the numerical estimates we use again the BPHYSICS subroutine included in the SuFla code~\cite{sufla} which
incorporates all the ingredients that we have pointed out above.   
 
Finally, for completeness, we include below the present experimental 
upper bound for this observable \cite{CMSLHCb}, 
and the prediction within the SM \cite{Buras:2009if}: 
\vspace{0.5cm}
{\noindent \begin{equation}
\bmm_{\rm exp} < 1.1 \times 10^{-8}\,\,\,\, (95\% ~{\rm CL}) 
\label{bsmumu-exp}
\end{equation}

\noindent \begin{equation}
\bmm_{\rm SM}= (3.6\pm 0.4)\times 10^{-9}
\label{bsmumu-SM}
\end{equation}

%%%%%%%%%%%%%%%%%%%%%%%%%%%%%%%%%%%%%%%%%%%%%%%%%%%%%%%%%%%%%%%%%%%%%%%%%%%%%%

\subsection{\boldmath{\dmbs}} 

The relevant effective Hamiltonian for $B_s-{\bar B_s}$ mixing and, hence, for 
the $B_s/{\bar B_s}$ mass difference \dmbs\ is:

\begin{equation}
\mathcal{H}_{\rm eff}=
\frac{G_F^2}{16\pi^2}M_W^2 
\left(\VCKM^{tb*}{}\VCKM^{ts}\right)^2
\sum_{i}C_i O_i.
\label{Ham}
\end{equation}
In the SM the most relevant operator is:
\begin{align}
O^{VLL}&=
(\bar{b}^{\alpha}\ga_{\mu}P_L s^{\alpha})(\bar{b}^{\beta}\ga^{\mu}P_L s^{\beta}).
\label{SMOps}
\end{align}
Where we have now written explicitly the color indices.
 In scenarios beyond the SM, as the present NMFV MSSM, other operators are also
 relevant:
\begin{align}
O^{LR}_{1}&=(\bar{b}^{\alpha}\ga_{\mu}P_L s^{\alpha})(\bar{b}^{\beta}\ga^{\mu}P_R s^{\beta}),
&O^{LR}_{2}&=(\bar{b}^{\alpha}P_L s^{\alpha})(\bar{b}^{\beta}P_R s^{\beta}),\label{Ops1}\\
O^{SLL}_{1}&=(\bar{b}^{\alpha}P_L s^{\alpha})(\bar{b}^{\beta}P_L s^{\beta}),
&O^{SLL}_{2}&=(\bar{b}^{\alpha}\sigma_{\mu\nu}P_L s^{\alpha})(\bar{b}^{\beta}\sigma^{\mu\nu}P_L s^{\beta}),\label{Ops2}
\end{align}
and the corresponding operators $O^{VRR}$ and
$O^{SRR}_{i}$ that can be obtained by replacing $P_L \leftrightarrow P_R$
  in~\eqref{SMOps} and~\eqref{Ops2}.
The mass difference $\Delta M_{B_s}$ is then evaluated by taking the matrix
element
\begin{align}
\Delta M_{B_s}=2\lvert\langle\bar{B}_s\lvert\mathcal{H}_{\rm eff}\rvert B_s\rangle\rvert,
\label{delmb}
\end{align}
where $\langle\bar{B}_s\lvert\mathcal{H}_{\rm eff}\rvert B_s\rangle$ is given
by
\begin{align}
\langle\bar{B}_s\lvert\mathcal{H}_{\rm eff}\rvert B_s\rangle=&
\frac{G_F^2}{48\pi^2}M_W^2 m_{B_s} f^2_{B_s}
\left(\VCKM^{tb*} \VCKM^{ts}\right)^2
\sum_{i}P_i C_i\left(\mu_W\right).
\label{matel}
\end{align}
Here $m_{B_s}$ is the $B_s$ meson mass, and
$f_{B_s}$ is the $B_s$ decay constant.
The coefficients $P_i$ contain the effects due to RG running between
the electroweak scale $\mu_W$ and $m_b$ as well as the relevant hadronic matrix element.
 We use the coefficients $P_i$ from the lattice calculation \cite{Becirevic:2001xt}:   
\begin{align}
P^{VLL}_1=&0.73,
&P^{LR}_1=&-1.97,
&P^{LR}_2=&2.50,
&P^{SLL}_1=&-1.02,
&P^{SLL}_2=&-1.97.
\label{pcoef}
\end{align}
The coefficients
$P^{VRR}_1$, etc.,~may be obtained from those above by simply exchanging 
$L \leftrightarrow R$.

In the present context of SUSY-NMFV, again with no preference for large $\tb$ 
values, and besides the SM loop contributions, there are in general three types of one-loop diagrams that contribute 
to the previous $C_i$ Wilson coefficients for $B_s-{\bar B_s}$ mixing: 
1) Box diagrams, 2) $Z$-penguin diagrams and 3) double Higgs-penguin diagrams.
In our numerical estimates we have included what are known to be the dominant 
contributions to these three types of diagrams in scenarios with non-minimal flavor violation (for a review see, for instance, \cite{Foster:2005wb}): gluino contributions to box
diagrams, chargino contributions to box and Z-penguin diagrams, and 
chargino and gluino contributions to double $H$-penguin diagrams. As in the previous observables, 
the total prediction for  $\Delta M_{B_s}$ includes, of course, the SM contribution.
 
Regarding the resummation of large $\tb$ effects we have followed again the effective Lagrangian formalism, generalized to NMFV-SUSY scenarios \cite{Isidori:2002qe}, as in the previous cases of $B \to X_s \ga$ and $B_s \to \mu^+ \mu^-$. It should be noted that, in the case of $\Delta M_{B_s}$, the resummation of large $\tb$ effects is very relevant for the double $H$-penguin contributions, which grow very fast with $\tb$. 

\begin{figure}[tp]
\begin{center}
\psfig{file=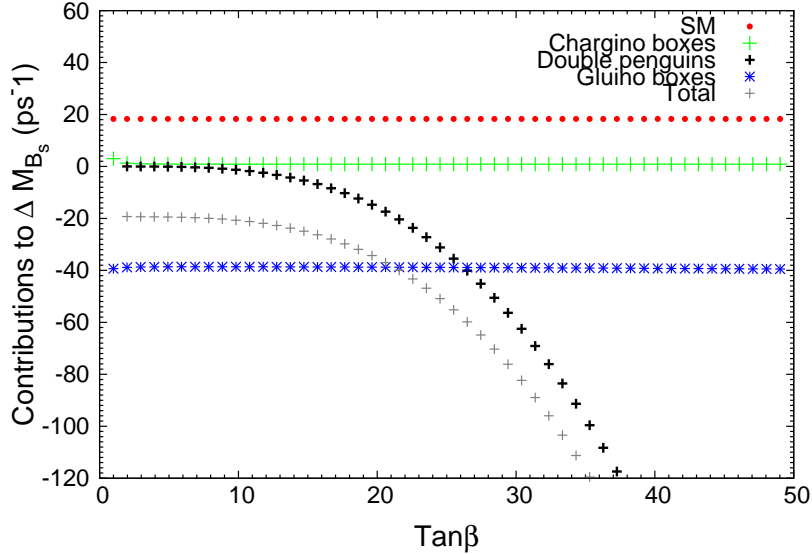,scale=0.45,angle=270,clip=}
\caption{Relevant contributions to $\Delta M_{B_s}$ in NMFV-SUSY scenarios as a function of $\tb$. They include: SM, Double Higgs penguins, gluino boxes and chargino boxes. The total prediction for $\Delta M_{B_s}$ should be understood here as $\Delta M_{B_s}=|{\rm Total}|$. The parameters are set to $\delta^{LL}_{23}=\delta^{RR}_{sb}=0.1$,$m_{\tilde q}=500 \gev$, $A_t= -m_{\tilde q}$, 
$m_{\tilde g}=\sqrt{2}m_{\tilde q}$, $\mu=m_{\tilde q}$, and $m_A= 300 \gev$. The other flavor changing deltas are set to zero.}
\label{fig:deltabs-allcontributions}
\end{center}
\end{figure} 
For the numerical estimates we have modified the BPHYSICS subroutine included in the SuFla code~\cite{sufla} which incorporates all the ingredients that we have pointed out above, except the contributions from gluino boxes. These contributions are known to be very important for $B_s-{\bar B_s}$ mixing in SUSY scenarios with non-minimal flavor violation \cite{Gabbiani:1996hi,Becirevic:2001jj,Foster:2005wb} and therefore they must be included into our analysis of $\Delta M_{B_s}$.  Concretely, we have incorporated them into the BPHYSICS subroutine by adding the full one-loop formulas for the gluino boxes of \cite{Baek:2001kh} to the other above quoted contributions that were already included in the code. In order to illustrate the relevance of these gluino contributions in our analysis of $\Delta M_{B_s}$, we show in Fig.\ref{fig:deltabs-allcontributions} each separate contribution as a function of $\tb$ in a particular example with $\delta^{LL}_{23}=\delta^{RR}_{sb}=0.1$, that we have chosen for comparison with \cite{Foster:2005wb}. The other flavor changing deltas are set to zero, and the other relevant MSSM parameters are set to $m_{\tilde q}=500 \gev$, $A_t= -m_{\tilde q}$, 
$m_{\tilde g}=\sqrt{2}m_{\tilde q}$, $\mu=m_{\tilde q}$, and $m_A= 300 \gev$, as in Fig.24 of \cite{Foster:2005wb}.   We clearly see in Fig.\ref{fig:deltabs-allcontributions} that it is just in the very large $\tb$ region where double Higgs- penguins dominate. For moderate and low $\tb$ values, $\tb \leq 20$ (which is a relevant region for our numerical analysis,
see below) the gluino boxes fully dominates the SUSY corrections to
$\Delta M_{B_s}$ and compete with the SM contributions. Our numerical
estimate in this plot is in complete agreement with the results in
\cite{Foster:2005wb} (see, in particular, Fig.24 of this reference) which
analyzed and compared in full detail these corrections.    
Finally, for completeness, we include below the experimental measurement of this observable \cite{Nakamura:2010zzi}\footnote{We have again added the various contributions to the experimental error in quadrature.}, and its prediction within the SM (using NLO expression of \cite{Buras:1990fn}
 and error estimate of \cite{Golowich:2011cx}):
%\vspace{0.5cm}
\noindent \begin{align}
\label{deltams-exp}
{\dmbs}_{\rm exp} &= (117.0 \pm 0.8) \times 10^{-10} \mev~, \\
{\dmbs}_{\rm SM} &= (117.1^{+17.2}_{-16.4}) \times 10^{-10} \mev~.
\label{deltams-SM}
\end{align} 
 
%%%%%%%%%%%%%%%%%%%%%%%%%%%%%%%%%%%%%%%%%%%%%%%%%%%%%%%%%%%%%%%%%%%%%%%%%%%%%%%

\subsection{Numerical results on \boldmath{$B$} observables}

In the following of this section we present our numerical results for
the three $B$ observables in the NMFV-SUSY scenarios  
and a discussion on the allowed values for the flavor changing deltas,
$\deXYij$. 

The predictions for \bsg, \bmm and \dmbs\ 
versus the various $\deXYij$, for the six selected SPS
points, are displayed respectively in  \reffis{figbsgamma}, \ref{figbmumu} and \ref{figdeltams}.  
For this analysis, we have assumed that just one at a time
of these deltas is not vanishing. Results for two non-vanishing deltas
will be shown later.  The following 7 flavor changing deltas are
considered: $\delta^{LL}_{23}$, $\delta^{LR}_{ct}$, $\delta^{LR}_{sb}$,
$\delta^{RL}_{ct}$, 
$\delta^{RL}_{sb}$, $\delta^{RR}_{ct}$ and $\delta^{RR}_{sb}$; and we
have explored delta values within the interval $-1<\deXYij<1$. In all
plots, the predictions for $\deXYij=0$ represent our estimate of the  
corresponding observable in the MFV case. This will allow us to compare
easily the results in the two kind of scenarios, NMFV and MFV. It should be noted
also, that some  of the predicted lines in these plots do not expand
along the full interval $-1<\deXYij<1$, and they are restricted to a smaller 
interval; for some regions of
the parameter space a too large delta value can generate very large corrections to any of the masses, and the mass squared turns negative. 
These problematic
points are consequently not shown in our plots.

We have also included in the right vertical axis of these figures, for
comparison, the respective SM prediction in (\ref{bsgamma-SM}),
(\ref{bsmumu-SM}), and (\ref{deltams-SM}). The error bars displayed are
the corresponding SM uncertainties as explained below. 
The shadowed horizontal bands in the case of \bsg\ and
\dmbs\ are their corresponding experimental measurements 
 in (\ref{bsgamma-exp}), and (\ref{deltams-exp}), expanded with
3${\sigma}_{\rm exp}$ errors. In the case of \bmm\ the shadowed area 
corresponds to the allowed region by the upper bound in (\ref{bsmumu-exp}).    

The main conclusions extracted from these figures for the three $B$
observables are summarized as follows: 

\begin{itemize}
\item \bsg:
\begin{itemize}
\item[-] Sensitivity to the various deltas:

We find strong sensitivity to $\delta^{LR}_{sb}$, $\delta^{RL}_{sb}$, $\delta^{LL}_{23}$, $\delta^{RR}_{sb}$ and $\delta^{LR}_{ct}$, in all the studied points, for both high and low $\tb$ values.
The order found from the highest to the lowest sensitivity is, generically:  1) $\delta^{LR}_{sb}$ and $\delta^{RL}_{sb}$ the largest, 2)   $\delta^{LL}_{23}$ the next, 3) $\delta^{LR}_{ct}$ and $\delta^{RR}_{sb}$ the next to next, and 4) slight sensitivity to $\delta^{RR}_{ct}$ and $\delta^{RL}_{ct}$. 

\item[-] Comparing the predictions with the experimental data:

If we focus on the five most relevant deltas, $\delta^{LR}_{sb}$,
$\delta^{RL}_{sb}$, $\delta^{LL}_{23}$, $\delta^{RR}_{sb}$ and
$\delta^{LR}_{ct}$, we see clearly that tiny deviations from zero (i.e.,
deviations from MFV) in these deltas, and specially in the first three,
produce sizeable shifts  
in  \bsg, and in many cases take it out the experimental
allowed band.    
Therefore, it is obvious from these plots that \bsg\ sets stringent bounds on the deltas (when varying just one delta), which are particularly tight on  $\delta^{LR}_{sb}$, $\delta^{RL}_{sb}$, $\delta^{LL}_{23}$, $\delta^{RR}_{sb}$, and  $\delta^{LR}_{ct}$, indeed for all the studied SPS points. There are just two exceptions, where the predicted central values of \bsg\ are already outside the experimental band 
in the MFV case (all deltas set to zero), and assuming one of these three most relevant deltas to be non-vanishing, the prediction 
moves inside the experimental band. This happens, for instance, in  the points SPS4 and SPS1b that have the largest $\tb$ values of 50 and 30 respectively.  Interestingly, it means that some points of the CMSSM, particularly those with large $\tb$ values, that would have been excluded in the context of MFV, can be recovered as compatible with data within NMFV-SUSY scenarios. 
\item[-] Intervals of $\deXYij$  allowed by data:

In order to conclude on the allowed delta intervals we have assumed that our
predictions of \bsg\ within SUSY scenarios have a somewhat
larger theoretical error $\Dtheo(\bsg)$ than the SM prediction 
$\Dtheo_{\rm SM}(\bsg)$
given in (\ref{bsgamma-SM}). As a very conservative value we use
$\Dtheo(\bsg) = 0.69 \times 10^{-4}$. 
A given $\deXYij$ value is then considered to be allowed
by data if the corresponding interval, defined by $\bsg \pm \Dtheo(\bsg)$,
intersects with the experimental band. It corresponds to 
adding linearly the experimental uncertainty and the MSSM theoretical
uncertainty. In total a predicted ratio in the
interval  
\noindent \begin{align}
\label{bsglinearerr}
2.08\times 10^{-4} < \bsg < 5.02\times 10^{-4},
\end{align} 
is regarded as allowed. 
Our results for these allowed intervals are summarized in table
\ref{tableintervals}. In this table we see again that the less constrained
parameters by \bsg\ are $\delta^{RL}_{ct}$,
and $\delta^{RR}_{ct}$. Therefore,
except for the excluded SPS4 case, these two deltas can be sizeable,
$|\deXYij|$ larger than ${\cal O}(0.1)$, and compatible with \bsg\ data.    
\end{itemize}
\end{itemize}

%%%%%%%%%%%%%%%%%%%%%%%%%% F I G U R E %%%%%%%%%%%%%%%%%%%%%%%%%%%%%%%%%%%%%%%%
\begin{figure}[h!] 
\centering
\hspace*{-8mm} 
{\resizebox{17.9cm}{!} 
{\begin{tabular}{cc} 
\includegraphics[width=13.3cm,height=16.5cm,angle=270]{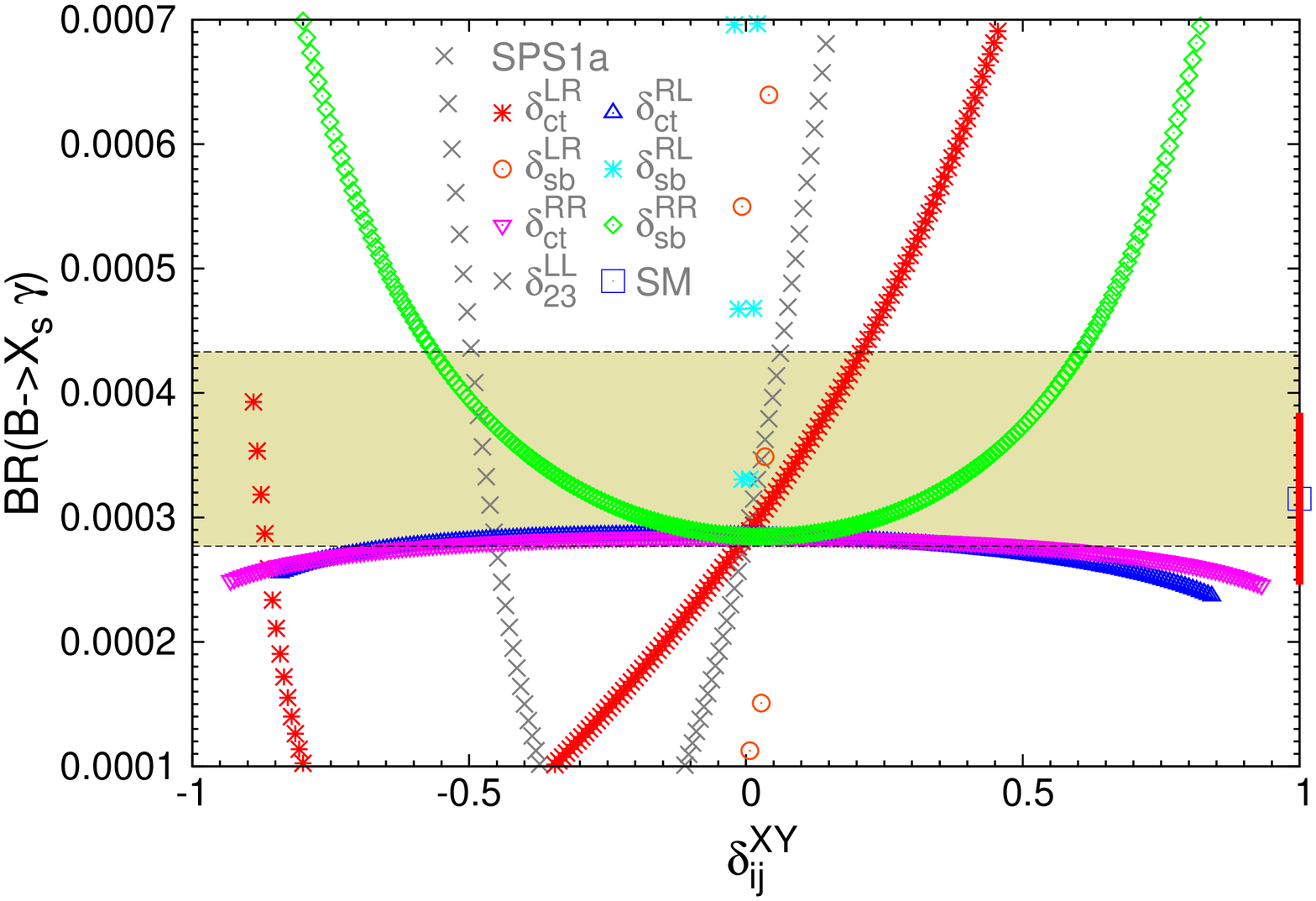}& 
\includegraphics[width=13.3cm,height=16.5cm,angle=270]{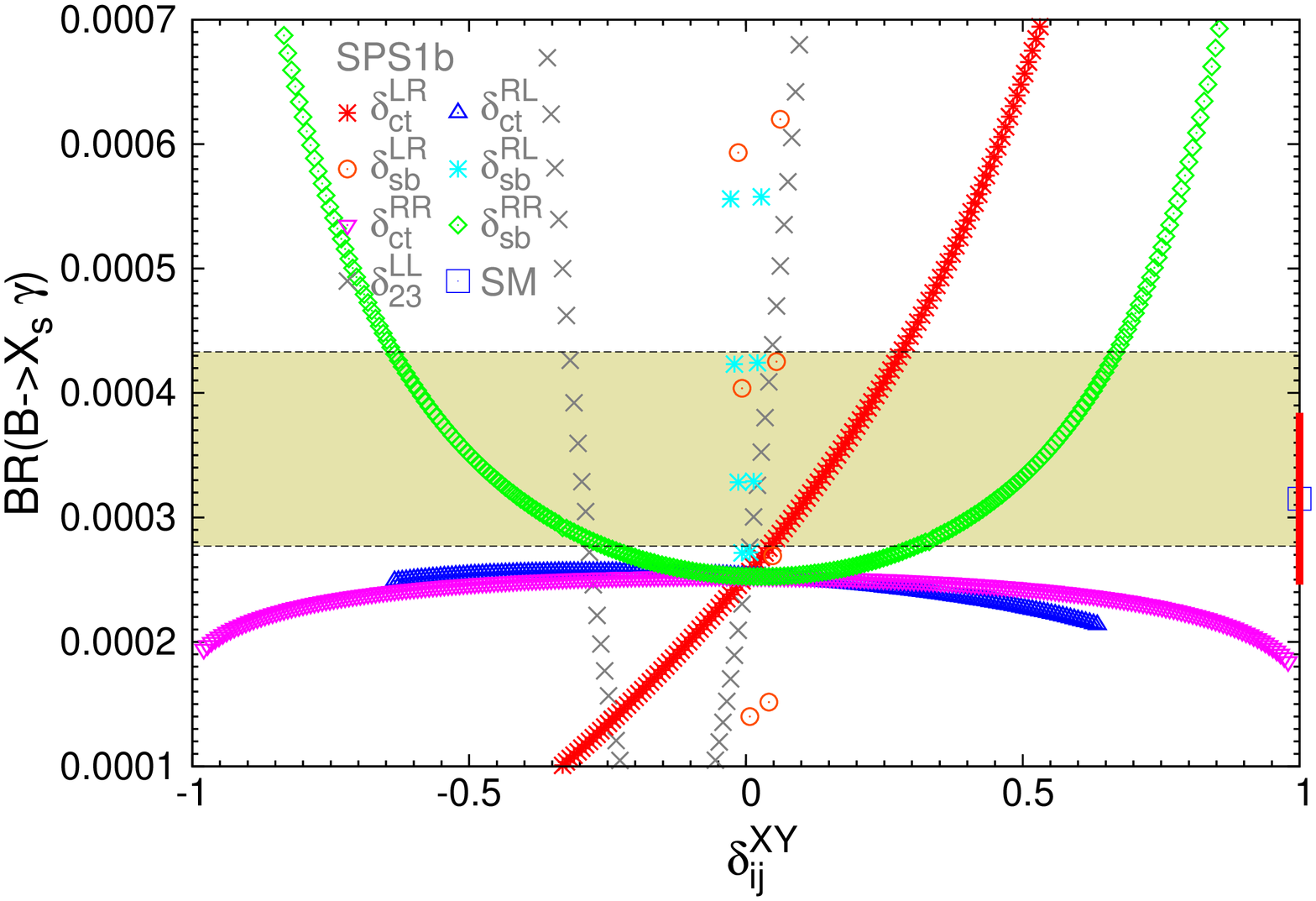}\\ 
\includegraphics[width=13.3cm,height=16.5cm,angle=270]{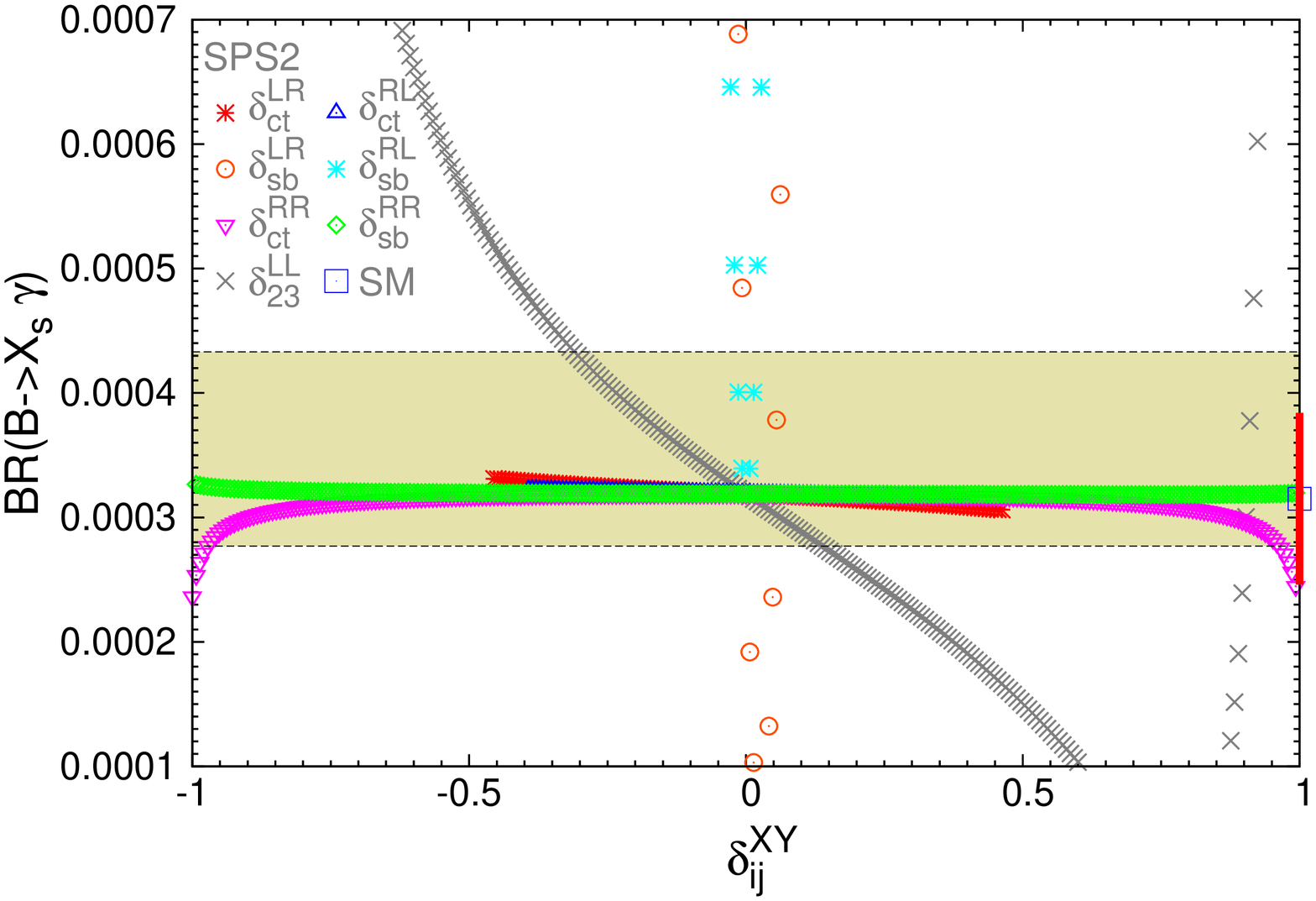}&
\includegraphics[width=13.3cm,height=16.5cm,angle=270]{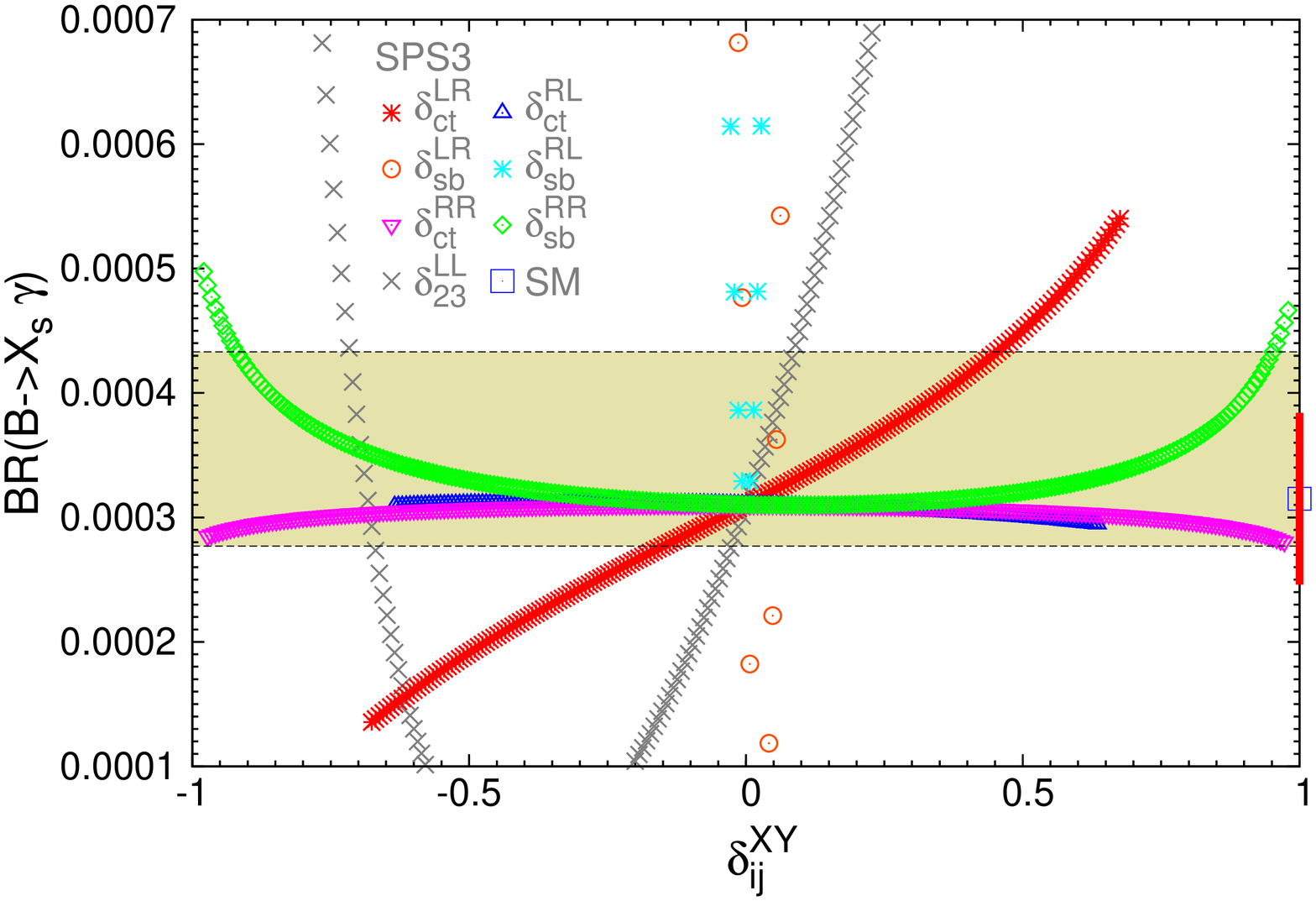}\\ 
\includegraphics[width=13.3cm,height=16.5cm,angle=270]{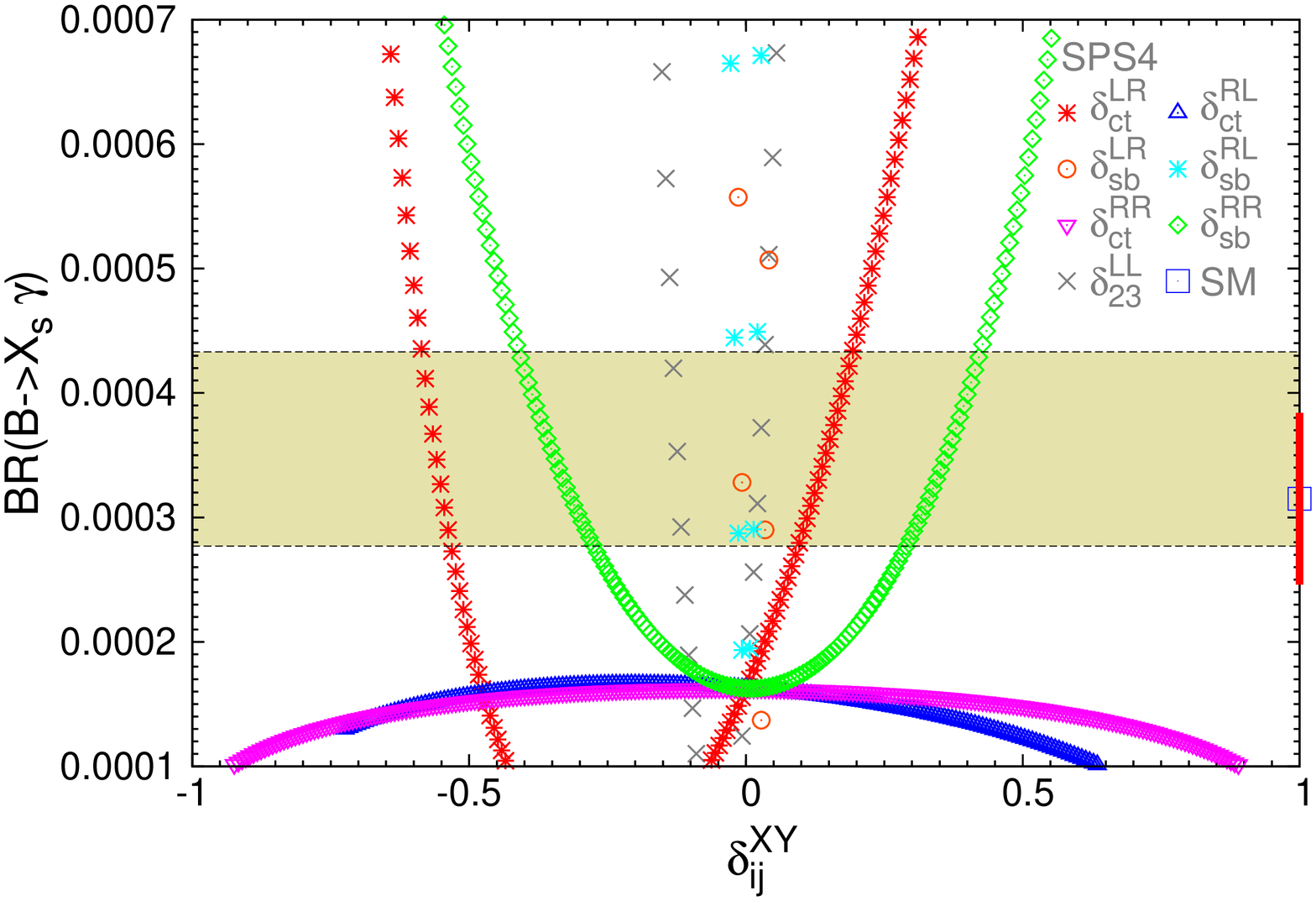}& 
\includegraphics[width=13.3cm,height=16.5cm,angle=270]{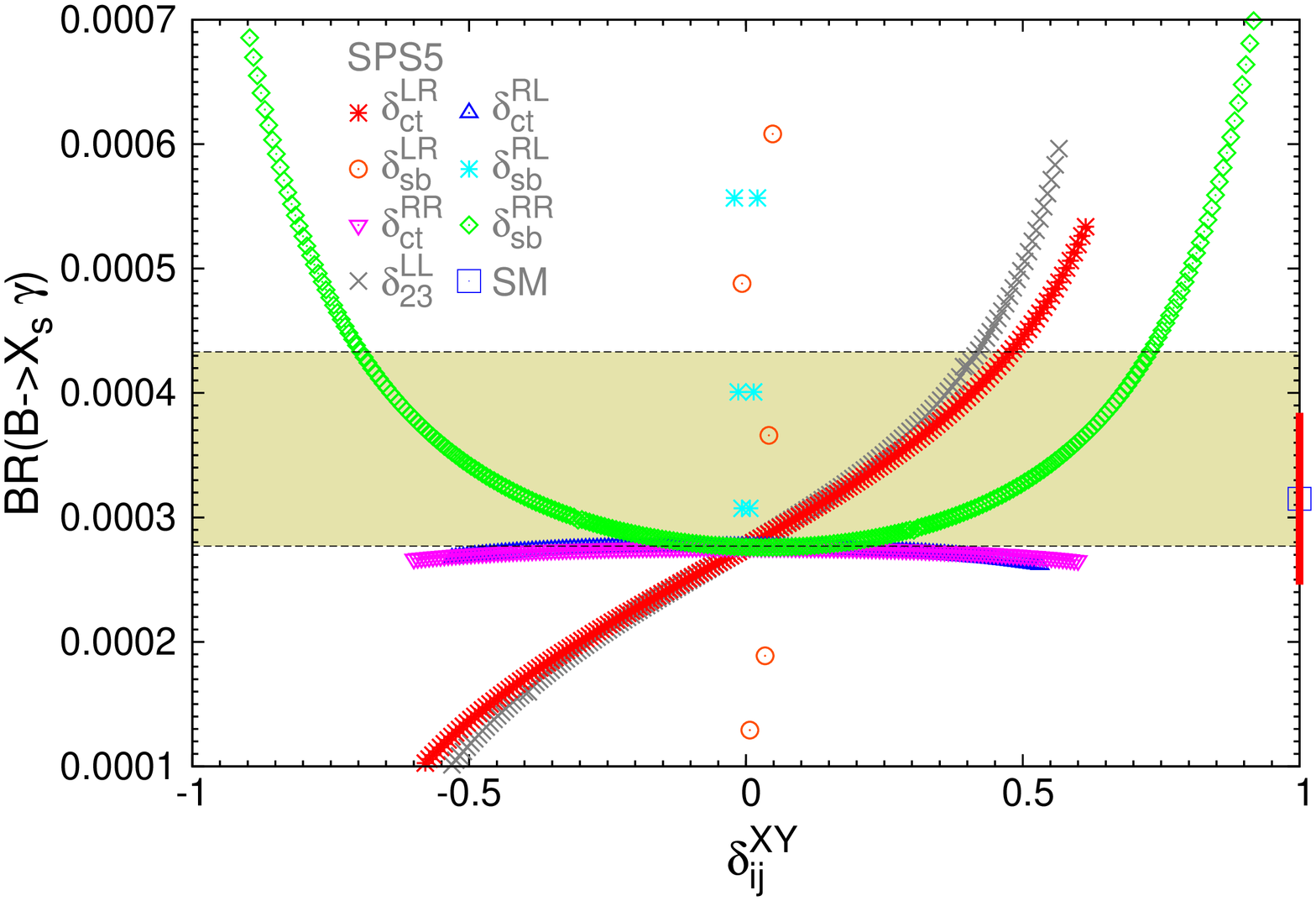}\\ 
\end{tabular}}}
\caption{Sensitivity to the NMFV deltas in \bsg\ for the SPSX points of table
  \ref{points}. The experimental allowed $3\sigma$ area is the horizontal
  colored band. The SM prediction and the theory uncertainty $\Dtheo(\bsg)$
  (red bar) is displayed on the right axis.}  
\label{figbsgamma}
\end{figure}
%%%%%%%%%%%%%%%%%%%%%%%%%%%%%%%%%%%%%%%%%%%%%%%%%%%%%%%%%%%%%%%%%%%%%%%%
%\clearpage
%\newpage

\begin{itemize}
\item
\bmm:

\begin{itemize}
\item[-] Sensitivity to the various deltas:

We find significant sensitivity to the NMFV deltas in scenarios with very large $\tb$, as it is the case of SPS4 and SPS1b. 
This sensitivity is clearly due to the Higgs-mediated contribution that, grows as $\tan^6\beta$. The largest sensitivity is to $\delta^{LL}_{23}$. In the case of SPS4, there is also significant sensitivity to $\delta^{LR}_{sb}$, $\delta^{RR}_{sb}$ and $\delta^{LR}_{ct}$. In the SPS1b scenario there is also found some sensitivity to $\delta^{LR}_{sb}$,  $\delta^{RR}_{ct}$, $\delta^{RR}_{sb}$ and $\delta^{LR}_{ct}$.
  
\item[-] Comparing the predictions with the experimental data:

\reffi{figbmumu} clearly shows that most of the $|\deXYij| \leq 1 $ explored values are allowed by experimental data on ${\rm BR}(B_s \to \mu^+ \mu^-)$. It is in the points with very large $\tb$, i.e SPS4 and SPS1b,  where there are some relevant differences between the MFV and the NMFV predictions. First, all predictions for MFV scenarios except for SPS4, are inside the experimental allowed area. In the case of SPS1b, the comparison of the NMFV predictions with data constraints specially $\delta^{LL}_{23}$, but also $\delta^{LR}_{sb}$,  $\delta^{RR}_{ct}$, $\delta^{RR}_{sb}$ and $\delta^{LR}_{ct}$. In the case of SPS4 some input non-vanishing values of  $\delta^{LL}_{23}$, $\delta^{LR}_{sb}$ or $\delta^{RR}_{sb}$ can place the prediction inside the experimental allowed area. In the case of the SPS1a and SPS3 scenarios some constraints for $\delta^{LL}_{23}$ can be found.
 
\item[-] Intervals of $\deXYij$  allowed by data:

As in the previous observable, we assume here that our predictions for 
\bmm\ have a slightly larger error as the SM prediction in
(\ref{bsmumu-SM}), where, however, the theory uncertainty is very small in
comparison with the experimental bound. We choose 
$\Dtheo(\bmm) = 0.12 \times 10^{-8}$. 
Then, a given $\deXYij$ value is allowed by data if the predicted interval,
defined by $\bmm + \Dtheo(\bmm)$, intersects
the experimental area. The upper line of the experimental area in this case is
set by the $95\% ~{\rm CL}$ upper bound given in (\ref{bsmumu-exp}).  It
implies that for a predicted ratio to be allowed it must fulfill: 
\noindent \begin{align}
\label{bsmmlinearerr}
\bmm < 1.22 \times 10^{-8}.
\end{align} 

The results for the allowed $\deXYij$ intervals are collected in table \ref{tableintervals}. We conclude from this table that, except for scenarios with large $\tanb \geq 30$, like SPS4 and SPS1b, the size of the deltas can be sizeable, $|\deXYij|$ larger than ${\cal O}(0.1)$, and compatible with \bmm\ data. 
\end{itemize}
\end{itemize}

%%%%%%%%%%%%%%%%%%%%%%%%%% F I G U R E %%%%%%%%%%%%%%%%%%%%%%%%%%%%%%%%%%%%%%%%
\begin{figure}[h!] 
\centering
\hspace*{-8mm} 
{\resizebox{17.9cm}{!} 
{\begin{tabular}{cc} 
\includegraphics[width=13.3cm,height=16.5cm,angle=270]{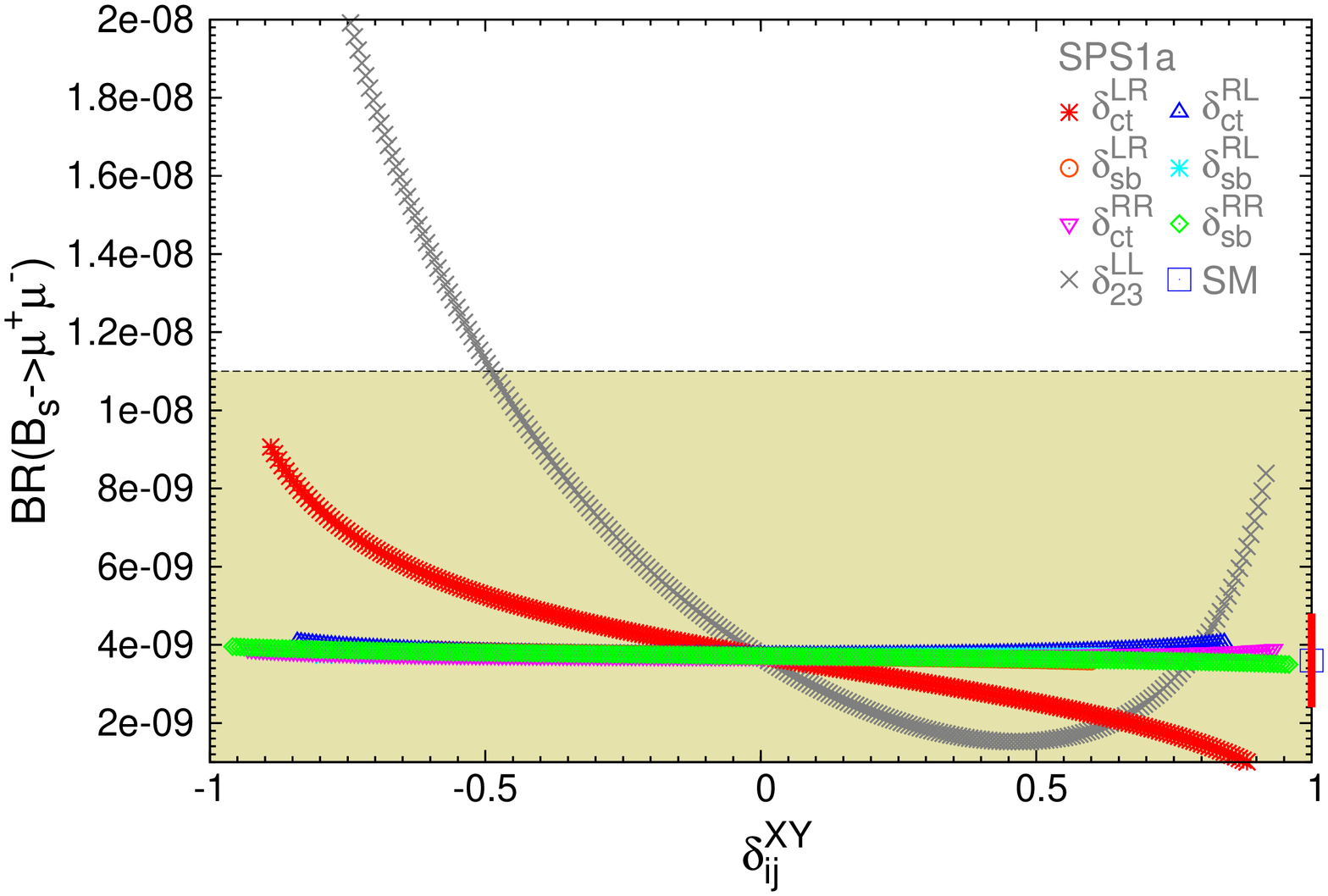}& 
\includegraphics[width=13.3cm,height=16.5cm,angle=270]{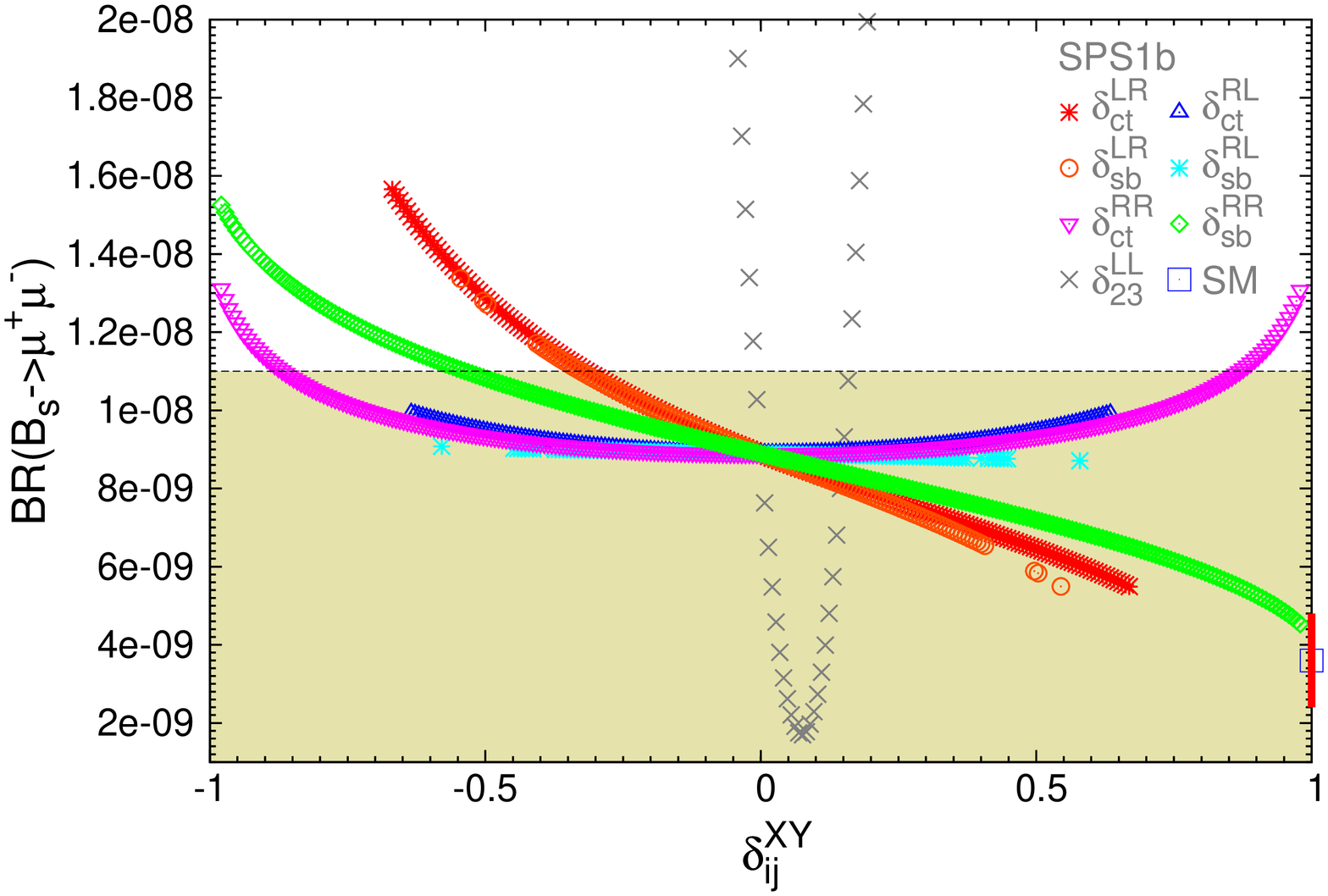}\\ 
\includegraphics[width=13.3cm,height=16.5cm,angle=270]{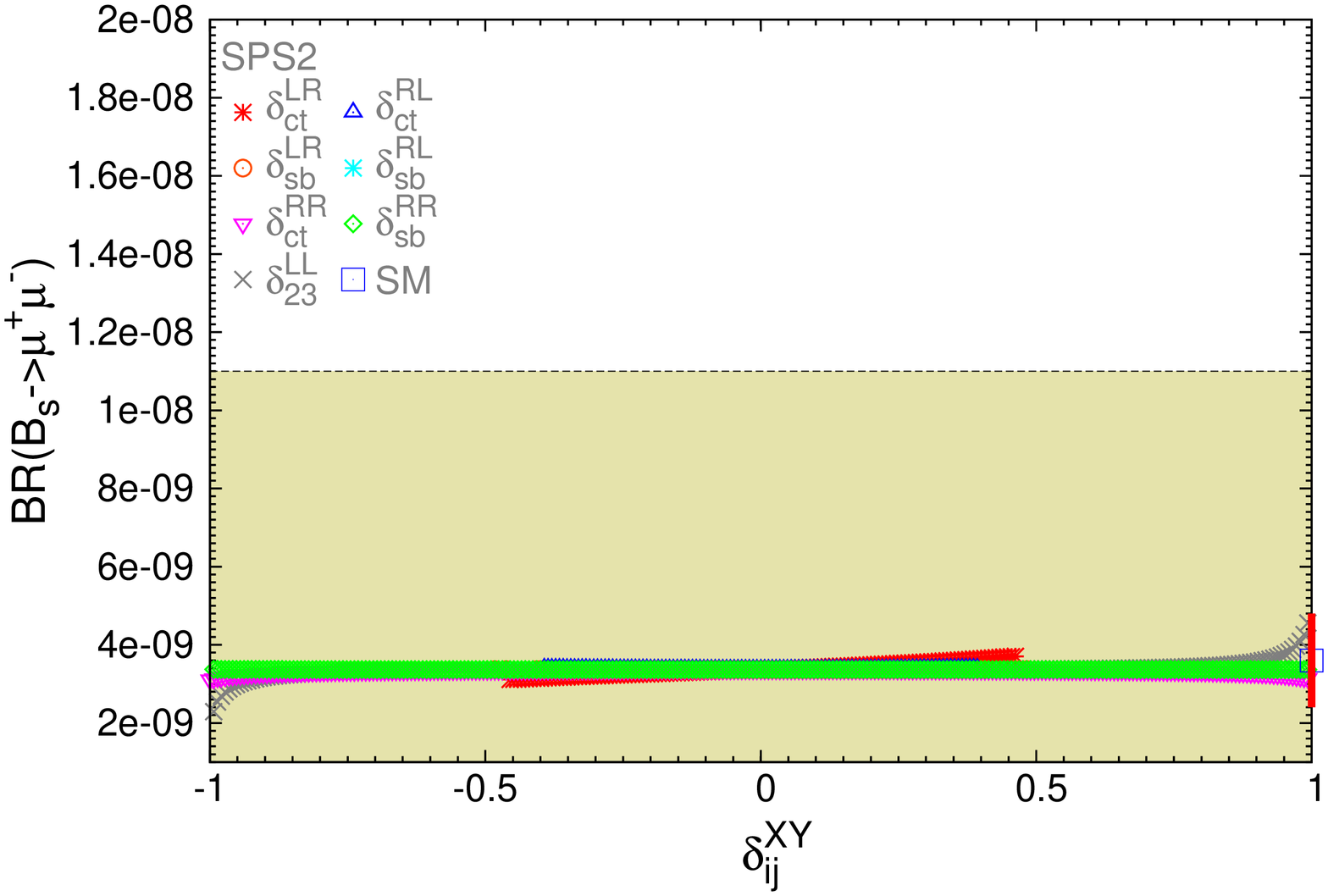}&
\includegraphics[width=13.3cm,height=16.5cm,angle=270]{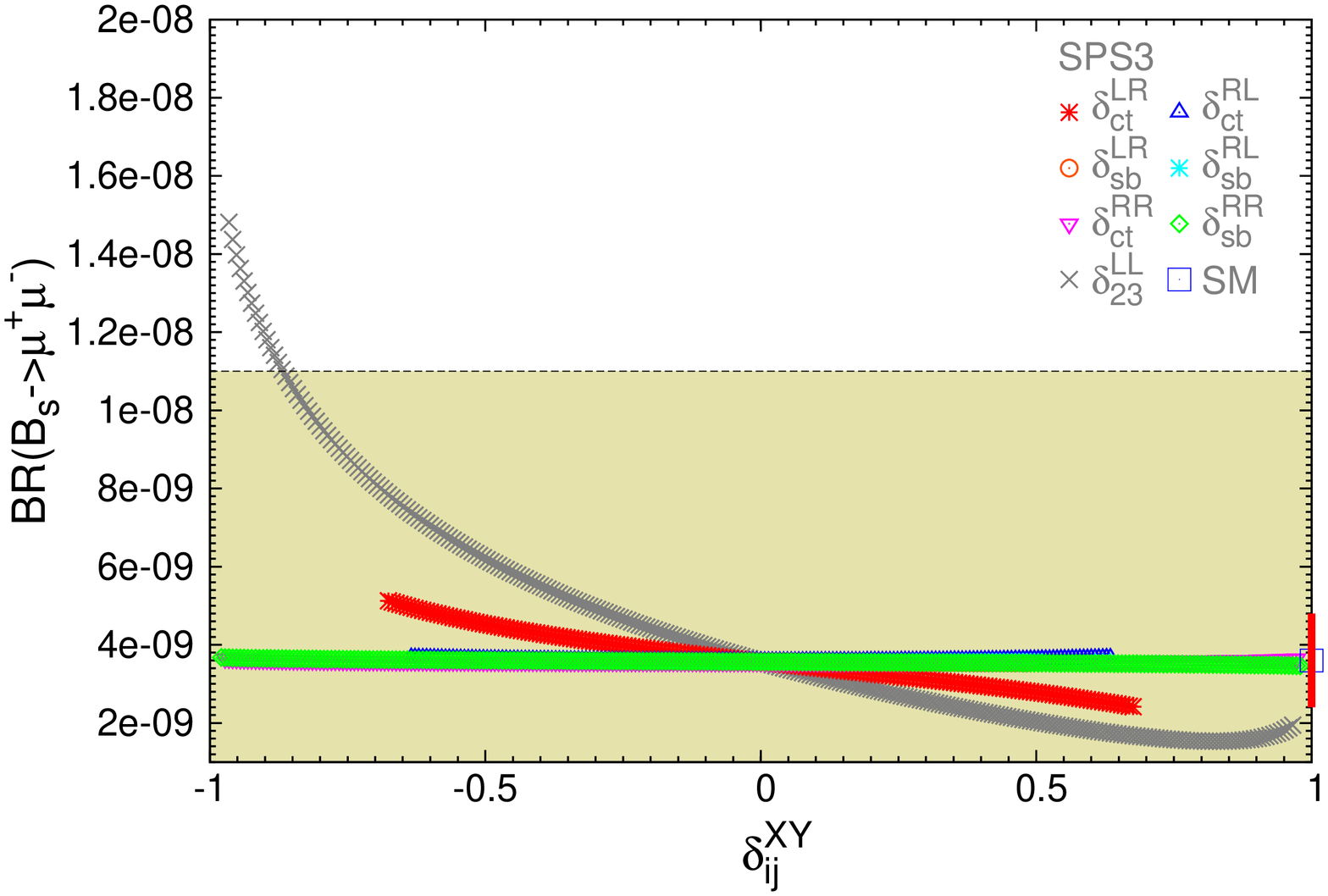}\\ 
\includegraphics[width=13.3cm,height=16.5cm,angle=270]{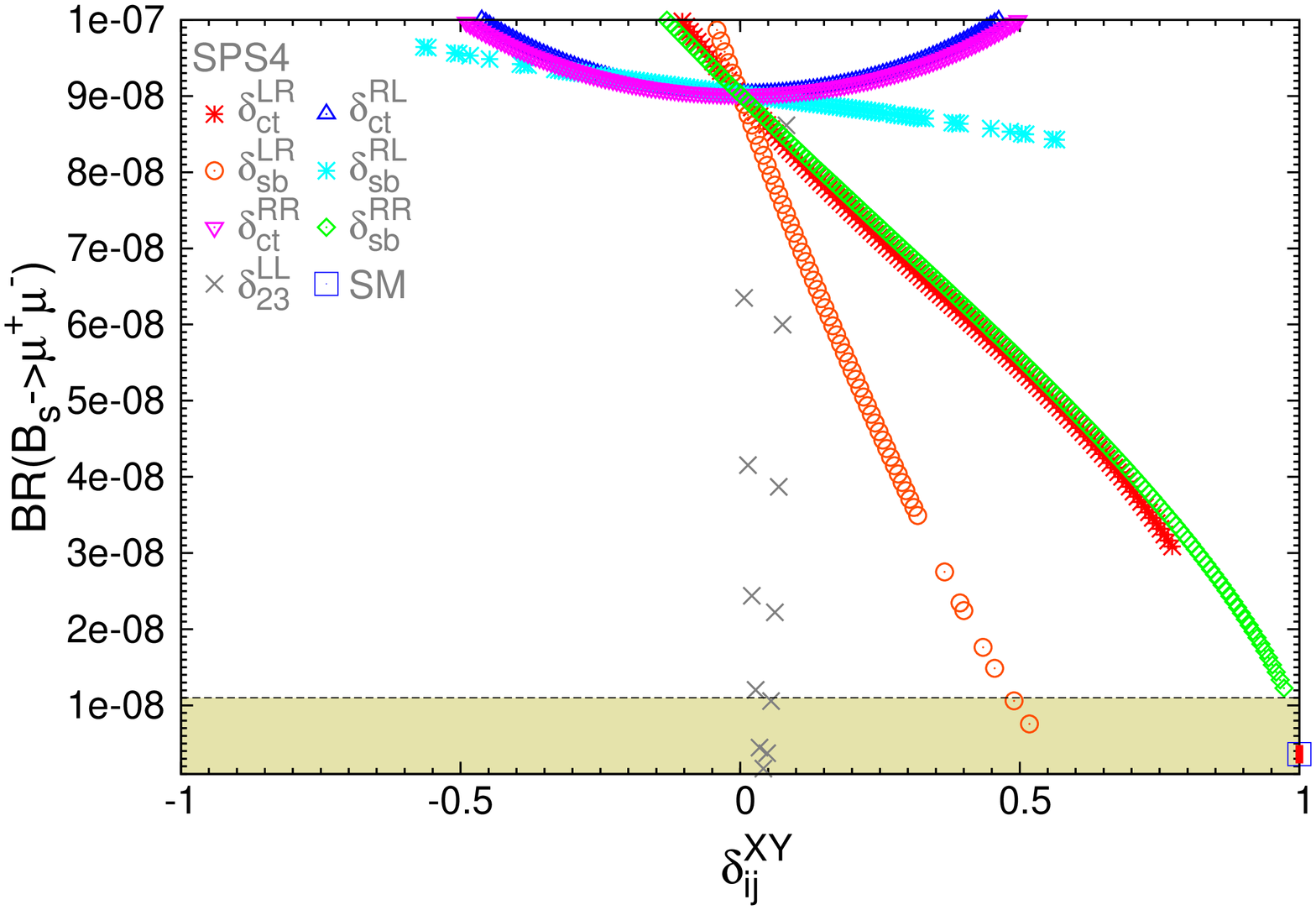}& 
\includegraphics[width=13.3cm,height=16.5cm,angle=270]{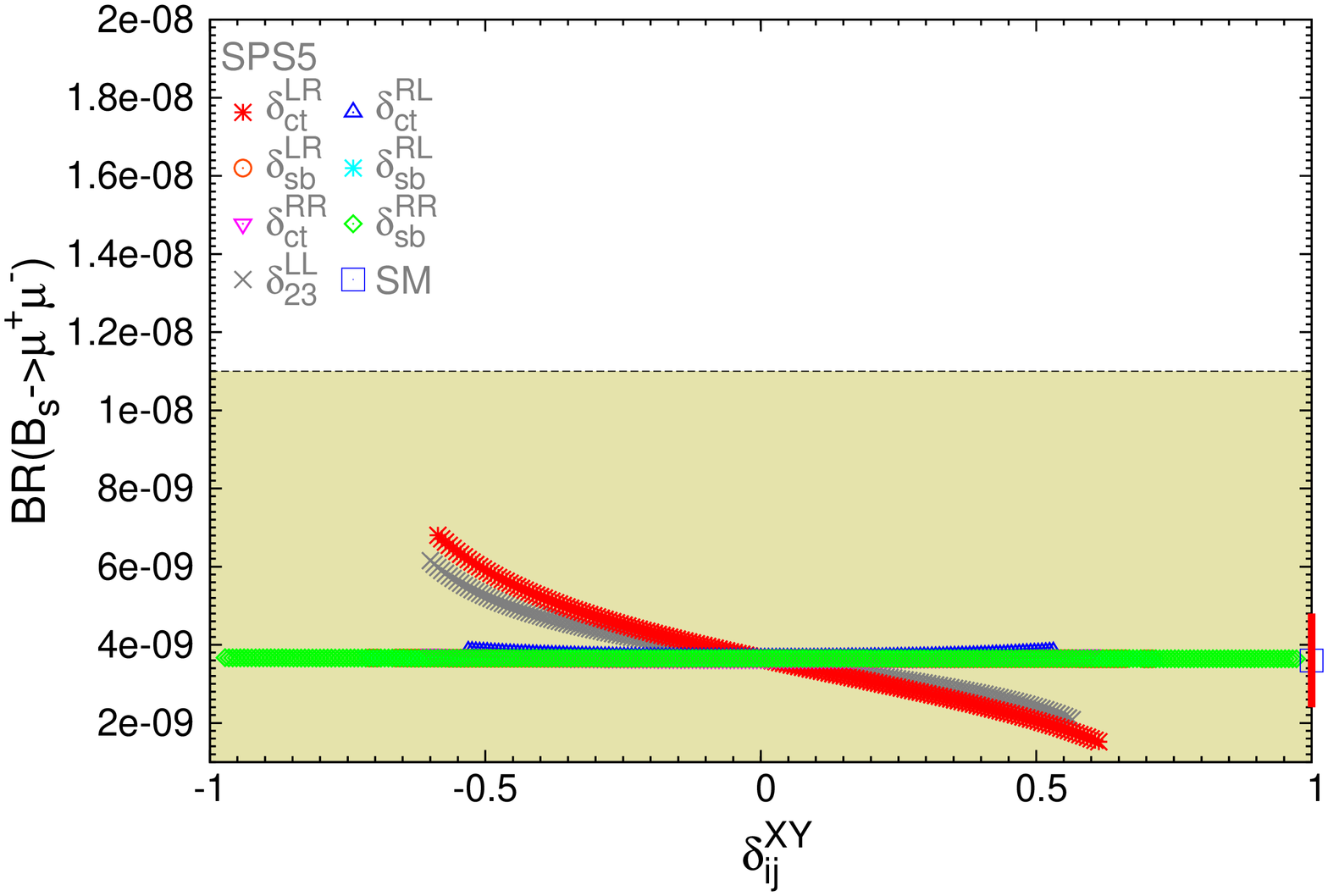}\\ 
\end{tabular}}}
\caption{Sensitivity to the NMFV deltas in ${\rm BR}(B_s \to \mu^+ \mu^-)$ for
  the SPSX points of table \ref{points}.  The experimental allowed region by
  the $95\% ~{\rm CL}$ bound is the horizontal colored area. The SM prediction
  and the theory uncertainty \Dtheo(\bmm) (red bar) is displayed on the right
  axis.}   
\label{figbmumu}
\end{figure}
%%%%%%%%%%%%%%%%%%%%%%%%%% F I G U R E %%%%%%%%%%%%%%%%%%%%%%%%%%%%%%%%%%%%%%%%
%\clearpage
%\newpage

\begin{itemize}
\item \dmbs:

\begin{itemize}
\item[-] Sensitivity to the various deltas:

Generically, we find strong sensitivity to most of the NMFV deltas in all the studied points, including those with large and low $\tb$ values. The pattern of the $\Delta M_{B_s}$ predictions as a function of the various
$\deXYij$ differs substantially for each SPS point. This is mainly because in this observable there are two large competing contributions, the double Higgs penguins and the gluino boxes, with very different behavior with  $\tb$, as we have seen in \reffi{fig:deltabs-allcontributions}.  
In the case of SPS4 with extremely large $\tb= 50$, the high sensitivity to all deltas is evident in this figure. In the case of SPS5 with low $\tb= 5$, there is important sensitivity to all deltas, except  $\delta^{RR}_{ct}$, $\delta^{LR}_{ct}$ and $\delta^{RL}_{ct}$. 
Generically, for all the studied points, we find the highest sensitivity to 1) $\delta^{LR}_{sb}$, $\delta^{RL}_{sb}$ and $\delta^{LL}_{23}$; 2)   $\delta^{RR}_{sb}$ the next, 3) $\delta^{LR}_{ct}$ the next to next; and 4)
the lowest sensitivity is to  $\delta^{RL}_{ct}$ and $\delta^{RR}_{ct}$. Consequently, these two later will be the less constrained ones.
 
\item[-] Comparing the predictions with the experimental data:

In this case, the experimental allowed $3 \sigma_{\rm exp}$ band is very narrow, and all the central predictions at $\deXYij=0$, i.e. for MFV scenarios,  lay indeed outside this band. However, if we assume again that our predictions suffer of a similar large uncertainty as the SM prediction, given in (\ref{deltams-SM}), then the MFV predictions are all compatible with data except for SPS4. It is also obvious from this figure that the predictions within NMFV, as compared to the MFV ones, lie quite generically outside the experimental allowed band, except for the above commented deltas with low sensitivity.  

\item[-] Intervals of $\deXYij$  allowed by data:

We consider again, that a given $\deXYij$ value is allowed by \dmbs\
data if the predicted interval $\dmbs \pm \Dtheo(\dmbs)$, intersects the  
experimental band. It corresponds to adding linearly the experimental uncertainty 
and the theoretical uncertainty. Given the present controversy on the realistic 
size of the theoretical error in the estimates of $\Dtheo(\dmbs)$ in the MSSM (see, for instance,~\cite{refLunghi}), 
we choose a very conservative value for the theoretical error in our estimates, 
considerably larger than the SM value in (\ref{deltams-SM}), of
$\Dtheo(\dmbs)=51 \times 10^{-10}$ MeV.  This  
implies that a predicted mass difference in the interval
\begin{align}
\label{deltabslinearerr}
63\times 10^{-10} < \dmbs {\rm (MeV)} < 168.6\times 10^{-10},
\end{align} 
is regarded as allowed.

The allowed intervals for the deltas that are obtained with this requirement are collected in table \ref{tableintervals}. As we have already commented, the restrictions on the b-sector parameters from $\Delta M_{B_s}$ are very strong, and in consequence, there are narrow intervals allowed for, 
$\delta^{LR}_{sb}$, $\delta^{RL}_{sb}$, and $\delta^{LL}_{23}$. In the case of 
 $\delta^{RR}_{sb}$ there are indeed sequences of very narrow allowed intervals, which in some cases reduce to just a single allowed value. The parameters that show a largest allowed interval, with sizeable $|\deXYij|$, larger than ${\cal O}(0.1)$, are $\delta^{RR}_{ct}$, $\delta^{RL}_{ct}$ and $\delta^{LR}_{ct}$. 
\end{itemize}

\end{itemize}

%%%%%%%%%%%%%%%%%%%%%%%%%% F I G U R E %%%%%%%%%%%%%%%%%%%%%%%%%%%%%%%%%%%%%%%%
\begin{figure}[h!] 
\centering
\hspace*{-8mm} 
{\resizebox{17.9cm}{!} 
{\begin{tabular}{cc} 
\includegraphics[width=13.3cm,height=16.5cm,angle=270]{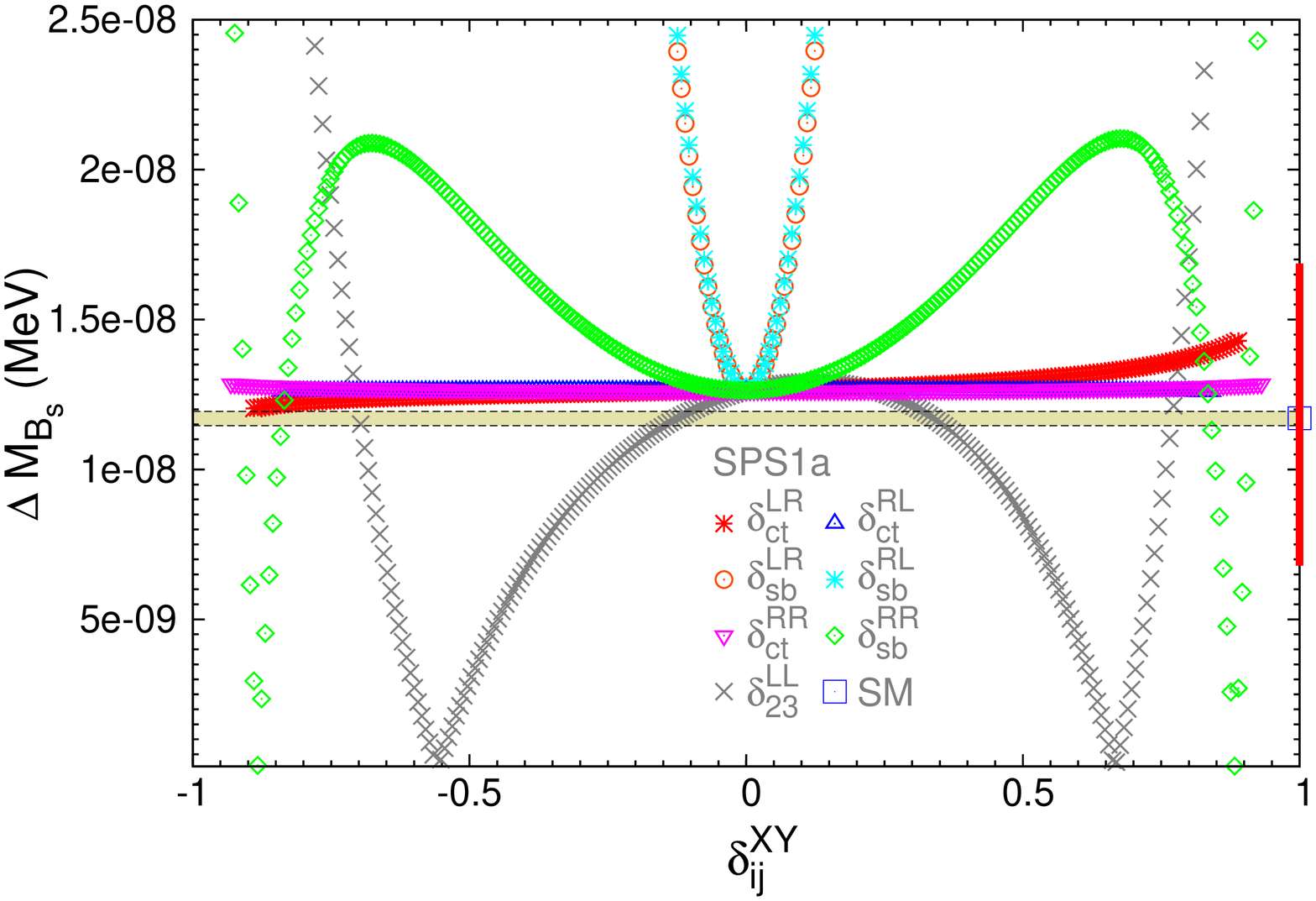}& 
\includegraphics[width=13.3cm,height=16.5cm,angle=270]{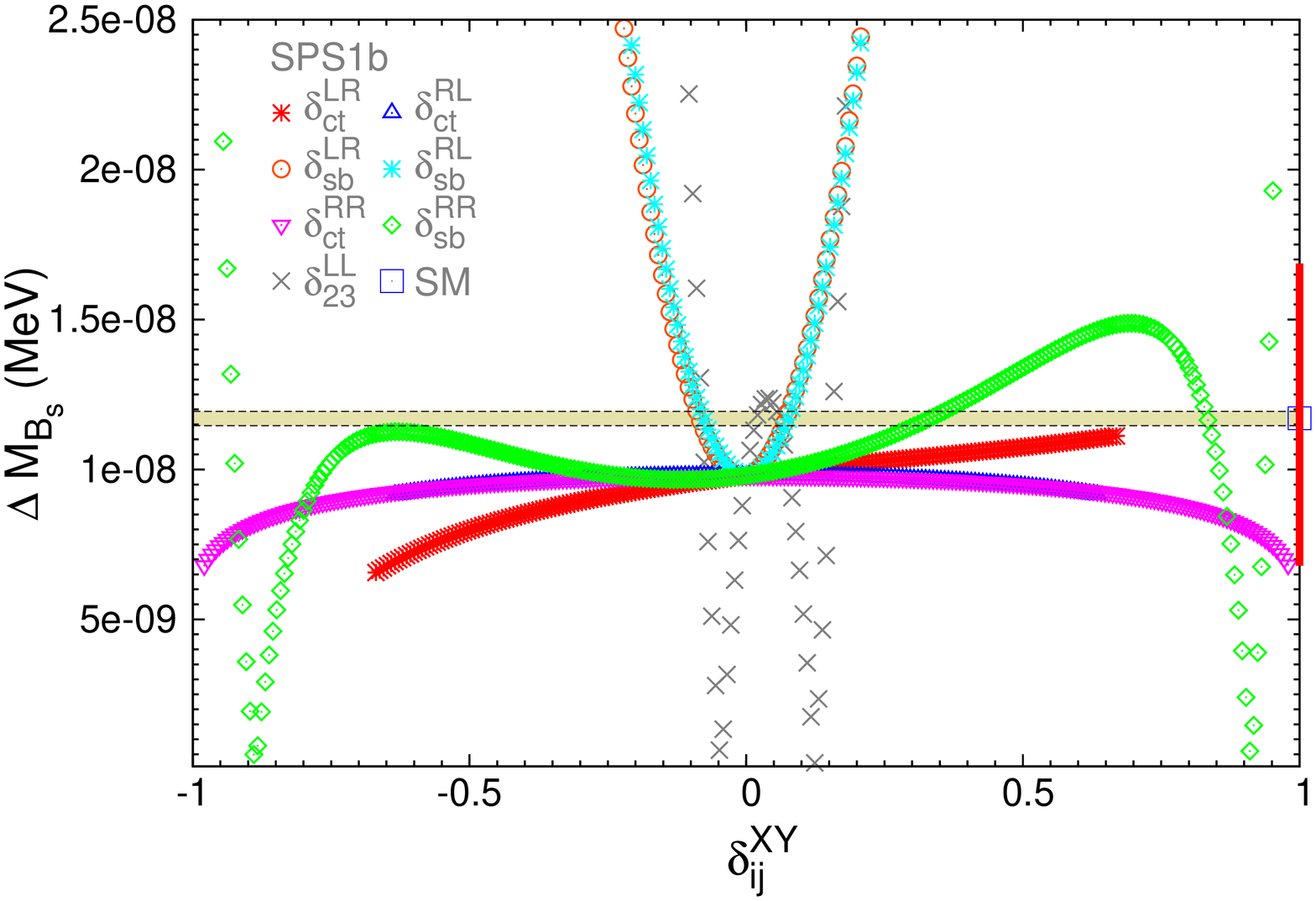}\\ 
\includegraphics[width=13.3cm,height=16.5cm,angle=270]{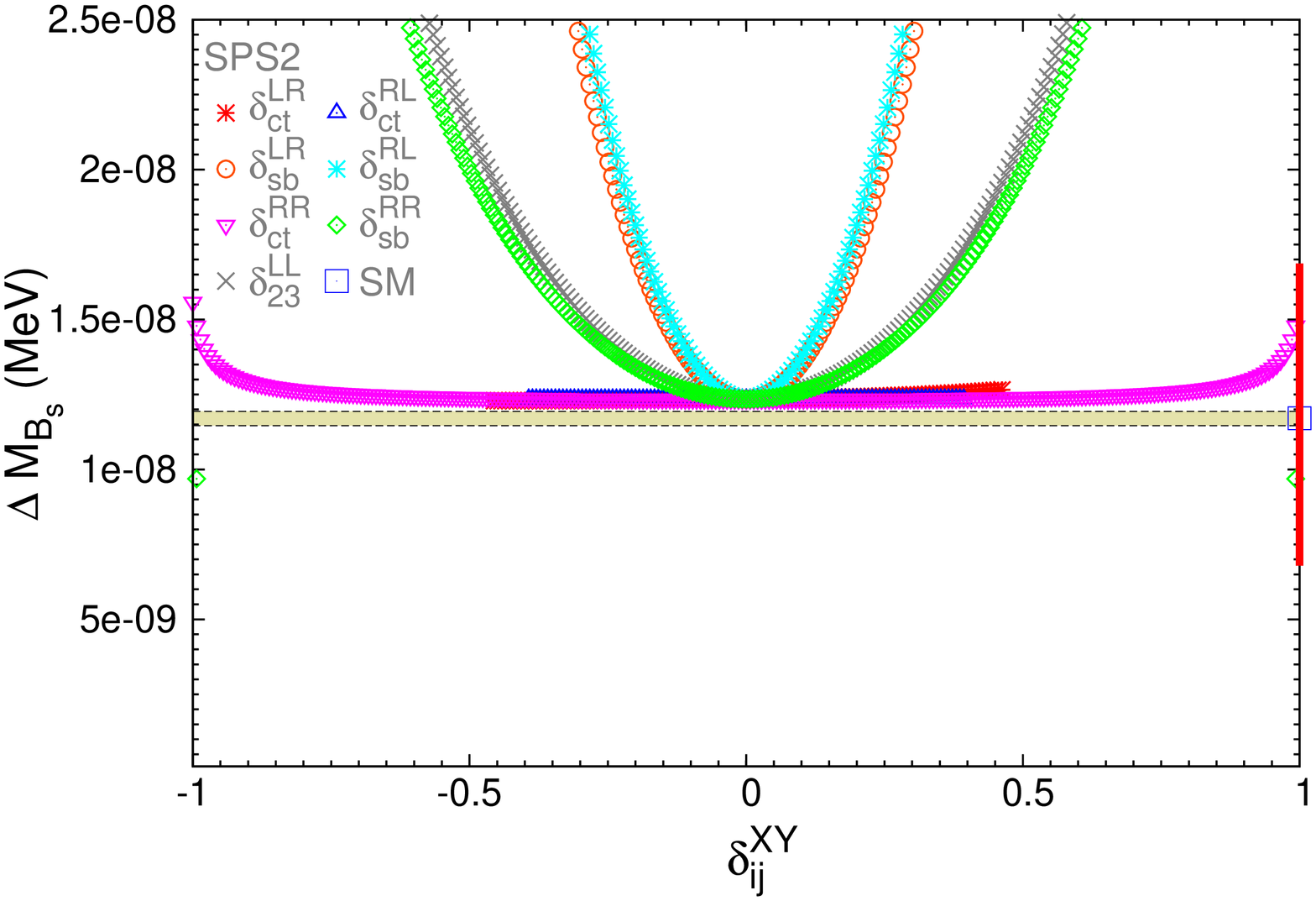}&
\includegraphics[width=13.3cm,height=16.5cm,angle=270]{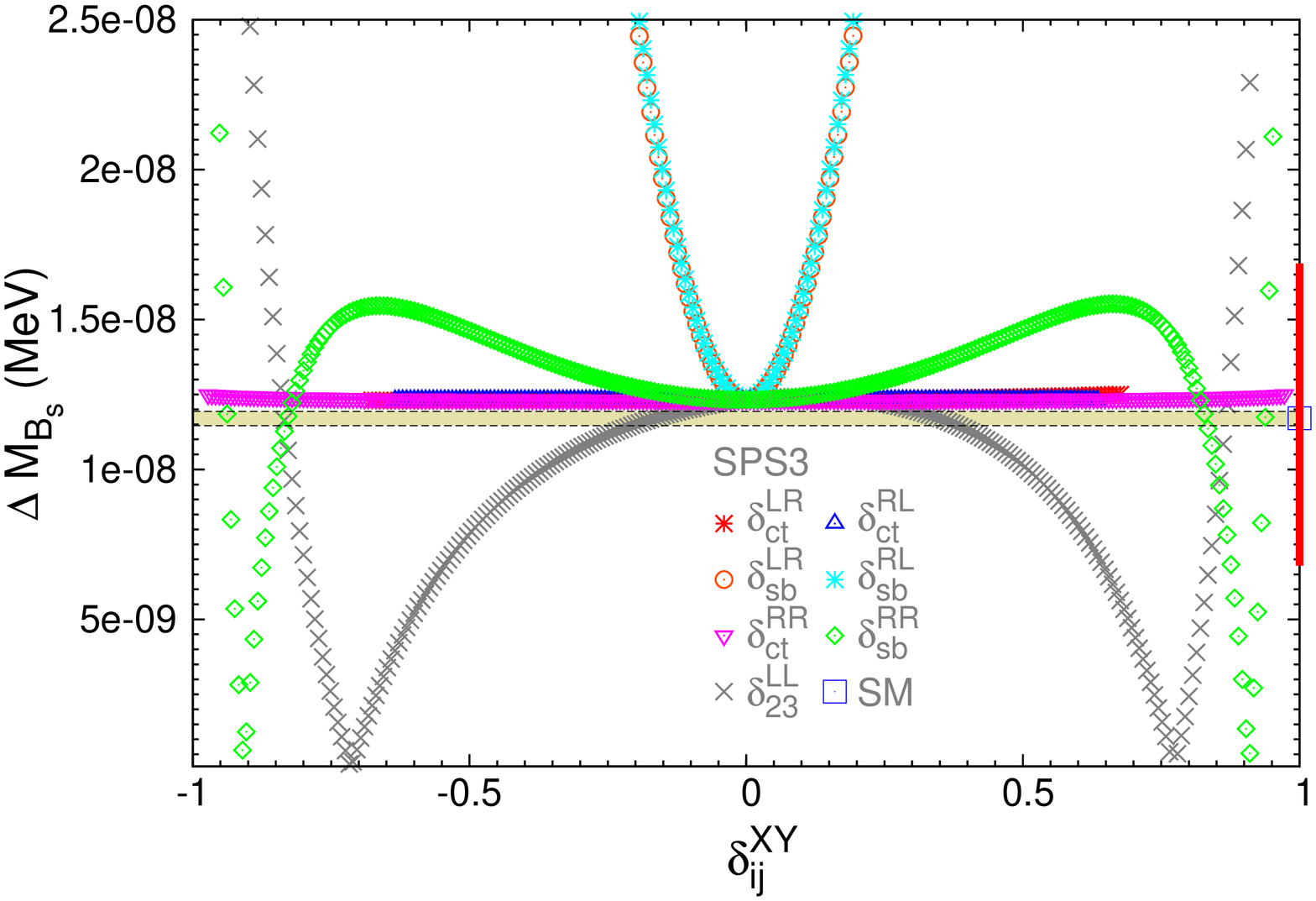}\\ 
\includegraphics[width=13.3cm,height=16.5cm,angle=270]{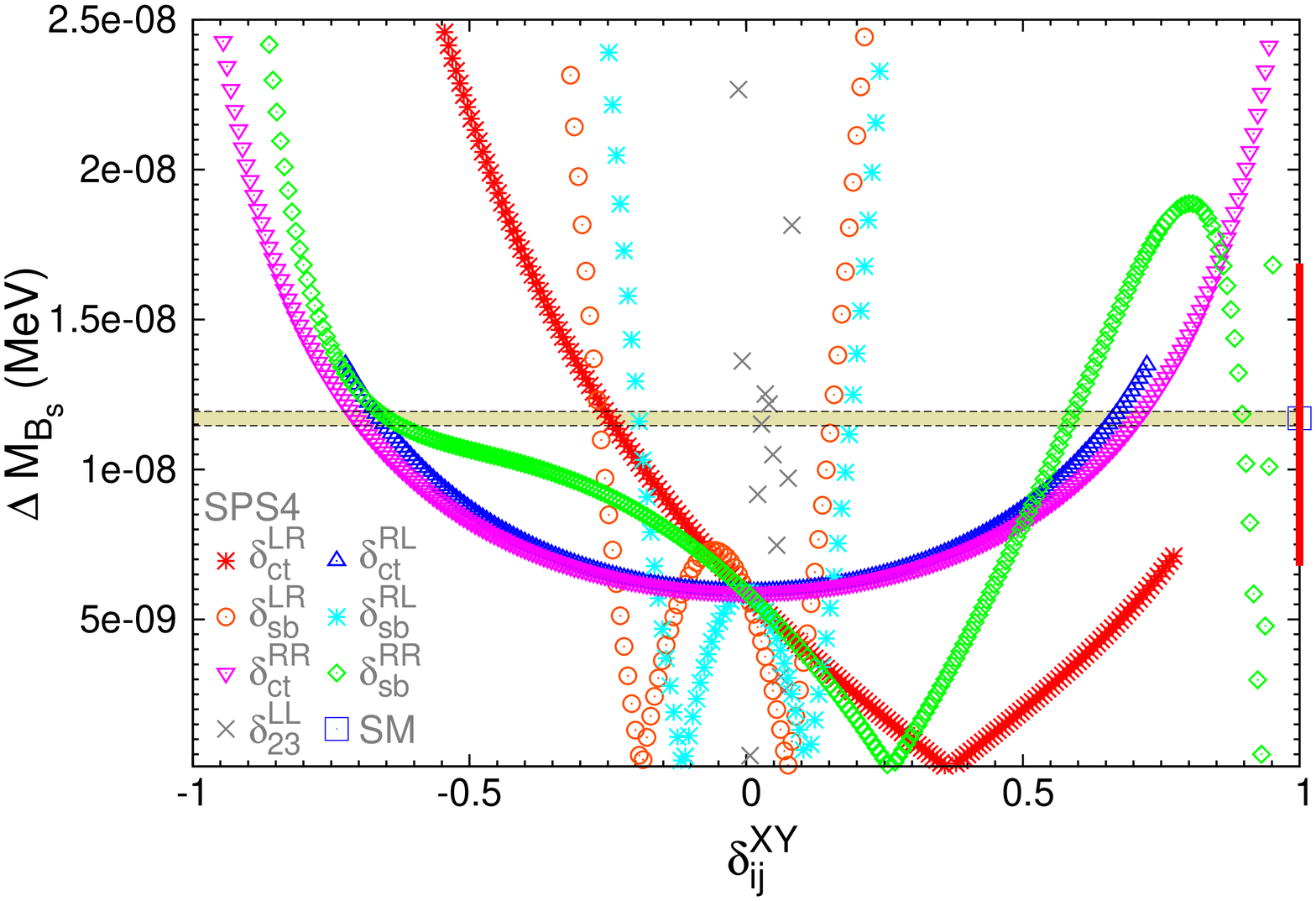}& 
\includegraphics[width=13.3cm,height=16.5cm,angle=270]{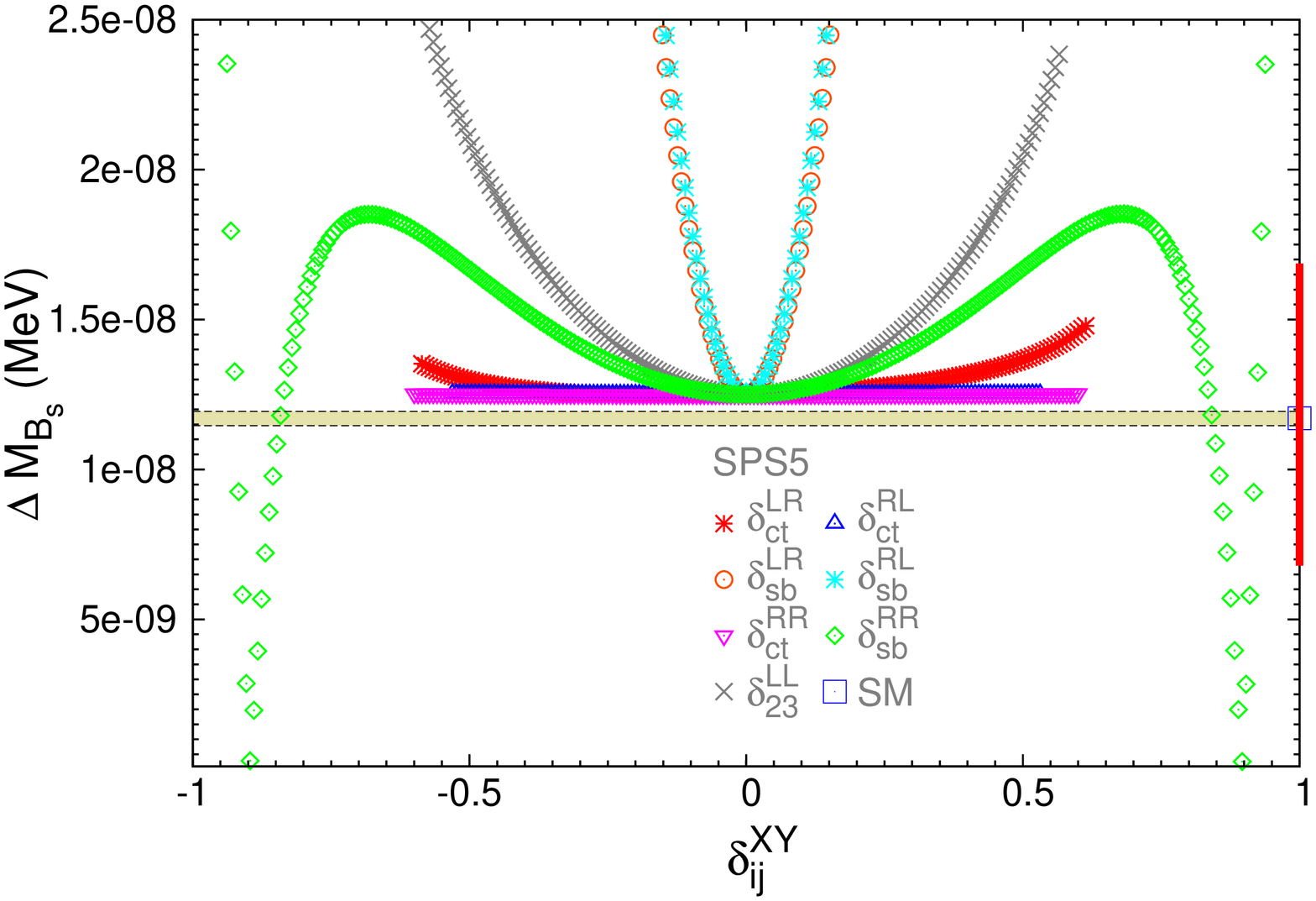}\\ 
\end{tabular}}}
\caption{Sensitivity to the NMFV deltas in $\Delta M_{B_s}$ for the SPSX
  points of table \ref{points}.  The experimental allowed $3\sigma_{\rm exp}$ area is
  the horizontal colored band. The SM prediction and the theory uncertainty
  \Dtheo(\dmbs) (red bar) is displayed on the right axis.}  
\label{figdeltams}
\end{figure}
%%%%%%%%%%%%%%%%%%%%%%%%%% F I G U R E %%%%%%%%%%%%%%%%%%%%%%%%%%%%%%%%%%%%%%%%
\clearpage
\newpage

\subsubsection*{Total Allowed $\deXYij$ Intervals}

\vspace{0.5cm}
We finally summarize in table \ref{tabdeltasummary} the total allowed
intervals for all the NMFV deltas, $\deXYij$, where now we have required
compatibility with the present data of the three considered $B$
observables, \bsg, ${\rm BR}(B_s \to \mu^+ \mu^-)$ and $\Delta
M_{B_s}$. It is obvious, from the previous discussion, that the most
restrictive observables are \bsg and $\Delta M_{B_s}$, leading to a
pattern of allowed delta intervals which is clearly the intersect of
their two corresponding intervals. The main conclusion from this table
is that, except for SPS4 (the point SPS4 is practically excluded), the
NMFV deltas in the top-sector can be sizeable $|\de_{ct}^{XY}|$  larger
than ${\cal O}(0.1)$ and still compatible with $B$ data. In particular,
$\delta^{RL}_{ct}$, and $\delta^{RR}_{ct}$ are the less constrained
parameters, and to a lesser extent also $\delta^{LR}_{ct}$. The
parameters on the bottom-sector are, in contrast, quite constrained. The
most tightly constrained are clearly  $\delta^{LR}_{sb}$ and
$\delta^{RL}_{sb}$, specially the first one with just some singular
allowed values: either positive and of the order of  $3-5 \times
10^{-2}$, or negative and with a small size of the order of $-7 \times
10^{-3}$; for the second the limits are around  $2 \times 10^{-2}$ for
both positive and negative values. $\delta^{RR}_{sb}$ is the less
constrained parameter in the bottom sector, with larger allowed
intervals of  $|\delta^{RR}_{sb}| \lsim 0.4-0.9$ depending on the
scenario. 

All SPS points are defined with a positive~$\mu$ value. We have
checked the effect of switching the sign of~$\mu$. While the numerical
results are changing somewhat, no qualitative change can be
observed. Consequently, confining ourselves to positive~$\mu$ does
not constitute a general restriction of our analysis. Similar
observations are made in the Higgs-sector analysis below.

The intervals allowed by $B$ data that we have presented above  will be
of interest for the following study in this work, where we will next
explore the size of the radiative corrections to the MSSM Higgs masses
within these NMFV-MSSM scenarios and we will require compatibility with
$B$ data. In the final analysis of these corrections, we will use the
constraints from $B$ data as extracted from two non-vanishing deltas. As
expected, these constraints vary significantly respect to the ones with
just one non-vanishing delta.   

\begin{table}[H]
\vspace*{-10mm}
\begin{sidewaystable}[H]
\begin{center}
\small
\resizebox{20cm}{!} {
\begin{tabular}{|c|c|c|c|c|} \hline
 & & ${\rm BR}(B\rightarrow X_s\gamma)$ & ${\rm BR}(B_s\rightarrow\mu^+\mu^-)$ & $\dmbs$  \\ \hline
$\delta^{LL}_{23}$ & \begin{tabular}{c} SPS1a \\ SPS1b \\ SPS2 \\ SPS3 \\ SPS4 \\ SPS5 \end{tabular} &  
\begin{tabular}{c}
(-0.51:-0.43) (-0.034:0.083) \\ (-0.33:-0.27)  (-0.014:0.062) \\ (-0.43:0.34)   (0.90:0.92) \\ (-0.73:-0.65) (-0.083:0.12) \\ (-0.14:-0.11) (0.0069:0.034) \\ (-0.26:0.50) \end{tabular}&  \begin{tabular}{c} (-0.53:0.92) \\ (-0.014:0.16) \\ (-0.99:0.99) \\ (-0.90:0.97) \\ (0.028:0.055) \\ (-0.60:0.57) \end{tabular}&  \begin{tabular}{c} (-0.73:-0.65) (-0.41:0.55) (0.73:0.79) \\ (-0.090:-0.069) (-0.021:0.097) (0.14:0.17) \\ (-0.37:0.37) \\ (-0.86:-0.79) (-0.56:0.66) (0.83:0.89) \\ (-0.0069)(0.021:0.055)(0.076) \\ (-0.37:0.39) 
 \end{tabular}  \\ \hline
$\delta^{LR}_{ct}$  & \begin{tabular}{c} SPS1a \\ SPS1b \\ SPS2 \\ SPS3 \\ SPS4 \\ SPS5 \end{tabular}    
& \begin{tabular}{c} 
(-0.89:-0.86) (-0.12:-0.097) (-0.062:0.28) \\ (-0.083:0.36) \\ (-0.46:0.46) \\ (-0.43:0.61) \\ (-0.61:-0.51) (0.041:0.23) \\ (-0.27:0.58) \end{tabular} & \begin{tabular}{c} (-0.89:0.89) \\ (-0.44:0.67) \\ (-0.46:0.46) \\ (-0.68:0.68) \\ excluded \\ (-0.59:0.61) \end{tabular} & \begin{tabular}{c} (-0.89:0.89) \\ (-0.67:0.67) \\ (-0.46:0.46) \\ (-0.68:0.68) \\ (-0.39:-0.021) (0.74:0.77) \\ (-0.59:0.61)
\end{tabular}  \\ \hline
$\delta^{LR}_{sb}$  & \begin{tabular}{c} SPS1a \\ SPS1b \\ SPS2 \\ SPS3 \\ SPS4 \\ SPS5 \end{tabular}    & 
\begin{tabular}{c} 
(0)(0.034) \\ (-0.0069:0) (0.048:0.055) \\ (-0.0069:0) (0.048:0.055) \\ (-0.0069:0) (0.048:0.055) \\ (-0.0069)(0.034) \\ (-0.0069:0) (0.041) \end{tabular} & \begin{tabular}{c} (-0.60:0.60) \\ (-0.43:0.54) \\ (-0.48:0.48) \\ (-0.61:0.61) \\ (0.49) \\ (-0.71:0.71) \end{tabular} & \begin{tabular}{c} (-0.076:0.076) \\ (-0.15:0.14) \\ (-0.19:0.19) \\ (-0.12:0.12) \\ (-0.29:-0.24) (-0.10:-0.014) (0.12:0.18) \\ (-0.090:0.090) 
\end{tabular}  \\ \hline
$\delta^{RL}_{ct}$  & \begin{tabular}{c} SPS1a \\ SPS1b \\ SPS2 \\ SPS3 \\ SPS4 \\ SPS5 \end{tabular}    & 
\begin{tabular}{c}
(-0.84:0.84) \\ (-0.63:0.63) \\ (-0.39:0.39) \\ (-0.63:0.63) \\ excluded \\ (-0.53:0.53) \end{tabular} 
& 
\begin{tabular}{c} 
(-0.84:0.84) \\ (-0.63:0.63) \\ (-0.39:0.39) \\ (-0.63:0.63) \\ excluded \\ (-0.53:0.53) \end{tabular} 
& 
\begin{tabular}{c}
(-0.84:0.84) \\ (-0.63:0.63) \\ (-0.39:0.39) \\ (-0.63:0.63) \\ (-0.72:-0.21) (0.21:0.72) \\ (-0.53:0.53)\end{tabular} \\ \hline
$\delta^{RL}_{sb}$  & \begin{tabular}{c} SPS1a \\ SPS1b \\ SPS2 \\ SPS3 \\ SPS4 \\ SPS5 \end{tabular}    &
\begin{tabular}{c} 
(-0.014:0.014) \\ (-0.021:0.021) \\ (-0.014:0.014) \\ (-0.021:0.021) \\ (-0.021:-0.014)(0.014:0.021) \\ (-0.014:0.014) \end{tabular} 
& 
\begin{tabular}{c}  (-0.71:0.71) \\ (-0.58:0.58) \\ (-0.55:0.55) \\ (-0.63:0.63) \\ excluded \\ (-0.72:0.72) \end{tabular} 
& 
\begin{tabular}{c}  (-0.069:0.069) \\ (-0.14:0.14) \\ (-0.17:0.17) \\ (-0.11:0.11) \\ (-0.21:-0.17) (0.16:0.21) \\ (-0.083:0.083)\end{tabular}  \\ \hline
$\delta^{RR}_{ct}$ & \begin{tabular}{c} SPS1a \\ SPS1b \\ SPS2 \\ SPS3 \\ SPS4 \\ SPS5 \end{tabular}   & \begin{tabular}{c} 
(-0.93:-0.67) (-0.64:0.93) \\ (-0.93:-0.61) (-0.56:0.90) \\ (-1.0:0.99) \\ (-0.97:0.97) \\ excluded \\ (-0.60:0.60) \end{tabular}     & \begin{tabular}{c} (-0.93:0.93) \\ (-0.95:0.94) \\ (-1.0:0.99) \\ (-0.97:0.97) \\ excluded \\ (-0.60:0.60) \end{tabular}     & \begin{tabular}{c} (-0.93:0.93) \\ (-0.98:0.98) \\ (-1.0:0.99) \\ (-0.98:0.97) \\ (-0.85:-0.22) (0.22:0.85) \\ (-0.60:0.60)
\end{tabular}   \\ \hline
$\delta^{RR}_{sb}$  & \begin{tabular}{c} SPS1a \\ SPS1b \\ SPS2 \\ SPS3 \\ SPS4 \\ SPS5 \end{tabular}    &
\begin{tabular}{c} 
(-0.65:0.68) \\ (-0.71:0.74) \\ (-0.99:0.99) \\ (-0.98:0.98) \\ (-0.45:-0.18) (0.19:0.46) \\ (-0.77:0.80) \end{tabular}  & \begin{tabular}{c} 
(-0.96:0.96) \\ (-0.73:0.98) \\ (-0.99:0.99) \\ (-0.98:0.98) \\ excluded \\ (-0.97:0.97) \end{tabular}  & \begin{tabular}{c} (-0.91:-0.90) (-0.86:-0.80) (-0.41:0.41) (0.81:0.86) (0.90:0.91) \\ (-0.94:-0.92) (-0.83:0.88)  (0.93:0.94) \\ (-0.99) (-0.39:0.39) (0.99) \\ (-0.94:-0.93) (-0.88:0.88) (0.93:0.94) \\ (-0.80:-0.028) (0.461:0.71) (0.86:0.91)  (0.94:0.95) \\ (-0.92) (-0.87:-0.78) (-0.51:0.51) (0.78:0.87)  (0.92)
\end{tabular}    \\ \hline
\end{tabular}}
\end{center}
\end{sidewaystable}
\caption{Allowed delta intervals by ${\rm BR}(B\rightarrow X_s\gamma)$, ${\rm BR}(B_s\rightarrow\mu^+\mu^-)$ and $\dmbs$. \label{tableintervals}}
\end{table}

%\newpage
\begin{table}[H]
%\hspace*{15mm} 
\begin{center}
\begin{tabular}{|c|c|c|} \hline
 & & Total allowed intervals \\ \hline
$\delta^{LL}_{23}$ & \begin{tabular}{c} SPS1a \\ SPS1b \\ SPS2 \\ SPS3 \\ SPS4 \\ SPS5 \end{tabular} &  
\begin{tabular}{c} 
(-0.034:0.083) \\ (-0.014:0.062) \\ (-0.37:0.34) \\ (-0.083:0.12) \\ (0.028:0.034) \\ (-0.26:0.39) \end{tabular} \\ \hline
$\delta^{LR}_{ct}$  & \begin{tabular}{c} SPS1a \\ SPS1b \\ SPS2 \\ SPS3 \\ SPS4 \\ SPS5 \end{tabular}    
& \begin{tabular}{c} 
(-0.89:-0.86) (-0.12:-0.097) (-0.062:0.28) \\ (-0.083:0.36) \\ (-0.46:0.46) \\ (-0.43:0.61) \\ excluded \\ (-0.27:0.58) \end{tabular}   \\ \hline
$\delta^{LR}_{sb}$  & \begin{tabular}{c} SPS1a \\ SPS1b \\ SPS2 \\ SPS3 \\ SPS4 \\ SPS5 \end{tabular}    & 
\begin{tabular}{c} 
(0)(0.034) \\ (-0.0069:0) (0.048:0.055) \\ (-0.0069:0) (0.048:0.055) \\ (-0.0069:0) (0.048:0.055) \\ excluded \\ (-0.0069:0) (0.041) \end{tabular}  \\ \hline
$\delta^{RL}_{ct}$  & \begin{tabular}{c} SPS1a \\ SPS1b \\ SPS2 \\ SPS3 \\ SPS4 \\ SPS5 \end{tabular}    & 
\begin{tabular}{c}
(-0.84:0.84) \\ (-0.63:0.63) \\ (-0.39:0.39) \\ (-0.63:0.63) \\ excluded \\ (-0.53:0.53) \end{tabular} 
  \\ \hline
$\delta^{RL}_{sb}$  & \begin{tabular}{c} SPS1a \\ SPS1b \\ SPS2 \\ SPS3 \\ SPS4 \\ SPS5 \end{tabular}    &
\begin{tabular}{c}  (-0.014:0.014) \\ (-0.021:0.021) \\ (-0.014:0.014) \\ (-0.021:0.021) \\ excluded \\ (-0.014:0.014) \end{tabular} 
  \\ \hline
$\delta^{RR}_{ct}$ & \begin{tabular}{c} SPS1a \\ SPS1b \\ SPS2 \\ SPS3 \\ SPS4 \\ SPS5 \end{tabular}   & \begin{tabular}{c} 
(-0.93:-0.67) (-0.64:0.93) \\ (-0.93:-0.61) (-0.56:0.90) \\ (-1.0:0.99) \\ (-0.97:0.97) \\ excluded \\ (-0.60:0.60) \end{tabular}    \\ \hline
$\delta^{RR}_{sb}$  & \begin{tabular}{c} SPS1a \\ SPS1b \\ SPS2 \\ SPS3 \\ SPS4 \\ SPS5 \end{tabular}    &
\begin{tabular}{c}  (-0.41:0.41)  \\ (-0.71:0.74) \\ (-0.99) (-0.39:0.39) (0.99) \\ (-0.94:-0.93) (-0.88:0.88) (0.93:0.94) \\ excluded \\ (-0.51:0.51) (0.78:0.80)
\end{tabular}    \\ \hline
\end{tabular}  
\end{center}
\caption{Total allowed delta intervals by ${\rm BR}(B\rightarrow X_s\gamma)$, ${\rm BR}(B_s\rightarrow\mu^+\mu^-)$ and $\dmbs$. \label{tabdeltasummary}}
\end{table}

%%%%%%%%%%%%%%%%%%%%%%%%%%%%%%%%%%%%%%%%%%%%%%%%%%%%%%%%%%%%%%%%%%%%%%%%%%%%%%%
%%%%%%%%%%%%%%%%%%%%%%%%%%%%%%%%%%%%%%%%%%%%%%%%%%%%%%%%%%%%%%%%%%%%%%%%%%%%%%%

\section{Radiative corrections to MSSM Higgs masses within NMFV scenarios}
\label{sec:mhiggs}
In this section we present our computation of the one-loop radiative corrections to MSSM Higgs boson masses within the NMFV scenarios. For completeness and definiteness, we first shortly review the relevant features of the MSSM Higgs sector at tree-level. Then we summarize the main one-loop renormalization issues that are involved in the computation and finally we present the analytical results for the one-loop corrected Higgs masses.  

\subsection{The Higgs boson sector at tree-level}
\label{sec:tree}

Contrary to the SM, in the MSSM two Higgs doublets
are required.
The  Higgs potential~\cite{hhg}
\BEA
V &=& m_{1}^2 |\cHe|^2 + m_{2}^2 |\cHz|^2 
      - m_{12}^2 (\epsilon_{ab} \cHe^a\cHz^b + \hc)  \non \\
  & & + \frac{1}{8}(g_1^2+g_2^2) \left[ |\cHe|^2 - |\cHz|^2 \right]^2
        + \frac{1}{2} g_2^2|\cHe^{\dag} \cHz|^2~,
\label{higgspot}
\EEA
contains $m_1, m_2, m_{12}$ as soft SUSY breaking parameters;
$g_2, g_1$ are the $SU(2)$ and $U(1)$ gauge couplings, and 
$\epsilon_{12} = -1$.

The doublet fields $H_1$ and $H_2$ are decomposed  in the following way:
\BEA
\cHe &=& \VL \cHe^0 \\[0.5ex] \cHe^- \VR \; = \; \VL v_1 
        + \ed{\wz}(\phi_1^0 - i\chi_1^0) \\[0.5ex] -\phi_1^- \VR~,  
        \non \\
\cHz &=& \VL \cHz^+ \\[0.5ex] \cHz^0 \VR \; = \; \VL \phi_2^+ \\[0.5ex] 
        v_2 + \ed{\wz}(\phi_2^0 + i\chi_2^0) \VR~.
\label{higgsfeldunrot}
\EEA
The potential (\ref{higgspot}) can be described with the help of two  
independent parameters (besides $g_1$ and $g_2$): 
$\Tb = v_2/v_1$ and $M_A^2 = -m_{12}^2(\Tb+\CTb)$,
where $M_A$ is the mass of the $\cp$-odd Higgs boson~$A$.

The diagonalization of the bilinear part of the Higgs potential,
i.e.\ of the Higgs mass matrices, is performed via the orthogonal
transformations 
\BEA
\label{hHdiag}
\VL H \\[0.5ex] h \VR &=& \ML \Ca & \Sa \\[0.5ex] -\Sa & \Ca \MR 
\VL \phi_1^0 \\[0.5ex] \phi_2^0~, \VR  \\
\label{AGdiag}
\VL G \\[0.5ex] A \VR &=& \ML \Cb & \Sbe \\[0.5ex] -\Sbe & \Cb \MR 
\VL \chi_1^0 \\[0.5ex] \chi_2^0 \VR~,  \\
\label{Hpmdiag}
\VL G^{\pm} \\[0.5ex] H^{\pm} \VR &=& \ML \Cb & \Sbe \\[0.5ex] -\Sbe & 
\Cb \MR \VL \phi_1^{\pm} \\[0.5ex] \phi_2^{\pm} \VR~.
\EEA
The mixing angle $\al$ is determined through
\BE
\al = {\rm arctan}\KKL 
  \frac{-(\MA^2 + \MZ^2) \Sbe \Cb}
       {\MZ^2 \CQb + \MA^2 \SQb - m^2_{h,{\rm tree}}} \KKR~, ~~
 -\frac{\pi}{2} < \al < 0~.
\label{alphaborn}
\end{equation}

One gets the following Higgs spectrum:
\BEA
\mbox{2 neutral bosons},\, {\cal CP} = +1 &:& h, H \non \\
\mbox{1 neutral boson},\, {\cal CP} = -1  &:& A \non \\
\mbox{2 charged bosons}                   &:& H^+, H^- \non \\
\mbox{3 unphysical Goldstone bosons}      &:& G, G^+, G^- .
\EEA

At tree level the mass matrix of the neutral $\cp$-even Higgs bosons
is given in the $\Pe$-$\Pz$-basis 
in terms of $\MZ$, $\MA$, and $\Tb$ by
\BEA
M_{\rm Higgs}^{2, {\rm tree}} &=& \ML \mpe^2 & \mpez^2 \\ 
                           \mpez^2 & \mpz^2 \MR \non\\
&=& \ML \MA^2 \SQb + \MZ^2 \CQb & -(\MA^2 + \MZ^2) \Sbe \Cb \\
    -(\MA^2 + \MZ^2) \Sbe \Cb & \MA^2 \CQb + \MZ^2 \SQb \MR,
\label{higgsmassmatrixtree}
\EEA
which by diagonalization according to \refeq{hHdiag} yields the
tree-level Higgs boson masses
\BE
M_{\rm Higgs}^{2, {\rm tree}} 
   \stackrel{\al}{\longrightarrow}
   \ML m_{H,{\rm tree}}^2 & 0 \\ 0 &  m_{h,{\rm tree}}^2 \MR~,
\end{equation}
where
\BE
\label{rMSSM:mtree}
(m_{H,h}^2)_{\rm tree}=
\edz\left[\MA^2+\MZ^2 \pm\sqrt{(\MA^2+\MZ^2)^2-
4\MZ^2\MA^2\cos^2 2\be}\right] ~.
 \end{equation}

The charged Higgs boson mass is given by
\BE
\label{rMSSM:mHp}
m^{2}_{H^{\pm},{\rm tree}} = \MA^2 + \MW^2~.
\end{equation}
The masses of the gauge bosons are given in analogy to the SM:
\BE
M_W^2 = \frac{1}{2} g_2^2 (v_1^2+v_2^2) ;\qquad
M_Z^2 = \frac{1}{2}(g_1^2+g_2^2)(v_1^2+v_2^2) ;\qquad M_\ga=0.
\end{equation}

%%%%%%%%%%%%%%%%%%%%%%%%%%%%%%%%%%%%%%%%%%%%%%%%%%%%%%%%%%%%%%%%%%%%%%%%%%%%%%%

\subsection{The Higgs boson sector at one-loop}
\label{sec:renrMSSM}

In order to calculate one-loop corrections to the Higgs boson
masses, the renormalized Higgs boson
self-energies are needed. Here we follow the procedure used in
\citeres{mhiggsf1lC,mhcMSSMlong} (and references therein) and review it for
completeness. The parameters appearing in the Higgs
potential, see \refeq{higgspot}, are renormalized as follows:
\begin{align}
\label{rMSSM:PhysParamRenorm}
  \MZ^2 &\to \MZ^2 + \dMZsq,  & \tadh &\to \tadh +
  \dtadh, \\ 
  \MW^2 &\to \MW^2 + \dMWsq,  & \tadH &\to \tadH +
  \dtadH, \notag \\ 
  M_{\rm Higgs}^2 &\to M_{\rm Higgs}^2 + \de M_{\rm Higgs}^2, & 
  \tanb &\to \tanb (1+\dtanb). \notag 
%  \mHp^2 &\to \mHp^2 + \de\mHp^2 \notag
\end{align}
$M_{\rm Higgs}^2$ denotes the tree-level Higgs boson mass matrix given
in \refeq{higgsmassmatrixtree}. $\tadh$ and $\tadH$ are the tree-level
tadpoles, i.e.\ the terms linear in $h$ and $H$ in the Higgs potential.

The field renormalization matrices of both Higgs multiplets
can be set up symmetrically, 
\begin{align}
\label{rMSSM:higgsfeldren}
  \begin{pmatrix} h \\[.5em] H \end{pmatrix} \to
  \begin{pmatrix}
    1+\tfrac{1}{2} \dZ{hh} & \tfrac{1}{2} \dZ{hH} \\[.5em]
    \tfrac{1}{2} \dZ{hH} & 1+\tfrac{1}{2} \dZ{HH} 
  \end{pmatrix} \cdot
  \begin{pmatrix} h \\[.5em] H \end{pmatrix}~.
\end{align}

\noindent
For the mass counter term matrices we use the definitions
\begin{align}
  \delta M_{\rm Higgs}^2 =
  \begin{pmatrix}
    \dmhsq  & \dmhHsq \\[.5em]
    \dmhHsq & \dmHsq  
  \end{pmatrix}~.
\end{align}
The renormalized self-energies, $\hSi(p^2)$, can now be expressed
through the unrenormalized self-energies, $\Si(p^2)$, the field
renormalization constants and the mass counter terms.
This reads for the $\cp$-even part,
\begin{subequations}
\label{rMSSM:renses_higgssector}
\begin{align}
\ser{hh}(p^2)  &= \se{hh}(p^2) + \dZ{hh} (p^2-\mhtree^2) - \dmhsq, \\
\ser{hH}(p^2)  &= \se{hH}(p^2) + \dZ{hH}
(p^2-\tfrac{1}{2}(\mhtree^2+\mHtree^2)) - \dmhHsq, \\ 
\ser{HH}(p^2)  &= \se{HH}(p^2) + \dZ{HH} (p^2-\mHtree^2) - \dmHsq~.
\end{align}
\end{subequations}

Inserting the renormalization transformation into the Higgs mass terms
leads to expressions for their counter terms which consequently depend
on the other counter terms introduced in~(\ref{rMSSM:PhysParamRenorm}). 

For the $\cp$-even part of the Higgs sectors, these counter terms are:
\begin{subequations}
\label{rMSSM:HiggsMassenCTs}
\begin{align}
\dmhsq &= \de\MA^2 \cos^2(\alpha-\beta) + \delta \MZ^2 \sin^2(\alpha+\beta) \\
&\quad + \tfrac{e}{2 \MZ \sw \cw} (\dtadH \cos(\alpha-\beta)
\sin^2(\alpha-\beta) + \dtadh \sin(\alpha-\beta)
(1+\cos^2(\alpha-\beta))) \notag \\ 
&\quad + \dtanb \sinb \cosb (\MA^2 \sin 2 (\alpha-\beta) + \MZ^2 \sin
2 (\alpha+\beta)), \notag \\ 
\dmhHsq &= \tfrac{1}{2} (\de\MA^2 \sin 2(\alpha-\beta) - \dMZsq \sin
2(\alpha+\beta)) \\ 
&\quad + \tfrac{e}{2 \MZ \sw \cw} (\dtadH \sin^3(\alpha-\beta) -
\dtadh \cos^3(\alpha-\beta)) \notag \\ 
&\quad - \dtanb \sinb \cosb (\MA^2 \cos 2 (\alpha-\beta) + \MZ^2 \cos
2 (\alpha+\beta)), \notag \\ 
\dmHsq &= \de\MA^2 \sin^2(\alpha-\beta) + \dMZsq \cos^2(\alpha+\beta) \\
&\quad - \tfrac{e}{2 \MZ \sw \cw} (\dtadH \cos(\alpha-\beta)
(1+\sin^2(\alpha-\beta)) + \dtadh \sin(\alpha-\beta)
\cos^2(\alpha-\beta)) \notag \\ 
&\quad - \dtanb \sinb \cosb (\MA^2 \sin 2 (\alpha-\beta) + \MZ^2 \sin
2 (\alpha+\beta))~. \notag 
\end{align}
\end{subequations}

\bigskip
For the field renormalization we chose to give each Higgs doublet one
renormalization constant,
\begin{align}
\label{rMSSM:HiggsDublettFeldren}
  \cHe \to (1 + \tfrac{1}{2} \dZ{\cHe}) \cHe, \quad
  \cHz \to (1 + \tfrac{1}{2} \dZ{\cHz}) \cHz~.
\end{align}
This leads to the following expressions for the various field
renormalization constants in \refeq{rMSSM:higgsfeldren}:
\begin{subequations}
\label{rMSSM:FeldrenI_H1H2}
\begin{align}
  \dZ{hh} &= \sinasq \dZ{\cHe} + \cosasq \dZ{\cHz}, \\[.2em]
  \dZ{hH} &= \sina \cosa (\dZ{\cHz} - \dZ{\cHe}), \\[.2em]
  \dZ{HH} &= \cosasq \dZ{\cHe} + \sinasq \dZ{\cHz}~.
%  \dZ{H^-H^+} &= \sinbsq \dZ{\cHe} + \cosbsq \dZ{\cHz}~.
\end{align}
\end{subequations}
The counter term for $\tb$ can be expressed in terms of the vacuum
expectation values as
\begin{equation}
\de\tb = \frac{1}{2} \KL \dZ{\cHz} - \dZ{\cHe} \KR +
\frac{\de v_2}{v_2} - \frac{\de v_1}{v_1}~,
\end{equation}
where the $\de v_i$ are the renormalization constants of the $v_i$:
\begin{equation}
v_1 \to \KL 1 + \dZ{\cHe} \KR \KL v_1 + \de v_1 \KR, \quad
v_2 \to \KL 1 + \dZ{\cHz} \KR \KL v_2 + \de v_2 \KR~.
\end{equation}
Similarly for the charged Higgs sector, the renormalized self-energy is expressed in terms of the unrenormalized one and the corresponding counter-terms as:
\noindent \begin{equation}
\hat{\Sigma}_{H^{-}H^{+}}\left(p^{2}\right)=\Sigma_{H^{-}H^{+}}\left(p^{2}\right)+\delta Z_{H^{-}H^{+}}\left(p^{2}-m^{2}_{H^{\pm},{\rm tree}} \right)-\delta m_{H^{\pm}}^{2},\end{equation}
where, 
\noindent \begin{equation}
\delta m_{H^{\pm}}^{2}=\delta M_{A}^{2}+\delta M_{W}^{2}\end{equation}
and,
\noindent \begin{equation}
\delta Z_{H^{-}H^{+}}=\sin^{2}\beta \, \,\dZ{\cHe}  
+\cos^{2}\beta \,\,\dZ{\cHz}. \end{equation}

The renormalization conditions are fixed by an appropriate
renormalization scheme. For the mass counter terms on-shell conditions
are used, resulting in:
\begin{align}
\label{rMSSM:mass_osdefinition}
  \dMZsq = \re \se{ZZ}(\MZ^2), \quad \dMWsq = \re \se{WW}(\MW^2),
  \quad \de\MA^2 = \re \se{AA}(\MA^2). 
\end{align}
For the gauge bosons $\Si$ denotes the transverse part of the self-energy. 
Since the tadpole coefficients are chosen to vanish in all orders,
their counter terms follow from $T_{\{h,H\}} + \de T_{\{h,H\}} = 0$: 
\begin{align}
  \dtadh = -{\tadh}, \quad \dtadH = -{\tadH}~. 
\end{align}
For the remaining renormalization constants for $\de\tb$, $\dZ{\cHe}$
and $\dZ{\cHz}$ the most convenient
choice is a \drbar\ renormalization of $\de\tb$, $\dZ{\cHe}$
and $\dZ{\cHz}$, 
\begin{subequations}
\label{rMSSM:deltaZHiggsTB}
\begin{align}
  \dZ{\cHe} &= \dZ{\cHe}^{\drbarm}
       \; = \; - \KKL \re \Sip_{HH \; |\al = 0} \KKR^{\rm div}, \\[.5em]
  \dZ{\cHz} &= \dZ{\cHz}^{\drbarm} 
       \; = \; - \KKL \re \Sip_{hh \; |\al = 0} \KKR^{\rm div}, \\[.5em]
  \dtanb &= -\edz (\dZ{\cHz} - \dZ{\cHe}) = \dtanb^{\drbarm}~.
\end{align}
\end{subequations}
The corresponding renormalization scale, $\mudim$, is set to 
$\mudim = \mt$ in all numerical evaluations. 

Finally, in the 
  Feynman diagrammatic (FD) approach that we are following here, the higher-order corrected 
$\cp$-even Higgs boson masses are derived by finding the
poles of the $(h,H)$-propagator 
matrix. The inverse of this matrix is given by
\BE
\left(\Delta_{\rm Higgs}\right)^{-1}
= - i \ML p^2 -  \mHtree^2 + \hSi_{HH}(p^2) &  \hSi_{hH}(p^2) \\
     \hSi_{hH}(p^2) & p^2 -  \mhtree^2 + \hSi_{hh}(p^2) \MR~.
\label{higgsmassmatrixnondiag}
\end{equation}
Determining the poles of the matrix $\Delta_{\rm Higgs}$ in
\refeq{higgsmassmatrixnondiag} is equivalent to solving
the equation
\begin{equation}
\left[p^2 - \mhtree^2 + \hSi_{hh}(p^2) \right]
\left[p^2 - \mHtree^2 + \hSi_{HH}(p^2) \right] -
\left[\hSi_{hH}(p^2)\right]^2 = 0\,.
\label{eq:proppole}
\end{equation}
Similarly, in the case of the charged Higgs sector, the corrected Higgs mass is derived by the position of the pole in the charged Higgs propagator, which is defined by: 
\noindent \begin{equation}
p^{2}-m^{2}_{H^{\pm},{\rm tree}} +
\hat{\Sigma}_{H^{-}H^{+}}\left(p^{2}\right)=0.
\label{eq:proppolech}
\end{equation}

\subsection{Analytical results of Higgs mass corrections in NMFV-SUSY}
\label{sec:analytical-results}
Following the previously detailed prescription for the computation of the one-loop corrected Higgs boson masses, one finds the analytical results for these masses in terms of the renormalized self-energies which, in turn, are written in terms of the unrenormalized self-energies and tadpoles. To shorten the presentation of these analytical results, it is convenient to report just on these unrenormalized self-energies and tadpoles. 

The relevant one-loop corrections have been evaluated 
with the help of 
\fa~\cite{feynarts} and  \fc~\cite{formcalc}.
For completeness the new Feynman rules included in the model file are listed
in the Appendix A.
All the results for the unrenormalized self-energies and tadpoles are collected in
Appendix B. We have shown explicitly just the relevant contributions
for the present study of the radiative corrections to the Higgs boson masses within NMFV scenarios, namely, the one-loop contributions from quarks
and squarks. The corresponding generic Feynman-diagrams for the
Higgs bosons self-energies, gauge boson
self-energy diagrams and tadpole diagrams are collected in the \reffi{figfdall} in Appendix B.
It should also be noticed that the contributions from the squarks are the only ones that differ from the usual ones of the MSSM with MFV.  
It should be noted also that the corrections from flavor mixing, which are the subject of our interest here,  are implicit in both the
$\VCKM$, and in the values of the rotation matrices, $R^{\tilde u}$, $R^{\tilde d}$, and the squark masses, $m_{\tilde u_i}$, $m_{\tilde d_i}$ ($i=1,..,6$) that appear in these formulas of the unrenormalized self-energies and tadpoles and that have been introduced in section \ref{sec:nmfv}. 

Finally, it is worth mentioning that we have checked the finiteness in our analytical results for the renormalized Higgs self-energies. It is obviously expected, but it is not a trivial check in the present scenarios with three generations of quarks and squarks and with flavor mixing. We have also checked that the analytical results of the self-energies in Appendix B agree with the formulas in \fh~\cite{feynhiggs,mhiggslong,mhiggsAEC,mhcMSSMlong}. Each one of the terms contained in the Appendix B was compared with the corresponding term in \fh. During this process and the check of the finiteness, discrepancies were found with the charged Higgs part of \fh, leading to an updated version of the code\footnote{
We especially thank T.~Hahn for his efforts put into this update.}.

%%%%%%%%%%%%%%%%%%%%%%%%%%%%%%%%%%%%%%%%%%%%%%%%%%%%%%%%%%%%%%%%%%%%%%%%%%%%%%
\section{Numerical analysis}
\label{sec:numanal}
In this section we present our numerical results for the radiative corrections to the Higgs boson
 masses from from flavor mixing within NMFV-SUSY scenarios. Since all one-loop corrections in the present NMFV scenario are common to the MSSM except for the corrections from squarks, which depend on the $\deXYij$ values,  we will focus just on the results of these corrections as a function of the flavor mixing parameters,  and present the differences with respect to the predictions within the MSSM. Correspondingly, we define: 
\begin{equation}
 \Dmphi (\deXYij) \equiv 
 \mphi^{\rm NMFV}(\deXYij) - \mphi^{\rm MSSM}, \quad \phi =h,\, H,\, H^{\pm}, 
\end{equation}
where $\mphi^{\rm NMFV}(\deXYij)$ and $\mphi^{\rm MSSM}$ have been calculated at the
one-loop level.
 It should be noted that  $\mphi^{\rm NMFV}(\deXYij=0) = \mphi^{\rm MSSM}$ and,
 therefore, by construction, $\Dmphi(\deXYij=0) = 0$, and $\Dmphi$ gives the
size of the one-loop NMFV contributions to $\mphi$. The numerical calculation of $\mphi^{\rm NMFV}(\deXYij)$ and $\mphi^{\rm MSSM}$ has been done 
with (the updated version of) \fh~\cite{feynhiggs,mhiggslong,mhiggsAEC,mhcMSSMlong}, which solves the eqs. \ref{eq:proppole} and \ref{eq:proppolech} for finding the positions of the poles of the propagator matrix. Previous results for $\De \mh (\de^{LL}_{23})$ can be found in~\cite{mhNMFVearly}. 

%%%%%%%%%%%%%%%%%%%%%%%%%%%%%%%%%%%%%%%%%%%%%%%%%%%%%%%%%%%%%%%%%%%%%%%%%%%%%%

\subsection{\boldmath{$\Dmphi$} versus one \boldmath{$\deXYij\neq 0$}}
\label{sec:numanalonedelta}

We show in figs. \ref{figdeltamh0}, \ref{figdeltamH0} and
\ref{figdeltamHp} our numerical results for  
$\Dmh$, $\DmH$ and $\DmHp$, respectively, as functions of the seven
considered flavor changing deltas, $\de^{LL}_{23}$, $\de^{LR}_{ct}$,
$\de^{LR}_{sb}$, $\de^{RL}_{ct}$, $\de^{RL}_{sb}$, $\de^{RR}_{ct}$ and
$\de^{RR}_{sb}$,  
which we vary in the interval $-1 \leq \deXYij \leq 1$. In these plots
we have chosen the same six SPS points of table \ref{points}, as for the
previous study of  constraints from $B$ physics in \ref{sec:nmfv}. We do
not take the experimental bounds into account here, since we just want
to show the general behavior of the masses with the deltas. The
experimental bounds will be taken into account in the next subsection.   
As before we have checked the impact of switching the sign of~$\mu$
  and found a small quantitative but no qualitative effect.

The main conclusions from these figures are the following:
\begin{itemize}
\item[-] General features: 

All mass corrections, $\Dmh$, $\DmH$ and $\DmHp$, are symmetric $\deXYij \to - \deXYij$, as expected. This feature is obviously different than in the previous plots of the $B$ observables. The sign of the mass corrections can be both positive and negative, depending on the particular delta value.  The size of the Higgs mass corrections, can be very large in some  
$\deXYij \neq 0$ regions, reaching values even larger than 10 GeV  
in some cases, at the central region with not very large delta values, 
$|\deXYij|< 0.5$. In fact, the restrictions from $B$ physics in this central region is crucial to get a reliable estimate of these effects.  

For low $\tb$, where the restrictions from $B$ physics to the deltas are less severe, the Higgs mass corrections are specially relevant. Particularly, $\Dmh$ turns out to be negative and large for $\tb =5$ (SPS5) for all deltas, except $\de^{RR}_{sb}$. For instance, at
$|\deXYij|\simeq 0.5$, the mass correction $\Dmh$ for SPS5 
is negative and $\gsim 5 \gev$ in all flavor changing deltas except $\de^{RR}_{sb}$ where the correction is negligible. In the case of $\DmH$ and $\DmHp$ the size of the correction at low $\tb$  is smaller, $\lsim 2 \gev$ in the central region, except for $\de^{LR}_{sb}$ and $\de^{RL}_{sb}$ that can also generate large  corrections $\gsim 5 \gev$. 

 In the cases with large 
$\tb$ (SPS4 and SPS1b), we also find large mass corrections but, as already said, they are much more limited by $B$ constraints. In particular,
for SPS4 all deltas are excluded, except for a very narrow window in $\de^{LL}_{23}$ (see table \ref{tabdeltasummary}).

In the cases with moderate $\tb=10$ (SPS1a, SPS2 and SPS3), we find large corrections $|\Dmh| \gsim 5 \gev$ in the central region 
of  $\de^{LR}_{sb}$, $\de^{RL}_{sb}$, $\de^{LR}_{ct}$ and $\de^{RL}_{ct}$. The other Higgs bosons get large corrections $|\DmH|,|\DmH|\gsim 5 \gev$ in the deltas central region only for  
$\de^{LR}_{sb}$ and $\de^{RL}_{sb}$.
   
\item[-] Sensitivity to the various deltas:
 
We find very strong sensitivity in the three mass corrections $\Dmh$, $\DmH$ and $\DmHp$, to $\de^{LR}_{sb}$ and $\de^{RL}_{sb}$ for all the seven considered SPS points. 

In the case of $\Dmh$ there is also an important sensitivity to 
$\de^{LR}_{ct}$ and $\de^{RL}_{ct}$ in all the considered points. The
strong sensitivity to $LR$ and $RL$ parameters can be understood due to
the relevance of the $A$-terms in these Higgs mass corrections. It can
be noticed in the Feynman rules (i.e. see the coupling of two squarks
and one/two Higgs bosons in Appendix A) that the $A$-terms enter
directly into the couplings, and in some cases,
as in the couplings of down-type squarks to the CP-odd Higgs boson,
enhanced by $\tb$. Therefore, considering the relationship between the
$A$-terms and these $LR$ and $RL$ parameters as is shown in
eq. \ref{deltasdefs}, the strong sensitivity to these parameters can be
understood. A similar strong sensitivity to $\de^{LR}_{ct}$ in $\Dmh$
has been found in \cite{Cao1}. 

In SPS5 there is a noticeable sensitivity to all deltas except
$\de^{RR}_{sb}$. In other points, the effects of $\de^{LL}_{23}$,
$\de^{RR}_{ct}$  on $\Dmh$ are only appreciated at the large delta
region, close to $\pm 1$. For instance, in SPS2,  $\Dmh = -5 \gev$
for $\de^{RR}_{ct}= \pm 1$.

In the case of $\DmH$, apart from $\de^{LR}_{sb}$ and $\de^{RL}_{sb}$, there is only noticeable  sensitivity to other deltas in  SPS5. The same comment applies to $\DmHp$.
\end{itemize}

%%%%%%%%%%%%%%%%%%%%%%%%%%%%%%%%%%%%%%%%%%%%%%%%%%%%%%%%%%%%%%%%%%%%%%%%%%%%%

%%%%%%%%%%%%%%%%%%%%%%%%%% F I G U R E %%%%%%%%%%%%%%%%%%%%%%%%%%%%%%%%%%%%%%%%
\begin{figure}[h!] 
\centering
\hspace*{-10mm} 
{\resizebox{17.9cm}{!} 
{\begin{tabular}{cc} 
\includegraphics[width=13.3cm,height=16.5cm,angle=270]{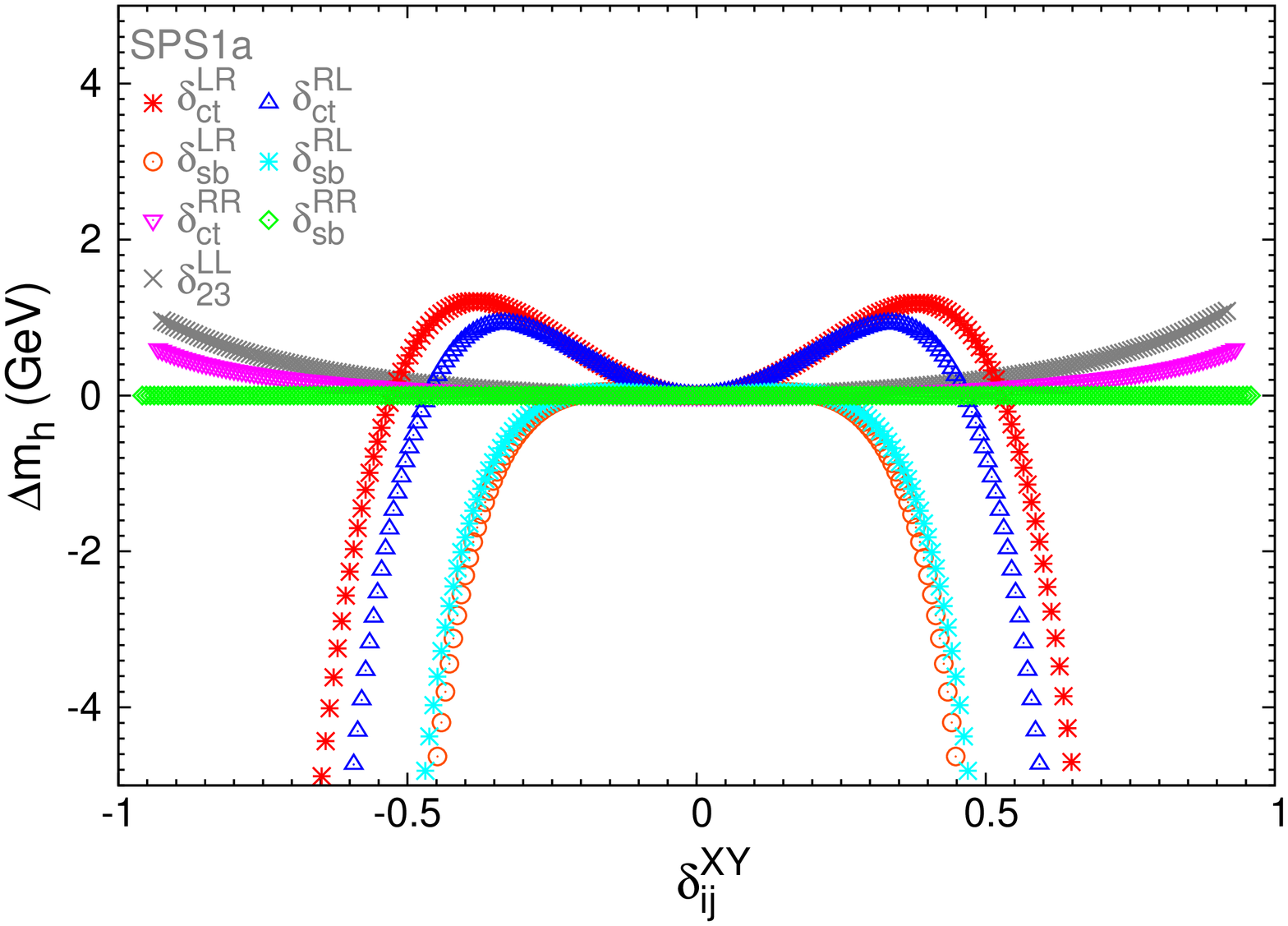}& 
\includegraphics[width=13.3cm,height=16.5cm,angle=270]{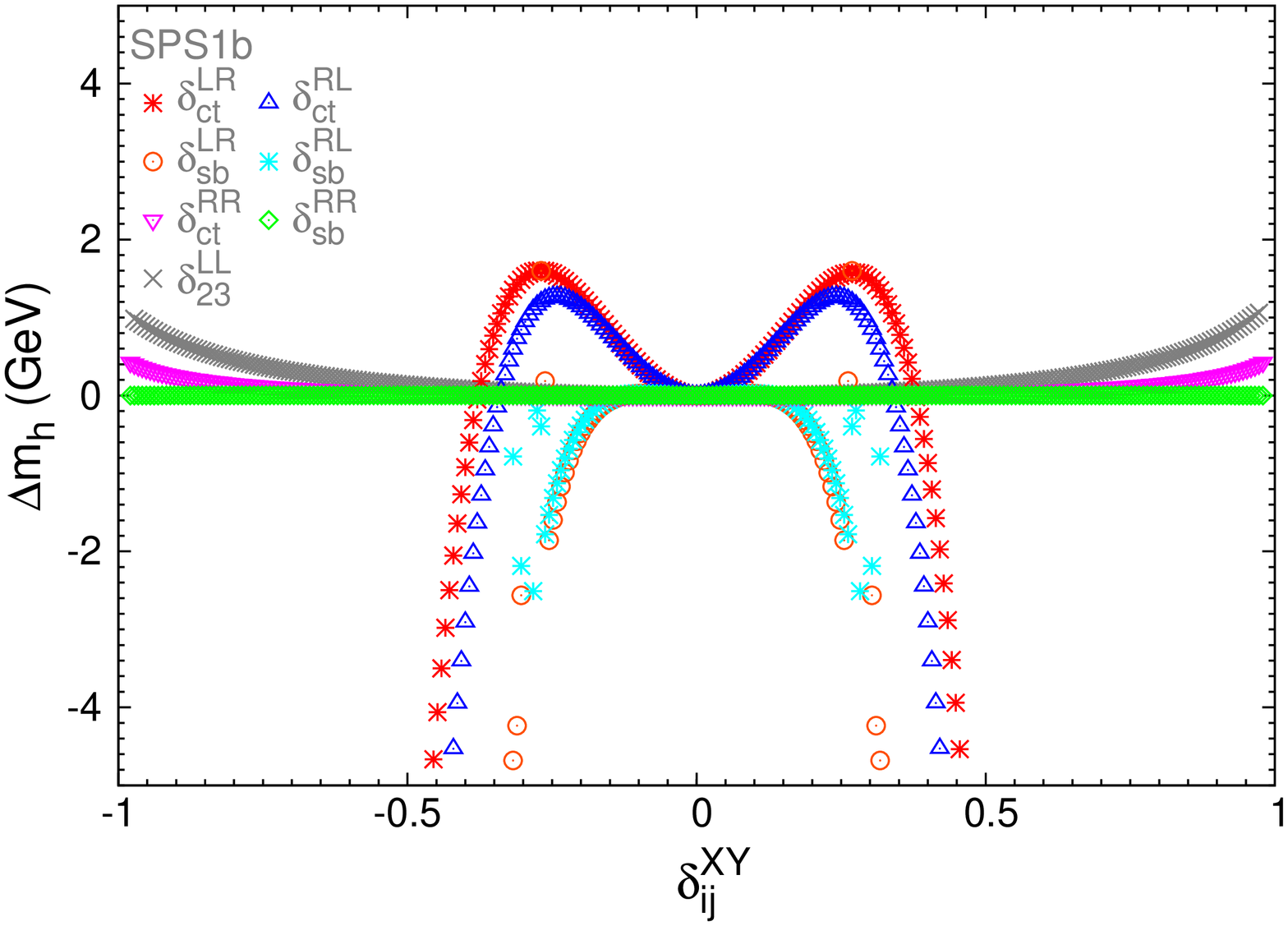}\\ 
\includegraphics[width=13.3cm,height=16.5cm,angle=270]{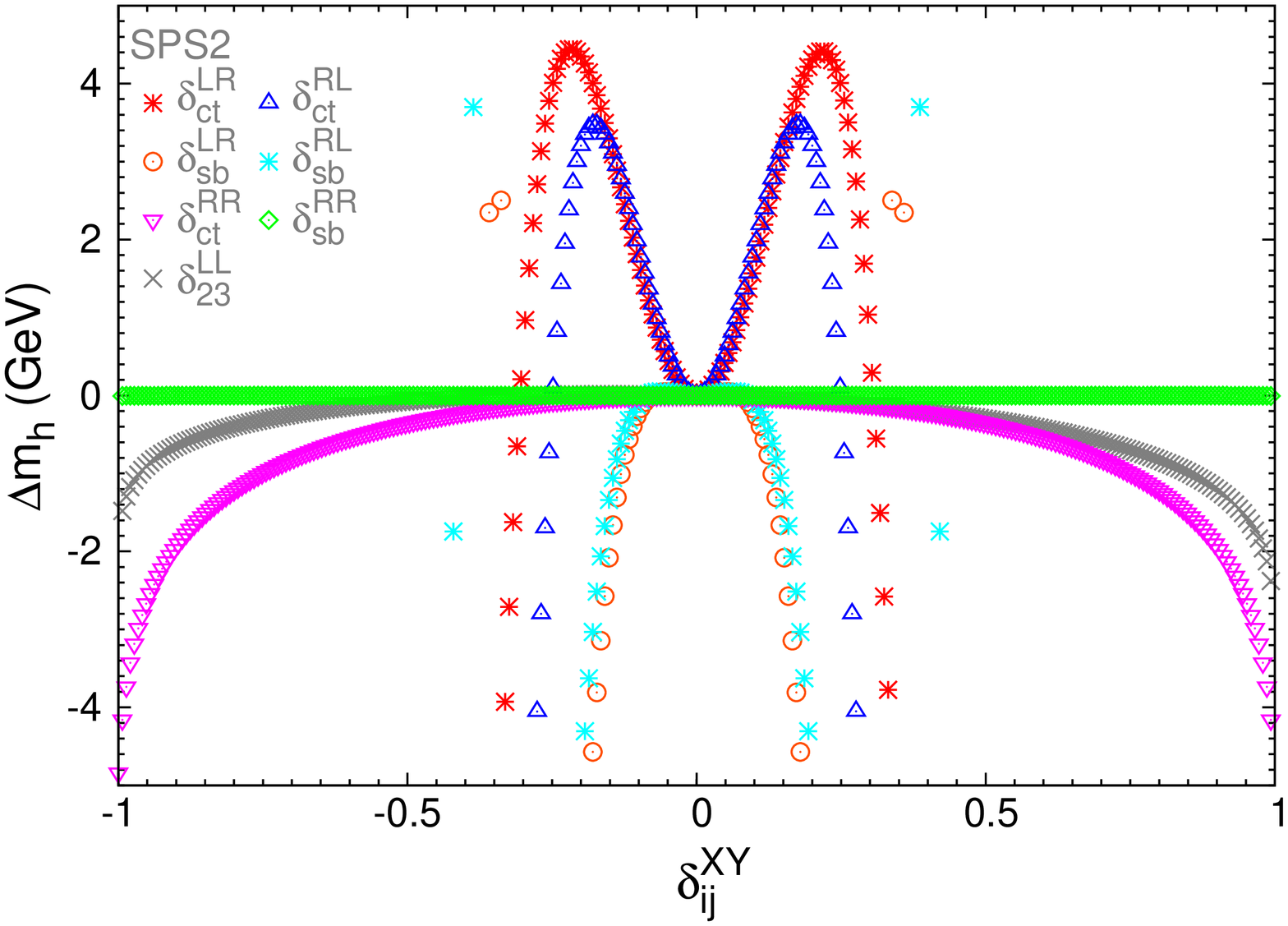}&
\includegraphics[width=13.3cm,height=16.5cm,angle=270]{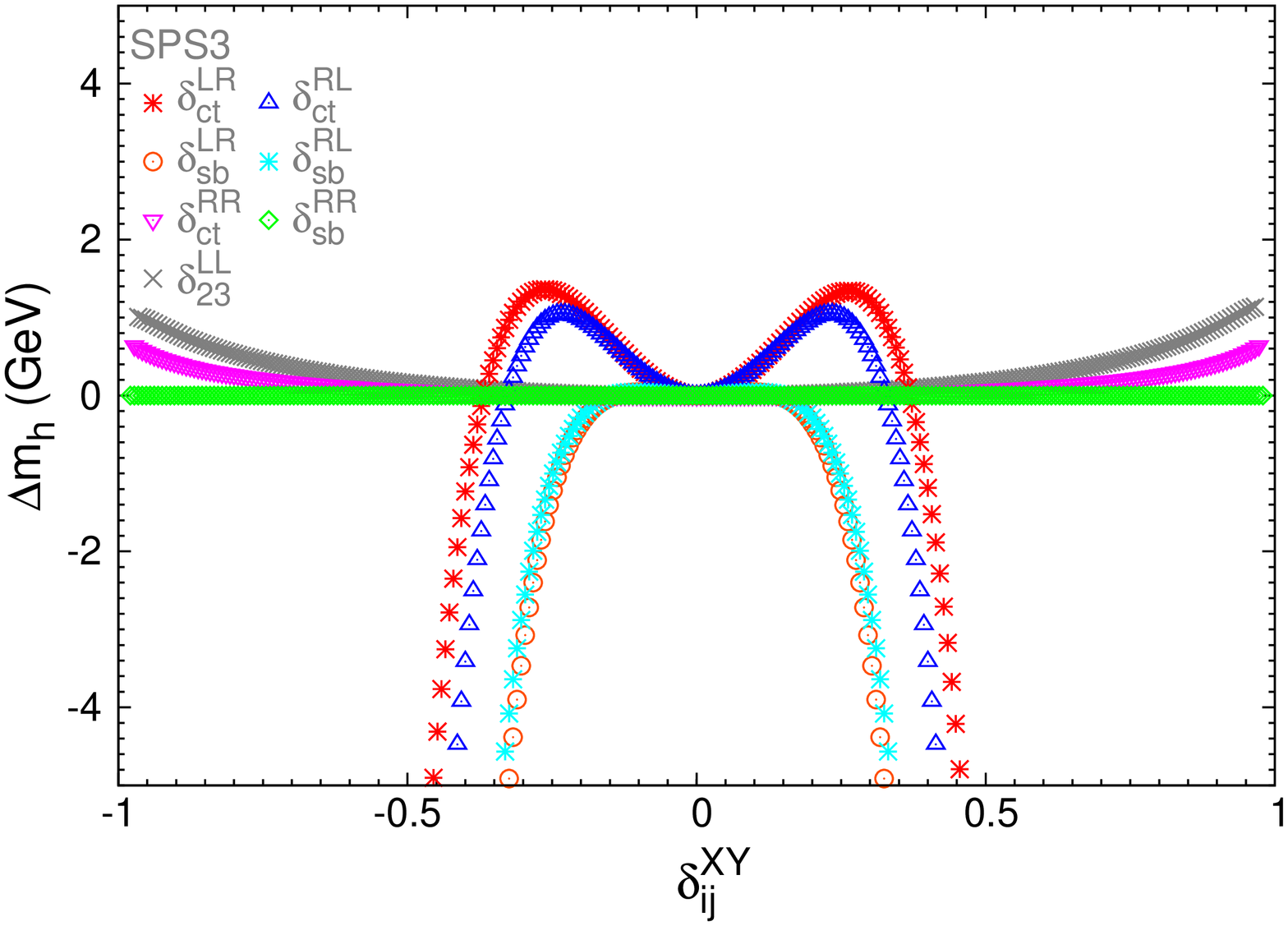}\\ 
\includegraphics[width=13.3cm,height=16.5cm,angle=270]{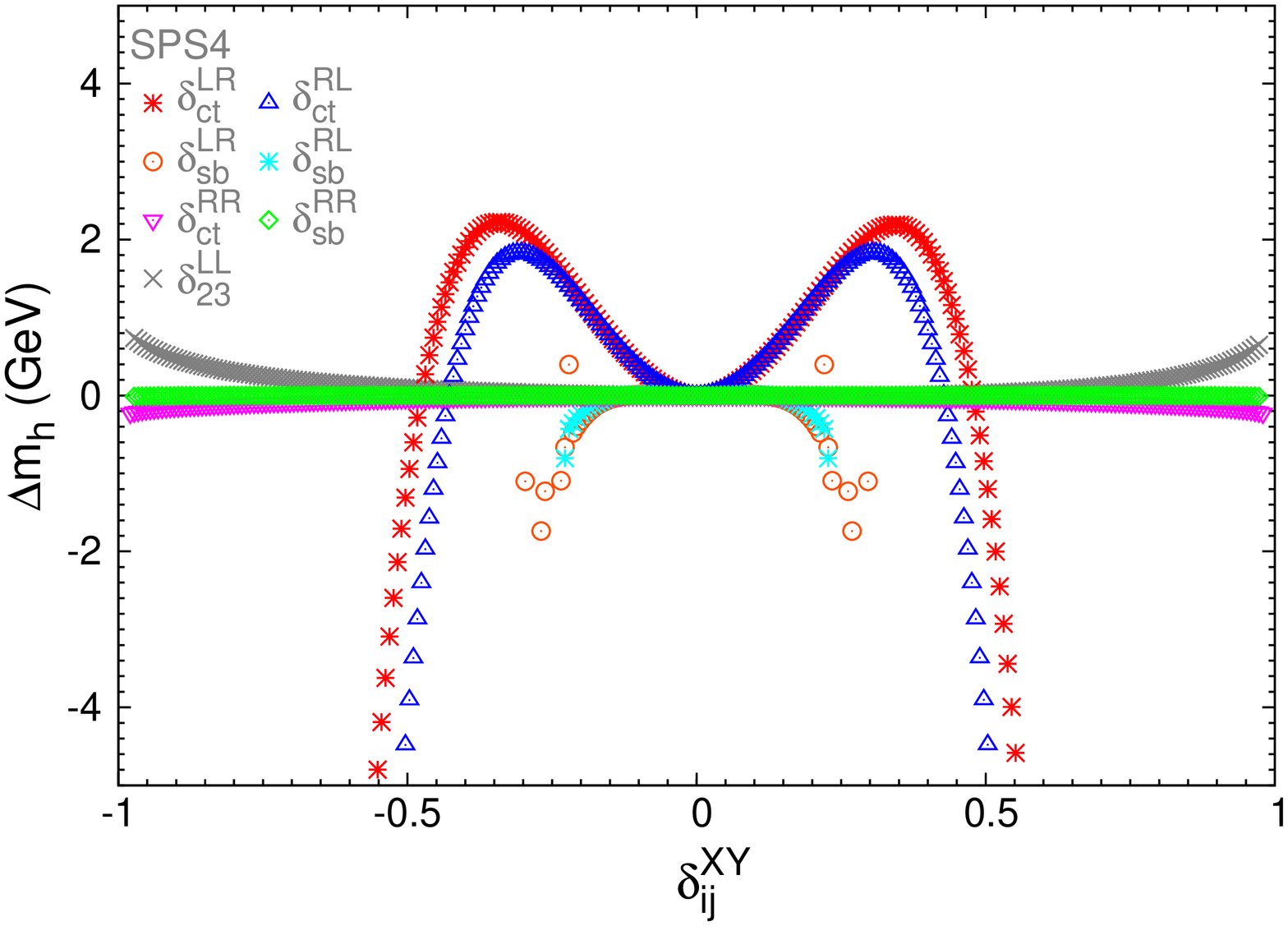}& 
\includegraphics[width=13.3cm,height=16.5cm,angle=270]{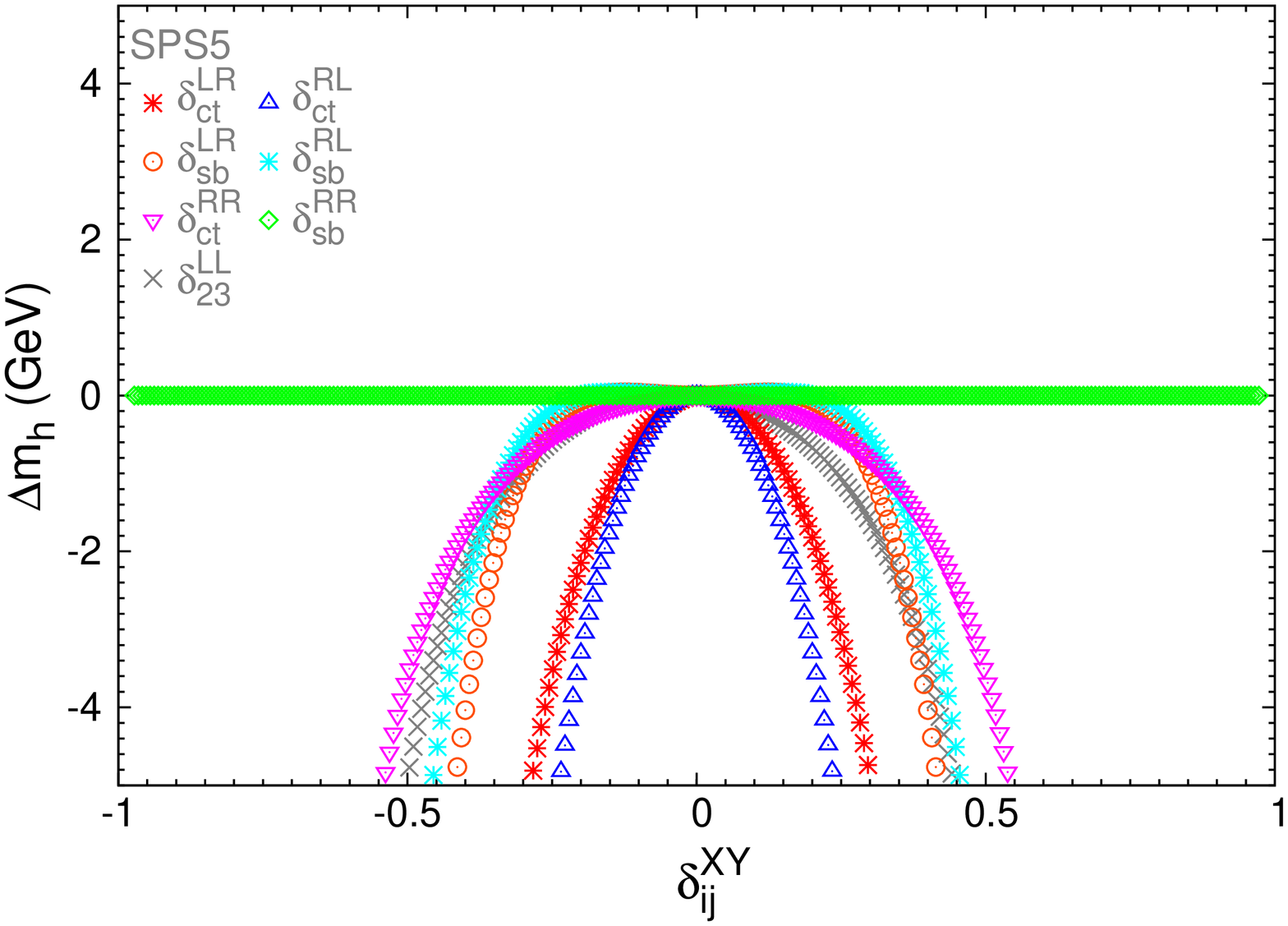}\\ 
\end{tabular}}}
\caption{Sensitivity to the NMFV deltas in $\Dmh$ for the SPSX points of table \ref{points}.}
 \label{figdeltamh0}
\end{figure}
%%%%%%%%%%%%%%%%%%%%%%%%%% F I G U R E %%%%%%%%%%%%%%%%%%%%%%%%%%%%%%%%%%%%%%%%
\clearpage
\newpage
%%%%%%%%%%%%%%%%%%%%%%%%%% F I G U R E %%%%%%%%%%%%%%%%%%%%%%%%%%%%%%%%%%%%%%%%
\begin{figure}[h!] 
\centering
\hspace*{-10mm} 
{\resizebox{17.9cm}{!} 
{\begin{tabular}{cc} 
\includegraphics[width=13.3cm,height=16.5cm,angle=270]{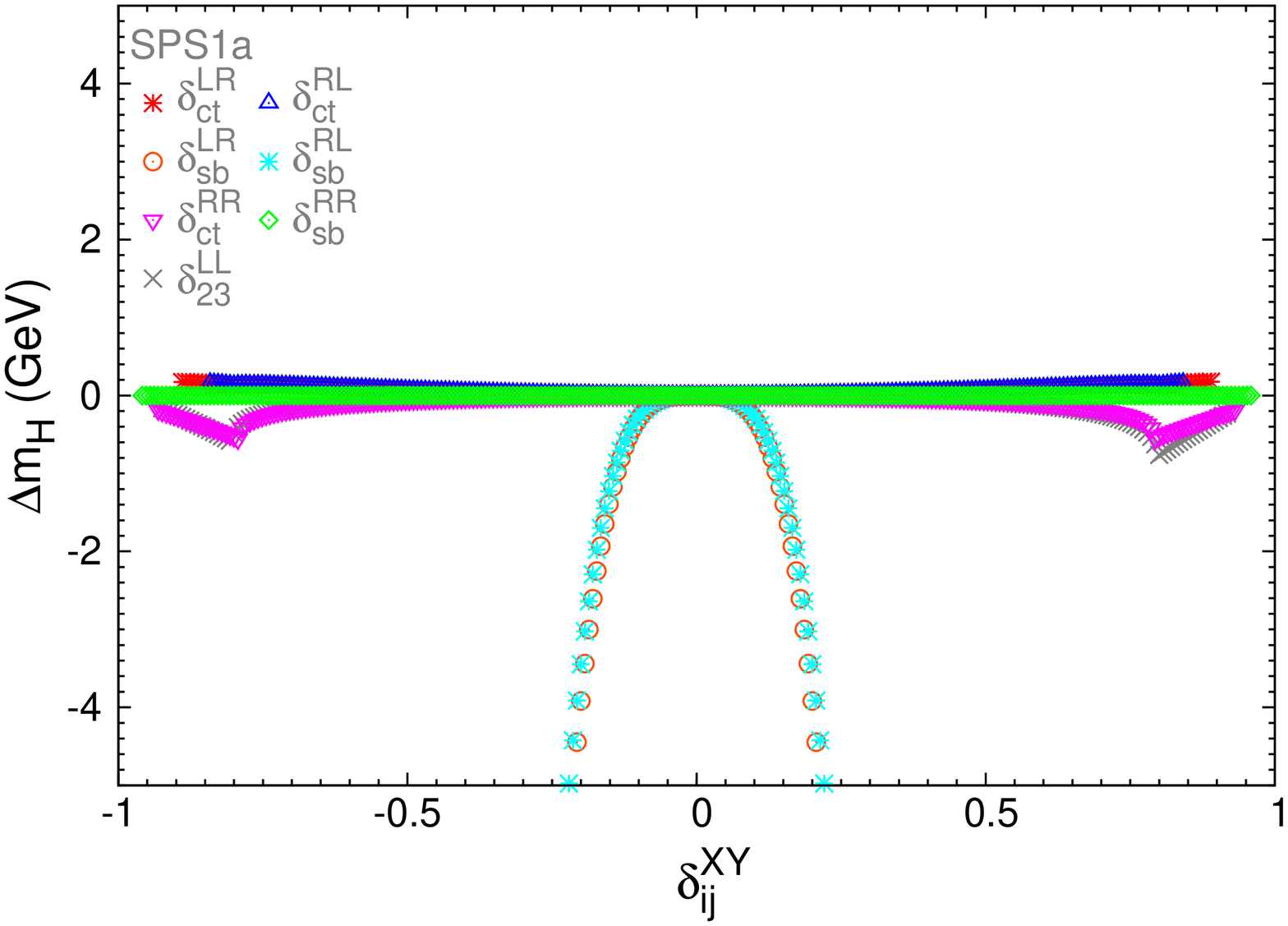}& 
\includegraphics[width=13.3cm,height=16.5cm,angle=270]{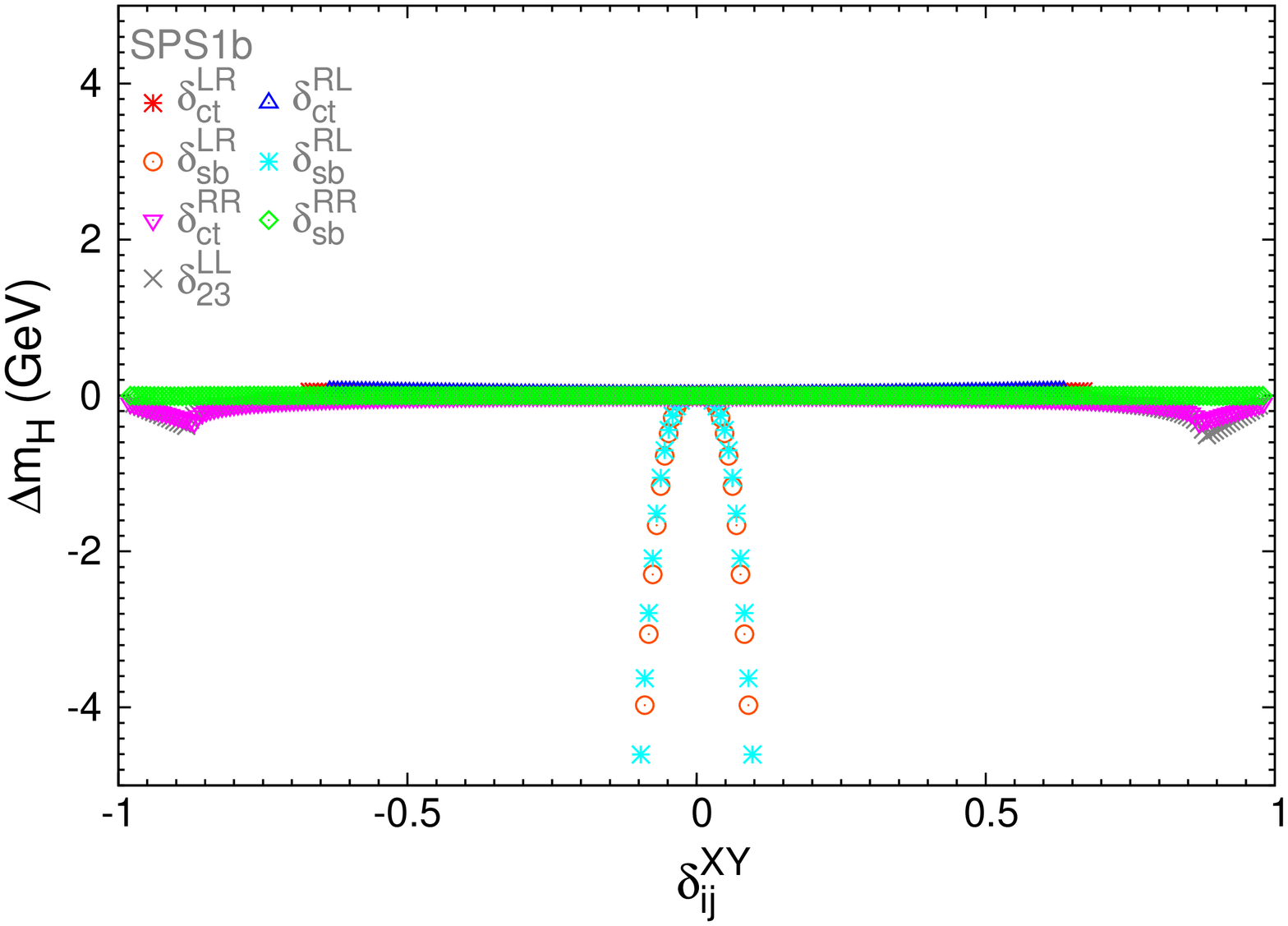}\\ 
\includegraphics[width=13.3cm,height=16.5cm,angle=270]{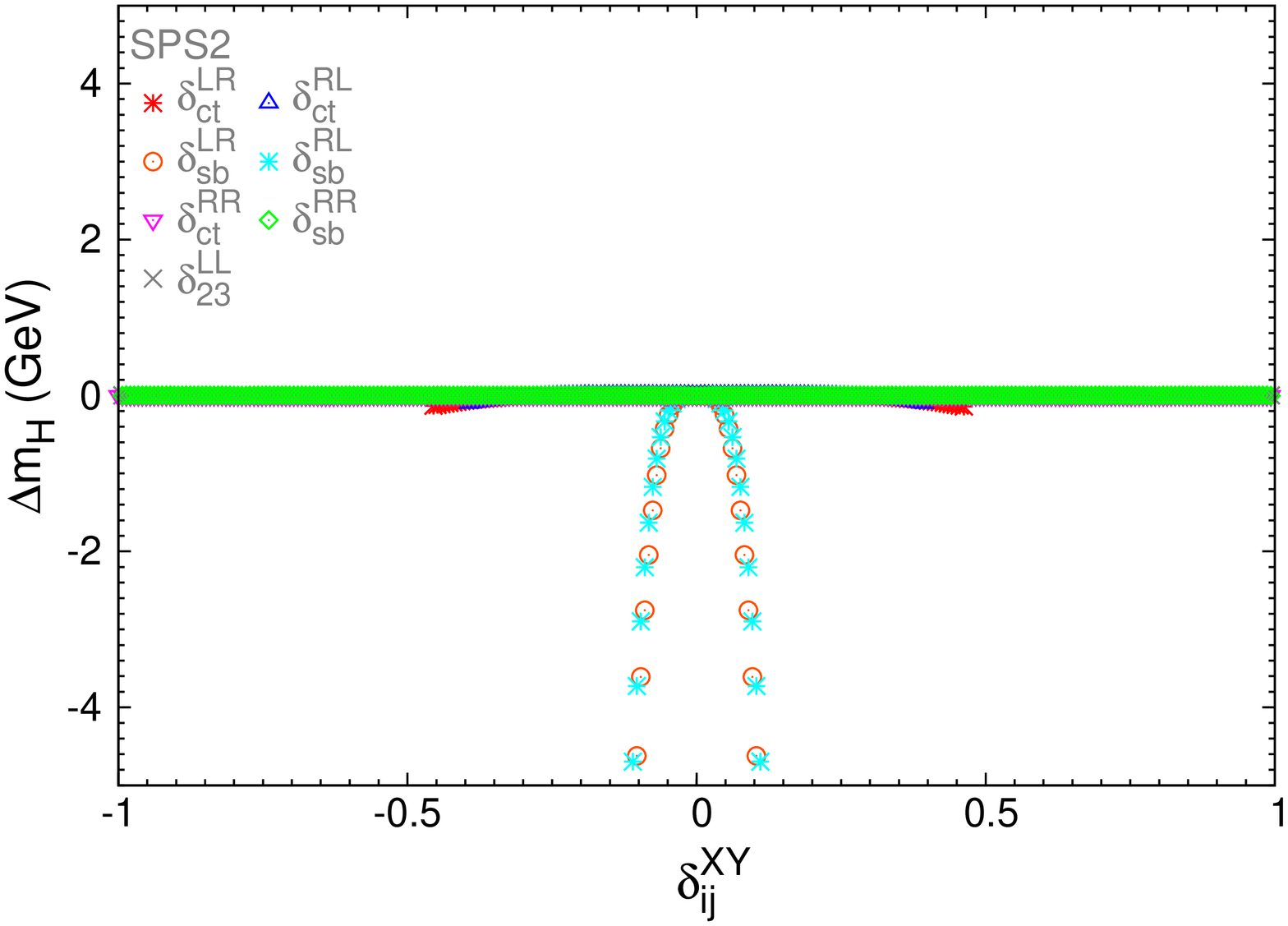}&
\includegraphics[width=13.3cm,height=16.5cm,angle=270]{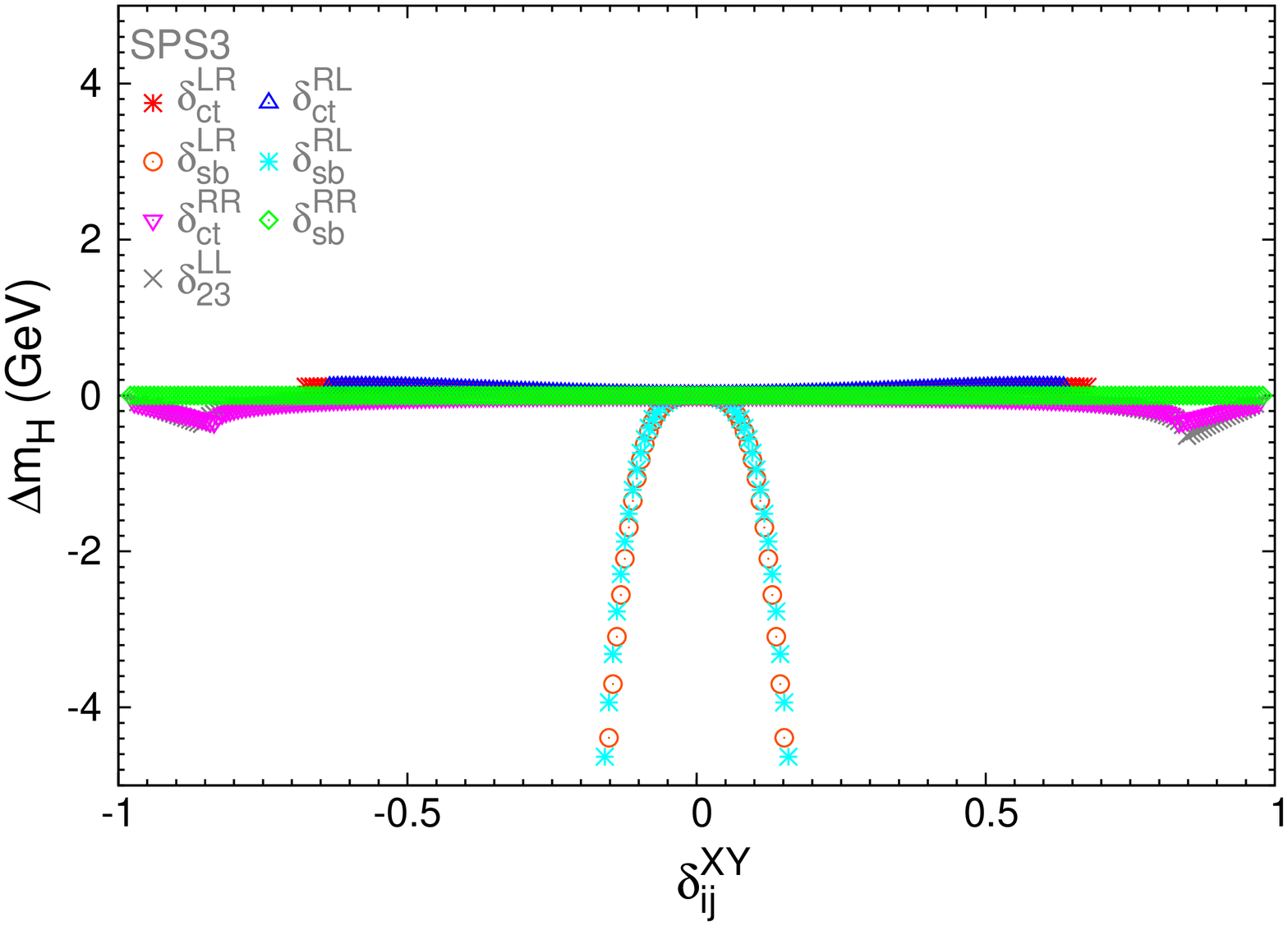}\\ 
\includegraphics[width=13.3cm,height=16.5cm,angle=270]{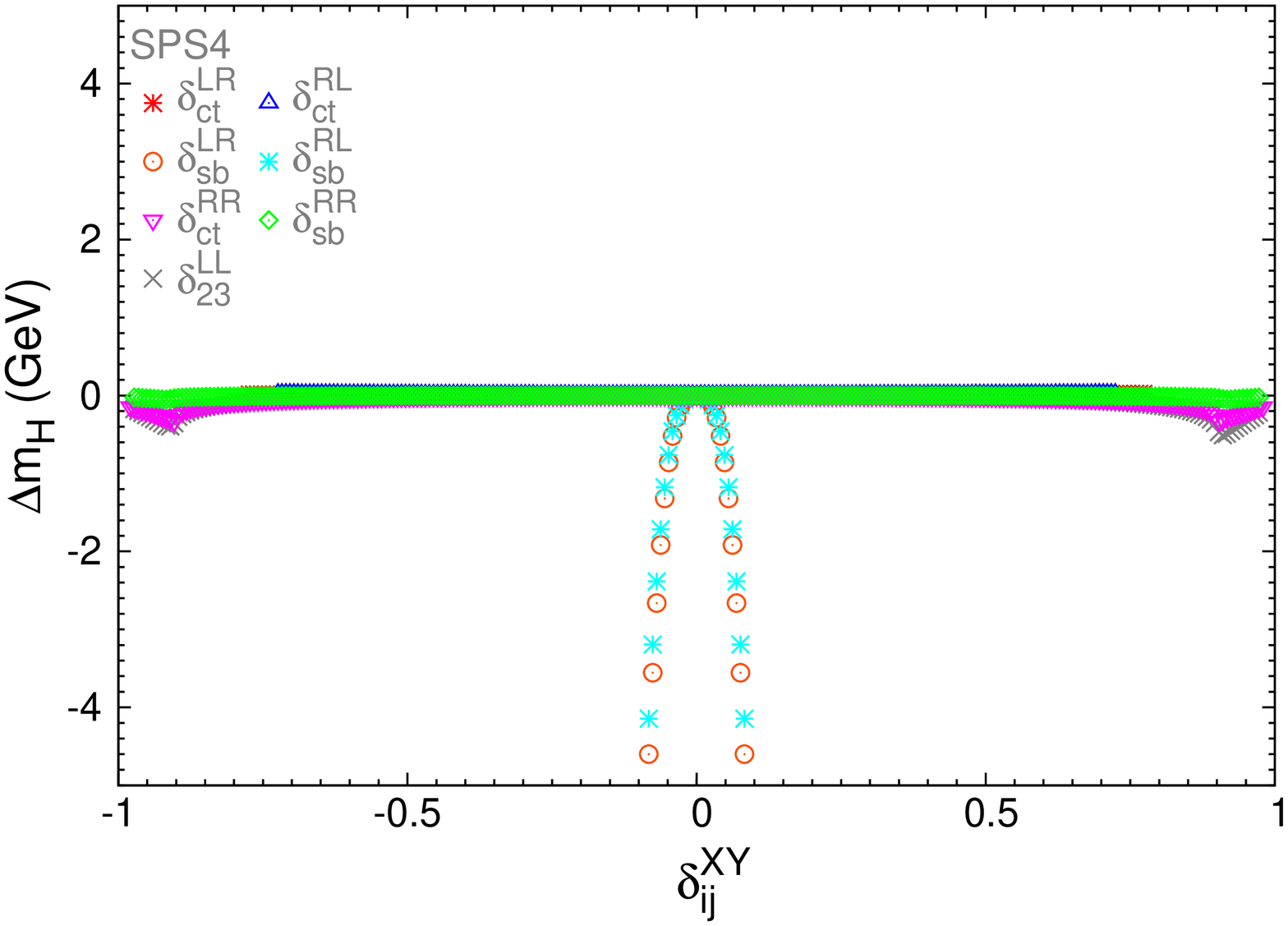}& 
\includegraphics[width=13.3cm,height=16.5cm,angle=270]{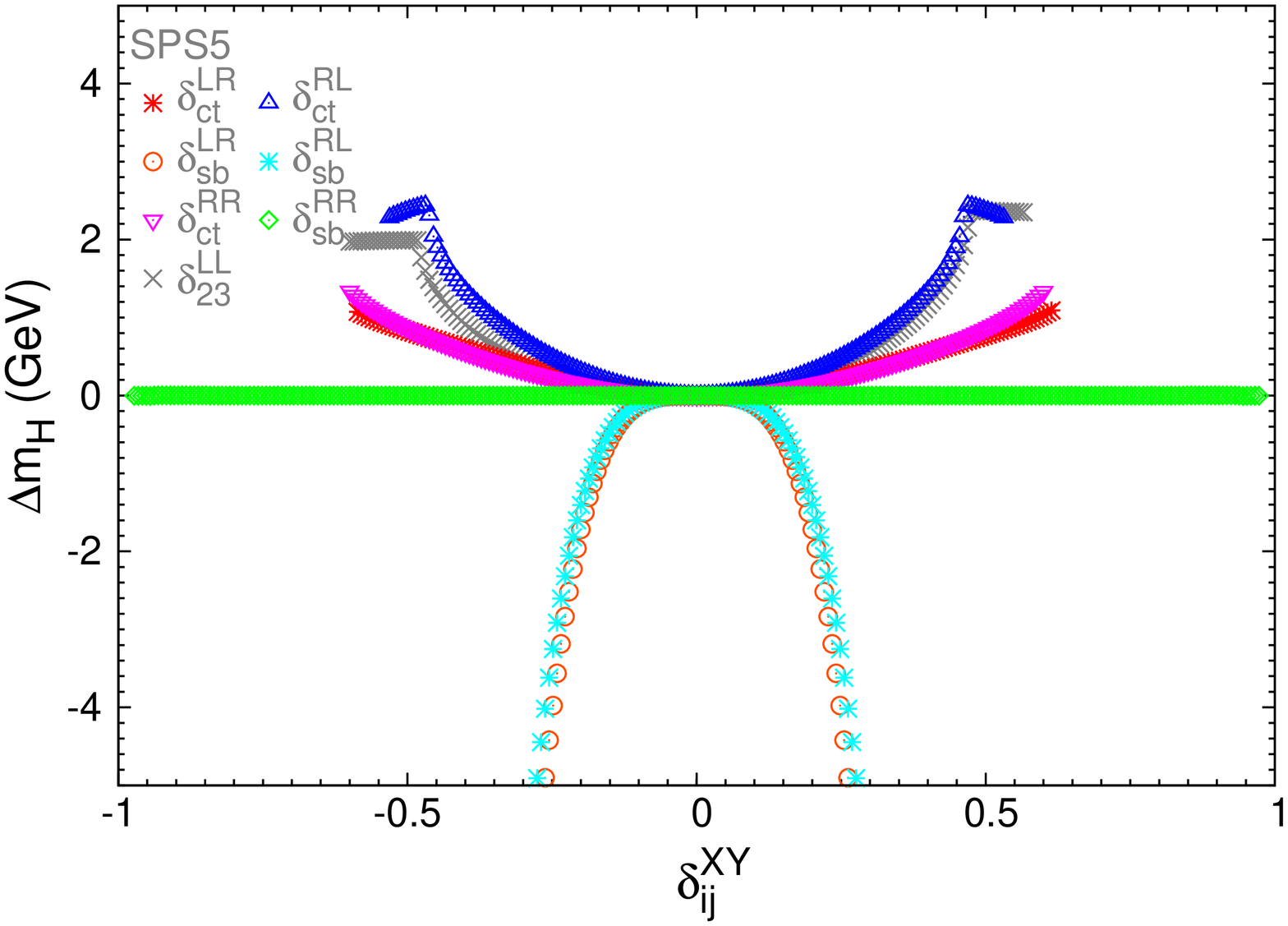}\\ 
\end{tabular}}}
\caption{Sensitivity to the NMFV deltas in $\DmH$ for the SPSX points of table \ref{points}.}
\label{figdeltamH0} 
\end{figure}
\clearpage
\newpage
%%%%%%%%%%%%%%%%%%%%%%%%%% F I G U R E %%%%%%%%%%%%%%%%%%%%%%%%%%%%%%%%%%%%%%%%
\begin{figure}[h!] 
\centering
\hspace*{-10mm} 
{\resizebox{17.9cm}{!} 
{\begin{tabular}{cc} 
\includegraphics[width=13.3cm,height=16.5cm,angle=270]{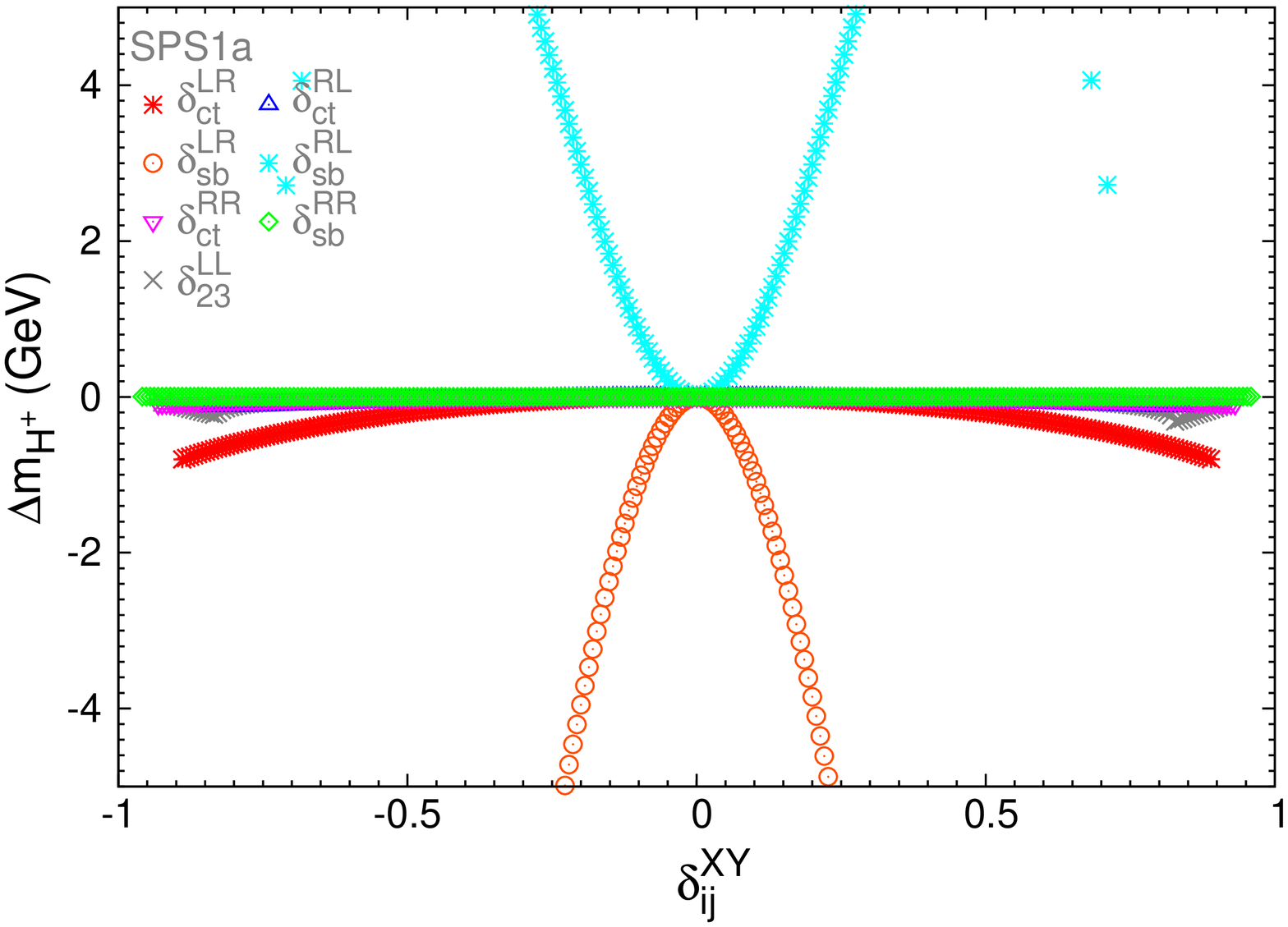}& 
\includegraphics[width=13.3cm,height=16.5cm,angle=270]{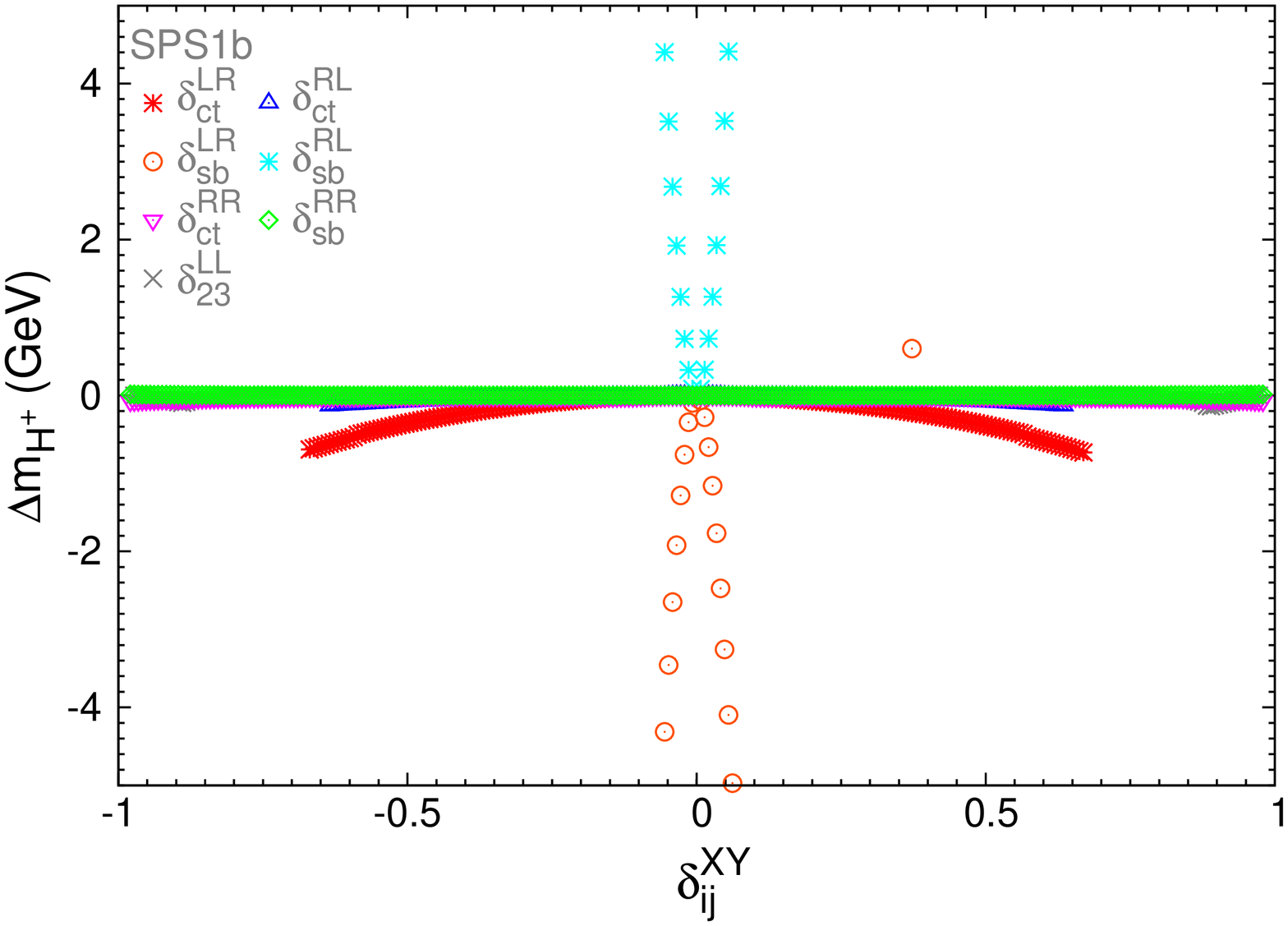}\\ 
\includegraphics[width=13.3cm,height=16.5cm,angle=270]{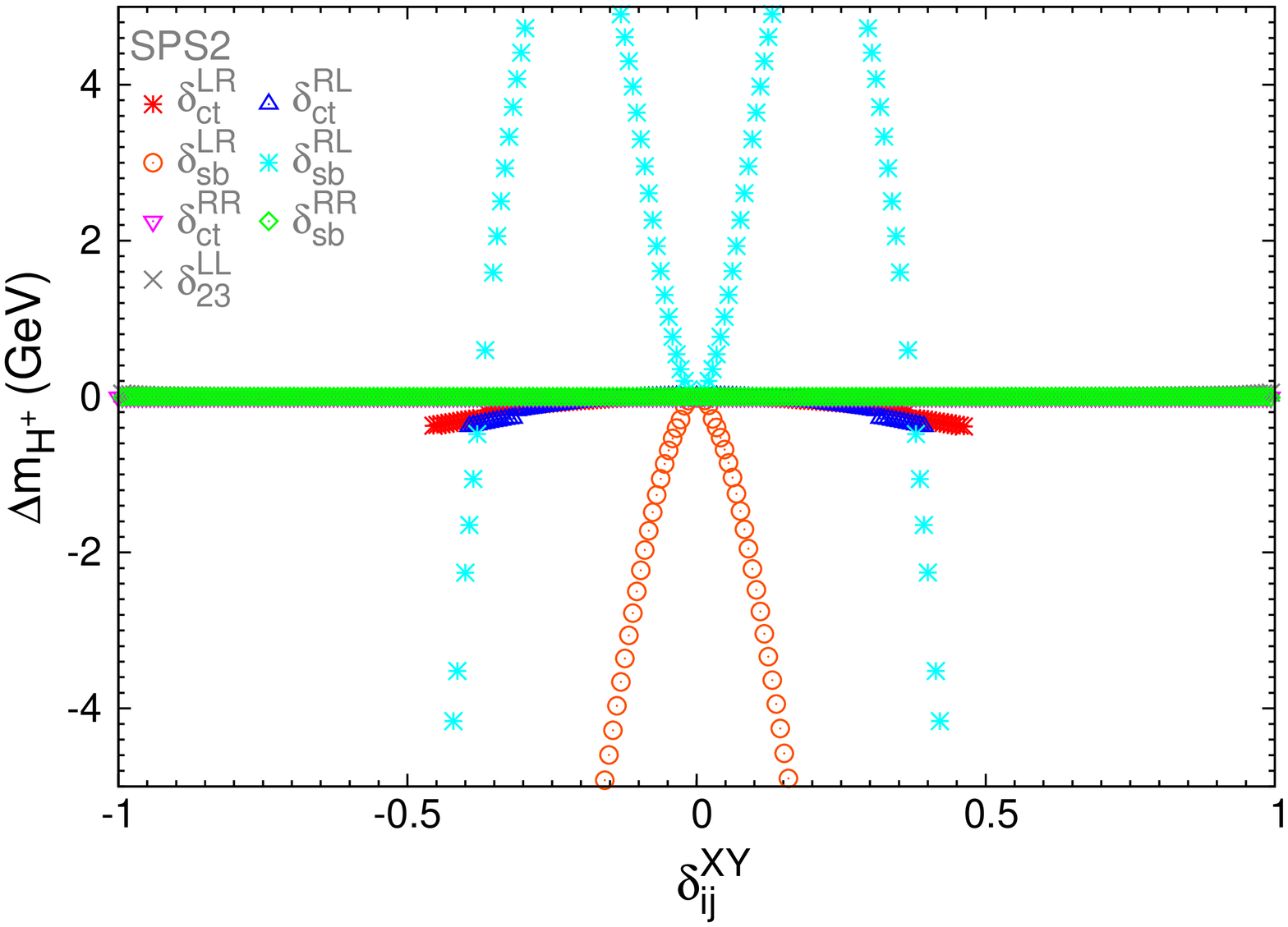}&
\includegraphics[width=13.3cm,height=16.5cm,angle=270]{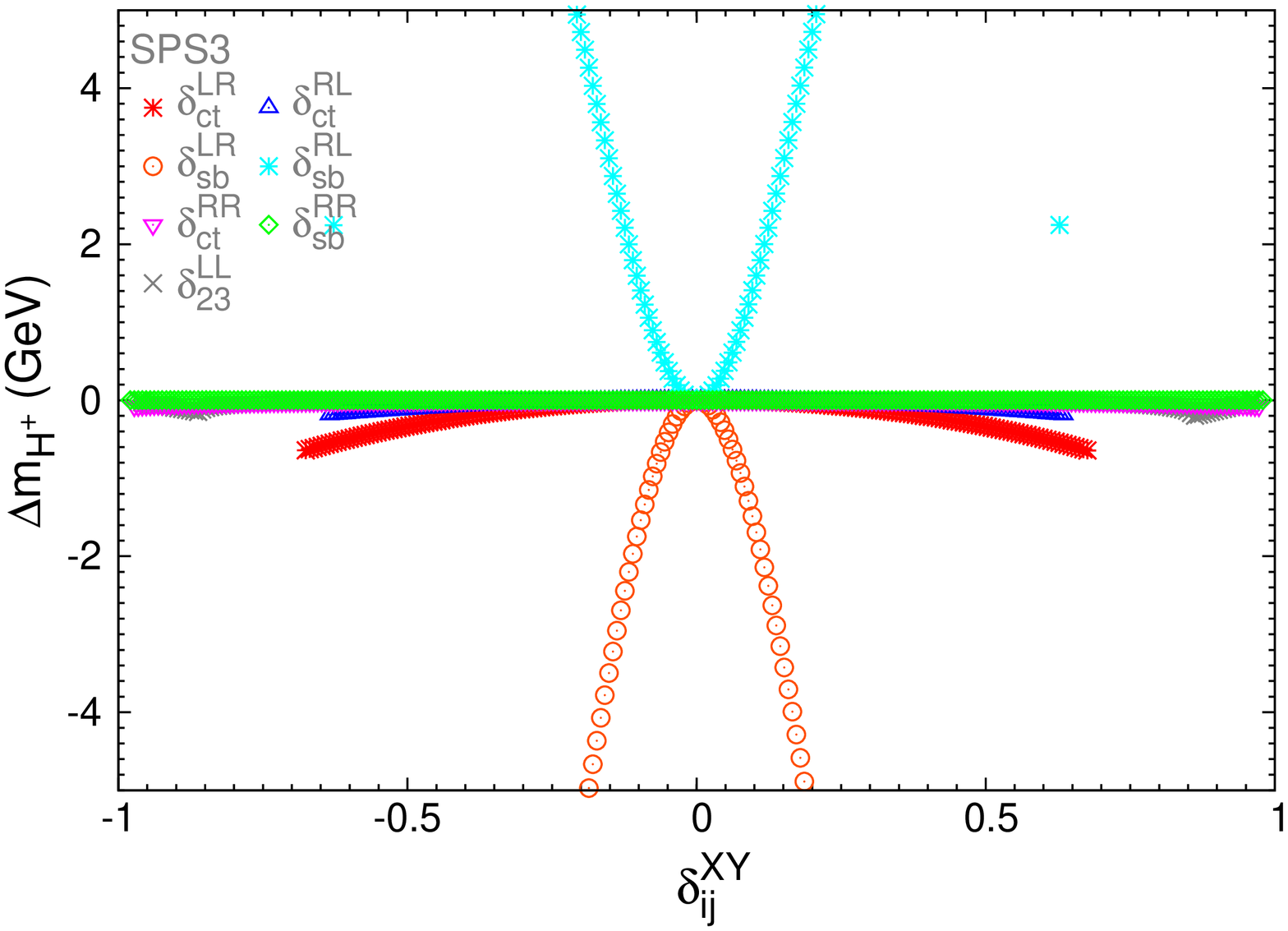}\\ 
\includegraphics[width=13.3cm,height=16.5cm,angle=270]{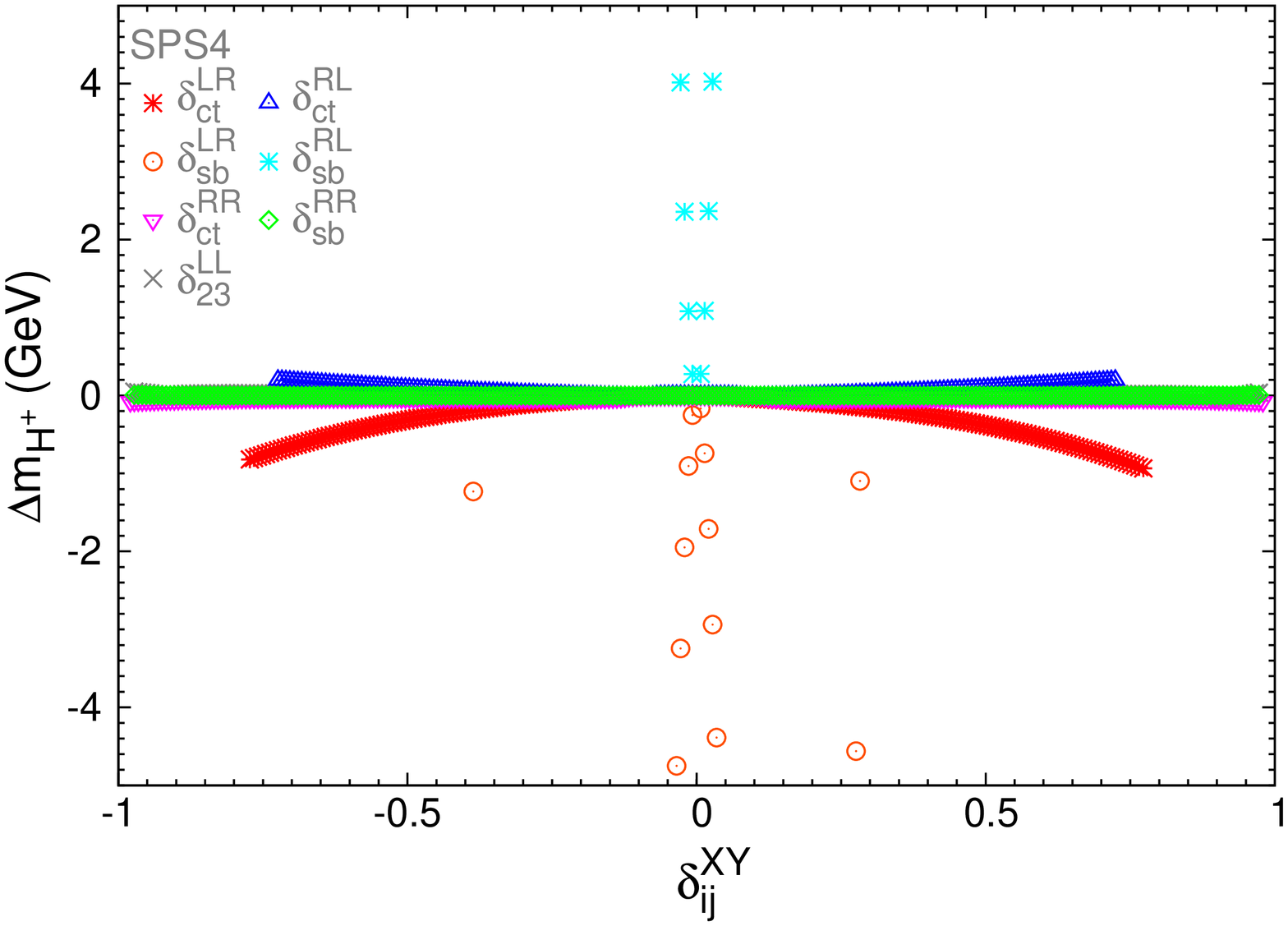}& 
\includegraphics[width=13.3cm,height=16.5cm,angle=270]{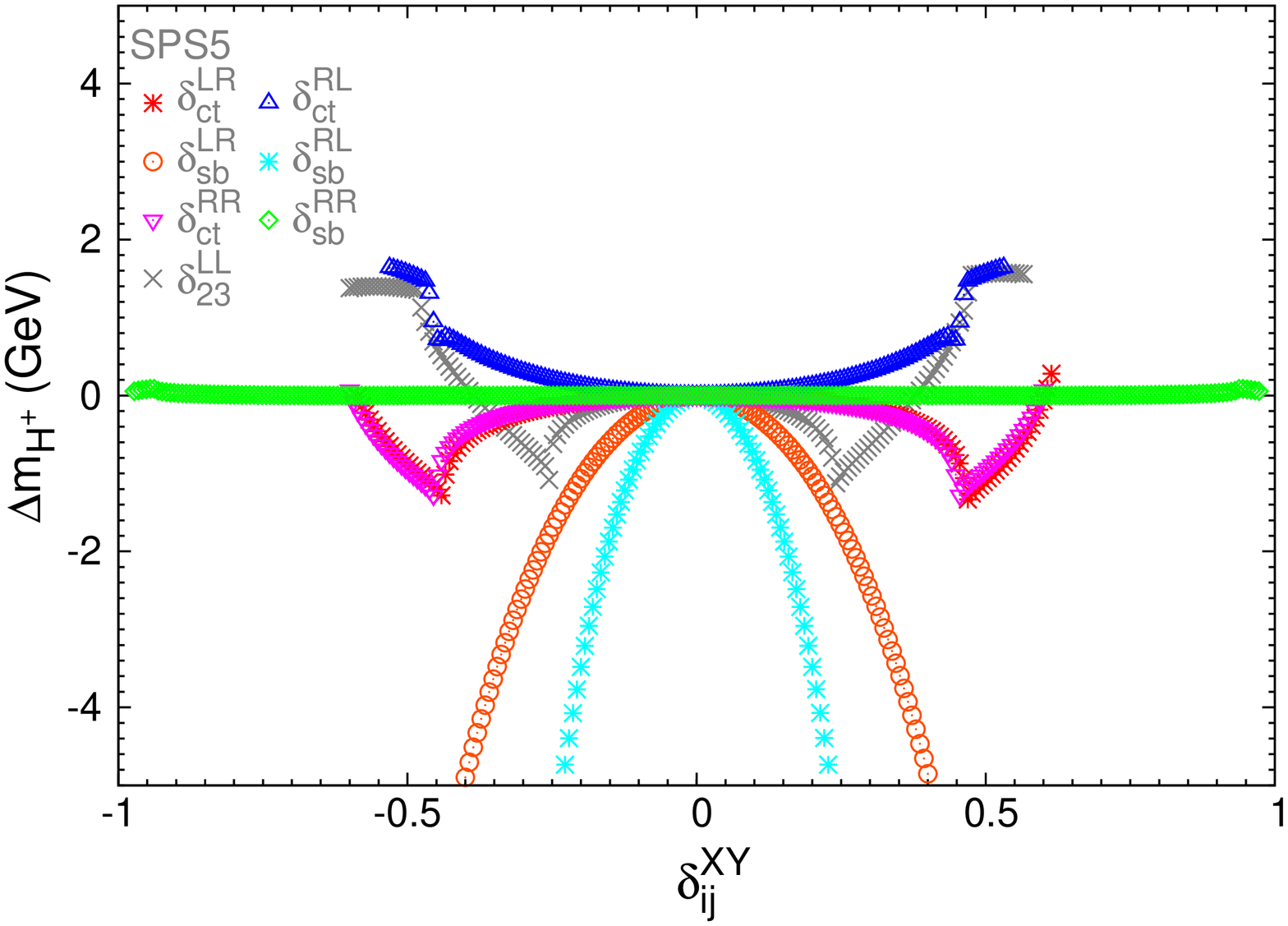}\\ 
\end{tabular}}}
\caption{Sensitivity to the NMFV deltas in $\DmHp$ for the SPSX points of table \ref{points}.}
 \label{figdeltamHp}
\end{figure}
\clearpage
\newpage
%%%%%%%%%%%%%%%%%%%%%%%%%% F I G U R E %%%%%%%%%%%%%%%%%%%%%%%%%%%%%%%%%%%%%%%%

\subsection{\boldmath{$\Dmh$} versus two \boldmath{$\deXYij\neq 0$}}

 Our previous results on the Higgs mass corrections show that the corrections to the lightest Higgs mass $\Dmh$ are negative in many of the studied cases and can be very large for some regions of the flavor changing deltas which are still allowed by present $B$ data. These negative and large
 mass corrections, can lead to a prediction for the corrected one-loop mass in these kind of NMFV-SUSY scenarios, $m_{h}^{\rm NMFV} \simeq m_{h}^{\rm MSSM} + \Dmh$, which are indeed too low and already excluded by present data~\cite{LEPHiggsSM,LEPHiggsMSSM}. Therefore, interestingly, the study of these mass corrections can be conclusive in the setting of additional restrictions on the size of some flavor changing deltas which otherwise are not bounded from present $B$ data.

In order to explore further the size of these 'dangerous' mass corrections, we have computed numerically the size of  $\Dmh$ as a function of two non-vanishing deltas and have looked for areas in these two dimensional plots that are allowed by $B$ data. We show in \reffis{figdoubledeltaBFP}, \ref{figdoubledeltaSPS3},  \ref{figdoubledeltaSPS2},  \ref{figdoubledeltaSPS5},  \ref{figdoubledeltaVHeavyS}, and \ref{figdoubledeltaHeavySLightH} the numerical results of the $\Dmh$ contour-lines in two dimensional plots, 
$(\de^{LL}_{23},\deXYij)$, for the respective points BFP, SPS2, SPS3, SPS5, VHeavyS and HeavySLightH of table \ref{points}.

%%%%%%%%%%%%%%%%%%%%%%%%%% F I G U R E %%%%%%%%%%%%%%%%%%%%%%%%%%%%%%%%%%%%%%%%
\begin{figure}[H]  
\centering 
%\hspace*{-10mm}  
{\resizebox{13cm}{!}  
{\begin{tabular}{cccc}  
\includegraphics[width=13.3cm]{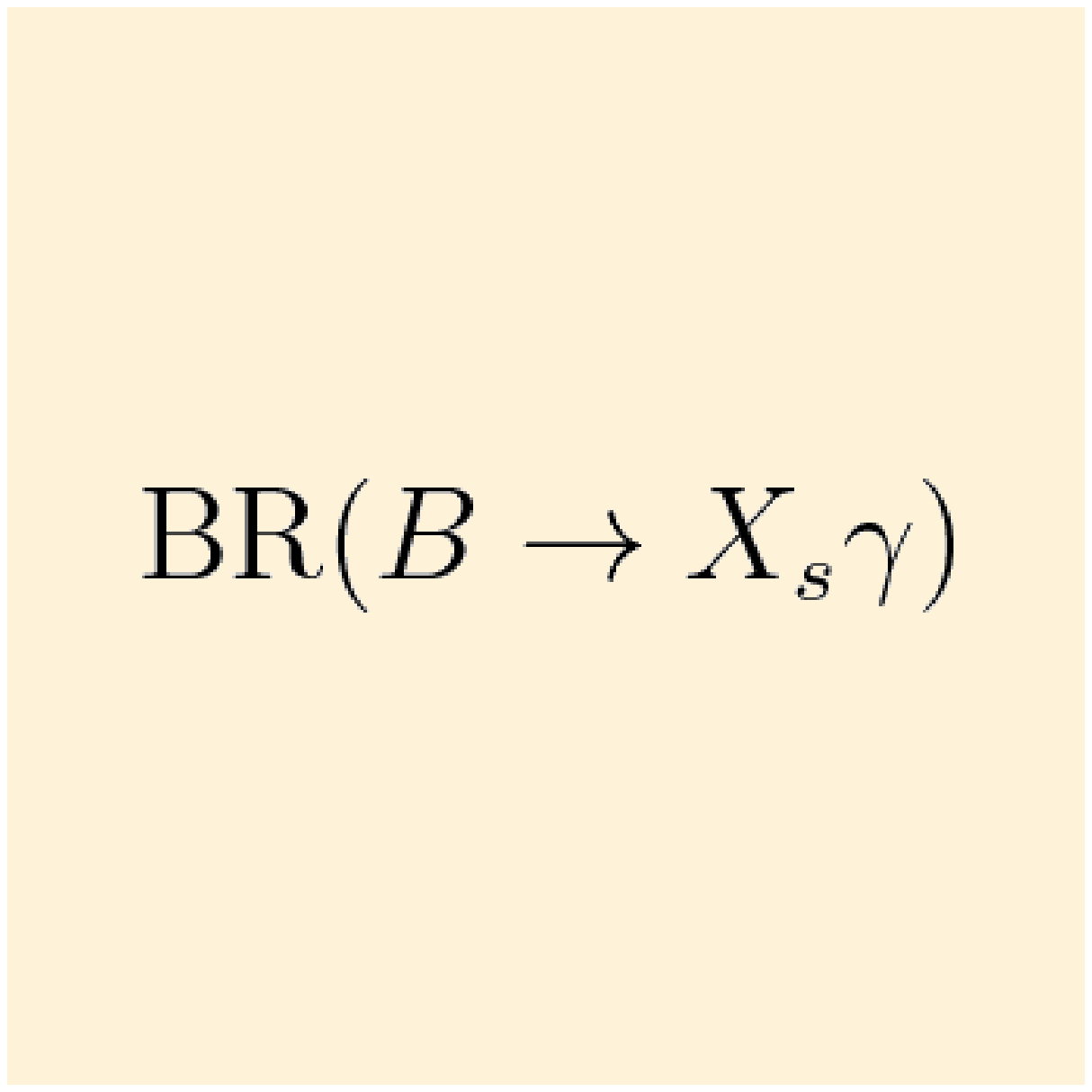}&  
\includegraphics[width=13.3cm]{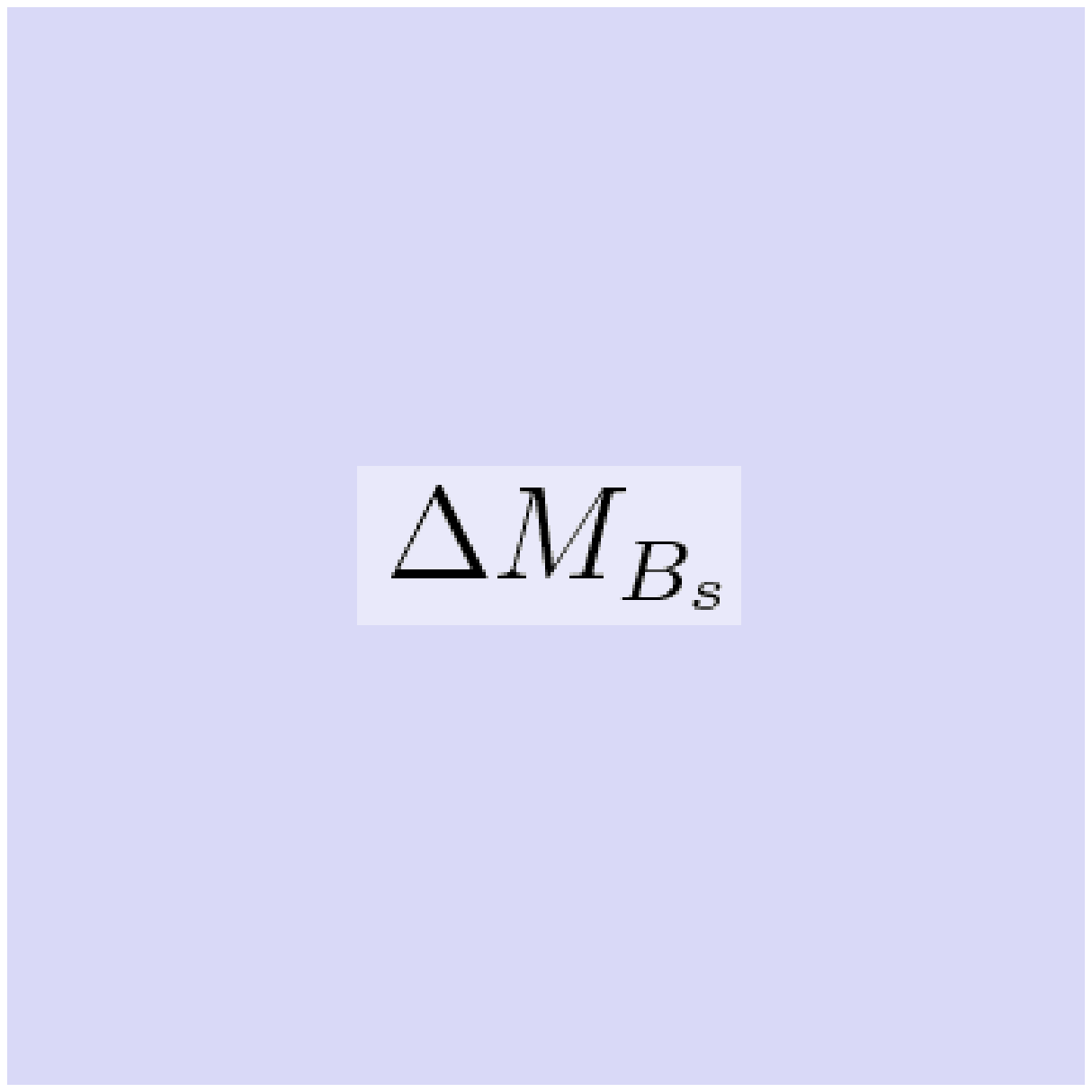}&  
\includegraphics[width=13.3cm]{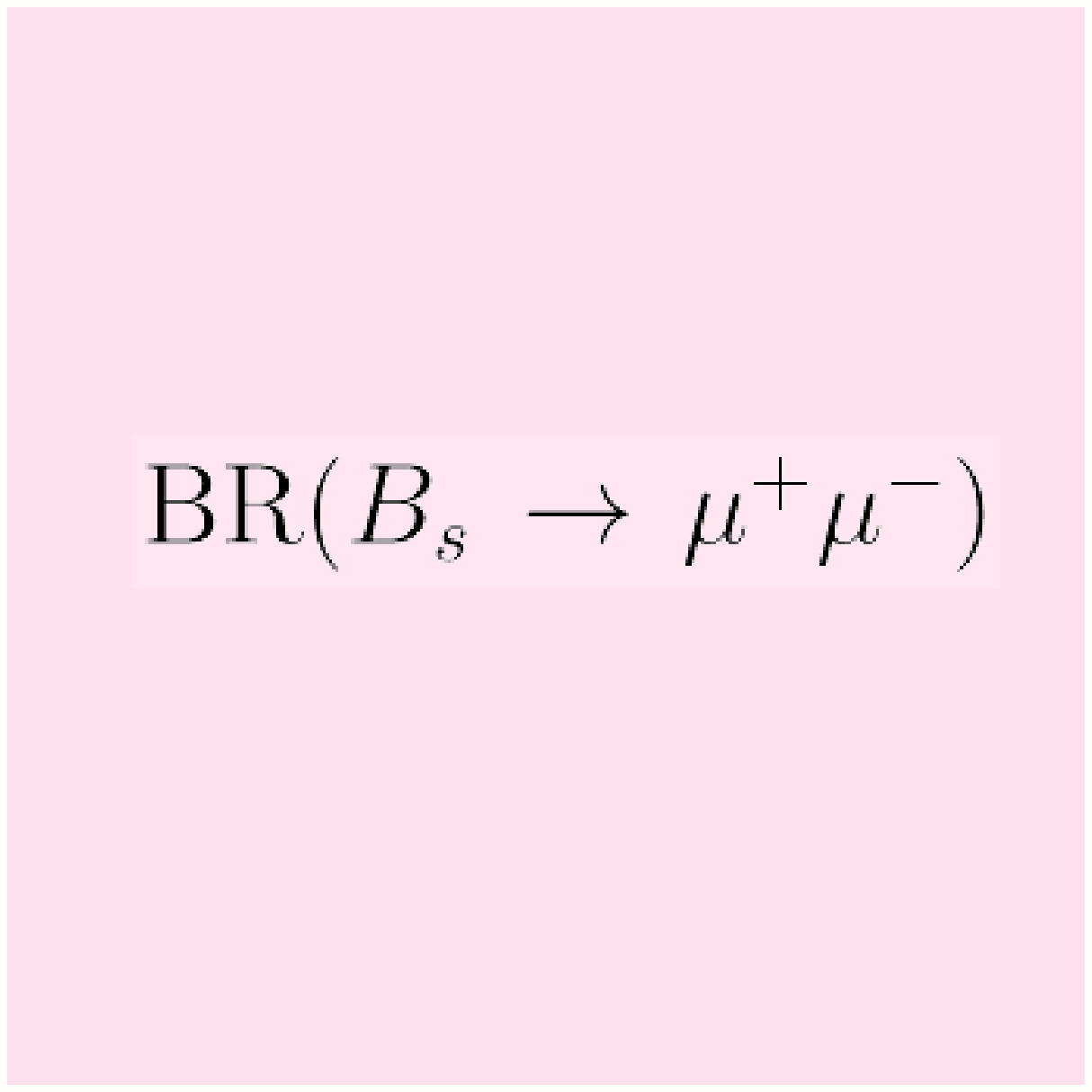}& 
\includegraphics[width=13.3cm]{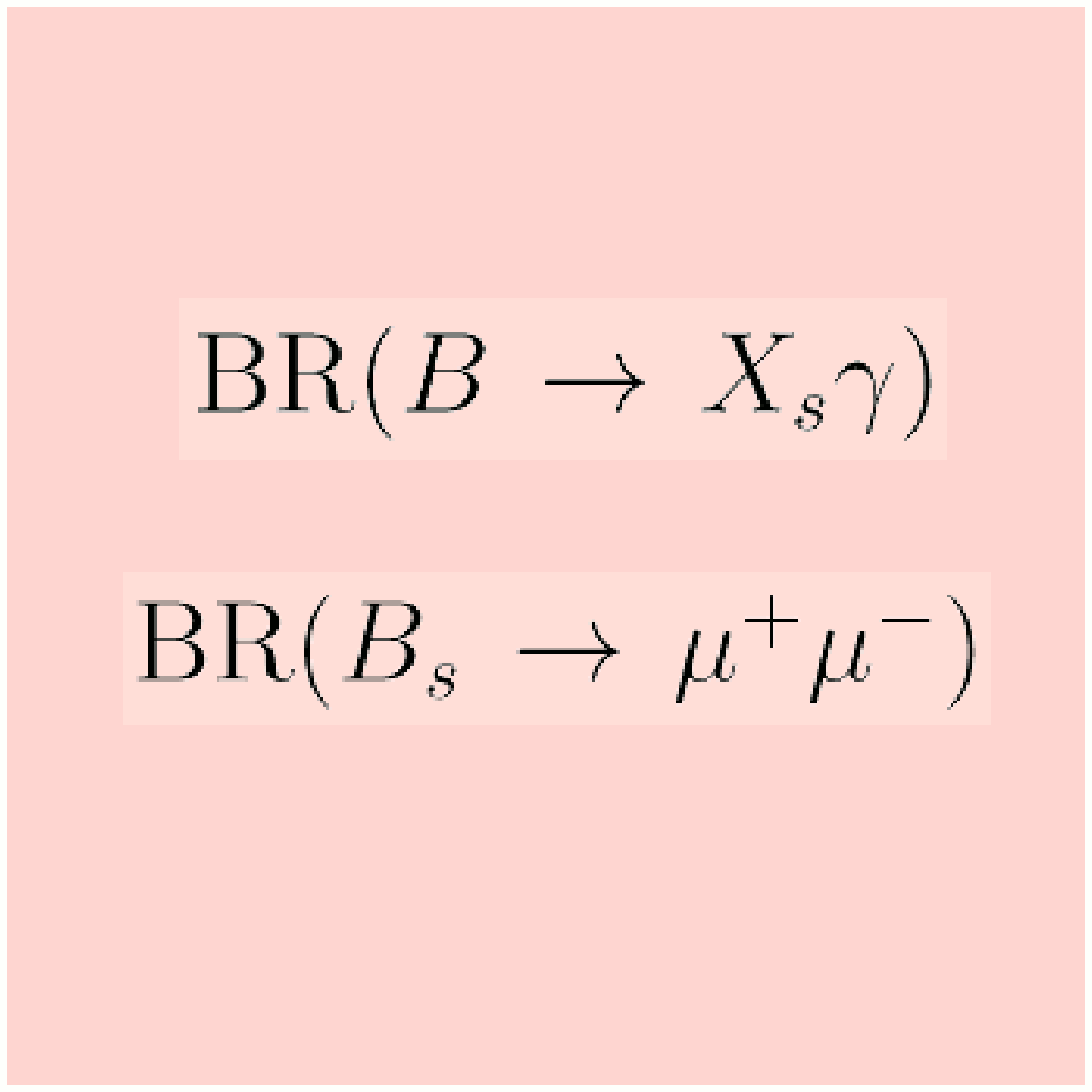}\\
\includegraphics[width=13.3cm]{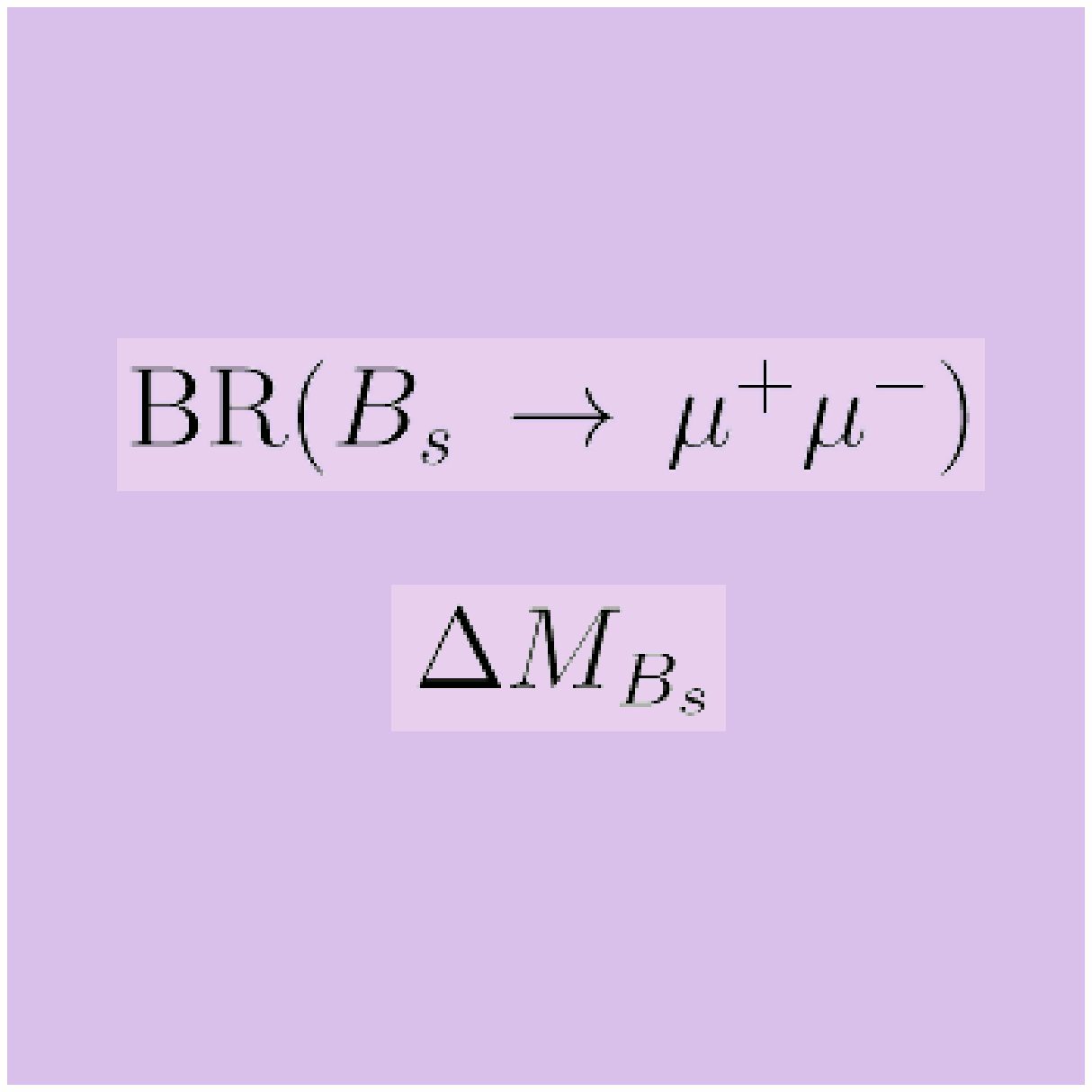}&  
\includegraphics[width=13.3cm]{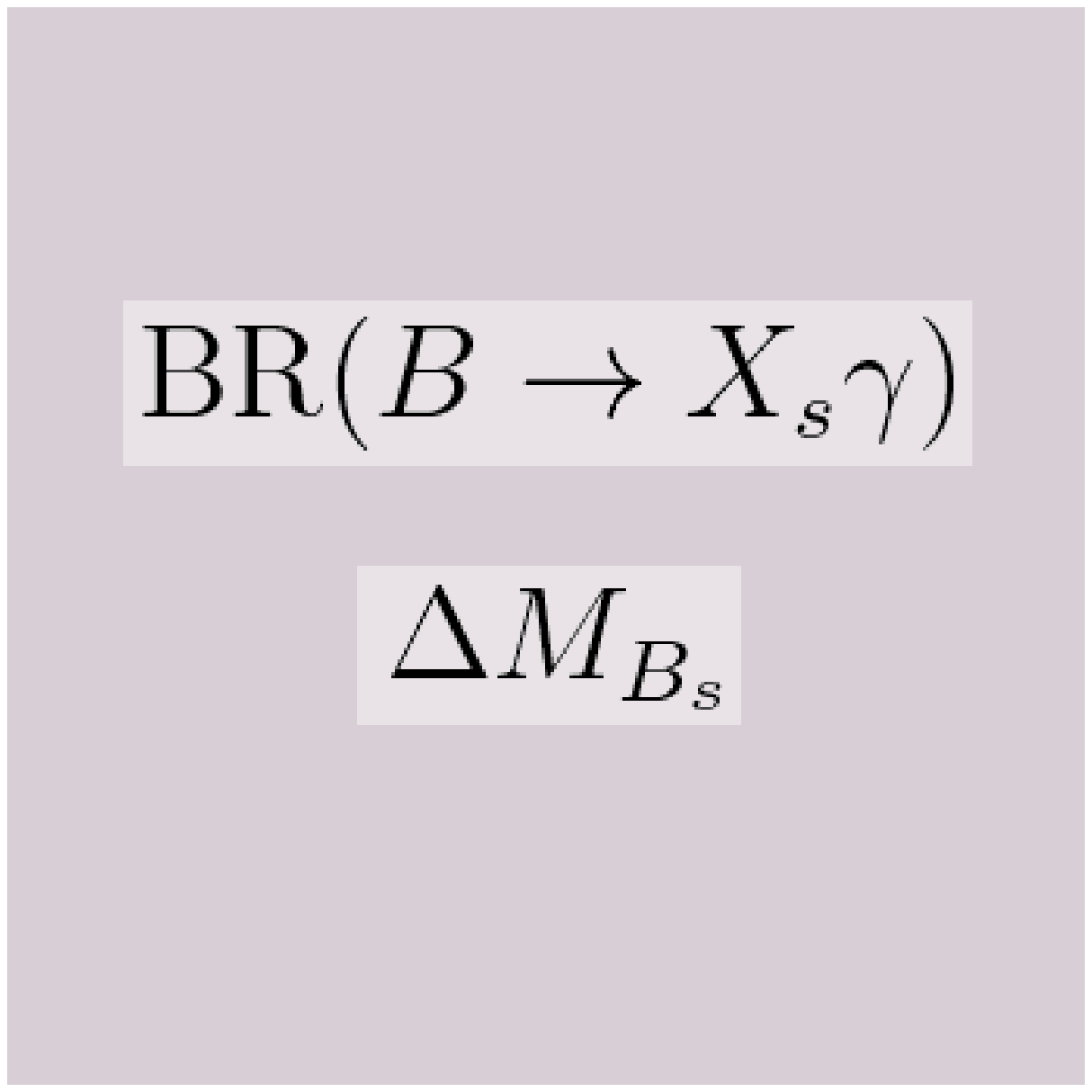}& 
\includegraphics[width=13.3cm]{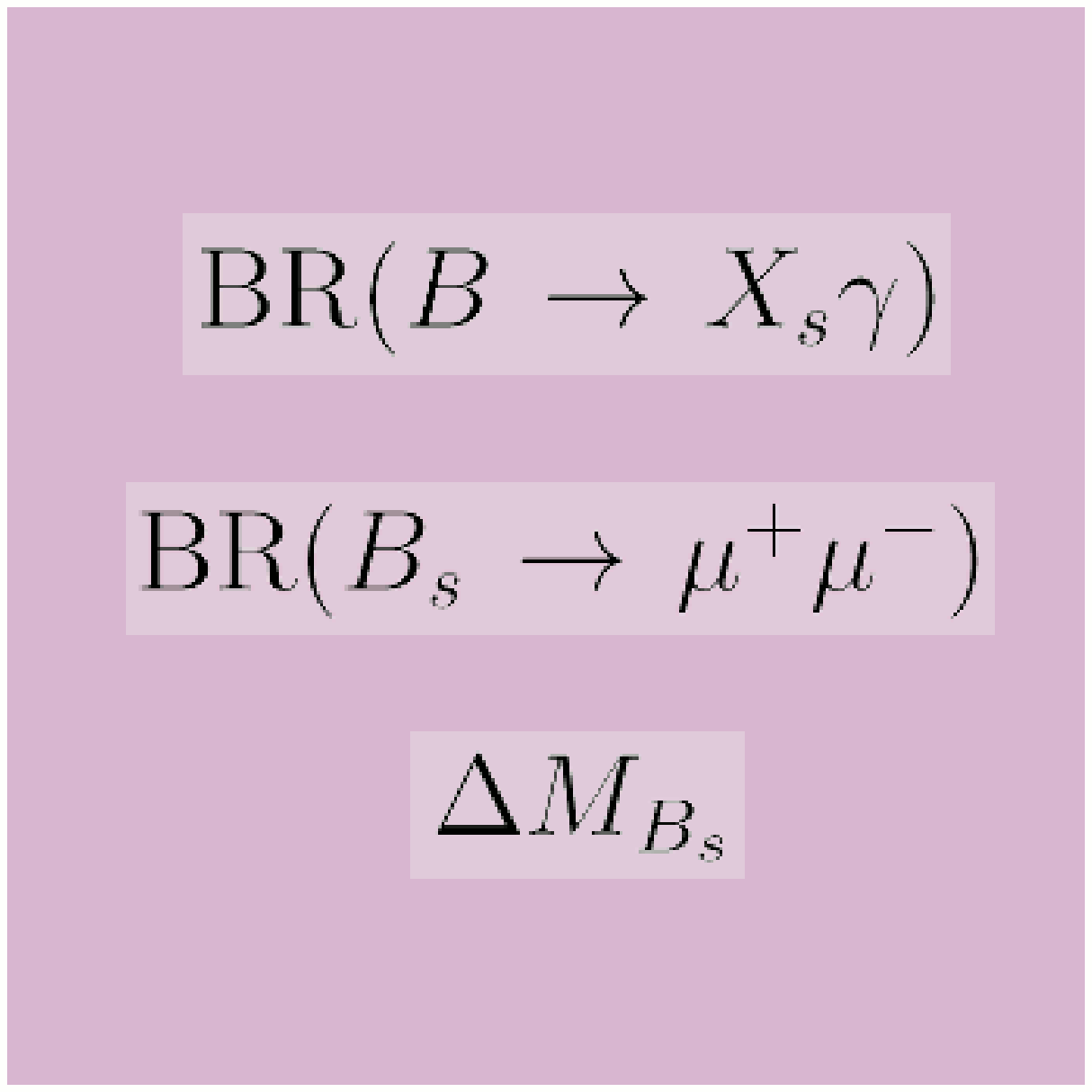}&
\includegraphics[width=13.3cm]{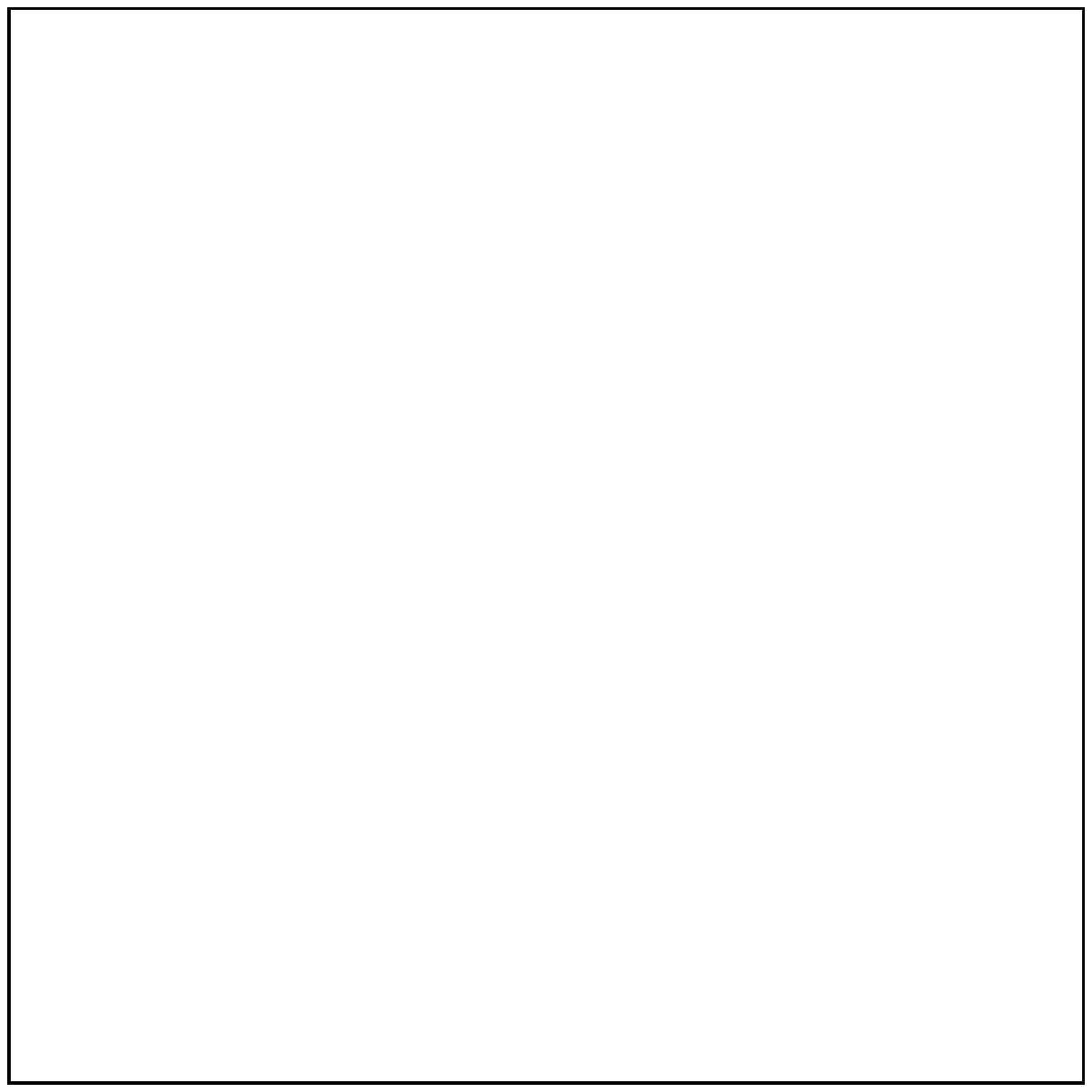}\\
\end{tabular}}} 
\caption{Legend for plots of Higgs mass corrections varying two deltas simultaneously displayed in figs. \ref{figdoubledeltaBFP}, \ref{figdoubledeltaSPS3}, \ref{figdoubledeltaSPS2}, \ref{figdoubledeltaSPS5}, \ref{figdoubledeltaVHeavyS} and \ref{figdoubledeltaHeavySLightH}. Each colored area represents the disallowed region by the specified observable/s inside each box. A white area placed at the central regions of the mentioned figures represents a region allowed by the three $B$ observables. A white area placed outside the colored areas represent regions of the parameter space that generate negative squared masses.  These problematic points are consequently not shown in our plots, as we did in the previous plots.
The discontinuous lines in those figures represent the contour lines for the $B$ observables corresponding to the maximum and minimum allowed values: dash-dot-dash for the upper bound of \bsg (eq.  (\ref{bsglinearerr})), dot-dash-dot for the lower bound of \bsg (eq.  (\ref{bsglinearerr})), dashed line for the upper bound of $\dmbs$ (eq. (\ref{deltabslinearerr})), a sequential three dotted line for  the lower bound of $\dmbs$ (eq. (\ref{deltabslinearerr})), and a dotted line for the upper bound of \bmm (eq. (\ref{bsmmlinearerr})).}  
\label{colleg} 
\end{figure} 
%%%%%%%%%%%%%%%%%%%%%%%%%% F I G U R E %%%%%%%%%%%%%%%%%%%%%%%%%%%%%%%%%%%%%%%%

 %%%%%%%%%%%%%%%%%%%%%%%%%% F I G U R E %%%%%%%%%%%%%%%%%%%%%%%%%%%%%%%%%%%%%%%%
\begin{figure}[h!] 
\centering
\hspace*{-10mm} 
{\resizebox{14.6cm}{!} 
{\begin{tabular}{cc} 
\includegraphics[width=13.3cm]{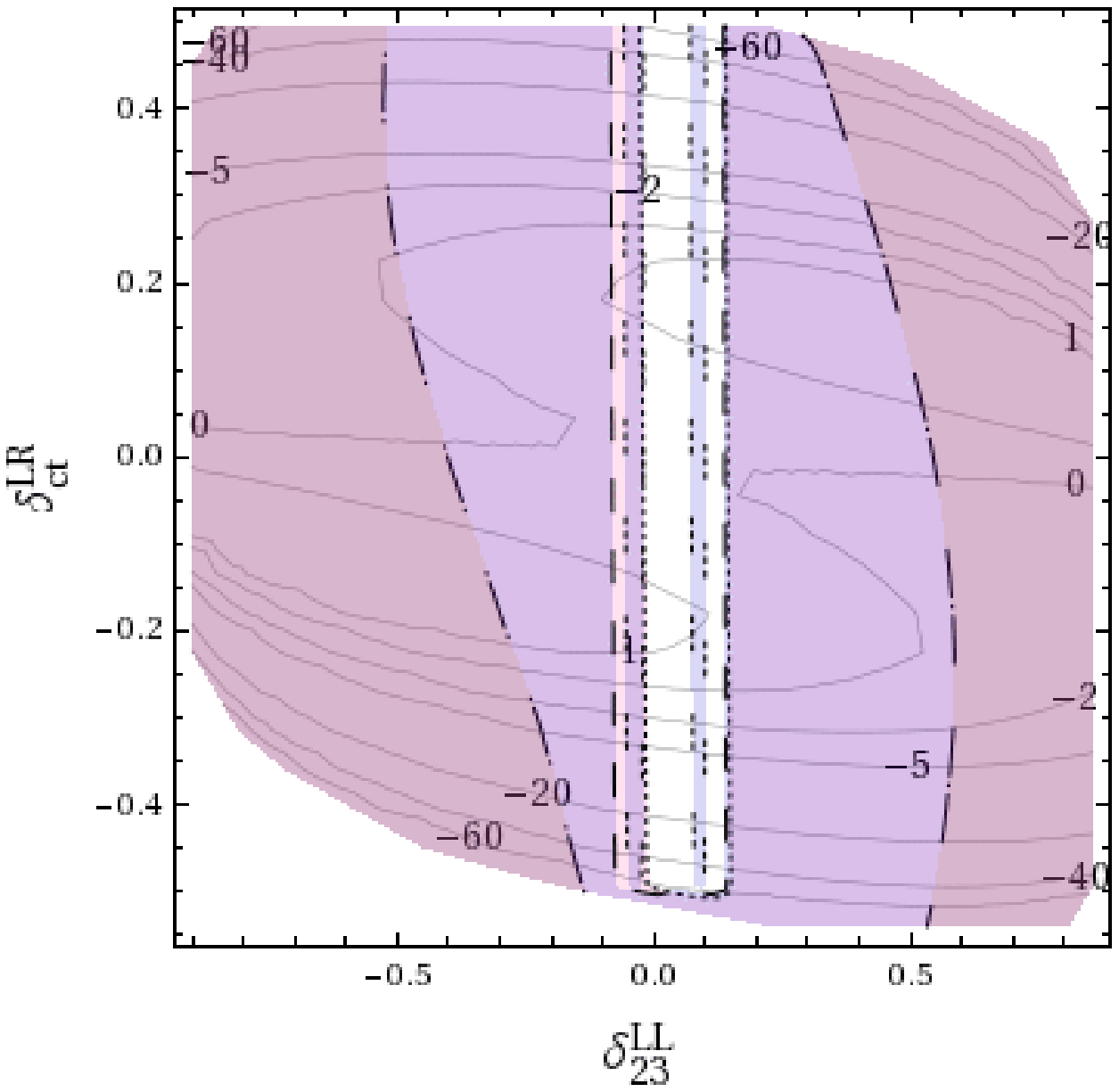}& 
\includegraphics[width=13.3cm]{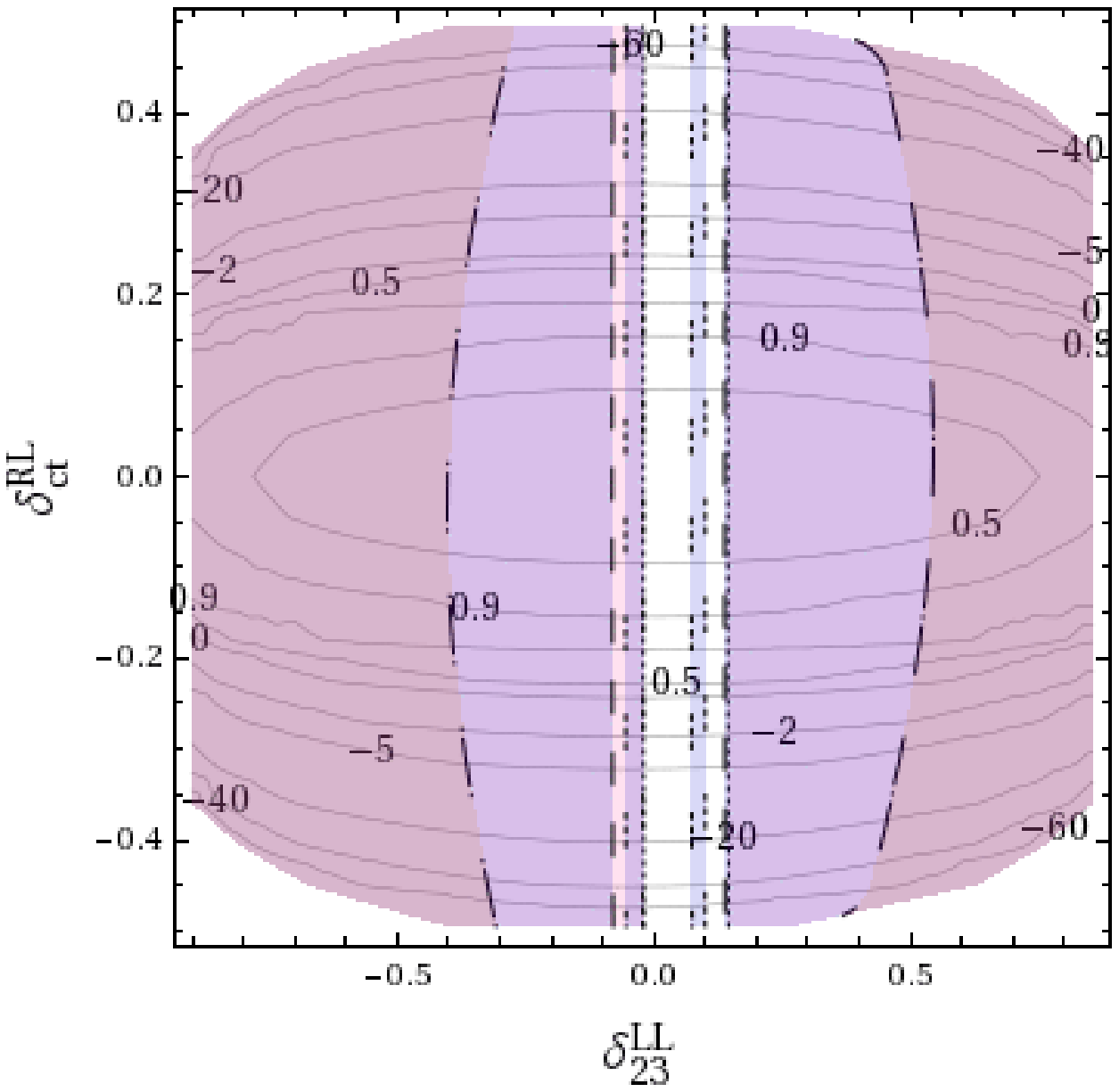}\\ 
\includegraphics[width=13.3cm]{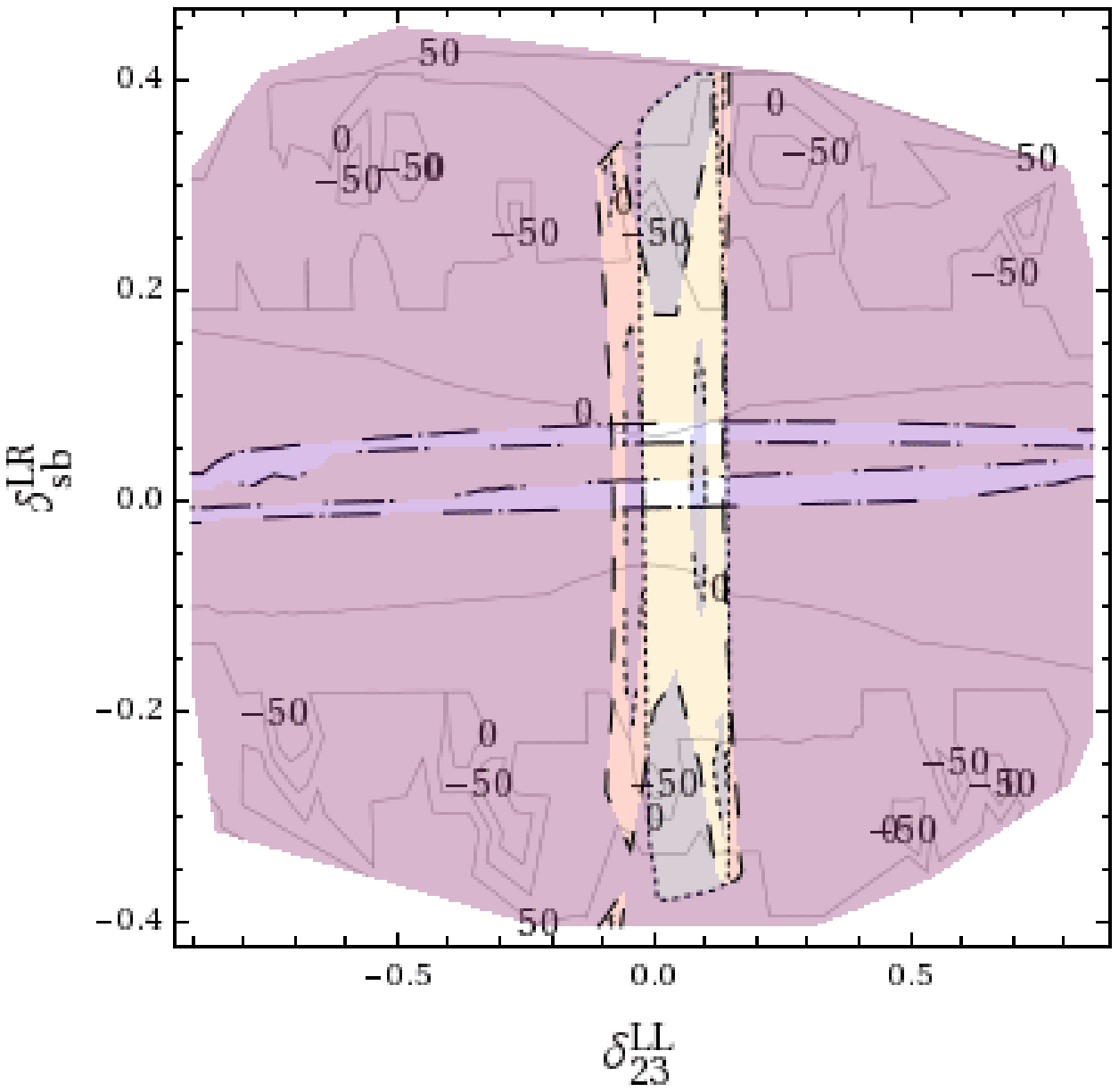}&
\includegraphics[width=13.3cm]{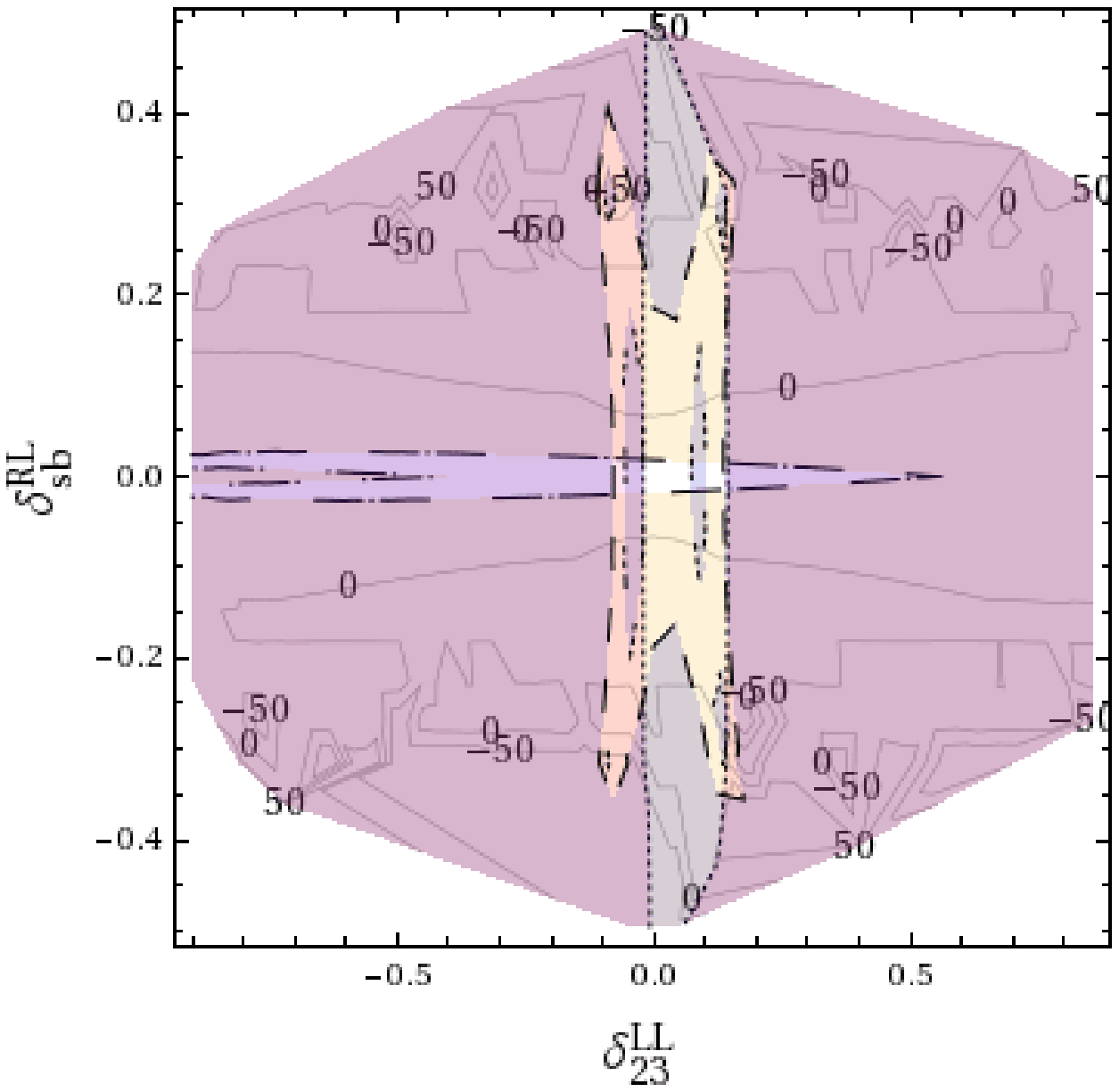}\\ 
\includegraphics[width=13.3cm]{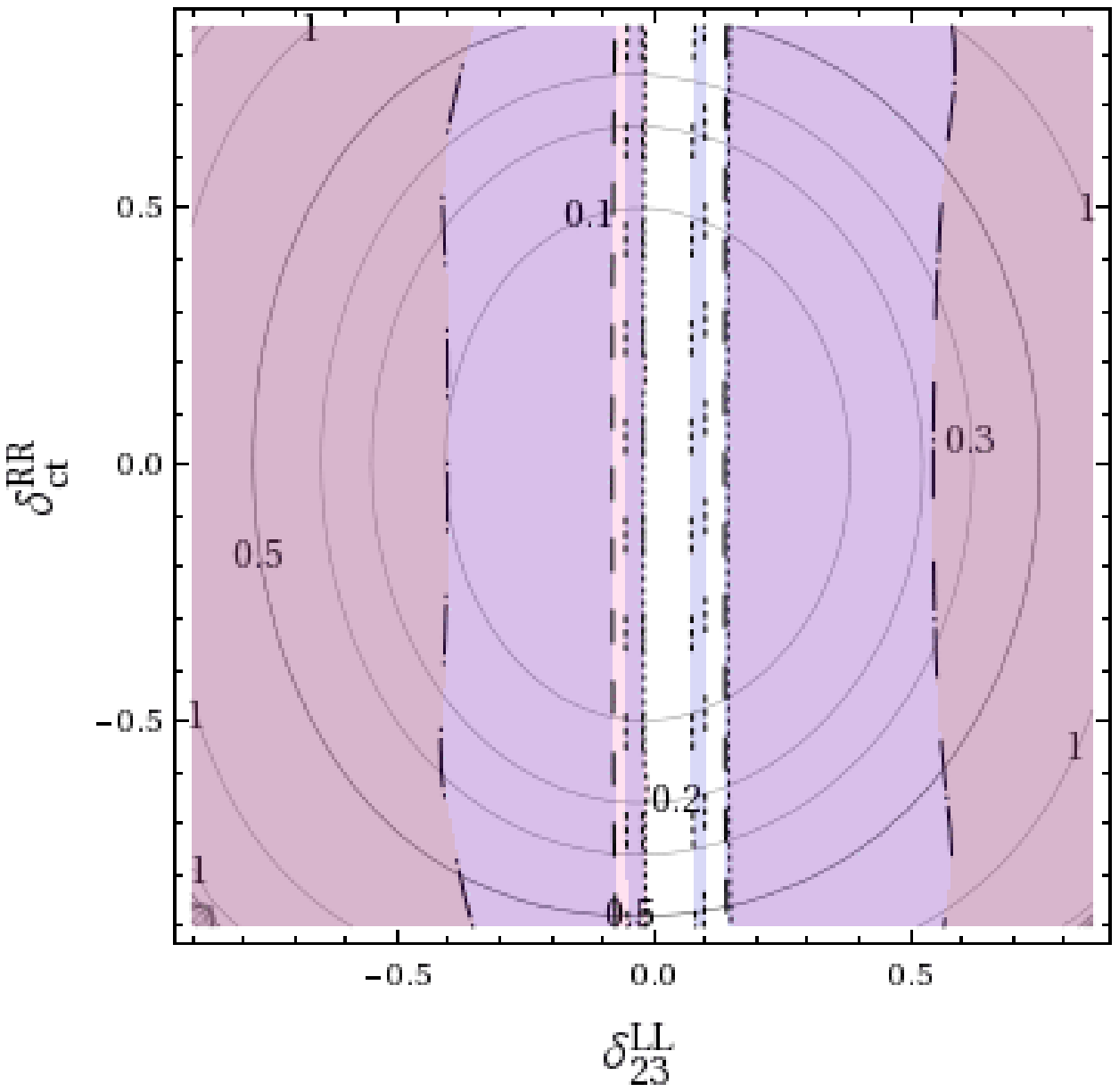}& 
\includegraphics[width=13.3cm]{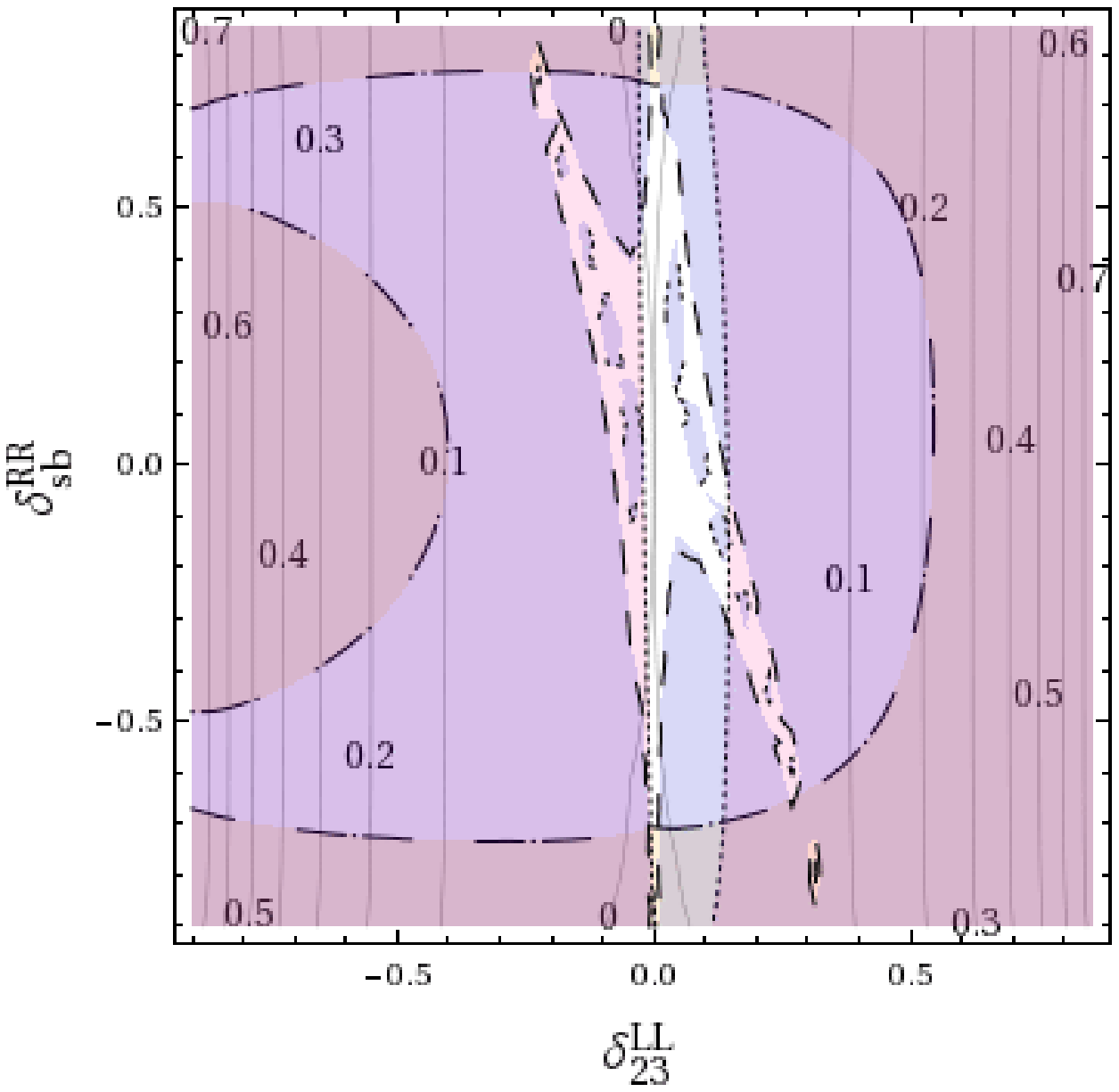}\\ 
\end{tabular}}}
\caption{$\Delta m_{h}$ (GeV) contour lines from our two deltas analysis for BFP. The color code for the allowed/disallowed areas by $B$ data is given in fig.\ref{colleg}.} 
 \label{figdoubledeltaBFP}
\end{figure}

%%%%%%%%%%%%%%%%%%%%%%%%%% F I G U R E %%%%%%%%%%%%%%%%%%%%%%%%%%%%%%%%%%%%%%%%
\clearpage
\newpage
%%%%%%%%%%%%%%%%%%%%%%%%%% F I G U R E %%%%%%%%%%%%%%%%%%%%%%%%%%%%%%%%%%%%%%%%
\begin{figure}[h!] 
\centering
\hspace*{-10mm} 
{\resizebox{14.6cm}{!} 
{\begin{tabular}{cc} 
\includegraphics[width=13.3cm]{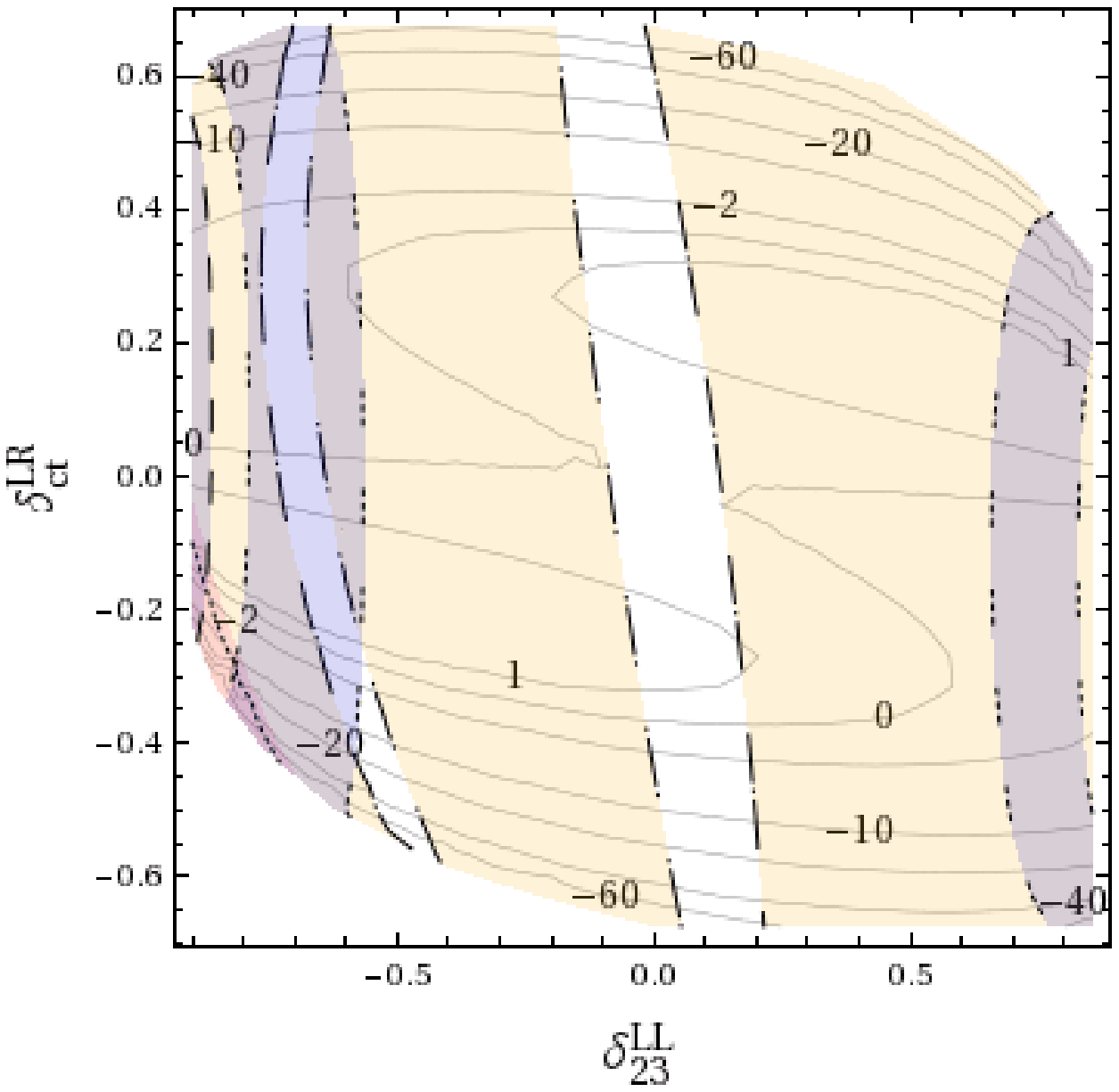}& 
\includegraphics[width=13.3cm]{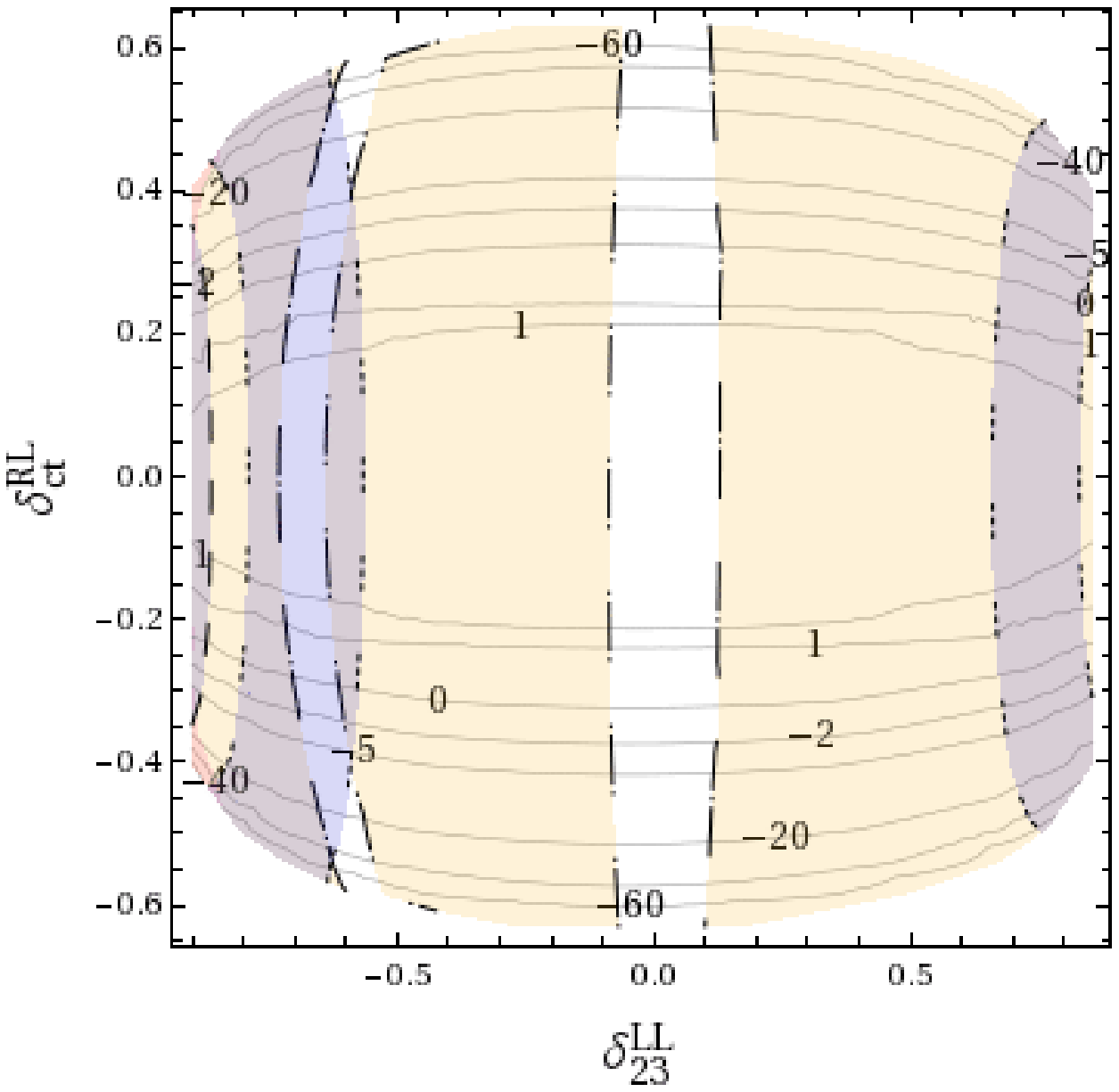}\\ 
\includegraphics[width=13.3cm]{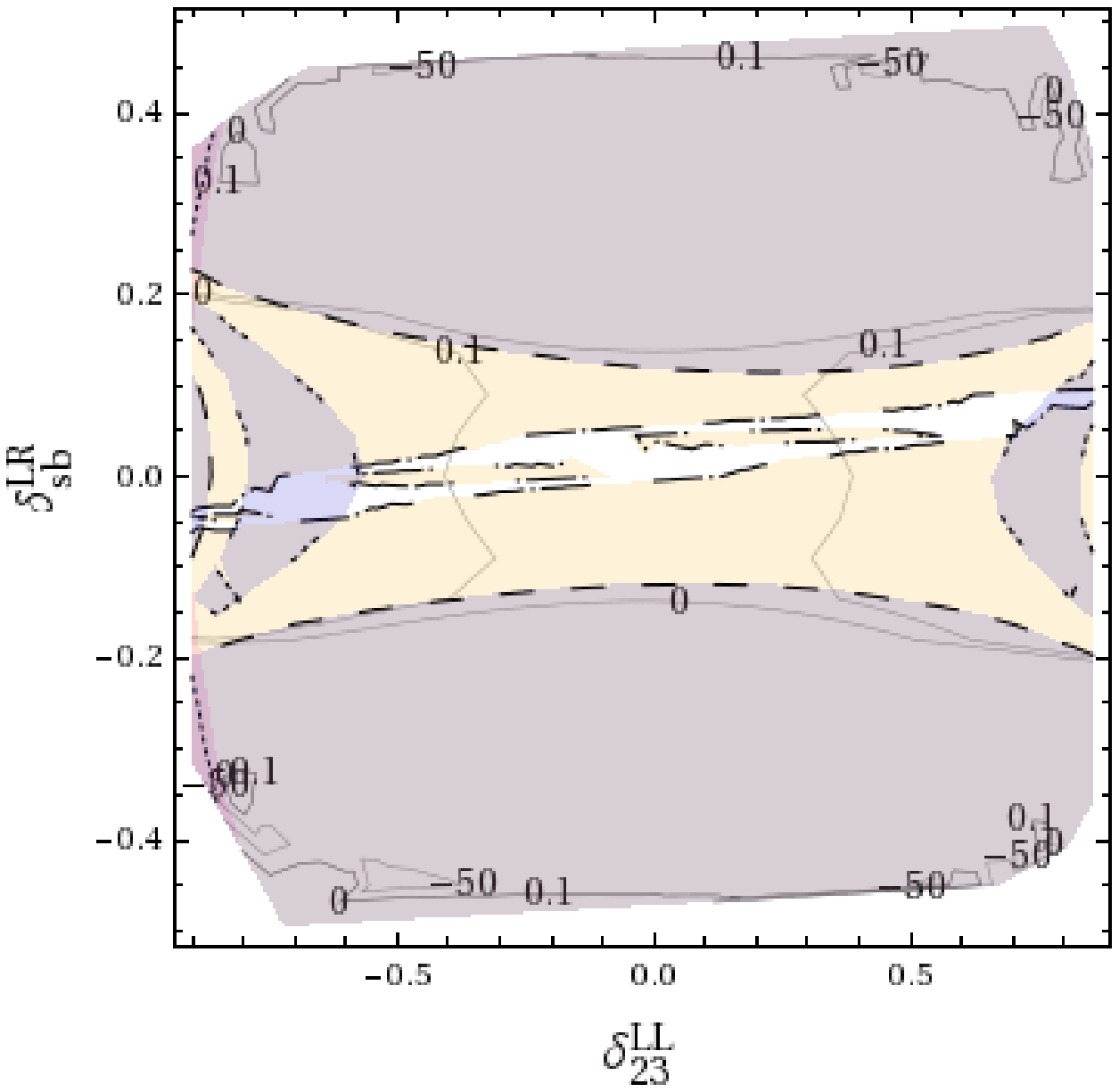}&
\includegraphics[width=13.3cm]{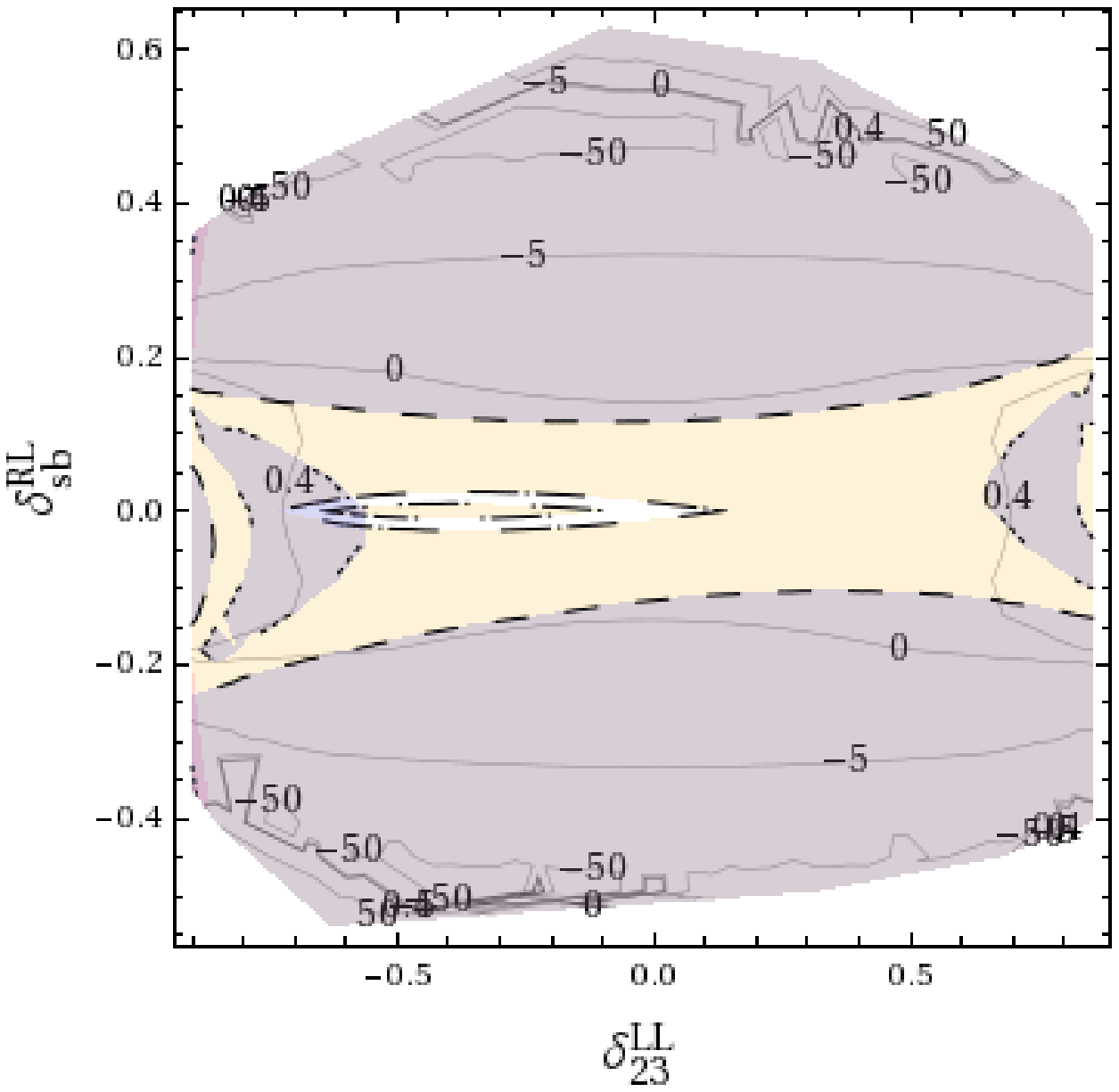}\\ 
\includegraphics[width=13.3cm]{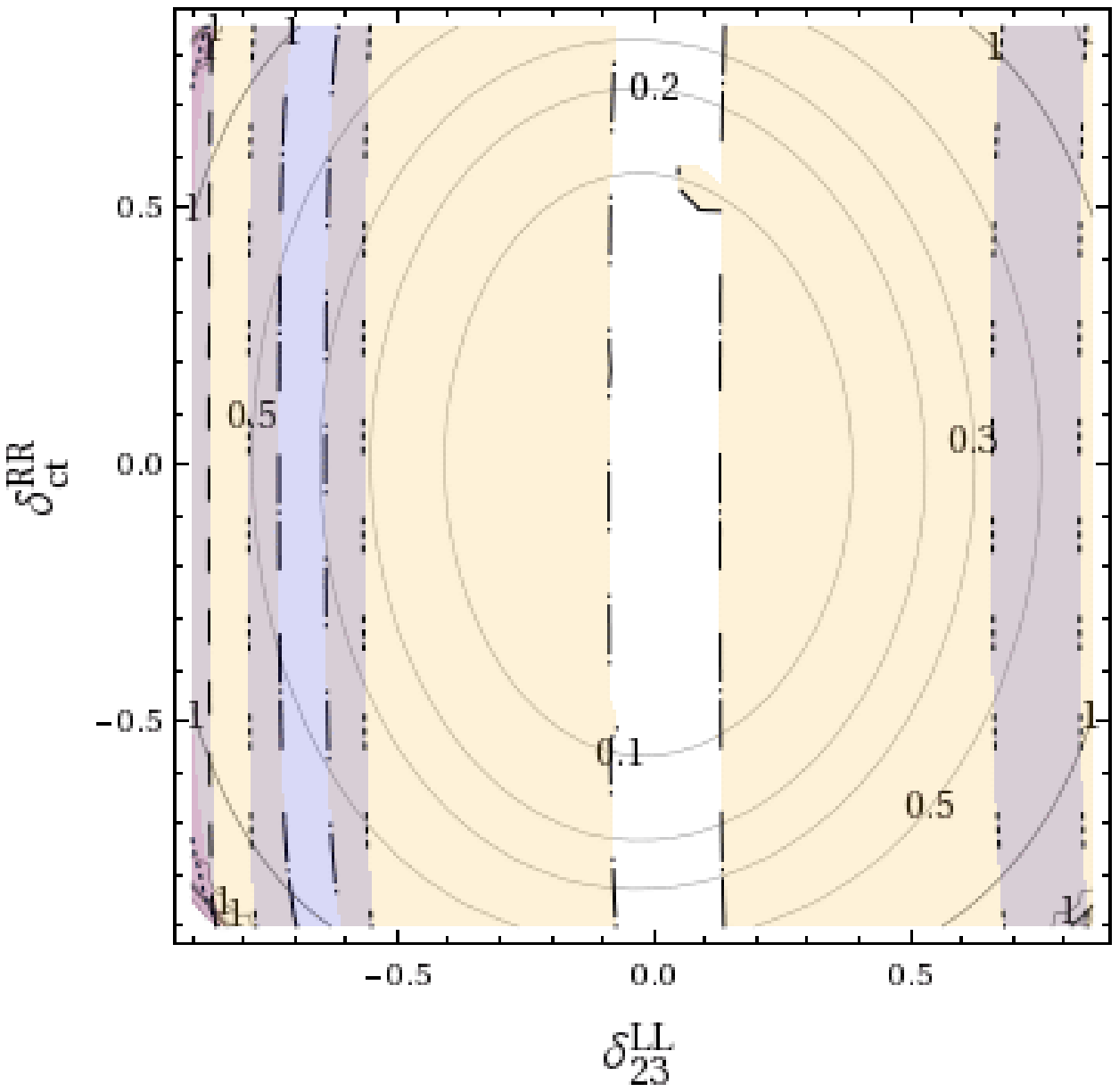}& 
\includegraphics[width=13.3cm]{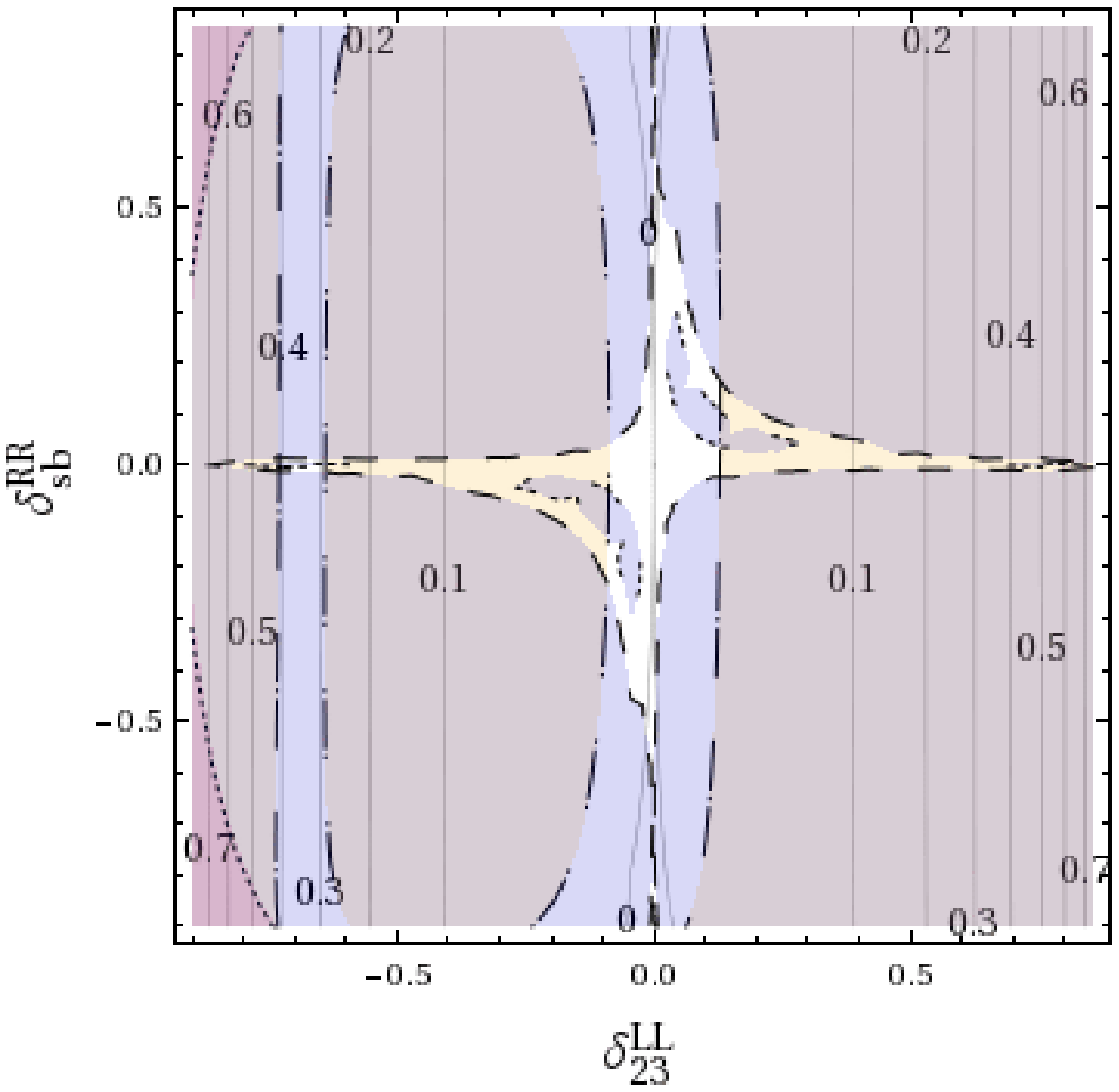}\\ 
\end{tabular}}}
\caption{$\Delta m_{h}$ (GeV) contour lines from our two deltas analysis for SPS3. The color code for the allowed/disallowed areas by $B$ data is given in fig.\ref{colleg}.} 
 \label{figdoubledeltaSPS3} 
\end{figure}
%%%%%%%%%%%%%%%%%%%%%%%%%% F I G U R E %%%%%%%%%%%%%%%%%%%%%%%%%%%%%%%%%%%%%%%%
\clearpage
\newpage
%%%%%%%%%%%%%%%%%%%%%%%%%% F I G U R E %%%%%%%%%%%%%%%%%%%%%%%%%%%%%%%%%%%%%%%%
\begin{figure}[h!] 
\centering
\hspace*{-10mm} 
{\resizebox{14.6cm}{!} 
{\begin{tabular}{cc} 
\includegraphics[width=13.3cm]{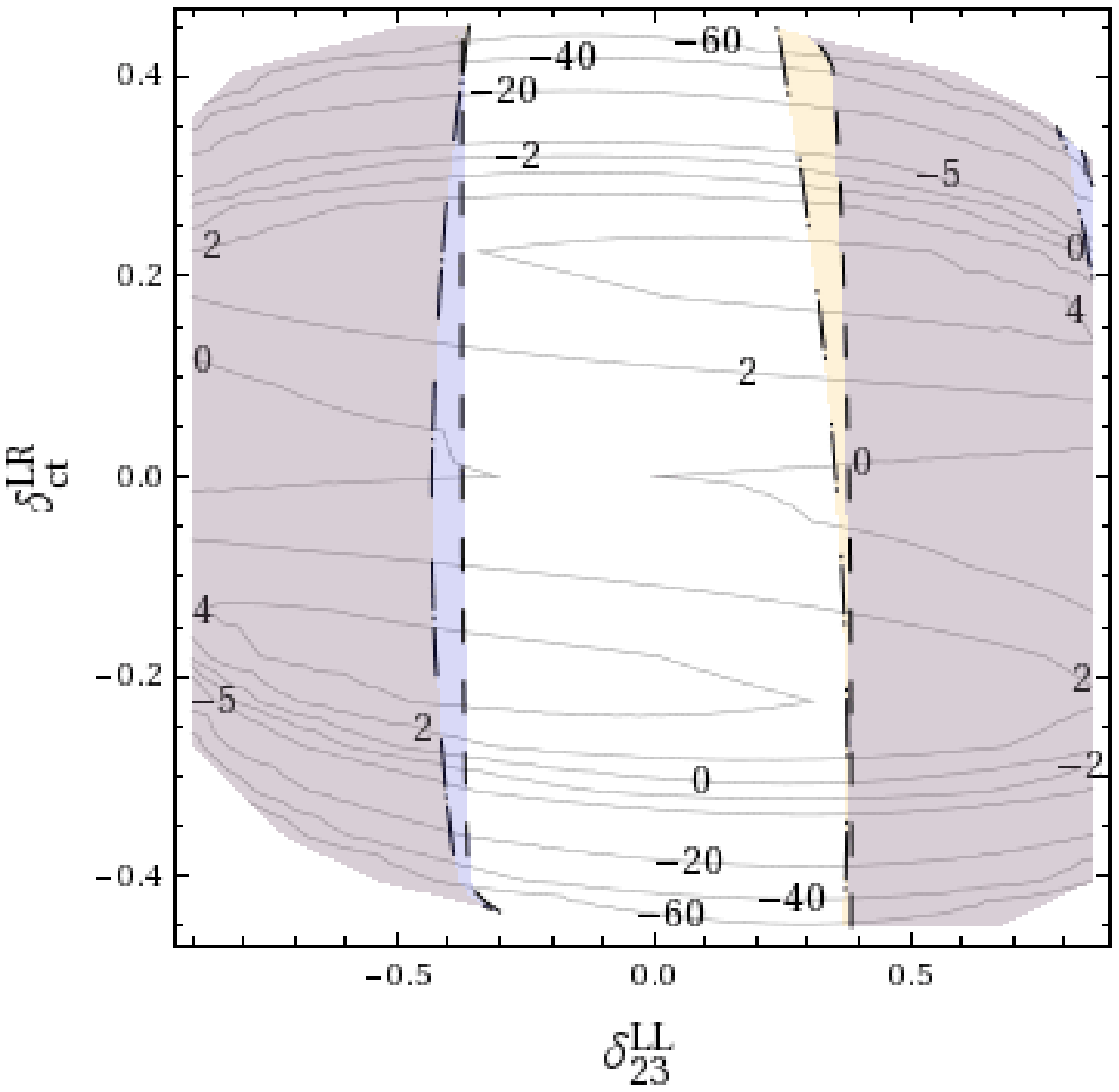}& 
\includegraphics[width=13.3cm]{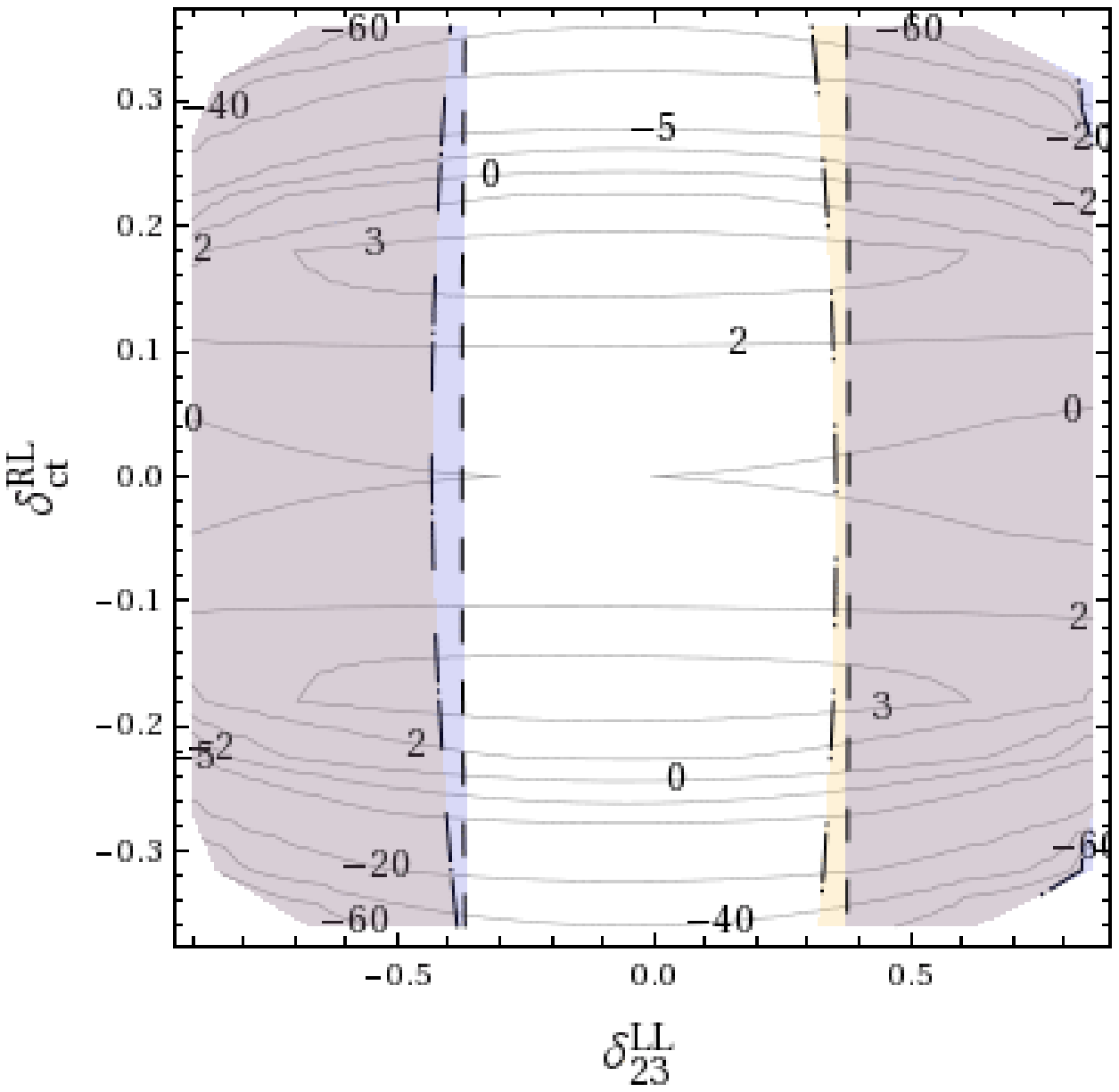}\\ 
\includegraphics[width=13.3cm]{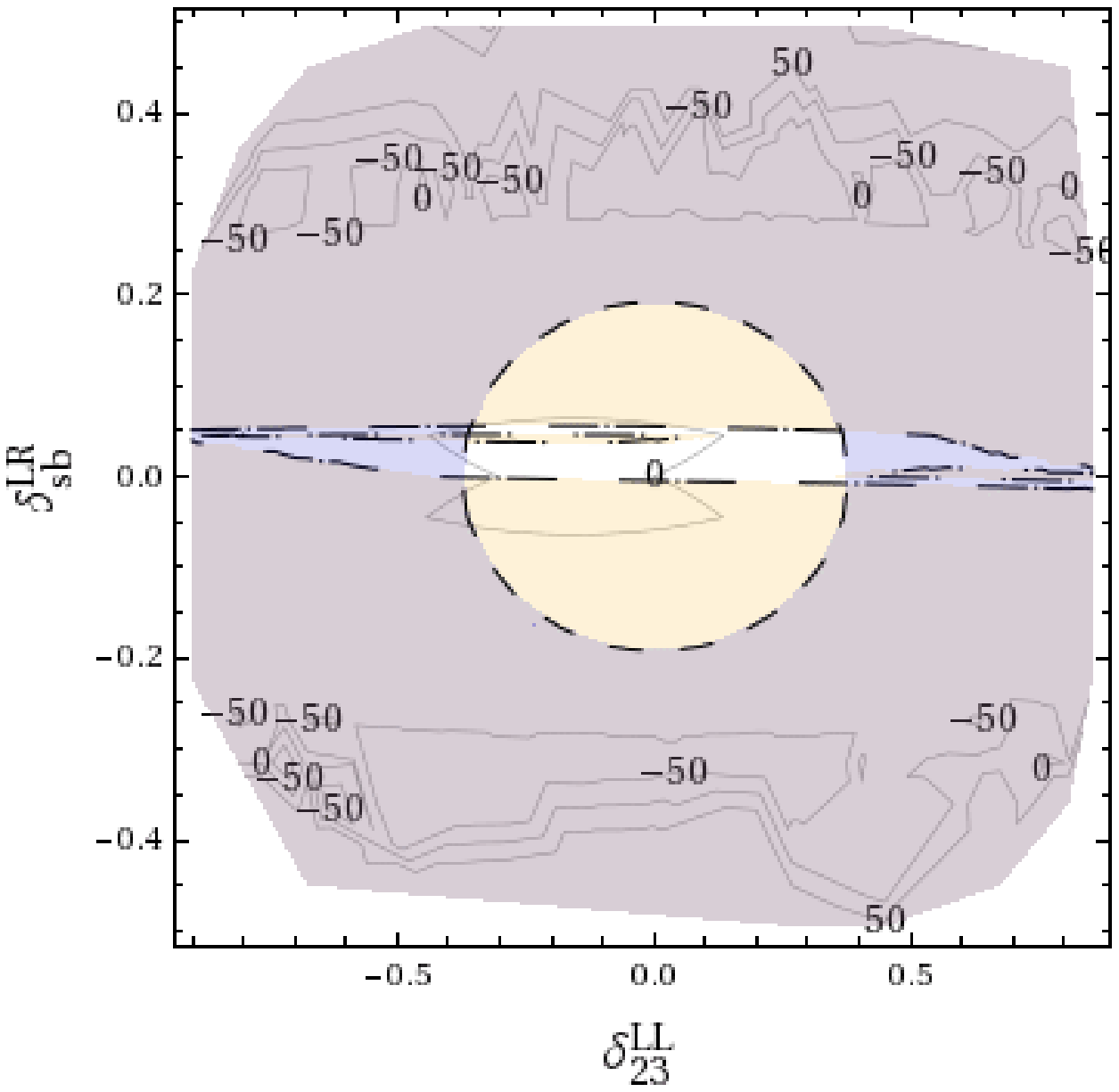}&
\includegraphics[width=13.3cm]{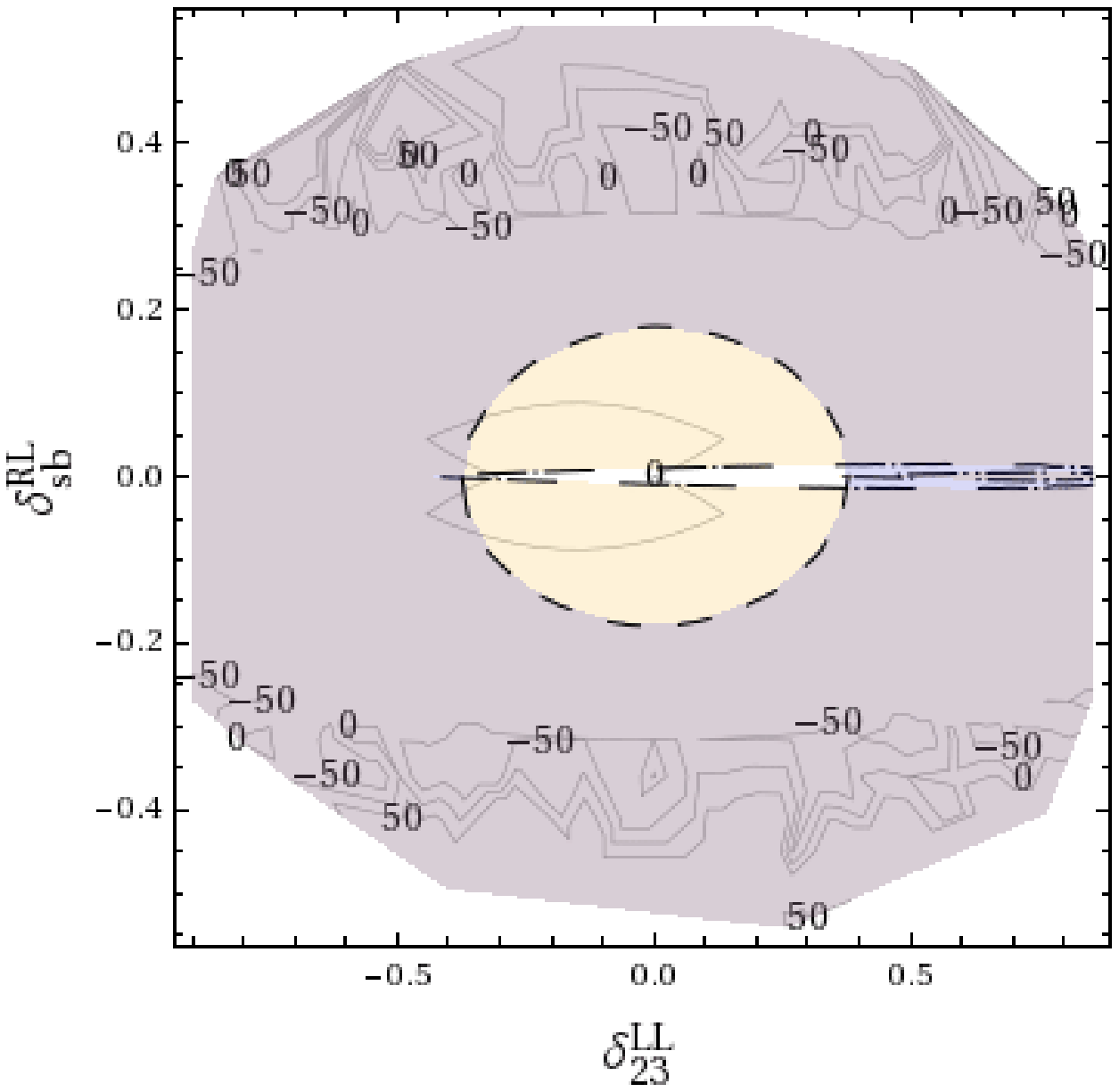}\\ 
\includegraphics[width=13.3cm]{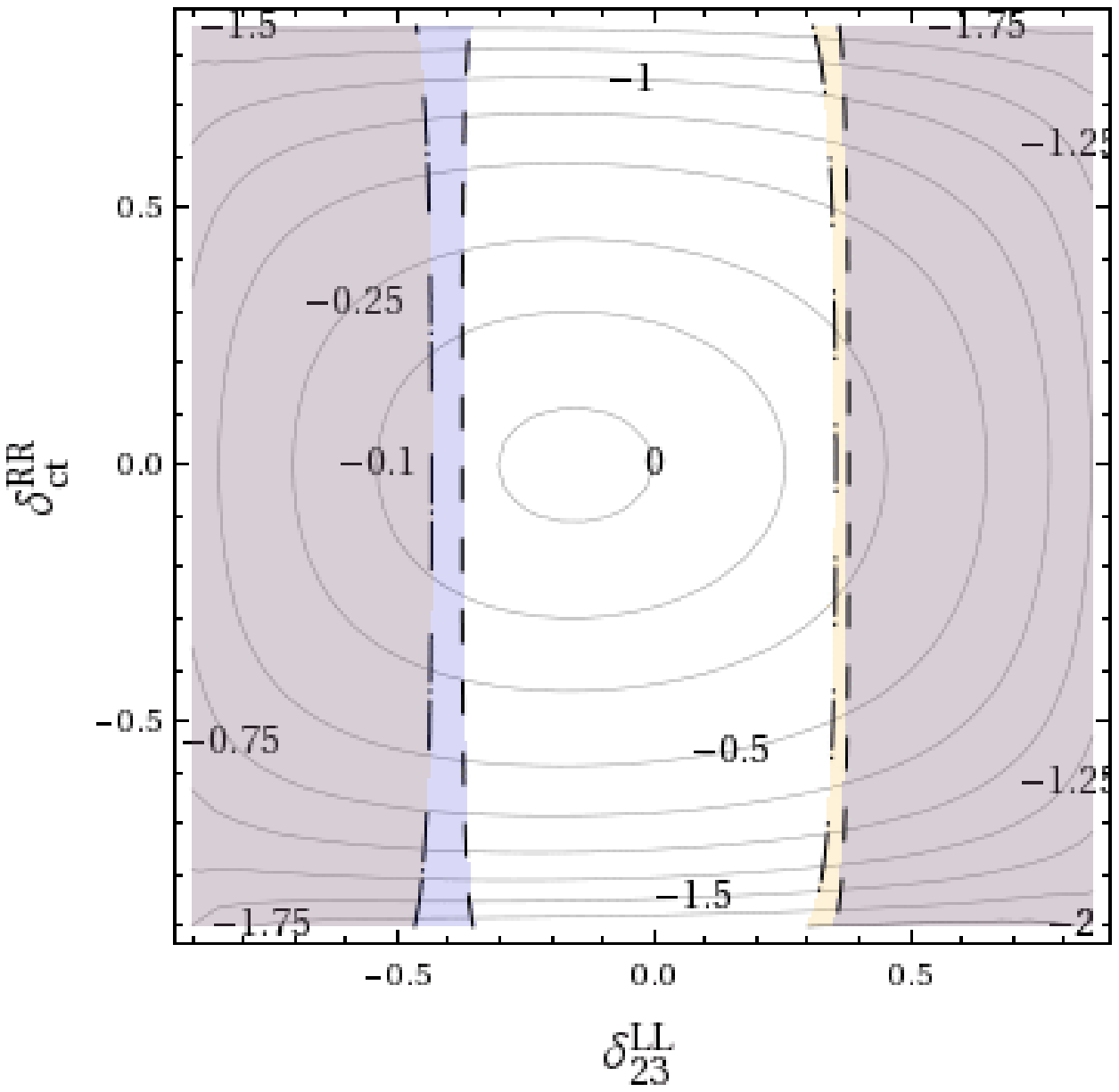}& 
\includegraphics[width=13.3cm]{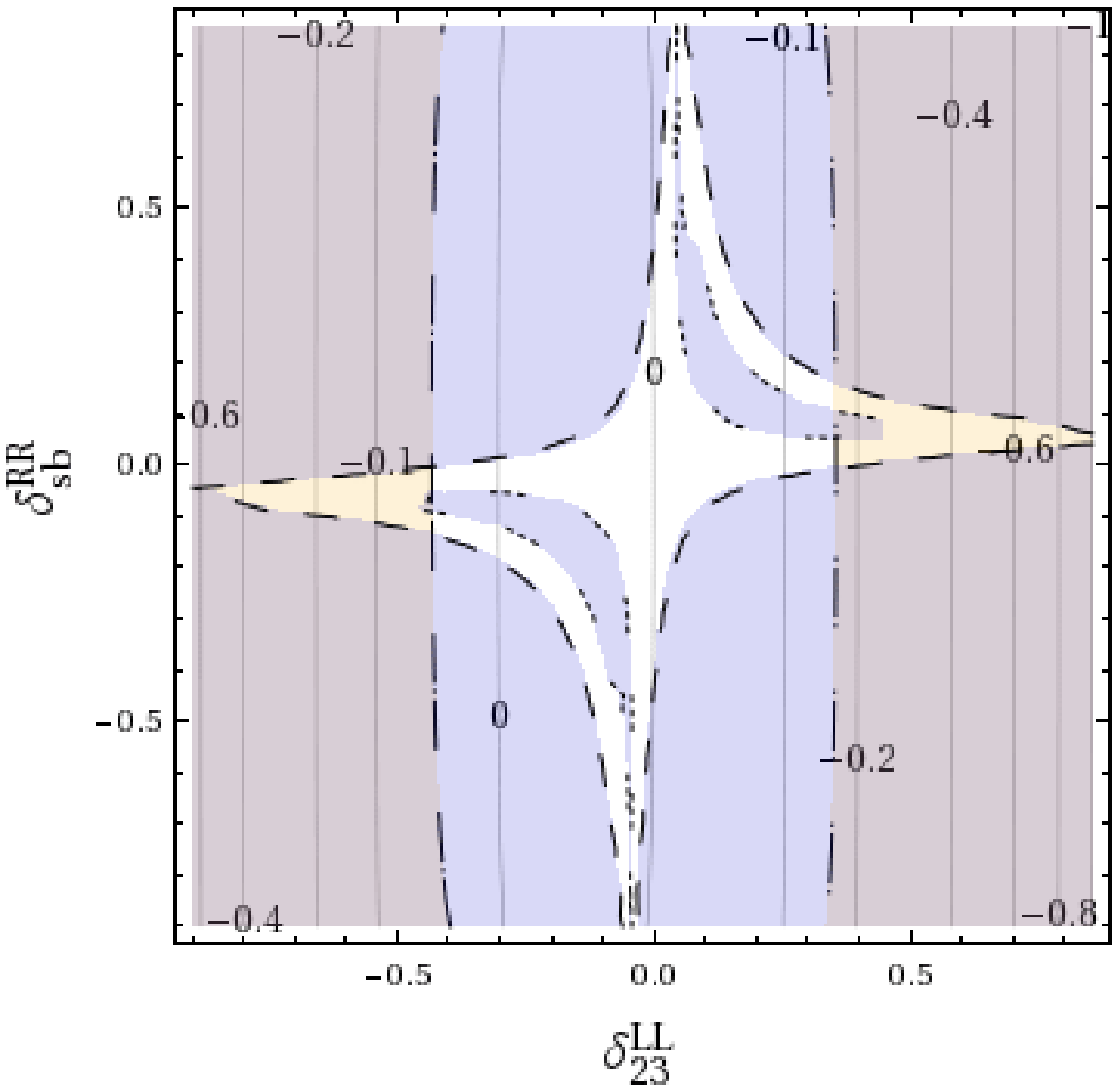}\\ 
\end{tabular}}}
\caption{$\Delta m_{h}$ (GeV) contour lines from our two deltas analysis for SPS2. The color code for the allowed/disallowed areas by $B$ data is given in fig.\ref{colleg}.} 
 \label{figdoubledeltaSPS2} 
\end{figure}
%%%%%%%%%%%%%%%%%%%%%%%%%% F I G U R E %%%%%%%%%%%%%%%%%%%%%%%%%%%%%%%%%%%%%%%%
\clearpage
\newpage

%%%%%%%%%%%%%%%%%%%%%%%%%% F I G U R E %%%%%%%%%%%%%%%%%%%%%%%%%%%%%%%%%%%%%%%%
\begin{figure}[h!] 
\centering
\hspace*{-10mm} 
{\resizebox{14.6cm}{!} 
{\begin{tabular}{cc} 
\includegraphics[width=13.3cm]{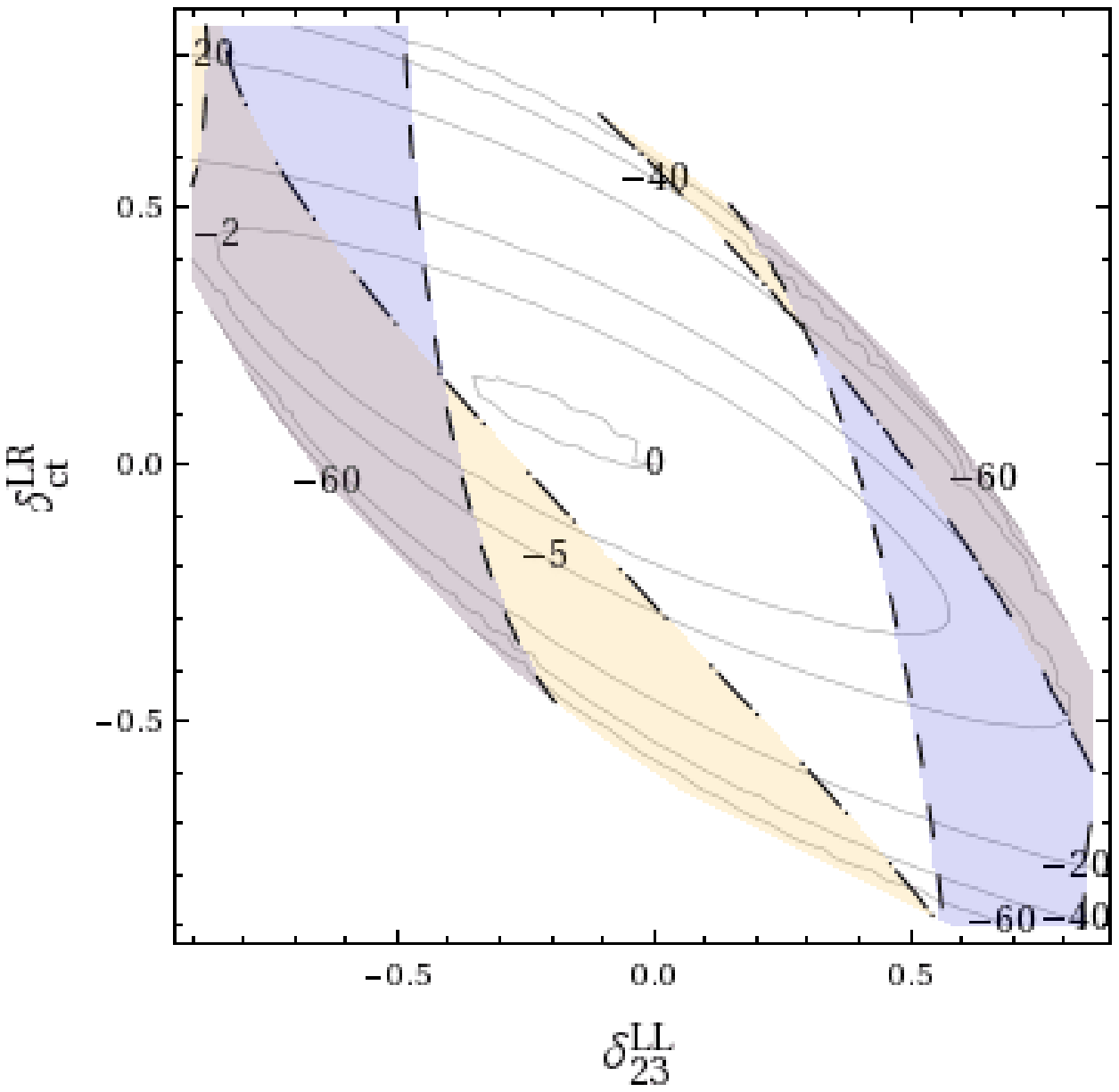}& 
\includegraphics[width=13.3cm]{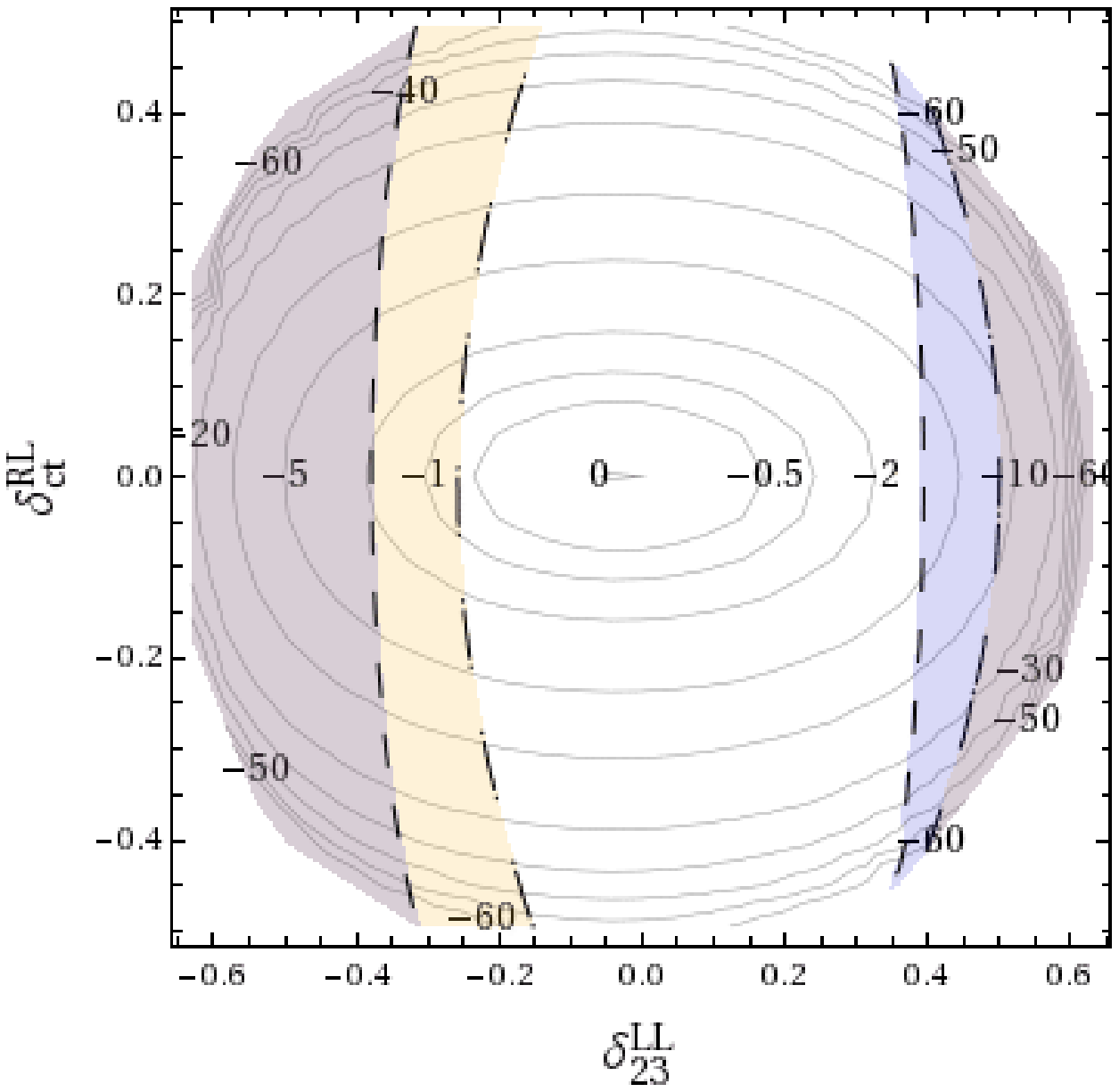}\\ 
\includegraphics[width=13.3cm]{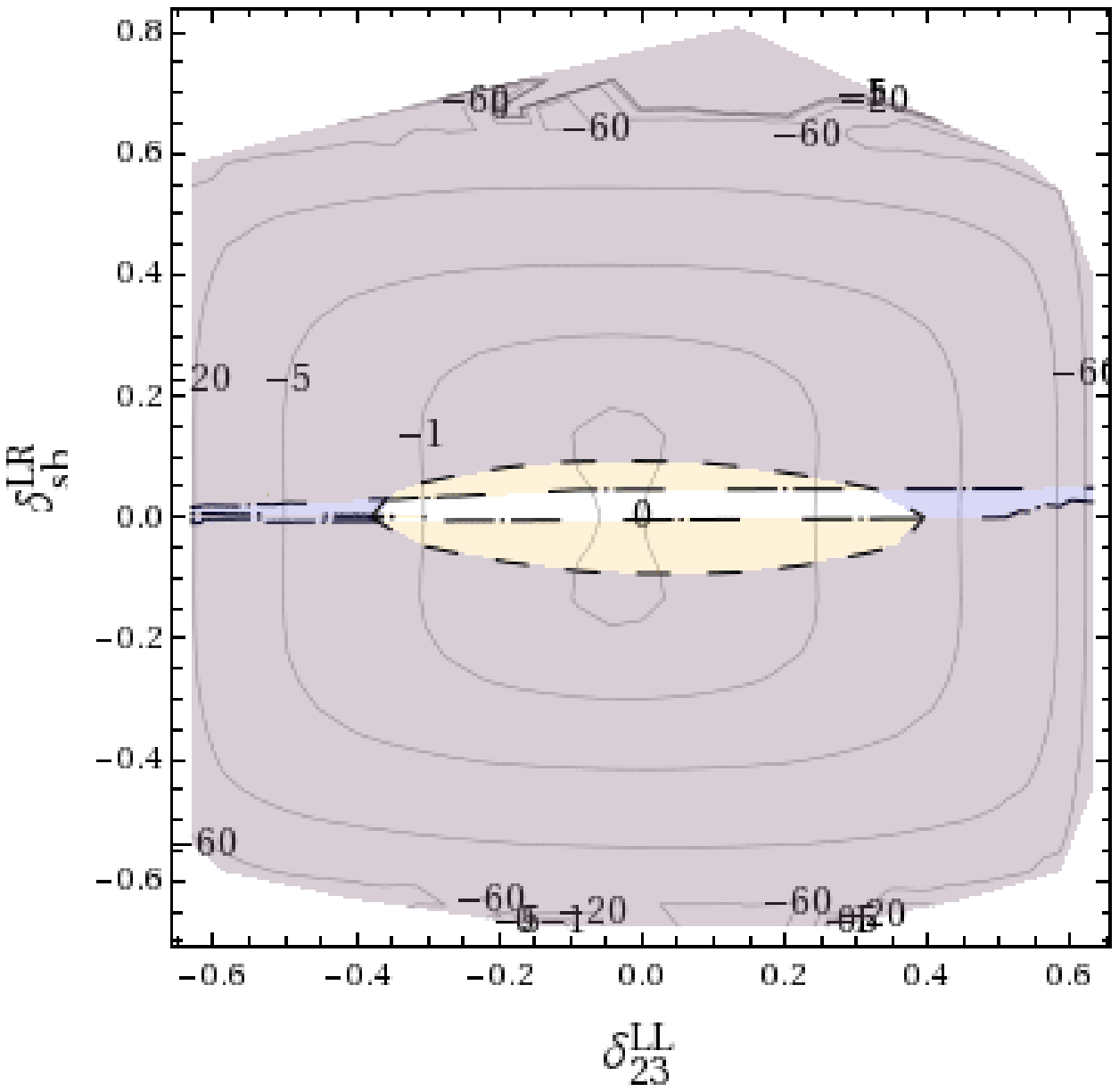}&
\includegraphics[width=13.3cm]{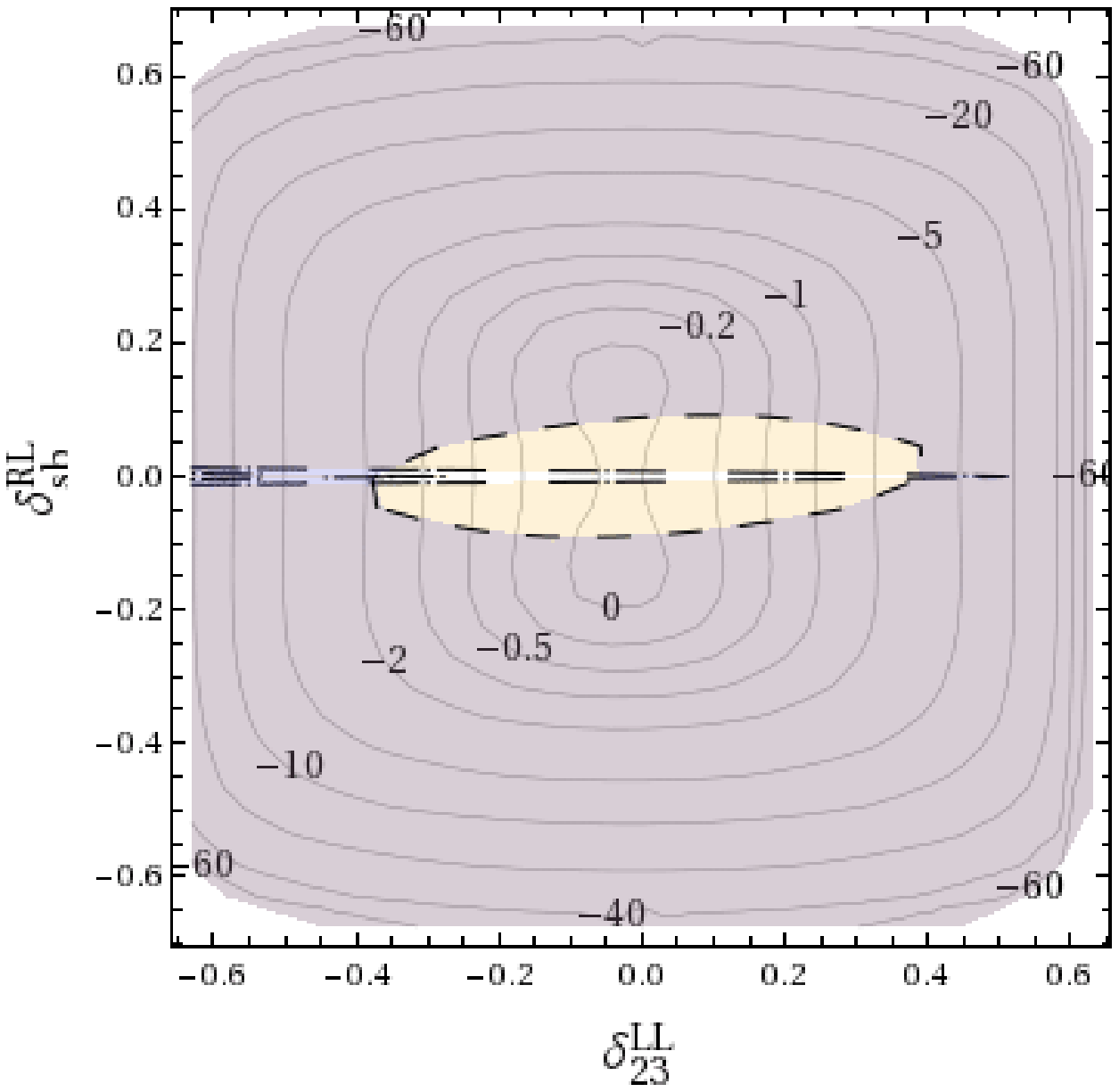}\\ 
\includegraphics[width=13.3cm]{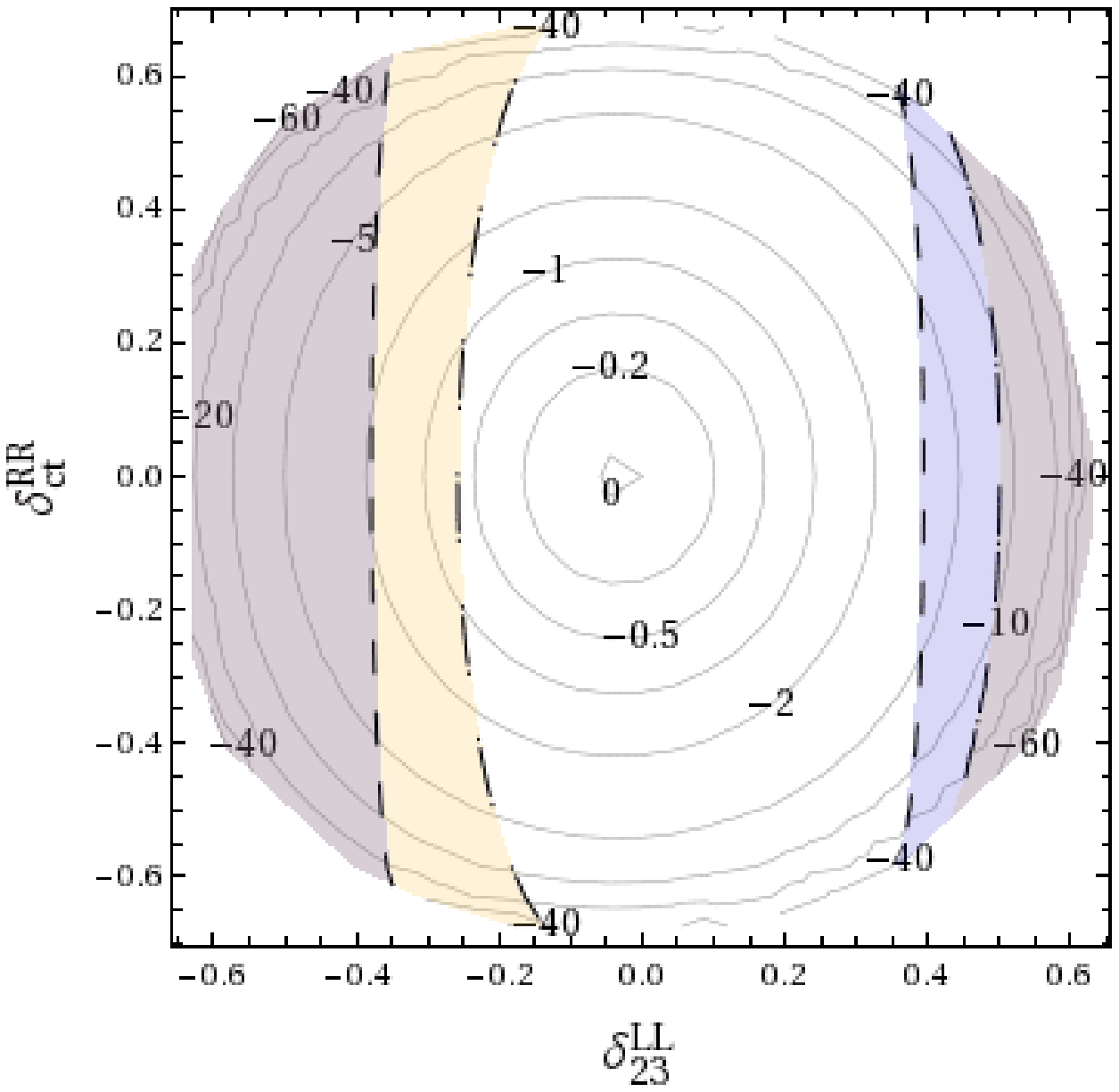}& 
\includegraphics[width=13.3cm]{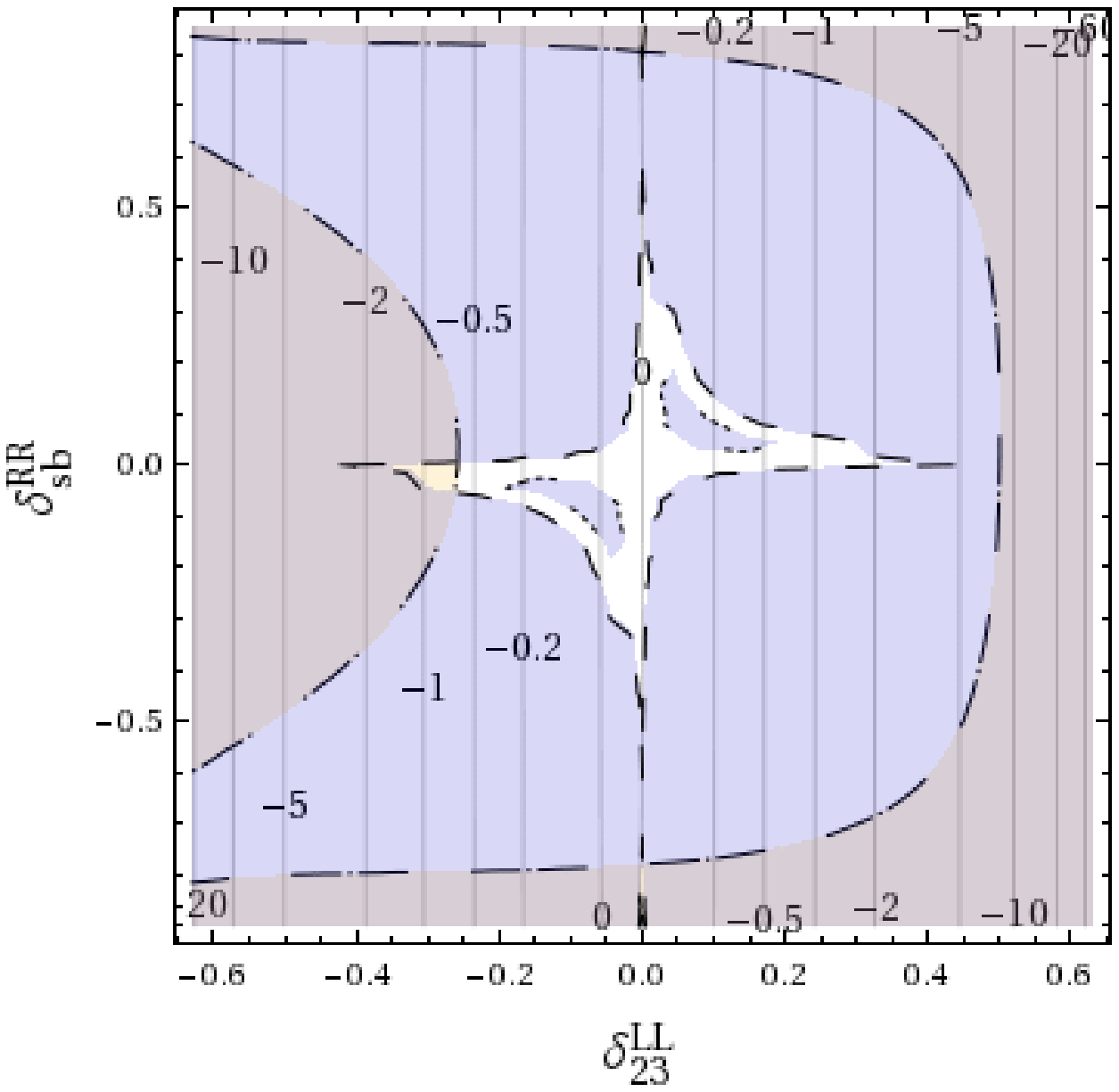}\\ 
\end{tabular}}}
\caption{$\Delta m_{h}$ (GeV) contour lines from our two deltas analysis for SPS5. The color code for the allowed/disallowed areas by $B$ data is given in fig.\ref{colleg}.} 
 \label{figdoubledeltaSPS5} 
\end{figure}
%%%%%%%%%%%%%%%%%%%%%%%%%% F I G U R E %%%%%%%%%%%%%%%%%%%%%%%%%%%%%%%%%%%%%%%%
\clearpage
\newpage
%%%%%%%%%%%%%%%%%%%%%%%%%% F I G U R E %%%%%%%%%%%%%%%%%%%%%%%%%%%%%%%%%%%%%%%%
\begin{figure}[h!] 
\centering
\hspace*{-10mm} 
{\resizebox{14.6cm}{!} 
{\begin{tabular}{cc} 
\includegraphics[width=13.3cm]{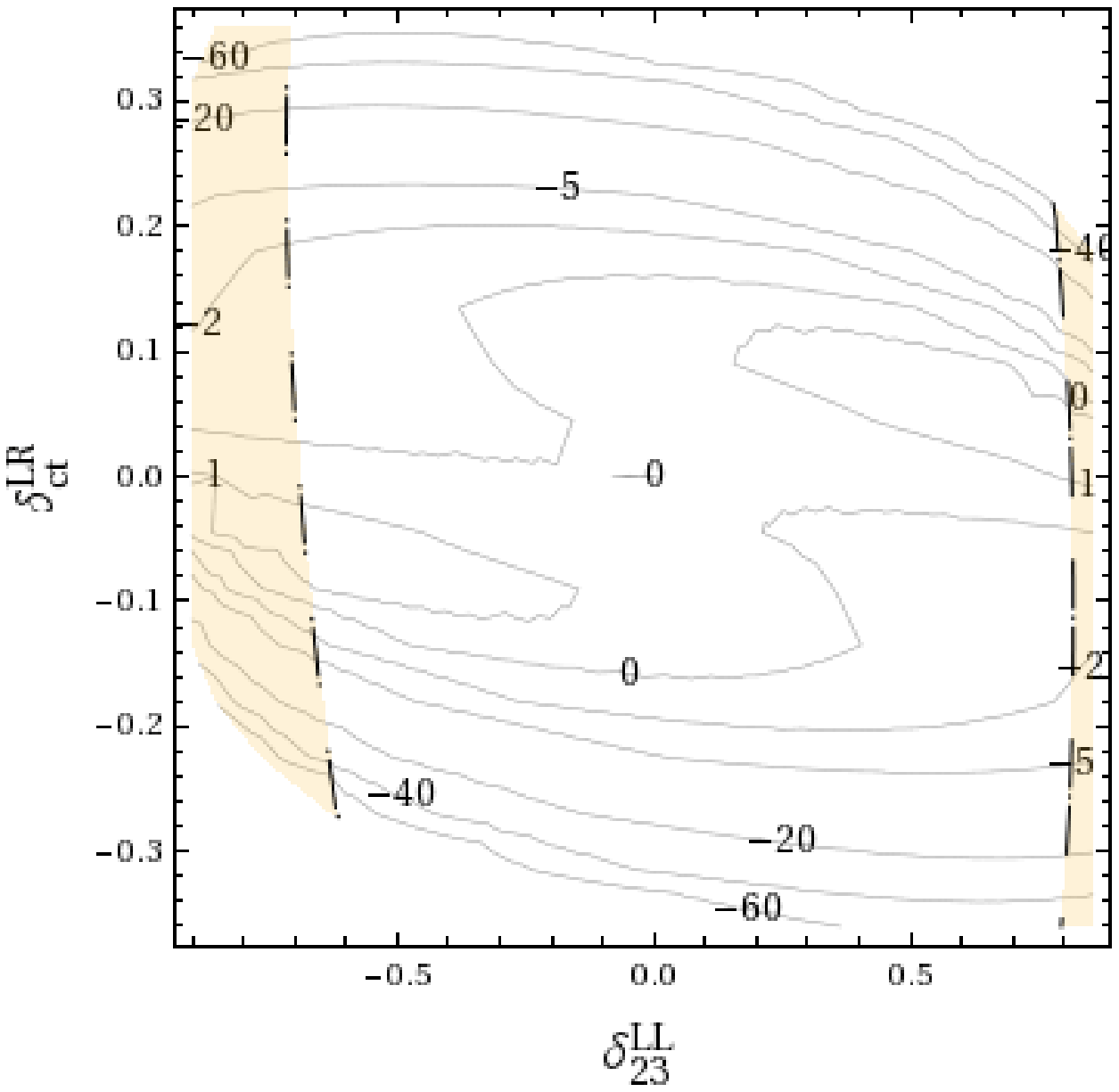}& 
\includegraphics[width=13.3cm]{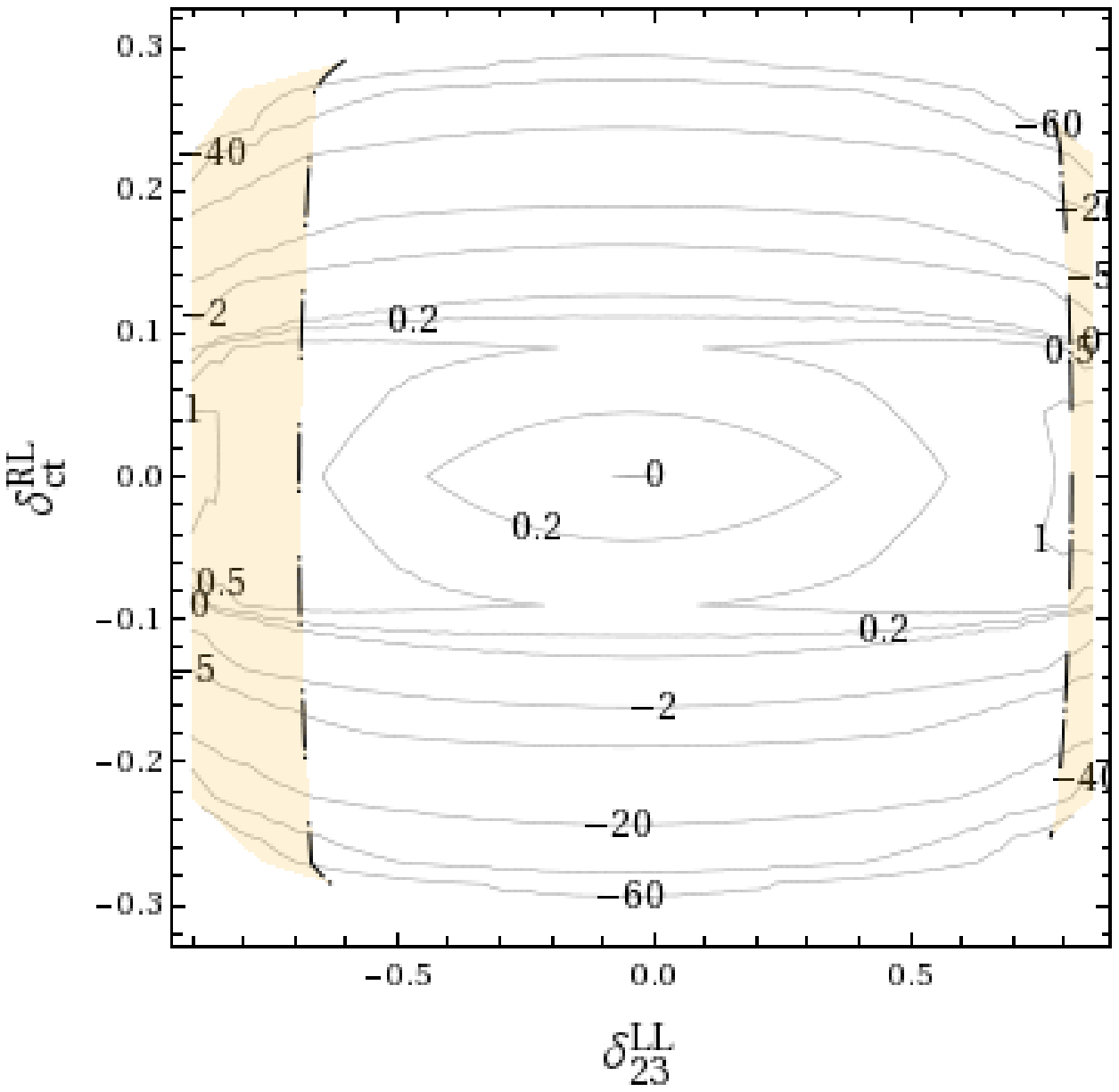}\\ 
\includegraphics[width=13.3cm]{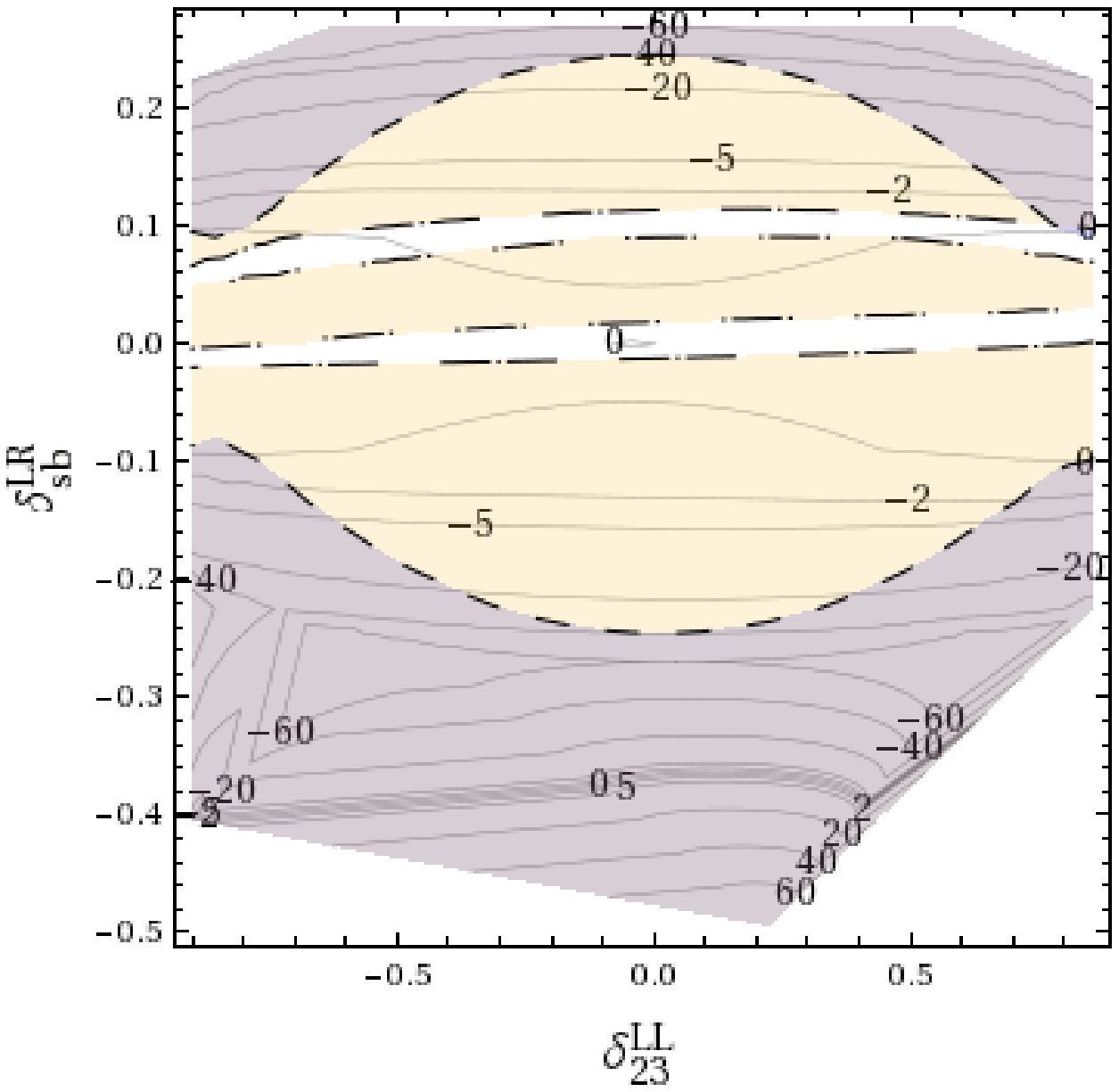}&
\includegraphics[width=13.3cm]{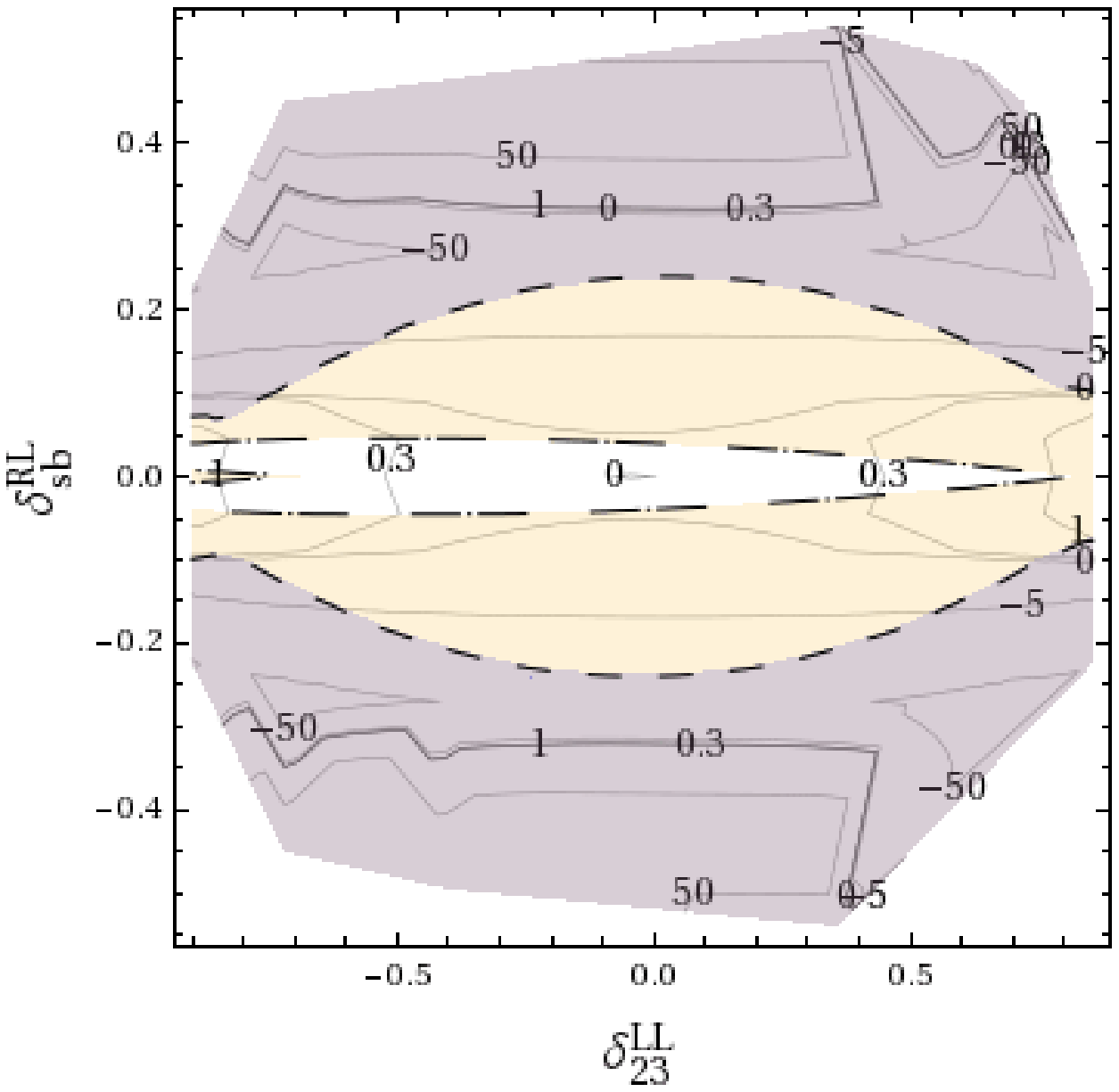}\\ 
\includegraphics[width=13.3cm]{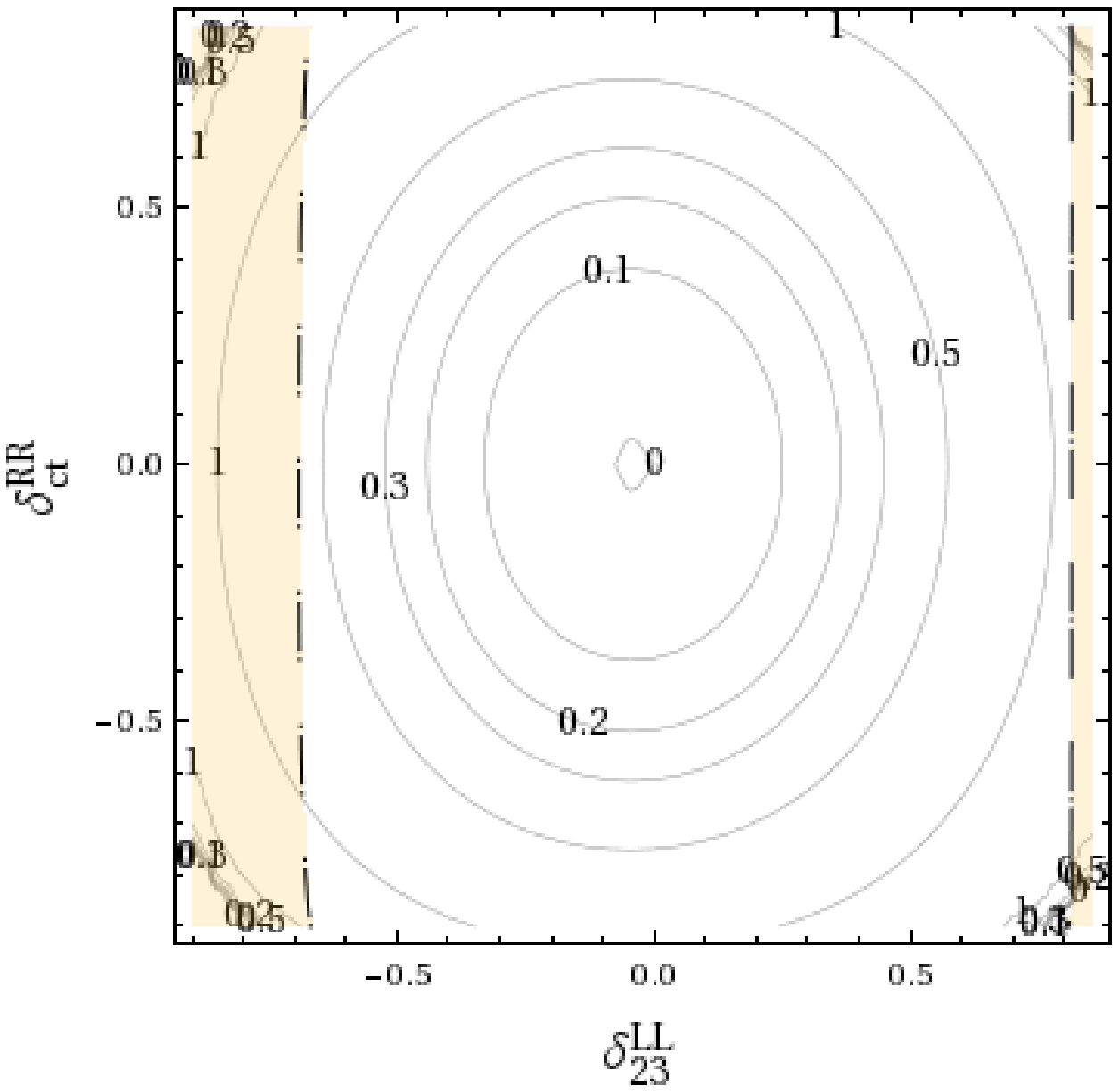}& 
\includegraphics[width=13.3cm]{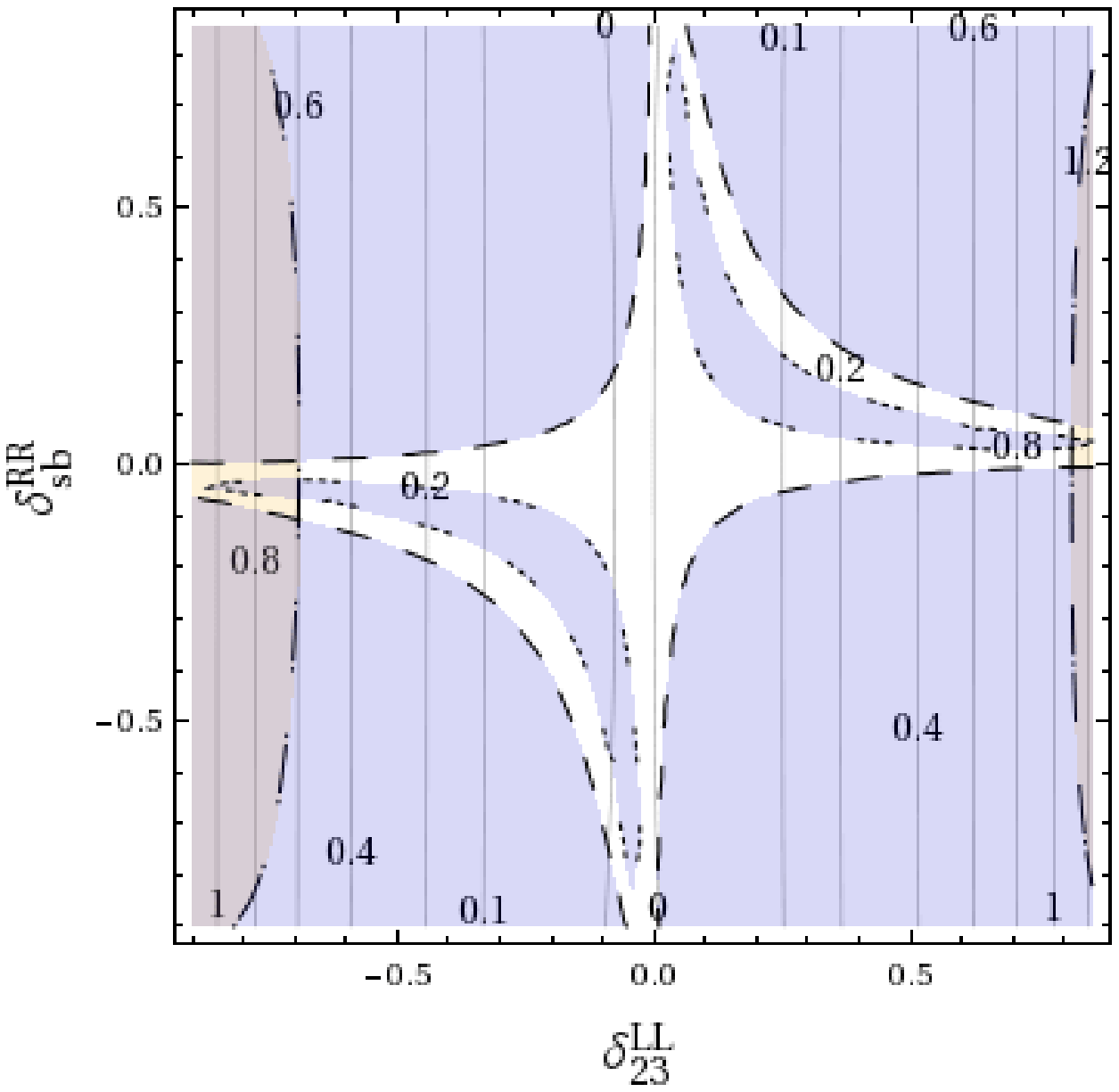}\\ 
\end{tabular}}}
\caption{$\Delta m_{h}$ (GeV) contour lines from our two deltas analysis for VHeavyS. The color code for the allowed/disallowed areas by $B$ data is given in fig.\ref{colleg}.} 
 \label{figdoubledeltaVHeavyS}
\end{figure}
%%%%%%%%%%%%%%%%%%%%%%%%%% F I G U R E %%%%%%%%%%%%%%%%%%%%%%%%%%%%%%%%%%%%%%%%
\clearpage
\newpage
%%%%%%%%%%%%%%%%%%%%%%%%%% F I G U R E %%%%%%%%%%%%%%%%%%%%%%%%%%%%%%%%%%%%%%%%
\begin{figure}[h!] 
\centering
\hspace*{-10mm} 
{\resizebox{14.4cm}{!} 
{\begin{tabular}{cc} 
\includegraphics[width=13.3cm]{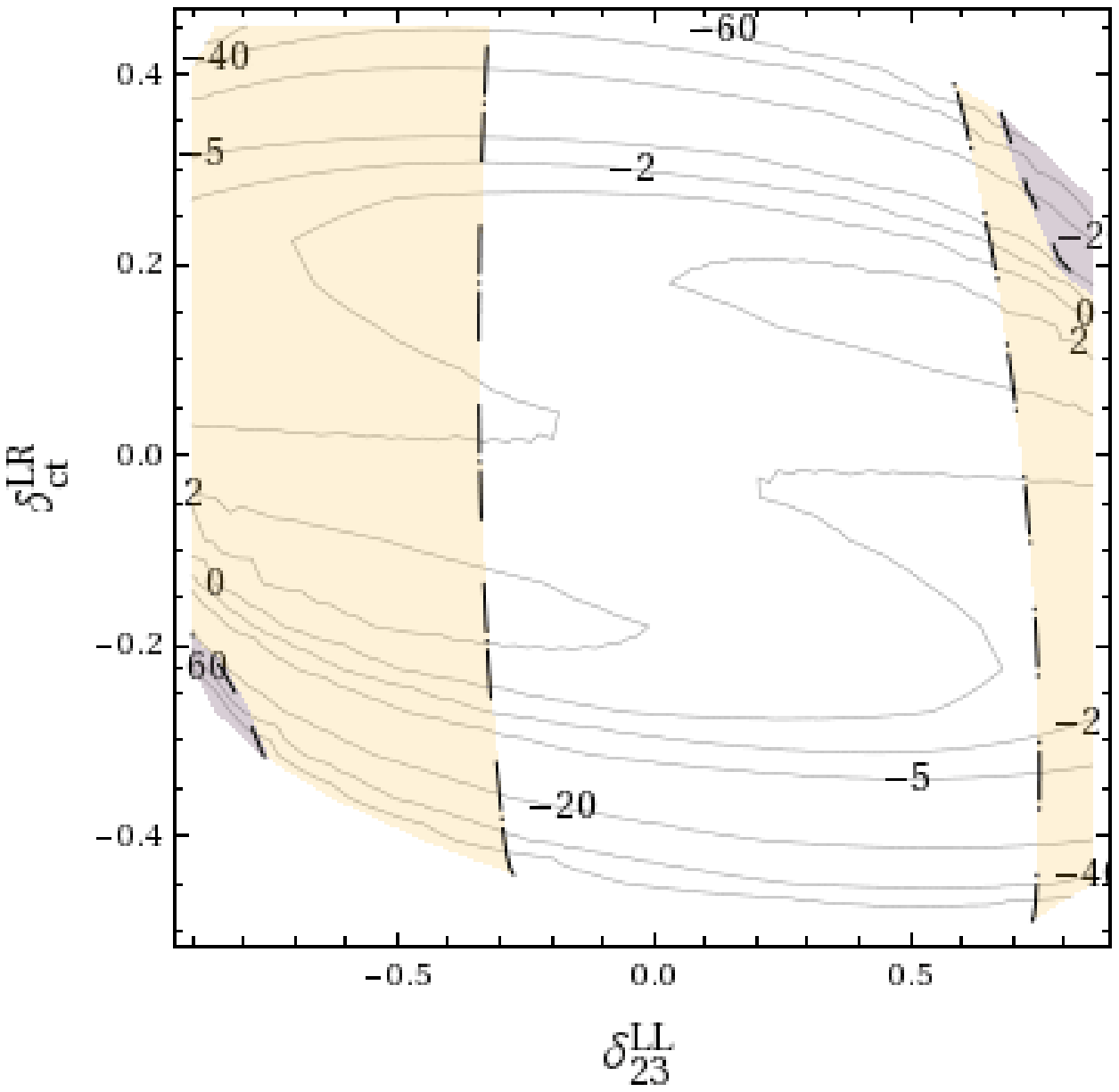}& 
\includegraphics[width=13.3cm]{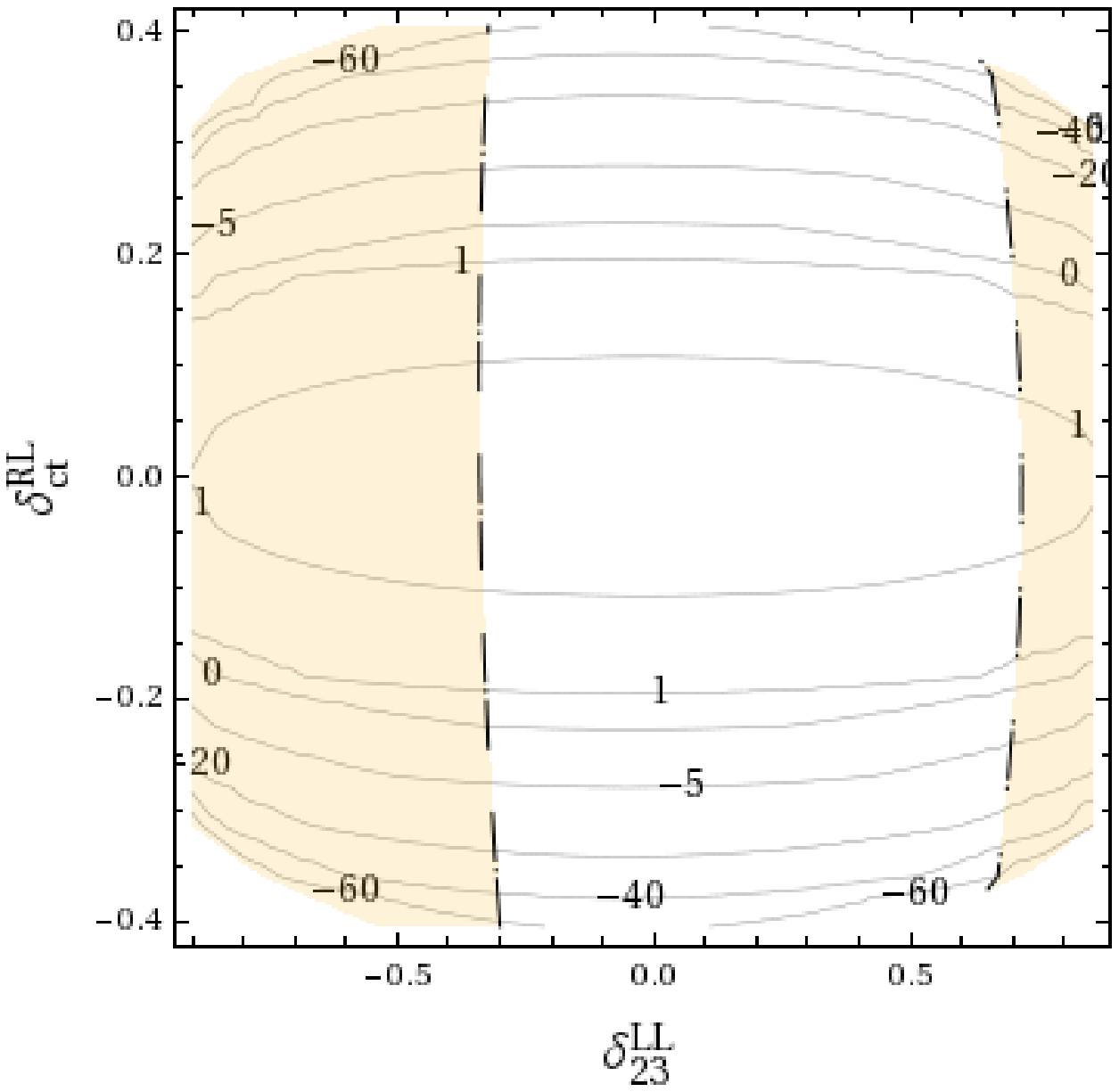}\\ 
\includegraphics[width=13.3cm]{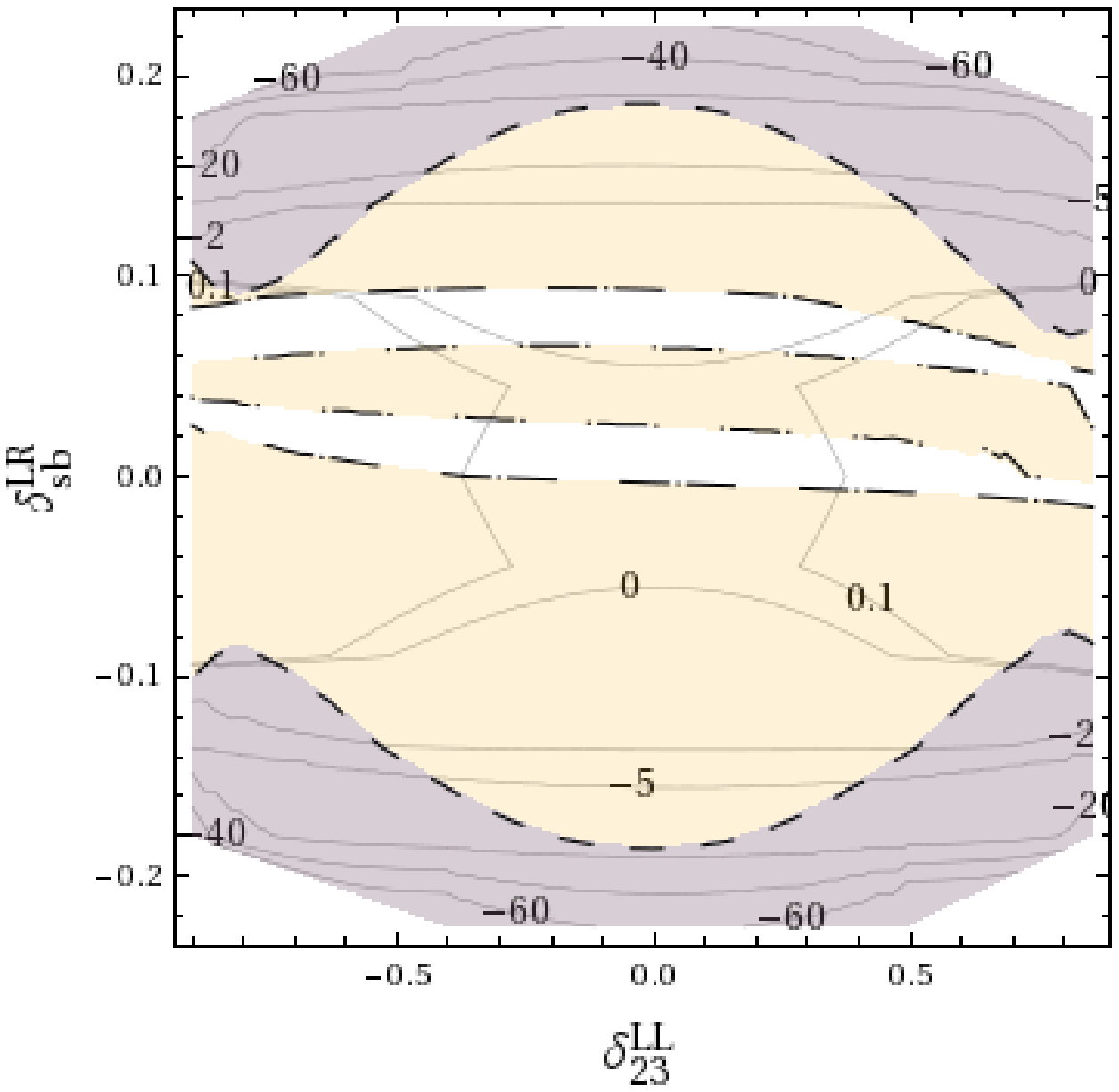}&
\includegraphics[width=13.3cm]{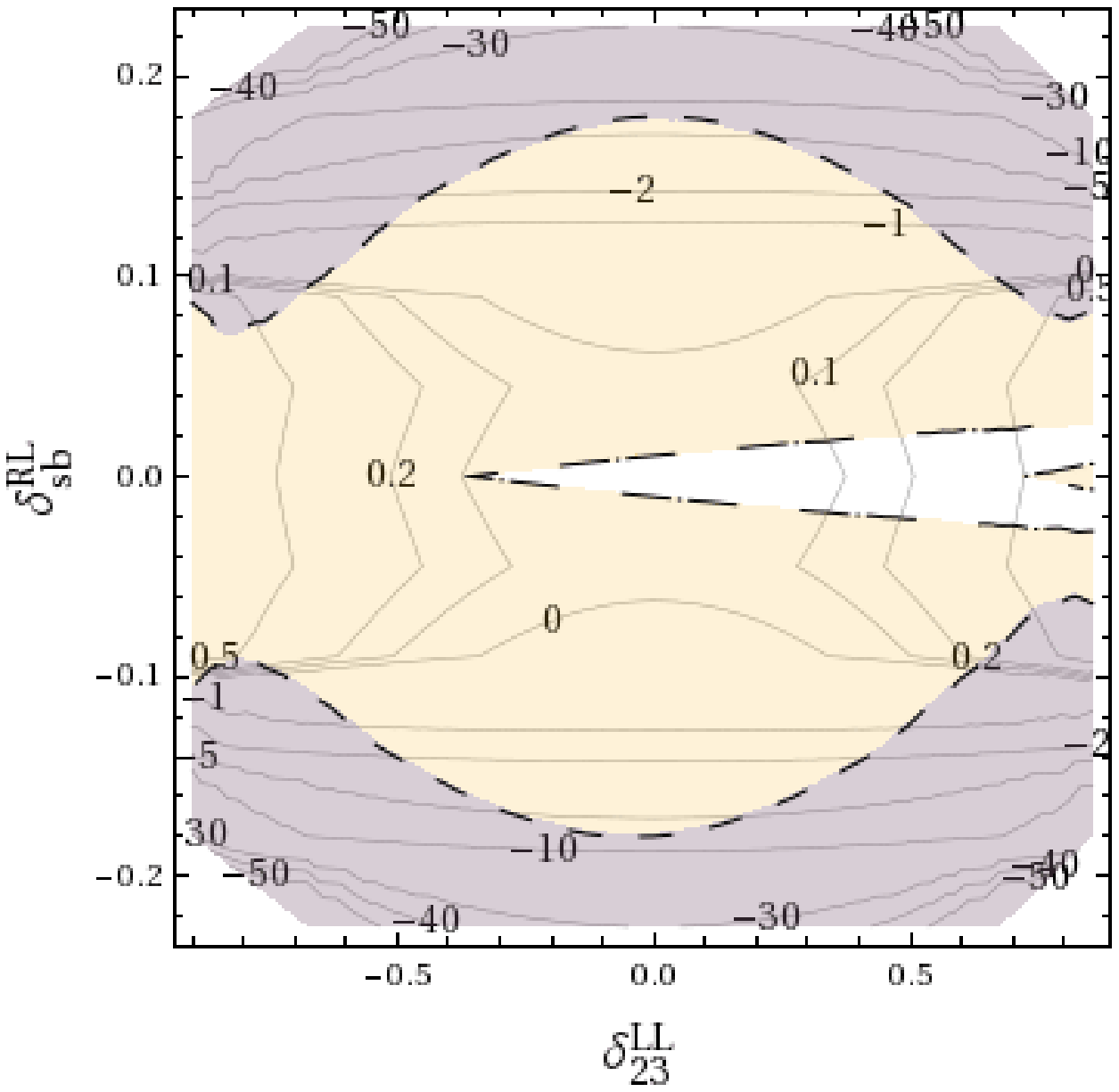}\\ 
\includegraphics[width=13.3cm]{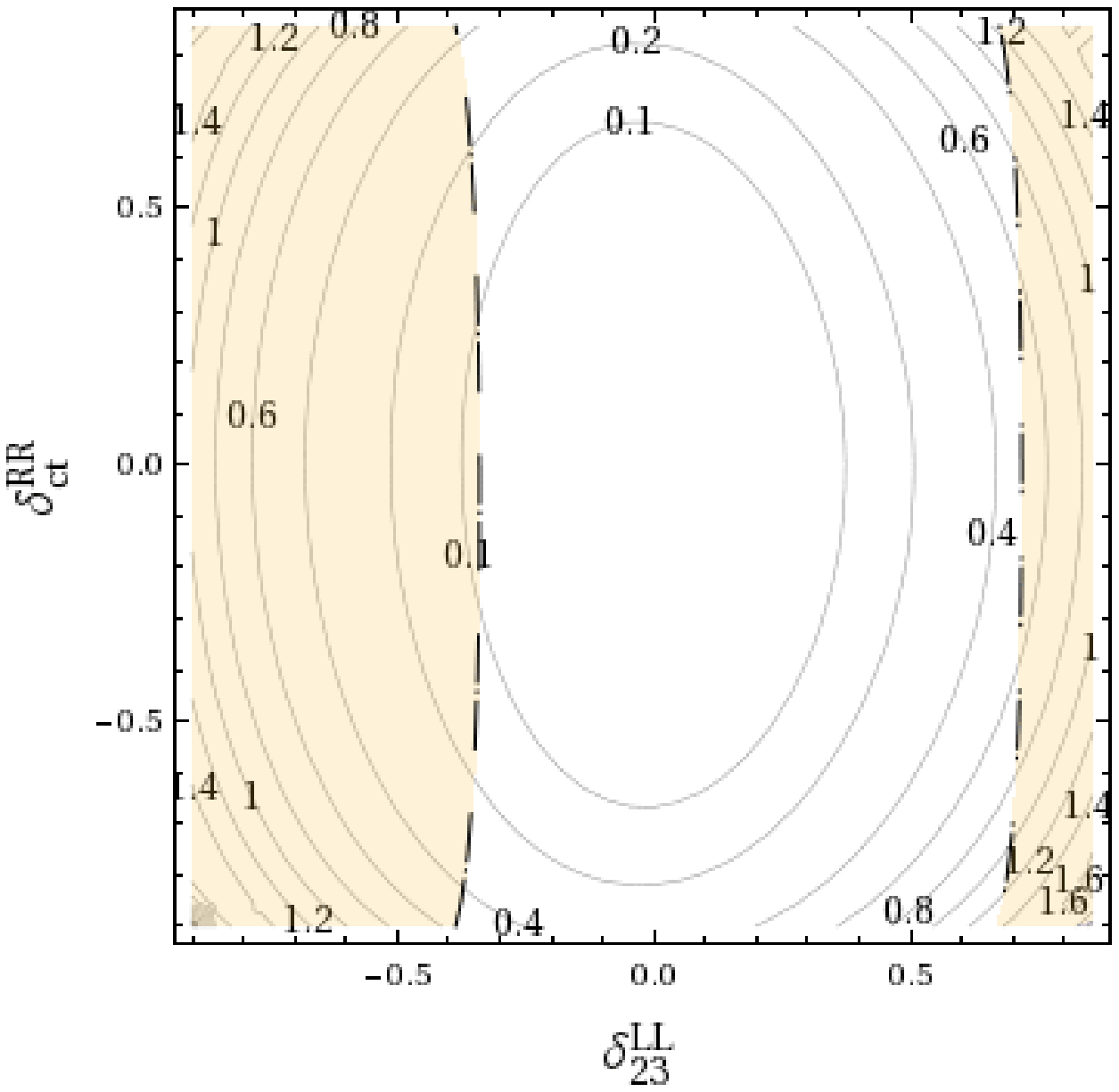}& 
\includegraphics[width=13.3cm]{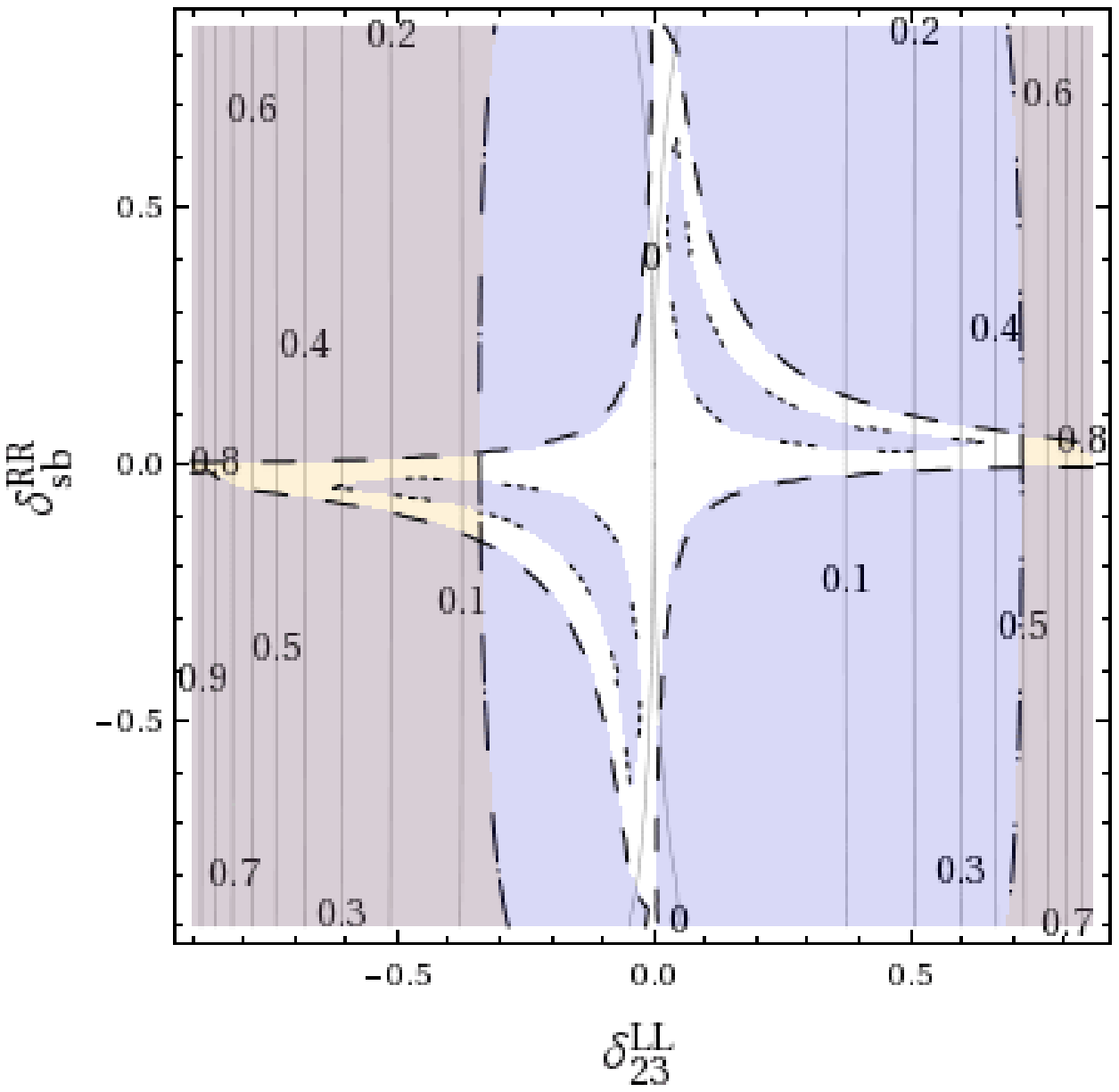}\\ 
\end{tabular}}}
\caption{$\Delta m_{h}$ (GeV) contour lines from our two deltas analysis for HeavySLightH. The color code for the allowed/disallowed areas by $B$ data is given in fig.\ref{colleg}.} 
\label{figdoubledeltaHeavySLightH}
\end{figure}
\clearpage
\newpage
%%%%%%%%%%%%%%%%%%%%%%%%%% F I G U R E %%%%%%%%%%%%%%%%%%%%%%%%%%%%%%%%%%%%%%%%
We have chosen in all plots $\de^{LL}_{23}$ as one of these non-vanishing deltas mainly because of two reasons. First, because it is one of the most frequently studied flavor changing parameters in the literature and, therefore, a convenient reference parameter. 
Second, because there are several well motivated SUSY scenarios, where this parameter gets the largest value, as we explained in section \ref{sec:nmfv}.

In these two-dimensional figures we have included the allowed/disallowed by
$B$ data areas that have been found by following the procedure explained in
section \ref{sec:Bphysics}, and the allowed intervals are given in
eqs.(\ref{bsglinearerr}), (\ref{bsmmlinearerr}) and (\ref{deltabslinearerr}). The
color code explaining the meaning of each colored area and the codes for the
discontinuous lines are given in fig.\ref{colleg}. Contour lines corresponding
to mass corrections above 60 GeV or below -60 GeV have not been represented. In several scenarios the plots involving $\de^{LR,RL}_{sb}$ 
show a seemingly abrupt behavior for $|\de^{LR,RL}_{sb}| \gsim 0.3$,
corresponding to extremely large (one-loop) corrections to $\mh$.
 In general, in the case of very large one-loop corrections, in order to get a more stable result further higher order corrections
would be required, as it is known from the higher-order corrections to
$\mh$ in the MFV case (see, e.g., \citere{mhiggsAEC}).
However, we cannot explore this possibility here. On the other hand, in order to understand the behavior of $\mh$ as a function of
$\de^{LR,RL}_{sb}$ a simple analytical formula would have to be
extracted from the general result. However, this is beyond the scope of
our paper.

The main conclusions from these two dimensional figures are summarized in the following:

The points that have been chosen in these plots are quite representative of all the different patterns found. The plots for SPS1a (not shown here) manifest  similar patterns as those of SPS3. The plots for SPS1b (not shown here) manifest similar patterns as those of BFP. The plots for SPS4 are not included because they do not manifest any allowed areas by $B$ data.

The largest mass corrections $\Dmh$ found, being allowed by $B$ data occur in plots $(\de^{LL}_{23},\de^{LR}_{ct} )$ and  $(\de^{LL}_{23},\de^{RL}_{ct} )$. This applies to all the studied points.
They can be as large as $(-50,-100)$ GeV at $\de^{LR}_{ct}$ or $\de^{RL}_{ct}$ close to the upper and lower horizontal bands in these plots where $\de^{LR}_{ct}$ or $\de^{RL}_{ct}$ are close to $\pm 0.5$. Again these large corrections from the $LR$ and $RL$ parameters are due to the $A$-terms, as we explained
at the end of section \ref{sec:numanalonedelta}.
Comparing the different plots, it can be seen that the size of the
allowed area by the $B$ data (the white area inside of the colored regions)
can be easily understood basically in terms of $\tb$, and the heaviness
of the SUSY and Higgs spectra. Generically, the plots  with largest
allowed regions and with largest Higgs mass corrections correspond to
scenarios with low $\tb = 5$ 
and heavy spectra. Consequently, the cases of VHeavyS and HeavySLightH
 are the most interesting ones, exhibiting very large radiative corrections,
 resulting from the heavy SUSY spectra. In the case of HeavySLightH
 the large corrections are not only found for $\Dmh$, but also, though to a 
lesser extent, for the other Higgs bosons,
 $\DmH$ and $\DmHp$ (not shown here). Consequently, in this scenario
 the deltas will be very restricted by the mass bounds, especially by
 $\mh$. 

There are also important corrections in the allowed areas of the two dimensional plots of   $(\de^{LL}_{23},\de^{RR}_{ct} )$ for some points, particularly for SPS5 (and to a lesser extent for SPS2). Here the corrections can be as large as -50 GeV in the upper and lower parts, i.e. for $\de^{RR}_{ct}$ close to $\pm 0.5$. In the case of SPS2 they can be up to -2 GeV for this same region.

As for the remaining two-dimensional plots they do not show relevant allowed areas where the mass corrections are interestingly large.

\section{Conclusions}
\label{sec:conclusions}

In this paper we have analyzed the 
one-loop corrections to the Higgs boson masses in the MSSM
with Non-Minimal Flavor Violation. 
We assume the flavor violation is being generated from the hypothesis of
general flavor mixing in the squark mass matrices, and these are parametrized
by a complete set of $\deXYij$ ($X,Y=L,R$; $i,j=t,c,u$ or $b,s,d$).

In the first step of the analysis we scanned over the NMFV parameters,
contrasting them with the experimental bounds on \bsg, \bmm\ and \dmbs.
We take into account the
most up-to-date evaluations in the NMFV MSSM for \bsg, \bmm\ and \dmbs,
as included in the BPHYSICS subroutine of the SuFla code~\cite{sufla}. 

For the evaluation of \dmbs\, we have added the
one-loop gluino boxes~\cite{Baek:2001kh} which are known to be very relevant
in the context of NMFV scenarios~\cite{Foster:2005wb,Gabbiani:1996hi,Becirevic:2001jj}. We have
estimated the size of these corrections and compared them with the other relevant contributions from
chargino boxes and double Higgs penguins for all values of $\tan \beta$ for the first time. And we have concluded
that gluino boxes dominate for moderate and low $\tan \beta \leq 20$ which is the 
interesting range for the present work.
In the final part of the $B$ physics analysis, we have evaluated in one-dimensional 
scans which intervals for the $\deXYij$
are still allowed in certain benchmark scenarios based on the SPS
points. 

In the second step we analyzed the one-loop contributions of NMFV to the
MSSM Higgs boson masses, focusing on the parameter space still allowed
by the experimental flavor constraints and by current limits from Higgs
boson searches. Here two relevant $\deXYij$ were varied simultaneously,
thus enlarging the allowed range for these parameters.
We found large corrections, mainly for the low  $\tan
\beta$ case, up to several tens of GeV for $\mh$ and somewhat smaller corrections for $\mH$ and $\mHp$. 
These corrections are specially relevant in the case of the light MSSM Higgs boson 
since they can be negative and up to two orders of magnitude larger than the anticipated LHC
precision. Consequently, these corrections must be taken into account in
any Higgs boson analysis in the NMFV MSSM framework. 
Conversely, in the case of a Higgs boson mass measurement these
corrections might be used to set further constraints on $\deXYij$. 
The present work clearly indicates that the flavor mixing parameters $\delta^{LR}_{ct}$ and  $\delta^{RL}_{ct}$ are
severely constrained by the present bounds on the lightest Higgs boson mass within the NMFV-MSSM scenarios.

%%%%%%%%%%%%%%%%%%%%%%%%%%%%%%%%%%%%%%%%%%%%%%%%%%%%%%%%%%%%%%%%%%%%%%%%%%%%%%

\vspace{-0.5em}
\subsection*{Acknowledgments}

We thank P. Paradisi and G. Isidori for kindly providing us the BPHYSICS
 subroutine and for helpful discussions. We thank T.~Hahn for invaluable
 help with FeynHiggs, as well as FeynArts 
 and FormCalc, and C.~Pena for helpful discussions on $B$-physics. We are indebted to P. Slavich for his valuable
comments and corrections regarding \bsg. We thank him for spotting the missing operators  $O'_{7,8}$  (eqs.(\ref{opo7p}) and (\ref{opo8p})) in a preliminary version of this paper. S.H.\ thanks A. Crivellin, L. Hofer and U. Nierste for interesting discussions.

The work of S.H.\ was supported in part by CICYT (grant FPA 2007--66387), in part by CICYT (grant FPA 2010--22163-C02-01) and by the Spanish MICINN's Consolider-Ingenio 2010 Program under grant MultiDark CSD2009-00064.
The work of M.H. and M.A.-C. was partially supported by CICYT (grant FPA2009-09017)
and  the Comunidad de Madrid project HEPHACOS, S2009/ESP-1473.
The work of S.P. was supported by a \textit{Ram{\'o}n y Cajal} contract 
from MEC (Spain) (PDRYC-2006-000930) and partially
by CICYT (grant FPA2009-09638), 
the Comunidad de Arag\'on project DCYT-DGA E24/2 and the Generalitat de Catalunya project 2009SGR502.
The work is also supported in part by  
the European Community's Marie-Curie Research
Training Network under contract MRTN-CT-2006-035505 `Tools and Precision Calculations for Physics Discoveries at Colliders' 
and also 
by the Spanish Consolider-Ingenio 2010 Programme CPAN (CSD2007-00042).

%%%%%%%%%%%%%%%%%%%%%%%%%%%%%%%%%%%%%%%%%%%%%%%%%%%%%%%%%%%%%%%%%%%%
\clearpage
\newpage
\begin{appendix}

\section*{Appendix A} 
%\subsection*{Appendix A}

We list the new Feynman rules of the NMFV scenario that are 
involved in the present computation.
The corresponding couplings to the Higgs boson $H$ are obtained
from the ones listened here for the lightest Higgs boson $h$ by replacing 

\begin{equation}
c_{\alpha}\rightarrow s_{\alpha}\quad;\quad s_{\alpha}\rightarrow-c_{\alpha}\quad;\quad s_{\alpha+\beta}\rightarrow-c_{\alpha+\beta}\quad;\quad c_{2\alpha}\rightarrow-c_{2\alpha}\label{eq:cambio}
\end{equation}

The notation used for the formulas is the following:
$s_{x}=\sin x;\, c_{x}=\cos x;\, \sw=\sin\theta_{W};\, \cw=\cos\theta_{W}=\frac{\MW}{\MZ};\, t_{\beta}=\tan\beta$.\\

{\it{1. Couplings of two squarks and one/two Higgs bosons}}\\

\begin{table}[H]
\begin{tabular}{ll}
\parbox[c]{1em}{\includegraphics{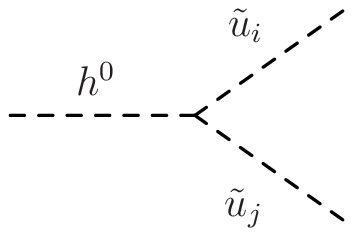}}
 & -$\sum_{k,l=1}^{3} \frac{ie}{6\MW\cw\sw s_{\beta}}
\left\{ R^{\tilde{u}\,*}_{i,k}\left\{ \delta_{kl} R^{\tilde{u}}_{j,l} 
\left(6\ca\cw {m_{{u}_{k}}^2}-\MW\MZ s_{\alpha+\beta} s_{\beta} 
(3-4\sw^2)\right)\right.
\right.$\vspace*{-0.6cm}\\
&\hspace*{3.0cm}$\left.+3\cw R^{\tilde{u}}_{j,3+l}(A^{u}_{k,l}\ca\, 
m_{{u}_{k}}+
\delta_{kl}\,m_{{u}_{k}}\,\mu^{*}\sa)
\right\}$\\
&\hspace*{3.0cm}$+R^{\tilde{u}\,*}_{i,3+k}\left\{
\delta_{kl} R^{\tilde{u}}_{j,3+l} 
\left(6\ca\cw {m_{{u}_{k}}^2}-4\MW\MZ s_{\alpha+\beta} s_{\beta}\sw^2
\right)\right.$\\
&\hspace*{3.0cm}$\left.\left.+
3\cw R^{\tilde{u}}_{j,l}(A^{u*}_{l,k}\ca\, m_{{u}_{l}}+
\delta_{kl}\,m_{{u}_{k}}\,\mu\sa)
\right\}
\right\}$\\
&\\
\parbox[c]{1em}{\includegraphics{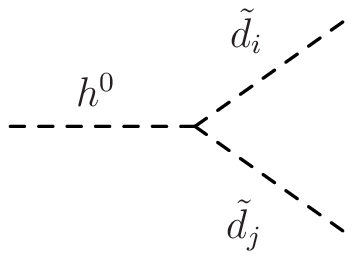}}
& $\sum_{k,l=1}^{3} \frac{ie}{6\MW\cw\sw c_{\beta}}
\left\{ R^{\tilde{d}\,*}_{i,k}\left\{ \delta_{kl} R^{\tilde{d}}_{j,l}
\left(6\sa\cw {m_{{d}_{k}}^2}-\MW\MZ s_{\alpha+\beta} c_{\beta} 
(3-2\sw^2)\right)\right.
\right.$\vspace*{-0.6cm}\\
&\hspace*{3.0cm}$\left.+3\cw R^{\tilde{d}}_{j,3+l}(A^{d}_{k,l}\sa\, 
m_{{d}_{k}}+
\delta_{kl}\,m_{{d}_{k}}\,\mu^{*}\ca)
\right\}$\\
&\hspace*{3.0cm}$+R^{\tilde{d}\,*}_{i,3+k}\left\{
\delta_{kl} R^{\tilde{d}}_{j,3+l} 
\left(6\sa\cw {m_{{d}_{k}}^2}-2\MW\MZ s_{\alpha+\beta} c_{\beta}\sw^2
\right)\right.$\\
&\hspace*{3.0cm}$\left.\left.+
3\cw R^{\tilde{d}}_{j,l}(A^{d*}_{l,k}\sa\, m_{{d}_{l}}+
\delta_{kl}\,m_{{d}_{k}}\,\mu\ca)
\right\}
\right\}$\\
&\\
\parbox[c]{1em}{\includegraphics{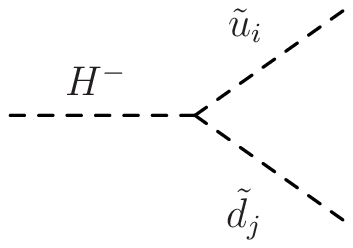}}
&$\sum_{k,l=1}^{3} \frac{ie}{\sqrt{2}\MW\sw t_{\beta}}
\left\{ R^{\tilde{u}\,*}_{i,3+k}\left\{ R^{\tilde{d}}_{j,l} 
\left(\sum_{n=1}^{3} A^{u*}_{n,k} \, m_{{u}_{n}} {V_{\rm{CKM}}^{*\,nl}}+
 m_{{u}_{k}}\,\mu {V_{\rm{CKM}}^{*\,kl}} t_{\beta}\right)\right.\right.$
\vspace*{-0.6cm}\\
&\hspace*{3.0cm}$\left.+ m_{{d}_{l}} m_{{u}_{k}} {V_{\rm{CKM}}^{*\,kl}}
R^{\tilde{d}}_{j,3+l}(1+{t_{\beta}}^{2})\right\}$\\
&\hspace*{3.0cm}$+R^{\tilde{u}\,*}_{i,k}\left\{ R^{\tilde{d}}_{j,3+l}t_{\beta} 
\left(\sum_{n=1}^{3} A^{d}_{n,l} \, m_{{d}_{n}} {V_{\rm{CKM}}^{*\,kn}} 
t_{\beta}+
 m_{{d}_{l}}\,\mu^{*} {V_{\rm{CKM}}^{*\,kl}}\right)\right.$\\
&\hspace*{3.0cm}$\left.\left.+ {V_{\rm{CKM}}^{*\,kl}}\, R^{\tilde{d}}_{j,l}
\left(m^{2}_{{u}_{k}}-t_{\beta}(\MW^2 s_{2\beta}-m^{2}_{{d}_{l}}t_{\beta})
\right)\right\}\right\}$\\
&\\
{\includegraphics{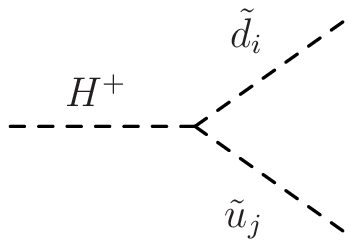}}
 &\put(10,30){$\sum_{k,l=1}^{3} \frac{ie}{\sqrt{2}\MW\sw t_{\beta}}
\left\{ R^{\tilde{d}\,*}_{i,3+l}\left\{ R^{\tilde{u}}_{j,k} t_{\beta}
\left(\sum_{n=1}^{3} A^{d*}_{n,l} \, m_{{d}_{n}} {V^{kn}_{\rm{CKM}}}t_{\beta}+
 m_{{d}_{l}}\,\mu {V^{kl}_{\rm{CKM}}}\right)\right.\right.$}
\vspace*{-0.6cm}\\
&\hspace*{3.0cm}$\left.+ m_{{d}_{l}} m_{{u}_{k}} {V^{kl}_{\rm{CKM}}}
R^{\tilde{u}}_{j,3+k}(1+{t_{\beta}}^{2})\right\}$\\
&\hspace*{3.0cm}$+R^{\tilde{d}\,*}_{i,l}\left\{ R^{\tilde{u}}_{j,3+k} 
\left(\sum_{n=1}^{3} A^{u}_{n,k} \, m_{{u}_{n}} {V^{nl}_{\rm{CKM}}}+
 m_{{u}_{k}}\,\mu^{*} {V^{kl}_{\rm{CKM}}}t_{\beta}\right)\right.$\\
&\hspace*{3.0cm}$\left.\left.+ {V^{kl}_{\rm{CKM}}}\, R^{\tilde{u}}_{j,k}
\left(m^{2}_{{u}_{k}}-t_{\beta}(\MW^2 s_{2\beta}-m^{2}_{{d}_{l}}t_{\beta})
\right)\right\}\right\}$\\
\end{tabular}
\end{table}

%\newpage
\begin{table}[H]
\begin{tabular}{ll}
\parbox[c]{1em}{\includegraphics{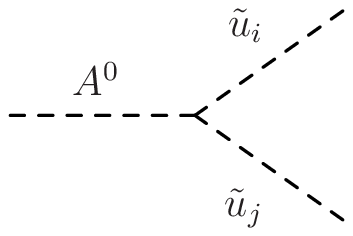}}
& -$\sum_{k,l=1}^{3} \frac{e}{2\MW\sw t_{\beta}}
\left\{ R^{\tilde{u}\,*}_{i,3+k} R^{\tilde{u}}_{j,l}
\left(A^{u\,*}_{l,k}\,m_{{u}_{l}}+\delta_{kl}\,m_{{u}_{k}}\,\mu\, t_{\beta}
\right)\right.$\vspace*{-0.6cm}\\
&\hspace*{3.0cm}$\left.-
R^{\tilde{u}\,*}_{i,k} R^{\tilde{u}}_{j,3+l}
\left(A^{u}_{k,l}{m_{u}}_{k}+\delta_{kl}\,m_{{u}_{k}}\,\mu^{*}\,t_{\beta}
\right)\right\}$\\
&\\
&\\
\parbox[c]{1em}{\includegraphics{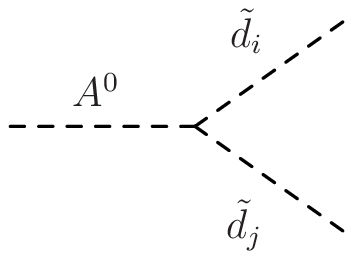}}
& -$\sum_{k,l=1}^{3} \frac{e}{2\MW\sw}
\left\{ R^{\tilde{d}\,*}_{i,3+k} R^{\tilde{d}}_{j,l}
\left(A^{d\,*}_{l,k}{m_{d}}_{l}t_{\beta}+\delta_{kl}\,m_{{d}_{k}}\mu
\right)\right.$\vspace*{-0.6cm}\\
&\hspace*{3.0cm}$\left.-
R^{\tilde{d}\,*}_{i,k} R^{\tilde{d}}_{j,3+l}
\left(A^{d}_{k,l}\,m_{{d}_{k}}\,t_{\beta}+
\delta_{kl}\,m_{{d}_{k}}\,\mu^{*}
\right)\right\}$\\
&\\
&\\
\parbox[c]{1em}{\includegraphics{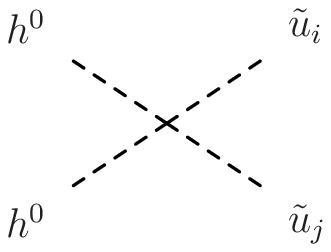}}
 & -$\sum_{k=1}^{3} \frac{ie^2}{12\MW^2\cw^2\sw^2 s^{2}_{\beta}}
\left\{ R^{\tilde{u}}_{i,k} R^{\tilde{u}\,*}_{j,k}
\left(6\ca^2\cw^2 {m_{{u}_{k}}^2}-c_{2\alpha}\MW^2 s^{2}_{\beta} 
(3-4\sw^2)\right)
\right.$\vspace*{-0.6cm}\\
&\hspace*{3.0cm}$\left.+2 R^{\tilde{u}}_{i,3+k} R^{\tilde{u}\,*}_{j,3+k}
\left(3\ca^2\cw^2 {m_{{u}_{k}}^2}-2c_{2\alpha}\MW^2 s^{2}_{\beta}\sw^2
\right)\right\}$\\
&\\
\parbox[c]{1em}{\includegraphics{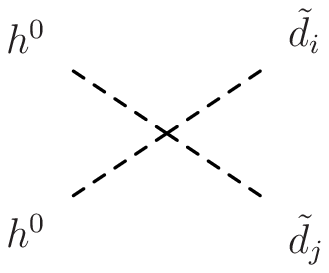}}
 &$-\sum_{k=1}^{3} \frac{ie^2}{12\MW^2\cw^2\sw^2 c^{2}_{\beta}}
\left\{ R^{\tilde{d}}_{i,k} R^{\tilde{d}\,*}_{j,k}
\left(6\sa^2\cw^2 {m_{{d}_{k}}^2}+c_{2\alpha}\MW^2 c^{2}_{\beta} 
(3-2\sw^2)\right)
\right.$\vspace*{-0.6cm}\\
&\hspace*{3.0cm}$\left.+2 R^{\tilde{d}}_{i,3+k} R^{\tilde{d}\,*}_{j,3+k}
\left(3\sa^2\cw^2 {m_{{d}_{k}}^2}+c_{2\alpha}\MW^2 c^{2}_{\beta}\sw^2
\right)\right\}$\\
&\\
\parbox[c]{1em}{\includegraphics{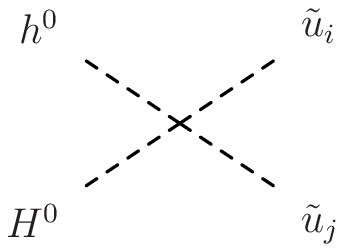}}
 & -$\sum_{k=1}^{3} \frac{ie^2 s_{2\alpha}}{12\MW^2\cw^2\sw^2 s^{2}_{\beta}}
\left\{ R^{\tilde{u}}_{i,k} R^{\tilde{u}\,*}_{j,k}
\left(3\cw^2 {m_{{u}_{k}}^2}-\MW^2 s^{2}_{\beta} 
(3-4\sw^2)\right)
\right.$\vspace*{-0.6cm}\\
&\hspace*{3.0cm}$\left.+R^{\tilde{u}}_{i,3+k} R^{\tilde{u}\,*}_{j,3+k}
\left(3\cw^2 {m_{{u}_{k}}^2}-4\MW^2 s^{2}_{\beta}\sw^2
\right)\right\}$\\
&\\
\parbox[c]{1em}{\includegraphics{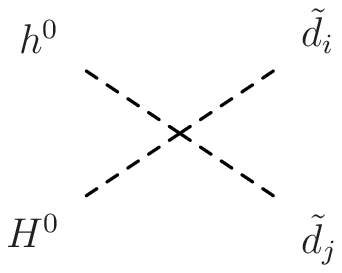}}
 &$\sum_{k=1}^{3} \frac{ie^2 s_{2\alpha}}{12\MW^2\cw^2\sw^2 c^{2}_{\beta}}
\left\{ R^{\tilde{d}}_{i,k} R^{\tilde{d}\,*}_{j,k}
\left(3\cw^2 {m_{{d}_{k}}^2}-\MW^2 c^{2}_{\beta} 
(3-2\sw^2)\right)
\right.$\vspace*{-0.6cm}\\
&\hspace*{3.0cm}$\left.+ R^{\tilde{d}}_{i,3+k} R^{\tilde{d}\,*}_{j,3+k}
\left(3\cw^2 {m_{{d}_{k}}^2}-2\MW^2 c^{2}_{\beta}\sw^2
\right)\right\}$\\
&\\
%\parbox[c]{1em}
{\includegraphics{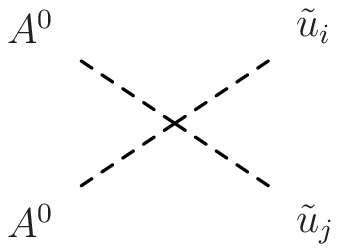}}
 &\put(10,30){$-\sum_{k=1}^{3} \frac{ie^2}{12\MW^2\cw^2\sw^2 t^{2}_{\beta}}
\left\{ R^{\tilde{u}}_{i,k} R^{\tilde{u}\,*}_{j,k}
\left(6\cw^2 {m_{{u}_{k}}^2}-c_{2\beta}\MW^2 t^{2}_{\beta} 
(3-4\sw^2)\right)
\right.$}\vspace*{-0.6cm}\\
&\hspace*{3.0cm}$\left.+2 R^{\tilde{u}}_{i,3+k} R^{\tilde{u}\,*}_{j,3+k}
\left(3\cw^2 {m_{{u}_{k}}^2}-2c_{2\beta}\MW^2 t^{2}_{\beta}\sw^2
\right)\right\}$\\
&\\
&\\
\end{tabular}
\end{table}

\begin{table}[H]
\begin{tabular}{ll}
\parbox[c]{1em}{\includegraphics{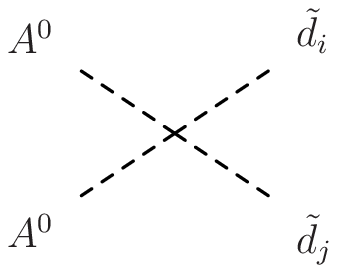}}
 &$-\sum_{k=1}^{3} \frac{ie^2}{12\MW^2\cw^2\sw^2}
\left\{ R^{\tilde{d}}_{i,k} R^{\tilde{d}\,*}_{j,k}
\left(6\cw^2 {m_{{d}_{k}}^2}t^{2}_{\beta}+c_{2\beta}\MW^2
(3-2\sw^2)\right)
\right.$\vspace*{-0.6cm}\\
&\hspace*{3.0cm}$\left.+2 R^{\tilde{d}}_{i,3+k} R^{\tilde{d}\,*}_{j,3+k}
\left(3\cw^2 {m_{{d}_{k}}^2}t^{2}_{\beta}+c_{2\beta}\MW^2 \sw^2
\right)\right\}$\\
&\\
\parbox[c]{1em}{\includegraphics{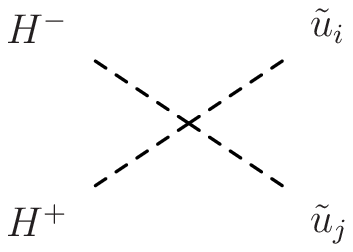}}
 &$-\sum_{k,l=1}^{3} \frac{ie^2}{12\MW^2\sw^2\cw^2 t^{2}_{\beta}}
\left\{ R^{\tilde{u}\,*}_{i,k}R^{\tilde{u}}_{j,l}t^{2}_{\beta}
\left(6\sum_{n=1}^{3} m^{2}_{{d}_{n}} {V_{\rm{CKM}}^{*\,kn}} {V^{ln}_{\rm{CKM}}}
\cw^2t^{2}_{\beta}\right.\right.$
\vspace*{-0.6cm}\\
&\hspace*{3.0cm}$\left.+c_{2\beta}\delta_{kl}\MW^2(1+2\cw^2)\right)$\\
&\hspace*{3.0cm}$\left.+2\delta_{kl}
R^{\tilde{u}\,*}_{i,3+k}R^{\tilde{u}}_{j,3+l}
\left(3\cw^2m^{2}_{{u}_{k}}-2c_{2\beta}\MW^2\sw^2t^{2}_{\beta}\right)
\right\}$\\
&\\
{\includegraphics{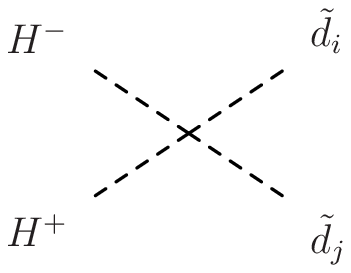}}
 &\put(10,30){$-\sum_{k,l=1}^{3} \frac{ie^2}{12\MW^2\sw^2\cw^2 t^{2}_{\beta}}
\left\{ R^{\tilde{d}\,*}_{i,k}R^{\tilde{d}}_{j,l}
\left(6\sum_{n=1}^{3} m^{2}_{{u}_{n}} {V^{nk}_{\rm{CKM}}} {V_{\rm{CKM}}^{*\,nl}}
\cw^2
\right.\right.$}
\vspace*{-0.6cm}\\
&\hspace*{3.0cm}$\left.+c_{2\beta}\delta_{kl}\MW^2\,t^{2}_{\beta}
(1-4\cw^2)\right)$\\
&\hspace*{3.0cm}$\left.+2\delta_{kl}
R^{\tilde{d}\,*}_{i,3+k}R^{\tilde{d}}_{j,3+l} t^{2}_{\beta}
\left(3\cw^2 t^{2}_{\beta} m^{2}_{{d}_{k}}+c_{2\beta}\MW^2\sw^2\right)
\right\}$\\
&\\
&\\
\end{tabular}
\end{table}

{\it{3. Couplings of two squarks and one/two gauge bosons}}\\

\begin{table}[H]
\begin{tabular}{ll}
\parbox[c]{1em}{\includegraphics{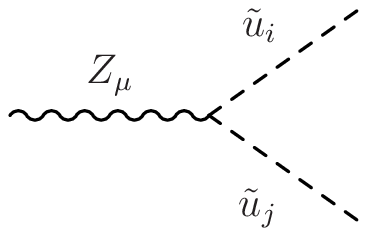}}
 & $\sum_{k=1}^{3} \frac{ie}{6\cw\sw}
\left( 4 R^{\tilde{u}\,*}_{i,3+k} R^{\tilde{u}}_{j,3+k}\sw^2-
R^{\tilde{u}\,*}_{i,k} R^{\tilde{u}}_{j,k} (3-4\sw^2)
\right)(p+p^{'})_{\mu}$\\
&\\
\parbox[c]{1em}{\includegraphics{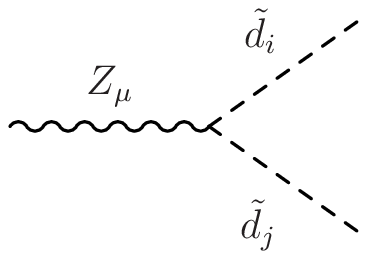}}
& $-\sum_{k=1}^{3} \frac{ie}{6\cw\sw}
\left( 2 R^{\tilde{d}\,*}_{i,3+k} R^{\tilde{d}}_{j,3+k}\sw^2-
R^{\tilde{d}\,*}_{i,k} R^{\tilde{d}}_{j,k} (3-2\sw^2)
\right)(p+p^{'})_{\mu}$\\
&\\
{\includegraphics{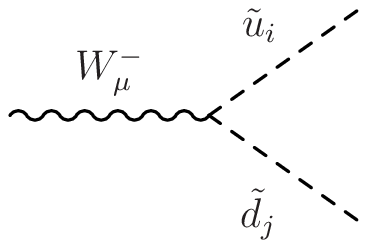}}
 &\put(10,30){$-\sum_{k,l=1}^{3} \frac{ie}{\sqrt{2}\sw}
 {V_{\rm{CKM}}^{*\,kl}}R^{\tilde{u}\,*}_{i,k} R^{\tilde{d}}_{j,l}
\,(p+p^{'})_{\mu}$}\\
&\\
\end{tabular}
\end{table}

\begin{table}[H]
\begin{tabular}{ll}
\parbox[c]{1em}{\includegraphics{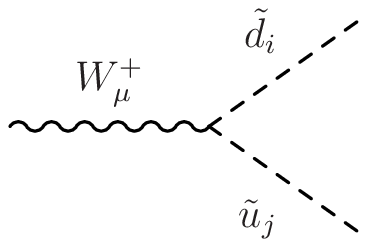}}
 & $-\sum_{k,l=1}^{3} \frac{ie}{\sqrt{2}\sw}
 {V_{\rm{CKM}}^{kl}}R^{\tilde{u}}_{j,k} R^{\tilde{d\,*}}_{i,l}\,(p+p^{'})_{\mu}$\\
&\\
\parbox[c]{1em}{\includegraphics{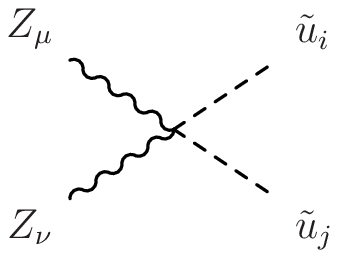}}
& $\sum_{k=1}^{3} \frac{i e^2}{18\cw^2\sw^2}
\left( R^{\tilde{u}\,*}_{i,k} R^{\tilde{u}}_{j,k} (3-4\sw^2)^2+
16 R^{\tilde{u}\,*}_{i,3+k} R^{\tilde{u}}_{j,3+k} \sw^4
\right)g_{\mu \nu}$\\
&\\
\parbox[c]{1em}{\includegraphics{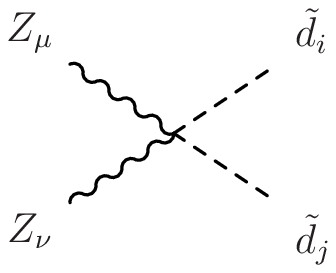}}
& {$\sum_{k=1}^{3} \frac{i e^2}{18\cw^2\sw^2}
\left( R^{\tilde{d}\,*}_{i,k} R^{\tilde{d}}_{j,k} (3-2\sw^2)^2+
4 R^{\tilde{d}\,*}_{i,3+k} R^{\tilde{d}}_{j,3+k} \sw^4
\right)g_{\mu \nu}$}\\
&\\
\parbox[c]{1em}{\includegraphics{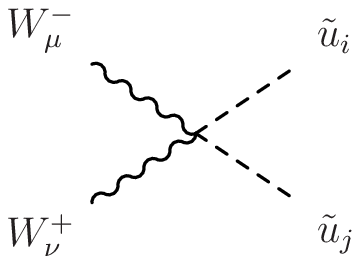}}
 & $\sum_{k=1}^{3} \frac{ie^2}{2\sw^2}
 R^{\tilde{u}\,*}_{i,k} R^{\tilde{u}}_{j,k}\,g_{\mu \nu}$\\
&\\
{\includegraphics{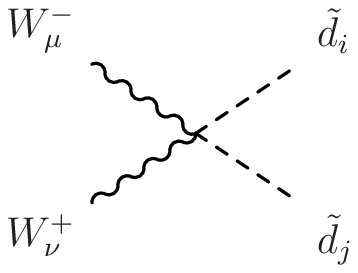}}
 &\put(10,30){$\sum_{k=1}^{3} \frac{ie^2}{2\sw^2}
 R^{\tilde{d\,*}}_{i,k} R^{\tilde{d}}_{j,k}\,g_{\mu \nu}$}\\
&\\
&\\
\end{tabular}
\end{table}

%%%%%%%%%%%%%%%%%%%%%%%%%%%%%%%%%%%%%%%%%%%%%%%%%%%%%%%%%%%%%%%%%%%%%%%%%%%%%%%
\clearpage
\newpage
\section*{Appendix B}

All the following Feynman diagrams have been calculated using FeynArts
3.5~\cite{feynarts} and FormCalc 6.0~\cite{formcalc}.
The notation used here is the same as in Appendix A. Furthermore we use the functions~\cite{a0b0c0}

\begin{equation}
\frac{i}{16\pi}A_{0}\left[m^{2}\right]\equiv\int\frac{\mu^{4-D}d^{D}k}{\left(2\pi\right)^{D}}\frac{1}{k^{2}-m^{2}}\end{equation}

\begin{equation}
\frac{i}{16\pi}B_{0}\left[p^{2},m_{1}^{2},m_{2}^{2}\right]\equiv\int\frac{\mu^{4-D}d^{D}k}{\left(2\pi\right)^{D}}\frac{1}{\left[k^{2}-m_{1}^{2}\right]\left[\left(k+p\right)^{2}-m_{2}^{2}\right]}\end{equation}

\begin{equation}
\frac{i}{16\pi}p^{2}B_{1}\left[p^{2},m_{1}^{2},m_{2}^{2}\right]\equiv\int\frac{\mu^{4-D}d^{D}k}{\left(2\pi\right)^{D}}\frac{pk}{\left[k^{2}-m_{1}^{2}\right]\left[\left(k+p\right)^{2}-m_{2}^{2}\right]}\end{equation}

The generic diagrams have been ordered according to its topologies, and the
particles involved in the internal loops (quarks $q$ or squarks $\tilde{q}$).
The diagrams can be found in fig \ref{figfdall}. The complete
self-energy can be expressed as a sum of three parts:
\begin{equation}
\Sigma_{\phi\phi^{\prime}}=\Sigma_{\phi\phi^{\prime}}^{2q}+\Sigma_{\phi\phi^{\prime}}^{2\tilde{q}}+\Sigma_{\phi\phi^{\prime}}^{1\tilde{q}}\qquad\Sigma_{VV}=\Sigma_{VV}^{2q}+\Sigma_{VV}^{2\tilde{q}}+\Sigma_{VV}^{1\tilde{q}}\qquad T_{\phi}=T_{\phi}^{q}+T_{\phi}^{\tilde{q}}\end{equation}
where $\phi,\,\phi^{\prime}=h,\, H,\, A,\, H^{\pm}$ and
$V=W,\, Z$. All the self-energies $\Sigma$ correspond to 
$\Sigma\left(p^{2}\right)$. The self-energies for $H$ are 
obtained by the replacements of eq.\ref{eq:cambio} on the results of $h$:

\begin{figure}[H]
\centering
\begin{tabular}{ccc}
\includegraphics[scale=0.33]{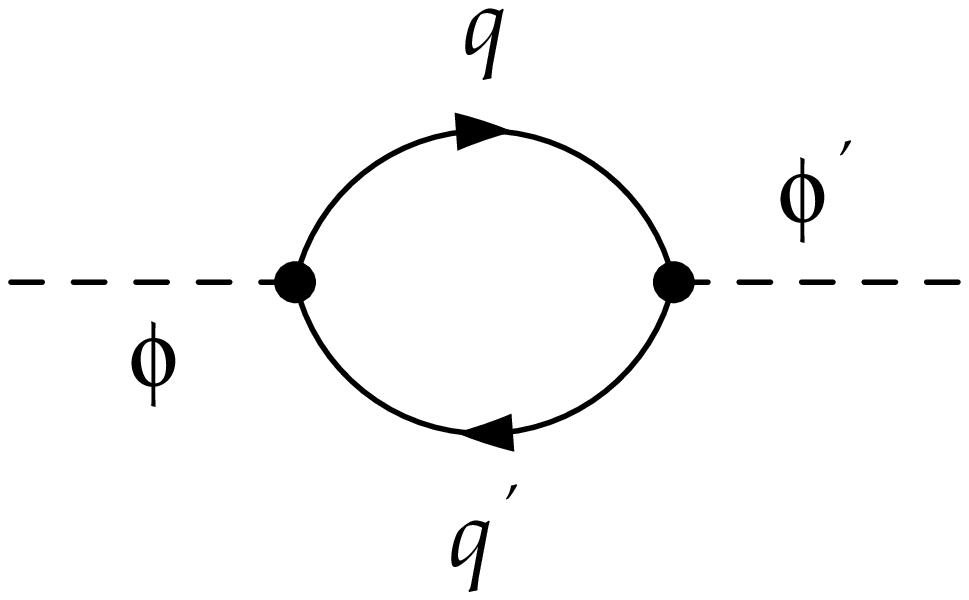}&
\includegraphics[scale=0.33]{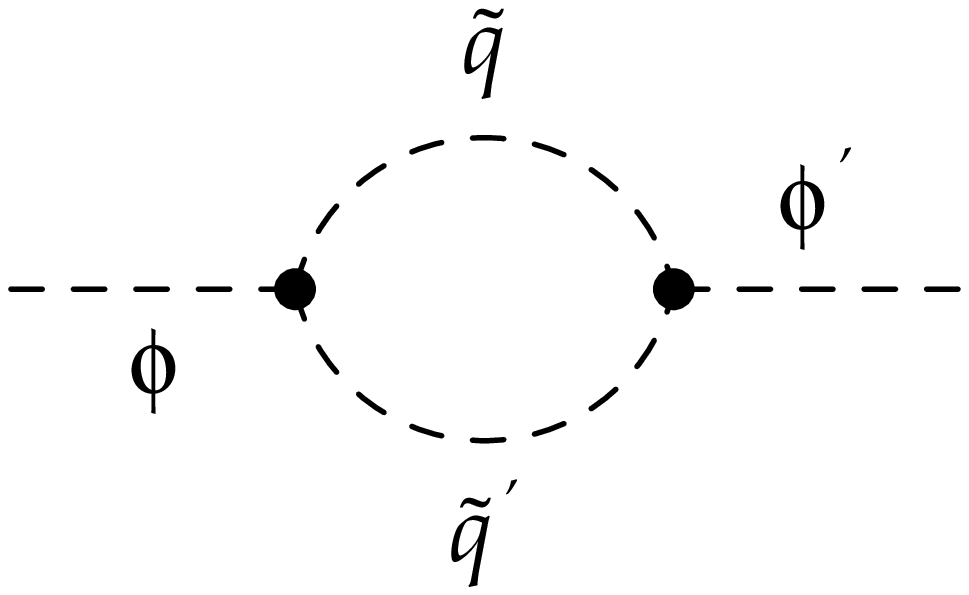}&
\includegraphics[scale=0.33]{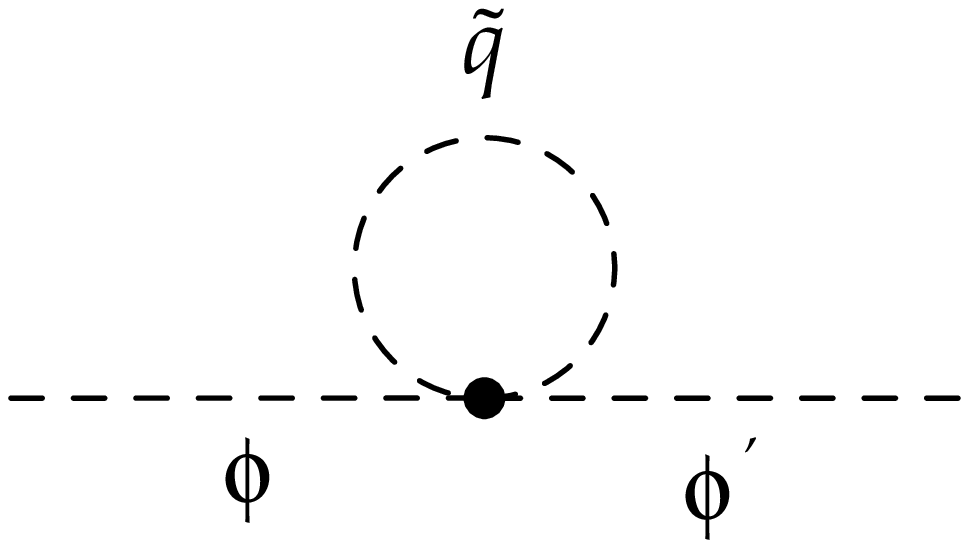}\\
\includegraphics[scale=0.33]{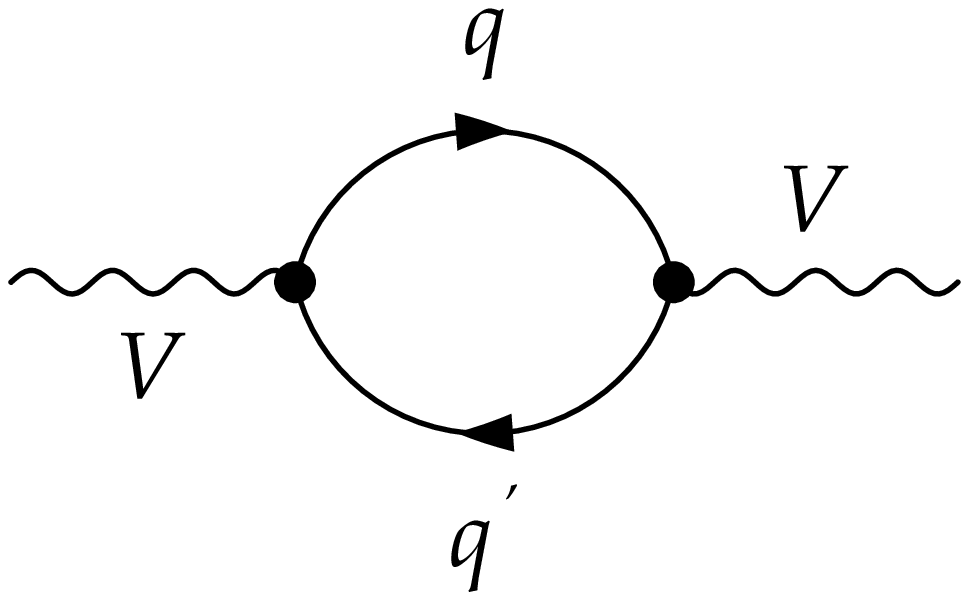}&
\includegraphics[scale=0.33]{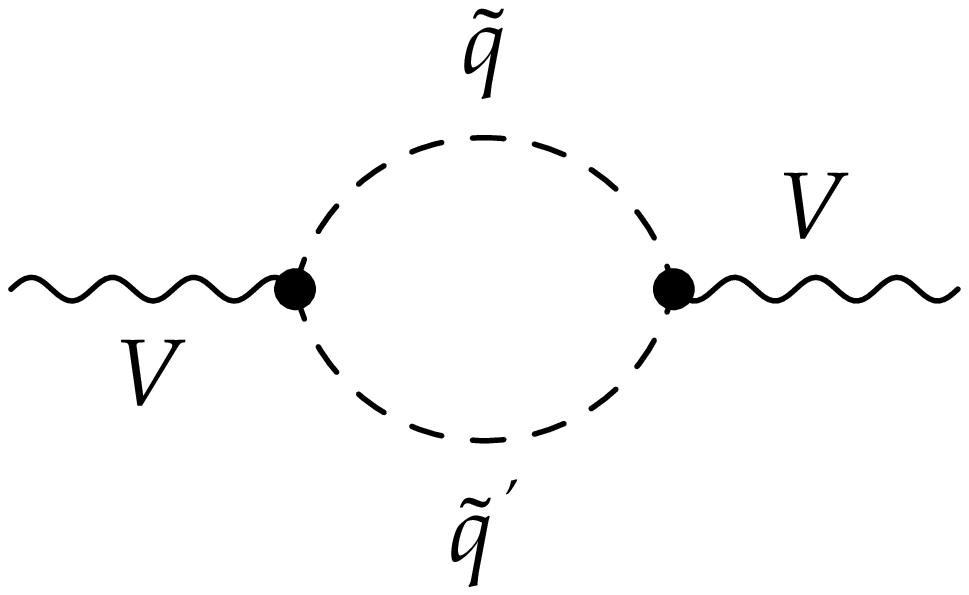}&
\includegraphics[scale=0.33]{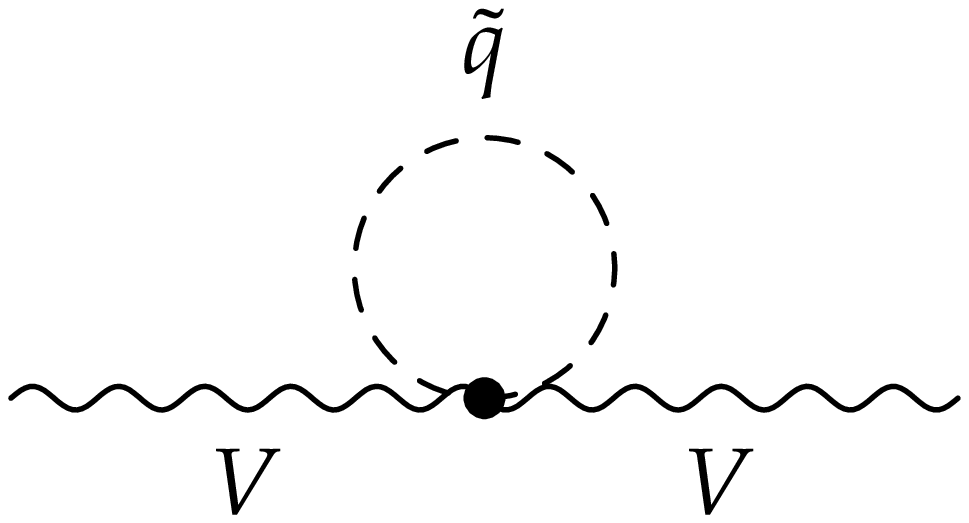}\\
\includegraphics[scale=0.33]{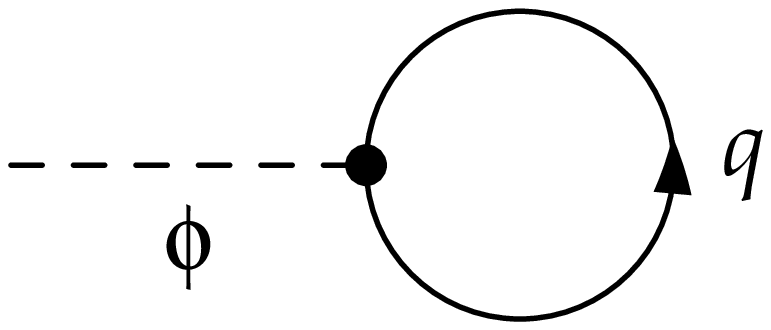}&
\includegraphics[scale=0.33]{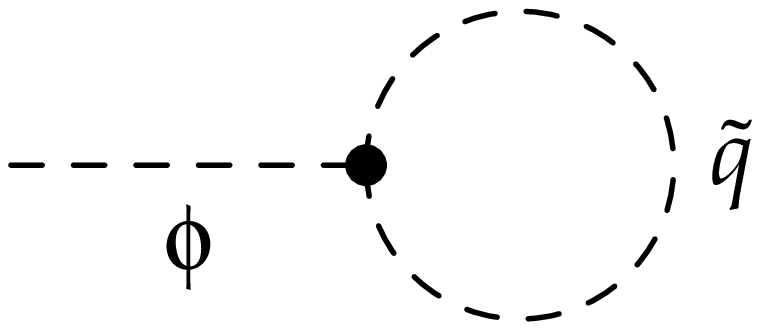}&
\end{tabular}
\caption{Different topologies for 
$\Sigma_{\phi\phi^{\prime}},\,\Sigma_{VV},\, T_{\phi}$}
\label{figfdall}
\end{figure}

\begin{itemize}
\item $h$
\end{itemize}
\begin{eqnarray}
\Sigma_{hh}^{2q} & = & -\sum_{i=1}^{3}\frac{3\alpha c_{\alpha}^{2}m_{u_{i}}^{2}}{4\pi M_{W}^{2}s_{\beta}^{2}s_{W}^{2}}\left\{ A_{0}\left[m_{u_{i}}^{2}\right]+p^{2}B_{1}\left[p^{2},m_{u_{i}}^{2},m_{u_{i}}^{2}\right]+2m_{u_{i}}^{2}B_{0}\left[p^{2},m_{u_{i}}^{2},m_{u_{i}}^{2}\right]\right\} \nonumber \\
 && -\sum_{i=1}^{3}\frac{3\alpha s_{\alpha}^{2}m_{d_{i}}^{2}}{4\pi M_{W}^{2}c_{\beta}^{2}s_{W}^{2}}\left\{ A_{0}\left[m_{d_{i}}^{2}\right]+p^{2}B_{1}\left[p^{2},m_{d_{i}}^{2},m_{d_{i}}^{2}\right]+2m_{d_{i}}^{2}B_{0}\left[p^{2},m_{d_{i}}^{2},m_{d_{i}}^{2}\right]\right\}\\
&&\nonumber \\
\Sigma_{hh}^{2\tilde{q}}&= & \sum_{m,n}^{6}\sum_{i,j,k,l}^{3}\frac{1}{48c_{W}^{2}M_{W}^{2}\pi s_{\beta}^{2}s_{W}^{2}}\alpha B_{0}\left[p^{2},m_{\tilde{u}_{m}}^{2},m_{\tilde{u}_{n}}^{2}\vphantom{m_{\tilde{d}_{l}}^{2}}\right]\nonumber \\
 && \times\left\{ \vphantom{\vphantom{\sum_{k=1}^{3}}}\delta_{i,j}\left(M_{W}m_{Z}s_{\alpha+\beta}s_{\beta}\left(-3+4s_{W}^{2}\right)+6c_{\alpha}c_{W}m_{u_{i}}^{2}\right)R_{n,j}^{\tilde{u}}R_{m,i}^{\tilde{u}*}\right.\nonumber \\
 && +3c_{W}\left(c_{\alpha}A_{i,j}^{u}+\mu^{*}s_{\alpha}\delta_{i,j}\right)m_{u_{i}}R_{n,3+j}^{\tilde{u}}R_{m,i}^{\tilde{u}*}\nonumber \\
&&+\left(3c_{\alpha}c_{W}A_{j,i}^{u*}m_{u_{j}}R_{n,j}^{\tilde{u}}+3c_{W}\mu s_{\alpha}\delta_{i,j}m_{u_{i}}R_{n,j}^{\tilde{u}}\right)R_{m,3+i}^{\tilde{u}*}\nonumber \\
 && \left.+2\delta_{i,j}\left(-2M_{W}m_{Z}s_{\alpha+\beta}s_{\beta}s_{W}^{2}+3c_{\alpha}c_{W}m_{u_{i}}^{2}\right)R_{n,3+j}^{\tilde{u}}R_{m,3+i}^{\tilde{u}*}\vphantom{\sum_{k=1}^{3}}\right\} \nonumber \\
 && \times\left\{ \vphantom{\vphantom{\sum_{k=1}^{3}}}\delta_{k,l}\left(M_{W}m_{Z}s_{\alpha+\beta}s_{\beta}\left(-3+4s_{W}^{2}\right)+6c_{\alpha}c_{W}m_{u_{k}}^{2}\right)R_{m,l}^{\tilde{u}}R_{n,k}^{\tilde{u}*}\right.\nonumber \\
 && +3c_{W}\left(c_{\alpha}A_{k,l}^{u}+\mu^{*}s_{\alpha}\delta_{k,l}\right)m_{u_{k}}R_{m,3+l}^{\tilde{u}}R_{n,k}^{\tilde{u}*}\nonumber \\
&&+\left(3c_{\alpha}c_{W}A_{l,k}^{u*}m_{u_{l}}R_{m,l}^{\tilde{u}}+3c_{W}\mu s_{\alpha}\delta_{k,l}m_{u_{k}}R_{m,l}^{\tilde{u}}\right)R_{n,3+k}^{\tilde{u}*}\nonumber \\
 && \left.+2\delta_{k,l}\left(-2M_{W}m_{Z}s_{\alpha+\beta}s_{\beta}s_{W}^{2}+3c_{\alpha}c_{W}m_{u_{k}}^{2}\right)R_{m,3+l}^{\tilde{u}}R_{n,3+k}^{\tilde{u}*}\vphantom{\vphantom{\sum_{k=1}^{3}}}\right\} \nonumber \\
 && +\sum_{m,n}^{6}\sum_{i,j,k,l}^{3}\frac{1}{48c_{W}^{2}M_{W}^{2}\pi c_{\beta}^{2}s_{W}^{2}}\alpha B_{0}\left[p^{2},m_{\tilde{d}_{m}}^{2},m_{\tilde{d}_{n}}^{2}\right]\nonumber \\
 && \times\left\{ \vphantom{\vphantom{\sum_{k=1}^{3}}}\delta_{i,j}\left(M_{W}m_{Z}s_{\alpha+\beta}c_{\beta}\left(-3+2s_{W}^{2}\right)+6s_{\alpha}c_{W}m_{d_{i}}^{2}\right)R_{n,j}^{\tilde{d}}R_{m,i}^{\tilde{d}*}\right.\nonumber \\
 && +3c_{W}\left(s_{\alpha}A_{i,j}^{d}+\mu^{*}c_{\alpha}\delta_{i,j}\right)m_{d_{i}}R_{n,3+j}^{\tilde{d}}R_{m,i}^{\tilde{d}*}\nonumber \\
&&+\left(3s_{\alpha}c_{W}A_{j,i}^{d*}m_{d_{j}}R_{n,j}^{\tilde{d}}+3c_{W}\mu c_{\alpha}\delta_{i,j}m_{d_{i}}R_{n,j}^{\tilde{d}}\right)R_{m,3+i}^{\tilde{d}*}\nonumber \\
 && \left.-2\delta_{i,j}\left(M_{W}m_{Z}s_{\alpha+\beta}c_{\beta}s_{W}^{2}-3s_{\alpha}c_{W}m_{d_{i}}^{2}\right)R_{n,3+j}^{\tilde{d}}R_{m,3+i}^{\tilde{d}*}\vphantom{\sum_{k=1}^{3}}\right\} \nonumber \\
 && \times\left\{ \vphantom{\sum_{k=1}^{3}}\left(\delta_{k,l}\left(M_{W}m_{Z}s_{\alpha+\beta}c_{\beta}\left(-3+2s_{W}^{2}\right)+6s_{\alpha}c_{W}m_{d_{k}}^{2}\right)R_{m,l}^{\tilde{d}}R_{n,k}^{\tilde{d}*}\right)\right.\nonumber \\
 && +3c_{W}\left(s_{\alpha}A_{k,l}^{d}+\mu^{*}c_{\alpha}\delta_{k,l}\right)m_{d_{k}}R_{m,3+l}^{\tilde{d}}R_{n,k}^{\tilde{d}*}\nonumber \\
&&+\left(3s_{\alpha}c_{W}A_{l,k}^{d*}m_{d_{l}}R_{m,l}^{\tilde{d}}+3c_{W}\mu c_{\alpha}\delta_{k,l}m_{d_{k}}R_{m,l}^{\tilde{d}}\right)R_{n,3+k}^{\tilde{d}*}\nonumber \\
 && \left.-2\delta_{k,l}\left(M_{W}m_{Z}s_{\alpha+\beta}c_{\beta}s_{W}^{2}-3s_{\alpha}c_{W}m_{d_{k}}^{2}\right)R_{m,3+l}^{\tilde{d}}R_{n,3+k}^{\tilde{d}*}\vphantom{\sum_{k=1}^{3}}\right\}\\
&&\nonumber \\
\Sigma_{hh}^{1\tilde{q}}&= & \sum_{l=1}^{6}\sum_{i=1}^{3}\frac{1}{16c_{W}^{2}M_{W}^{2}\pi s_{\beta}^{2}s_{W}^{2}}\alpha A_{0}\left[m_{\tilde{u}_{l}}^{2}\vphantom{m_{\tilde{d}_{l}}^{2}}\right]\left\{ \vphantom{R_{l,i}^{\tilde{d}}}R_{l,i}^{\tilde{u}}R_{l,i}^{\tilde{u}*}\left(c_{2\alpha}m_{w}^{2}s_{\beta}^{2}\left(-3+4s_{W}^{2}\right)+6c_{\alpha}^{2}c_{W}^{2}m_{u_{i}}^{2}\right)\right.\nonumber \\
 && \left.+2R_{l,3+i}^{\tilde{u}}R_{l,3+i}^{\tilde{u}*}\left(-2c_{2\alpha}M_{W}^{2}s_{\beta}^{2}s_{W}^{2}+3c_{\alpha}^{2}c_{W}^{2}m_{u_{i}}^{2}\right)\vphantom{R_{l,i}^{\tilde{d}}}\right\} \nonumber \\
 && -\sum_{l=1}^{6}\sum_{i=1}^{3}\frac{1}{16c_{W}^{2}M_{W}^{2}\pi c_{\beta}^{2}s_{W}^{2}}\alpha A_{0}\left[m_{\tilde{d}_{l}}^{2}\right]\left\{ R_{l,i}^{\tilde{d}}R_{l,i}^{\tilde{d}*}\left(c_{2\alpha}m_{w}^{2}c_{\beta}^{2}\left(-3+2s_{W}^{2}\right)-6s_{\alpha}^{2}c_{W}^{2}m_{d_{i}}^{2}\right)\right.\nonumber \\
 && \left.-2R_{l,3+i}^{\tilde{d}}R_{l,3+i}^{\tilde{d}*}\left(c_{2\alpha}M_{W}^{2}c_{\beta}^{2}s_{W}^{2}+3s_{\alpha}^{2}c_{W}^{2}m_{d_{i}}^{2}\right)\vphantom{R_{l,i}^{\tilde{d}}}\right\} 
\end{eqnarray}

\begin{itemize}
\item $hH$
\end{itemize}
\begin{eqnarray}
\Sigma_{hH}^{2q} & = & -\sum_{i=1}^{3}\frac{3\alpha c_{\alpha}s_{\alpha}m_{u_{i}}^{2}}{4\pi M_{W}^{2}s_{\beta}^{2}s_{W}^{2}}\left\{ A_{0}\left[m_{u_{i}}^{2}\right]+p^{2}B_{1}\left[p^{2},m_{u_{i}}^{2},m_{u_{i}}^{2}\right]+2m_{u_{i}}^{2}B_{0}\left[p^{2},m_{u_{i}}^{2},m_{u_{i}}^{2}\right]\right\} \nonumber \\
 &  & +\sum_{i=1}^{3}\frac{3\alpha c_{\alpha}s_{\alpha}m_{d_{i}}^{2}}{4\pi M_{W}^{2}c_{\beta}^{2}s_{W}^{2}}\left\{ A_{0}\left[m_{d_{i}}^{2}\right]+p^{2}B_{1}\left[p^{2},m_{d_{i}}^{2},m_{d_{i}}^{2}\right]+2m_{d_{i}}^{2}B_{0}\left[p^{2},m_{d_{i}}^{2},m_{d_{i}}^{2}\right]\right\}\,\,\\
\Sigma_{hH}^{2\tilde{q}}&= & \sum_{m,n}^{6}\sum_{i,j,k,l}^{3}\frac{1}{48c_{W}^{2}M_{W}^{2}\pi s_{\beta}^{2}s_{W}^{2}}\alpha B_{0}\left[p^{2},m_{\tilde{u}_{m}}^{2},m_{\tilde{u}_{n}}^{2}\vphantom{m_{\tilde{d}_{l}}^{2}}\right]\nonumber \\
 && \times\left\{ \vphantom{\vphantom{\sum_{k=1}^{3}}}\delta_{i,j}\left(M_{W}m_{Z}s_{\alpha+\beta}s_{\beta}\left(-3+4s_{W}^{2}\right)+6c_{\alpha}c_{W}m_{u_{i}}^{2}\right)R_{n,j}^{\tilde{u}}R_{m,i}^{\tilde{u}*}\right.\nonumber \\
 && +3c_{W}\left(c_{\alpha}A_{i,j}^{u}+\mu^{*}s_{\alpha}\delta_{i,j}\right)m_{u_{i}}R_{n,3+j}^{\tilde{u}}R_{m,i}^{\tilde{u}*}\nonumber \\
&&+\left(3c_{\alpha}c_{W}A_{j,i}^{u*}m_{u_{j}}R_{n,j}^{\tilde{u}}+3c_{W}\mu s_{\alpha}\delta_{i,j}m_{u_{i}}R_{n,j}^{\tilde{u}}\right)R_{m,3+i}^{\tilde{u}*}\nonumber \\
 && \left.+2\delta_{i,j}\left(-2M_{W}m_{Z}s_{\alpha+\beta}s_{\beta}s_{W}^{2}+3c_{\alpha}c_{W}m_{u_{i}}^{2}\right)R_{n,3+j}^{\tilde{u}}R_{m,3+i}^{\tilde{u}*}\vphantom{\vphantom{\sum_{k=1}^{3}}}\right\} \nonumber \\
 && \times\left\{ \vphantom{\vphantom{\sum_{k=1}^{3}}}\delta_{k,l}\left(M_{W}m_{Z}c_{\alpha+\beta}s_{\beta}\left(3-4s_{W}^{2}\right)+6s_{\alpha}c_{W}m_{u_{k}}^{2}\right)R_{m,l}^{\tilde{u}}R_{n,k}^{\tilde{u}*}\right.\nonumber \\
 && +3c_{W}\left(s_{\alpha}A_{k,l}^{u}-\mu^{*}c_{\alpha}\delta_{k,l}\right)m_{u_{k}}R_{m,3+l}^{\tilde{u}}R_{n,k}^{\tilde{u}*}\nonumber \\
&&+\left(3s_{\alpha}c_{W}A_{l,k}^{u*}m_{u_{l}}R_{m,l}^{\tilde{u}}-3c_{W}\mu c_{\alpha}\delta_{k,l}m_{u_{k}}R_{m,l}^{\tilde{u}}\right)R_{n,3+k}^{\tilde{u}*}\nonumber \\
 & &\left.+2\delta_{k,l}\left(2M_{W}m_{Z}c_{\alpha+\beta}s_{\beta}s_{W}^{2}+3s_{\alpha}c_{W}m_{u_{k}}^{2}\right)R_{m,3+l}^{\tilde{u}}R_{n,3+k}^{\tilde{u}*}\vphantom{\vphantom{\sum_{k=1}^{3}}}\right\} \nonumber \\
 && -\sum_{m,n}^{6}\sum_{i,j,k,l}^{3}\frac{1}{48c_{W}^{2}M_{W}^{2}\pi c_{\beta}^{2}s_{W}^{2}}\alpha B_{0}\left[p^{2},m_{\tilde{d}_{m}}^{2},m_{\tilde{d}_{n}}^{2}\right]\nonumber \\
 && \times\left\{ \vphantom{\vphantom{\sum_{k=1}^{3}}}\delta_{i,j}\left(M_{W}m_{Z}s_{\alpha+\beta}c_{\beta}\left(-3+2s_{W}^{2}\right)+6s_{\alpha}c_{W}m_{d_{i}}^{2}\right)R_{n,j}^{\tilde{d}}R_{m,i}^{\tilde{d}*}\right.\nonumber \\
 && +3c_{W}\left(s_{\alpha}A_{i,j}^{d}+\mu^{*}c_{\alpha}\delta_{i,j}\right)m_{d_{i}}R_{n,3+j}^{\tilde{d}}R_{m,i}^{\tilde{d}*}\nonumber \\
&&+\left(3s_{\alpha}c_{W}A_{j,i}^{d*}m_{d_{j}}R_{n,j}^{\tilde{d}}+3c_{W}\mu c_{\alpha}\delta_{i,j}m_{d_{i}}R_{n,j}^{\tilde{d}}\right)R_{m,3+i}^{\tilde{d}*}\nonumber \\
 && \left.-2\delta_{i,j}\left(M_{W}m_{Z}s_{\alpha+\beta}c_{\beta}s_{W}^{2}-3s_{\alpha}c_{W}m_{d_{i}}^{2}\right)R_{n,3+j}^{\tilde{d}}R_{m,3+i}^{\tilde{d}*}\vphantom{\vphantom{\sum_{k=1}^{3}}}\right\} \nonumber \\
 && \times\left\{ \vphantom{\vphantom{\sum_{k=1}^{3}}}\delta_{k,l}\left(M_{W}m_{Z}c_{\alpha+\beta}c_{\beta}\left(-3+2s_{W}^{2}\right)+6c_{\alpha}c_{W}m_{d_{k}}^{2}\right)R_{m,l}^{\tilde{d}}R_{n,k}^{\tilde{d}*}\right.\nonumber \\
 && +3c_{W}\left(c_{\alpha}A_{k,l}^{d}-\mu^{*}s_{\alpha}\delta_{k,l}\right)m_{d_{k}}R_{m,3+l}^{\tilde{d}}R_{n,k}^{\tilde{d}*}\nonumber \\
&&+\left(3c_{\alpha}c_{W}A_{l,k}^{d*}m_{d_{l}}R_{m,l}^{\tilde{d}}-3c_{W}\mu s_{\alpha}\delta_{k,l}m_{d_{k}}R_{m,l}^{\tilde{d}}\right)R_{n,3+k}^{\tilde{d}*}\nonumber \\
 && \left.-2\delta_{k,l}\left(2M_{W}m_{Z}c_{\alpha+\beta}c_{\beta}s_{W}^{2}-3c_{\alpha}c_{W}m_{d_{k}}^{2}\right)R_{m,3+l}^{\tilde{d}}R_{n,3+k}^{\tilde{d}*}\vphantom{\vphantom{\sum_{k=1}^{3}}}\right\}\\
\Sigma_{hH}^{1\tilde{q}}&= & \sum_{l=1}^{6}\sum_{i=1}^{3}\frac{1}{16c_{W}^{2}M_{W}^{2}\pi s_{\beta}^{2}s_{W}^{2}}\alpha A_{0}\left[m_{\tilde{u}_{l}}^{2}\vphantom{m_{\tilde{d}_{l}}^{2}}\right]\left\{ \vphantom{R_{l,i}^{\tilde{d}}}R_{l,i}^{\tilde{u}}R_{l,i}^{\tilde{u}*}\left(s_{2\alpha}m_{w}^{2}s_{\beta}^{2}\left(-3+4s_{W}^{2}\right)+3s_{2\alpha}c_{W}^{2}m_{u_{i}}^{2}\right)\right.\nonumber \\
 && \left.+R_{l,3+i}^{\tilde{u}}R_{l,3+i}^{\tilde{u}*}\left(-4s_{2\alpha}M_{W}^{2}s_{\beta}^{2}s_{W}^{2}+3s_{2\alpha}c_{W}^{2}m_{u_{i}}^{2}\right)\vphantom{R_{l,i}^{\tilde{d}}}\right\} \nonumber \\
 && -\sum_{l=1}^{6}\sum_{i=1}^{3}\frac{1}{16c_{W}^{2}M_{W}^{2}\pi c_{\beta}^{2}s_{W}^{2}}\alpha A_{0}\left[m_{\tilde{d}_{l}}^{2}\right]\left\{ R_{l,i}^{\tilde{d}}R_{l,i}^{\tilde{d}*}\left(s_{2\alpha}m_{w}^{2}c_{\beta}^{2}\left(-3+4s_{W}^{2}\right)+3s_{2\alpha}c_{W}^{2}m_{d_{i}}^{2}\right)\right.\nonumber \\
 && \left.+R_{l,3+i}^{\tilde{d}}R_{l,3+i}^{\tilde{d}*}\left(-2s_{2\alpha}M_{W}^{2}c_{\beta}^{2}s_{W}^{2}+3s_{2\alpha}c_{W}^{2}m_{d_{i}}^{2}\right)\vphantom{R_{l,i}^{\tilde{d}}}\right\}
\end{eqnarray}

\begin{itemize}
\item $A$
\end{itemize}
\begin{eqnarray}
\Sigma_{AA}^{2q} & = & -\sum_{i=1}^{3}\frac{3\alpha m_{u_{i}}^{2}}{4\pi M_{W}^{2}t_{\beta}^{2}s_{W}^{2}}\left\{ A_{0}\left[m_{u_{i}}^{2}\right]+p^{2}B_{1}\left[p^{2},m_{u_{i}}^{2},m_{u_{i}}^{2}\right]\right\} \nonumber \\
 &  & -\sum_{i=1}^{3}\frac{3\alpha t_{\beta}^{2}m_{d_{i}}^{2}}{4\pi M_{W}^{2}s_{W}^{2}}\left\{ A_{0}\left[m_{d_{i}}^{2}\right]+p^{2}B_{1}\left[p^{2},m_{d_{i}}^{2},m_{d_{i}}^{2}\right]\right\} \\
&&\nonumber \\
\Sigma_{AA}^{2\tilde{q}}&= & -\sum_{m,n}^{6}\sum_{i,j,k,l}^{3}\frac{3}{16M_{W}^{2}\pi t_{\beta}^{2}s_{W}^{2}}\alpha B_{0}\left[p^{2},m_{\tilde{u}_{m}}^{2},m_{\tilde{u}_{n}}^{2}\vphantom{m_{\tilde{d}_{l}}^{2}}\right]\nonumber \\
 && \times\left\{ \left(-A_{i,j}^{u}-\mu^{*}t_{\beta}\delta_{i,j}\right)m_{u_{i}}R_{n,3+j}^{\tilde{u}}R_{m,i}^{\tilde{u}*}+\left(A_{j,i}^{u*}m_{u_{j}}+\mu t_{\beta}\delta_{i,j}m_{u_{i}}\right)R_{n,j}^{\tilde{u}}R_{m,3+i}^{\tilde{u}*}\vphantom{R_{m,i}^{\tilde{d}*}}\right\} \nonumber \\
 & &\times\left\{ \left(-A_{k,l}^{u}-\mu^{*}t_{\beta}\delta_{k,l}\right)m_{u_{k}}R_{m,3+l}^{\tilde{u}}R_{n,k}^{\tilde{u}*}+\left(A_{l,k}^{u*}m_{u_{l}}+\mu t_{\beta}\delta_{k,l}m_{u_{k}}\right)R_{m,l}^{\tilde{u}}R_{n,3+k}^{\tilde{u}*}\vphantom{R_{m,i}^{\tilde{d}*}}\right\} \nonumber \\
 && -\sum_{m,n}^{6}\sum_{i,j,k,l}^{3}\frac{3}{16M_{W}^{2}\pi s_{W}^{2}}\alpha B_{0}\left[p^{2},m_{\tilde{d}_{m}}^{2},m_{\tilde{d}_{n}}^{2}\right]\nonumber \\
 && \times\left\{ \left(-t_{\beta}A_{i,j}^{d}-\mu^{*}\delta_{i,j}\right)m_{d_{i}}R_{n,3+j}^{\tilde{d}}R_{m,i}^{\tilde{d}*}+\left(t_{\beta}A_{j,i}^{d*}m_{d_{j}}+\mu\delta_{i,j}m_{d_{i}}\right)R_{n,j}^{\tilde{d}}R_{m,3+i}^{\tilde{d}*}\right\} \nonumber \\
 && \times\left\{ \left(-t_{\beta}A_{k,l}^{d}-\mu^{*}\delta_{k,l}\right)m_{d_{k}}R_{m,3+l}^{\tilde{d}}R_{n,k}^{\tilde{d}*}+\left(t_{\beta}A_{l,k}^{d*}m_{d_{l}}+\mu\delta_{k,l}m_{d_{k}}\right)R_{m,l}^{\tilde{d}}R_{n,3+k}^{\tilde{d}*}\right\}\\
&&\nonumber \\
\Sigma_{AA}^{1\tilde{q}}&= & \sum_{l=1}^{6}\sum_{i=1}^{3}\frac{1}{16c_{W}^{2}M_{W}^{2}\pi t_{\beta}^{2}s_{W}^{2}}\alpha A_{0}\left[m_{\tilde{u}_{l}}\vphantom{m_{\tilde{d}_{l}}^{2}}\right]\left\{ \vphantom{R_{l,i}^{\tilde{d}}}R_{l,i}^{\tilde{u}}R_{l,i}^{\tilde{u}*}\left(c_{2\beta}m_{w}^{2}t_{\beta}^{2}\left(-3+4s_{W}^{2}\right)+6c_{W}^{2}m_{u_{i}}^{2}\right)\right.\nonumber \\
 && \left.+2R_{l,3+i}^{\tilde{u}}R_{l,3+i}^{\tilde{u}*}\left(-2c_{2\beta}M_{W}^{2}t_{\beta}^{2}s_{W}^{2}+3c_{W}^{2}m_{u_{i}}^{2}\right)\vphantom{R_{l,i}^{\tilde{d}}}\right\} \nonumber \\
 && +\sum_{l=1}^{6}\sum_{i=1}^{3}\frac{1}{16c_{W}^{2}M_{W}^{2}\pi s_{W}^{2}}\alpha A_{0}\left[m_{\tilde{d}_{l}}\vphantom{m_{\tilde{d}_{l}}^{2}}\right]\left\{ R_{l,i}^{\tilde{d}}R_{l,i}^{\tilde{d}*}\left(c_{2\beta}m_{w}^{2}\left(3-2s_{W}^{2}\right)+6c_{W}^{2}t_{\beta}^{2}m_{d_{i}}^{2}\right)\right.\nonumber \\
 && \left.+2R_{l,3+i}^{\tilde{d}}R_{l,3+i}^{\tilde{d}*}\left(c_{2\beta}M_{W}^{2}s_{W}^{2}+3c_{W}^{2}t_{\beta}^{2}m_{d_{i}}^{2}\right)\vphantom{R_{l,i}^{\tilde{d}}}\right\} \end{eqnarray}

\begin{itemize}
\item $H^{\pm}$
\end{itemize}
\begin{align}
\Sigma_{H^{-}H^{+}}^{2q}=- & \sum_{i=1}^{3}\sum_{j=1}^{3}\frac{3\alpha}{4\pi M_{W}^{2}s_{W}^{2}}\left\{ \vphantom{\vphantom{\sum_{k=1}^{3}}}m_{u_{i}}^{2}\left(2m_{d_{j}}^{2}+m_{u_{i}}^{2}/t_{\beta}^{2}+m_{d_{j}}^{2}t_{\beta}^{2}\right)\VCKM^{i,j}\VCKM^{i,j*}B_{0}\left[p^{2},m_{d_{j}}^{2},m_{u_{i}}^{2}\right]\right.\nonumber \\
 & +\left(m_{u_{i}}^{2}/t_{\beta}^{2}+m_{d_{j}}^{2}t_{\beta}^{2}\right)\VCKM^{i,j}\VCKM^{i,j*}p^{2}B_{1}\left[p^{2},m_{u_{i}}^{2},m_{d_{j}}^{2}\right]\nonumber \\
 & \left.+A_{0}\left[m_{d_{j}}^{2}\right]\left(m_{u_{i}}^{2}\VCKM^{i,j}\VCKM^{i,j*}/t_{\beta}^{2}+m_{d_{j}}^{2}\VCKM^{i,j}\VCKM^{i,j*}t_{\beta}^{2}\right)\vphantom{\vphantom{\sum_{k=1}^{3}}}\right\} \end{align}

\begin{align}
\Sigma_{H^{-}H^{+}}^{2\tilde{q}}= & -\sum_{m,n}^{6}\sum_{i,j,k,l}^{3}\frac{3}{8M_{W}^{2}\pi t_{\beta}^{2}s_{W}^{2}}\alpha B_{0}\left[p^{2},m_{\tilde{u}_{m}}^{2},m_{\tilde{u}_{n}}^{2}\right]\left\{ \vphantom{\vphantom{\sum_{k=1}^{3}}}\right.\nonumber \\
 & \sum_{p,q}^{3}\left[\vphantom{\sum_{k=1}^{3}}\left(t_{\beta}^{2}A_{p,i}^{d}\VCKM^{k,p*}m_{d_{p}}R_{n,3+i}^{\tilde{d}}R_{m,k}^{\tilde{u}*}+A_{p,k}^{u*}\VCKM^{p,i*}m_{u_{p}}R_{n,i}^{\tilde{d}}R_{m,3+k}^{\tilde{u}*}\right)\right.\nonumber \\
 & \left.\times\left(t_{\beta}^{2}A_{q,j}^{d*}\VCKM^{l,q}m_{d_{q}}R_{m,l}^{\tilde{u}}R_{n,3+j}^{\tilde{d}*}+A_{q,l}^{u}\VCKM^{q,j}m_{u_{q}}R_{m,3+l}^{\tilde{u}}R_{n,j}^{\tilde{d}*}\right)\vphantom{\sum_{k=1}^{3}}\right]\nonumber \\
 & +\sum_{p}^{3}\left[\vphantom{\sum_{k=1}^{3}}\VCKM^{k,i*}\left\{ R_{n,i}^{\tilde{d}}\left(\left(-M_{W}^{2}s_{2\beta}t_{\beta}+m_{u_{k}}^{2}+t_{\beta}^{2}m_{d_{i}}^{2}\vphantom{R_{n,j}^{\tilde{d}*}}\right)R_{m,k}^{\tilde{u}*}+\mu t_{\beta}m_{u_{k}}R_{m,3+k}^{\tilde{u}*}\right)\right.\right.\nonumber \\
 & \left.+m_{d_{i}}R_{n,3+i}^{\tilde{d}}\left(\mu^{*}t_{\beta}R_{m,k}^{\tilde{u}*}+\left(1+t_{\beta}^{2}\right)m_{u_{k}}R_{m,3+k}^{\tilde{u}*}\right)\right\} \nonumber \\
 & \times\left\{ A_{p,l}^{u}\VCKM^{p,j}m_{u_{p}}R_{m,3+l}^{\tilde{u}}R_{n,j}^{\tilde{d}*}+t_{\beta}^{2}A_{p,j}^{d*}\VCKM^{l,p}m_{d_{p}}R_{m,l}^{\tilde{u}}R_{n,3+j}^{\tilde{d}*}\right\} \nonumber \\
 & +\VCKM^{l,j}\left\{ R_{m,l}^{\tilde{u}}\left(\left(-M_{W}^{2}s_{2\beta}t_{\beta}+m_{u_{l}}^{2}+t_{\beta}^{2}m_{d_{j}}^{2}\vphantom{R_{n,j}^{\tilde{d}*}}\right)R_{n,j}^{\tilde{d}*}+\mu t_{\beta}m_{d_{j}}R_{n,3+j}^{\tilde{d}*}\right)\right.\nonumber \\
 & \left.+m_{u_{l}}R_{m,3+l}^{\tilde{u}}\left(\mu^{*}t_{\beta}R_{n,j}^{\tilde{d}*}+\left(1+t_{\beta}^{2}\right)m_{d_{j}}R_{n,3+j}^{\tilde{d}*}\right)\right\} \nonumber \\
 & \left.\times\left\{ A_{p,k}^{u*}\VCKM^{p,i}m_{u_{p}}R_{n,i}^{\tilde{d}}R_{m,3+k}^{\tilde{u}*}+t_{\beta}^{2}A_{p,i}^{d}\VCKM^{k,p*}m_{d_{p}}R_{n,3+i}^{\tilde{d}}R_{m,k}^{\tilde{u}*}\right\} \vphantom{\sum_{k=1}^{3}}\right]\nonumber \\
 & +\left[\vphantom{\sum_{k=1}^{3}}\VCKM^{l,j}\VCKM^{k,i*}\left\{ R_{n,i}^{\tilde{d}}\left(\left(-M_{W}^{2}s_{2\beta}t_{\beta}+m_{u_{k}}^{2}+t_{\beta}^{2}m_{d_{i}}^{2}\right)R_{m,k}^{\tilde{u}*}+\mu t_{\beta}m_{u_{k}}R_{m,3+k}^{\tilde{u}*}\right)\right.\right.\nonumber \\
 & \left.+m_{d_{i}}R_{n,3+i}^{\tilde{d}}\left(\mu^{*}t_{\beta}R_{m,k}^{\tilde{u}*}+\left(1+t_{\beta}^{2}\right)m_{u_{k}}R_{m,3+k}^{\tilde{u}*}\right)\right\} \nonumber \\
 & \times\left\{ R_{m,l}^{\tilde{u}}\left(\left(-M_{W}^{2}s_{2\beta}t_{\beta}+m_{u_{l}}^{2}+t_{\beta}^{2}m_{d_{j}}^{2}\right)R_{n,j}^{\tilde{d}*}+\mu t_{\beta}m_{d{}_{j}}R_{n,3+j}^{\tilde{d}*}\right)\right.\nonumber \\
 & \left.\left.\left.+m_{u_{l}}R_{m,3+l}^{\tilde{u}}\left(\mu^{*}t_{\beta}R_{n,j}^{\tilde{d}*}+\left(1+t_{\beta}^{2}\right)m_{d_{j}}R_{n,3+j}^{\tilde{d}*}\right)\right\} \vphantom{\sum_{k=1}^{3}}\right]\vphantom{\vphantom{\sum_{k=1}^{3}}}\right\} \end{align}

\begin{align}
\Sigma_{H^{-}H^{+}}^{1\tilde{q}}= & \sum_{l=1}^{6}\sum_{i=1}^{3}\frac{1}{16c_{W}^{2}M_{W}^{2}\pi t_{\beta}^{2}s_{W}^{2}}\alpha A_{0}\left[m_{\tilde{u}_{l}}^{2}\vphantom{m_{\tilde{d}_{l}}^{2}}\right]\left\{ 2R_{l,3+i}^{\tilde{u}}R_{l,3+i}^{\tilde{u}*}\left(-2c_{2\beta}M_{W}^{2}t_{\beta}^{2}s_{W}^{2}+3c_{W}^{2}m_{u_{i}}^{2}\right)\vphantom{\vphantom{\sum_{k=1}^{3}}}\right.\nonumber \\
 & \left.+t_{\beta}^{2}R_{l,i}^{\tilde{u}}\left(R_{l,i}^{\tilde{u}*}c_{2\beta}m_{w}^{2}\left(1+2c_{W}^{2}\right)+\sum_{j=1}^{3}\sum_{k=1}^{3}6R_{l,j}^{\tilde{u}*}c_{W}^{2}t_{\beta}^{2}\VCKM^{i,k}\VCKM^{j,k*}m_{d_{k}}^{2}\right)\right\} \nonumber \\
 & +\sum_{l=1}^{6}\sum_{i=1}^{3}\frac{1}{16c_{W}^{2}M_{W}^{2}\pi t_{\beta}^{2}s_{W}^{2}}\alpha A_{0}\left[m_{\tilde{d}_{l}}^{2}\right]\left\{ \vphantom{\vphantom{\sum_{k=1}^{3}}}2R_{l,3+i}^{\tilde{d}}R_{l,3+i}^{\tilde{d}*}\left(c_{2\beta}M_{W}^{2}t_{\beta}^{2}s_{W}^{2}+3c_{W}^{2}t_{\beta}^{4}m_{d_{i}}^{2}\right)\right.\nonumber \\
 & \left.+R_{l,i}^{\tilde{d}}\left(R_{l,i}^{\tilde{d}*}c_{2\beta}m_{w}^{2}t_{\beta}^{2}\left(1-4c_{W}^{2}\right)+\sum_{j=1}^{3}\sum_{k=1}^{3}6R_{l,j}^{\tilde{d}*}c_{W}^{2}\VCKM^{k,j}\VCKM^{k,i*}m_{u_{k}}^{2}\right)\right\} \end{align}

\begin{itemize}
\item $Z$
\end{itemize}
\begin{eqnarray}
\Sigma_{ZZ}^{2q} & =- & \sum_{i=1}^{3}\frac{\left(9-24s_{W}^{2}+32s_{W}^{4}\right)\alpha}{36c_{W}^{2}\pi s_{W}^{2}}\left\{ A_{0}\left[m_{u_{i}}^{2}\right]+p^{2}B_{1}\left[p^{2},m_{u_{i}}^{2},m_{u_{i}}^{2}\right]\right\} \nonumber \\
 &  & -\sum_{i=1}^{3}\frac{\left(9+48s_{W}^{2}-64s_{W}^{4}\right)\alpha}{72c_{W}^{2}\pi s_{W}^{2}}\left\{ m_{u_{i}}^{2}B_{0}\left[p^{2},m_{u_{i}}^{2},m_{u_{i}}^{2}\right]\right\} \nonumber \\
 &  & -\sum_{i=1}^{3}\frac{\left(9-12s_{W}^{2}+8s_{W}^{4}\right)\alpha}{36c_{W}^{2}\pi s_{W}^{2}}\left\{ A_{0}\left[m_{d_{i}}^{2}\right]+p^{2}B_{1}\left[p^{2},m_{d_{i}}^{2},m_{d_{i}}^{2}\right]\right\} \nonumber \\
 &  & -\sum_{i=1}^{3}\frac{\left(9+24s_{W}^{2}-16s_{W}^{4}\right)\alpha}{72c_{W}^{2}\pi s_{W}^{2}}\left\{ m_{d_{i}}^{2}B_{0}\left[p^{2},m_{d_{i}}^{2},m_{d_{i}}^{2}\right]\right\}\\
&&\nonumber \\
\Sigma_{ZZ}^{2\tilde{q}}&=&- \sum_{m,n}^{6}\sum_{i,j}^{3}\frac{\alpha}{72c_{W}^{2}\pi s_{W}^{2}}\left\{ A_{0}\left[m_{\tilde{u}_{n}}^{2}\vphantom{m_{\tilde{d}_{l}}^{2}}\right]+2m_{\tilde{u}_{m}}^{2}B_{0}\left[p^{2},m_{\tilde{u}_{m}}^{2},m_{\tilde{u}_{n}}^{2}\vphantom{m_{\tilde{d}_{l}}^{2}}\right]\right.\nonumber \\
 && \left.+\left(p^{2}+m_{\tilde{u}_{m}}^{2}-m_{\tilde{u}_{n}}^{2}\right)B_{1}\left[p^{2},m_{\tilde{u}_{m}}^{2},m_{\tilde{u}_{n}}^{2}\vphantom{m_{\tilde{d}_{l}}^{2}}\right]-\frac{p^{2}}{3}+m_{\tilde{u}_{m}}^{2}-m_{\tilde{u}_{n}}^{2}\right\} \nonumber \\
 && \times\left\{ \left(-3+4s_{W}^{2}\right)R_{n,i}^{\tilde{u}}R_{m,i}^{\tilde{u}*}+4s_{W}^{2}R_{n,3+i}^{\tilde{u}}R_{m,3+i}^{\tilde{u}*}\right\}\nonumber \\ 
&& \,\,\,\left\{ \left(-3+4s_{W}^{2}\right)R_{m,j}^{\tilde{u}}R_{n,j}^{\tilde{u}*}+4s_{W}^{2}R_{m,3+j}^{\tilde{u}}R_{n,3+j}^{\tilde{u}*}\right\} \nonumber \\
 && -\sum_{m,n}^{6}\sum_{i,j}^{3}\frac{\alpha}{72c_{W}^{2}\pi s_{W}^{2}}\left\{ A_{0}\left[m_{\tilde{d}_{n}}^{2}\vphantom{m_{\tilde{d}_{l}}^{2}}\right]+2m_{\tilde{d}_{m}}^{2}B_{0}\left[p^{2},m_{\tilde{d}_{m}}^{2},m_{\tilde{d}_{n}}^{2}\vphantom{m_{\tilde{d}_{l}}^{2}}\right]\right.\nonumber \\
 && \left.+\left(p^{2}+m_{\tilde{d}_{m}}^{2}-m_{\tilde{d}_{n}}^{2}\right)B_{1}\left[p^{2},m_{\tilde{d}_{m}}^{2},m_{\tilde{d}_{n}}^{2}\vphantom{m_{\tilde{d}_{l}}^{2}}\right]-\frac{p^{2}}{3}+m_{\tilde{d}_{m}}^{2}-m_{\tilde{d}_{n}}^{2}\right\} \nonumber \\
 && \times\left\{ \left(-3+2s_{W}^{2}\right)R_{n,i}^{\tilde{d}}R_{m,i}^{\tilde{d}*}+2s_{W}^{2}R_{n,3+i}^{\tilde{d}}R_{m,3+i}^{\tilde{d}*}\right\}\nonumber \\ 
&&\,\,\,\left\{ \left(-3+2s_{W}^{2}\right)R_{m,j}^{\tilde{d}}R_{n,j}^{\tilde{d}*}+2s_{W}^{2}R_{m,3+j}^{\tilde{d}}R_{n,3+j}^{\tilde{d}*}\right\}\\
&&\nonumber \\
\Sigma_{ZZ}^{1\tilde{q}}&= & \sum_{l=1}^{6}\sum_{i=1}^{3}\frac{1}{24c_{W}^{2}\pi s_{W}^{2}}\alpha A_{0}\left[m_{\tilde{u}_{l}}^{2}\vphantom{m_{\tilde{d}_{l}}^{2}}\right]\left(\vphantom{R_{l,i}^{\tilde{d}}}R_{l,i}^{\tilde{u}}R_{l,i}^{\tilde{u}*}\left(-3+4s_{W}^{2}\right)^{2}+16s_{W}^{4}R_{l,3+i}^{\tilde{u}}R_{l,3+i}^{\tilde{u}*}\right)\nonumber \\
 && +\sum_{l=1}^{6}\sum_{i=1}^{3}\frac{1}{24c_{W}^{2}\pi s_{W}^{2}}\alpha A_{0}\left[m_{\tilde{d}_{l}}^{2}\right]\left(R_{l,i}^{\tilde{d}}R_{l,i}^{\tilde{d}*}\left(3-2s_{W}^{2}\right)^{2}+4s_{W}^{4}R_{l,3+i}^{\tilde{d}}R_{l,3+i}^{\tilde{d}*}\right)
\end{eqnarray}

\begin{itemize}
\item $W$
\end{itemize}
\begin{eqnarray}
\Sigma_{WW}^{2q} & =- & \sum_{i,l}^{3}\frac{\alpha}{4\pi s_{W}^{2}}\VCKM^{l,i}\VCKM^{l,i*}\left\{ 2A_{0}\left[m_{d_{i}}^{2}\right]+m_{u_{l}}^{2}B_{0}\left[p^{2},m_{d_{i}}^{2},m_{u_{l}}^{2}\right]\right.\nonumber \\
 &  & \left.+\left(m_{d_{l}}^{2}-m_{u_{l}}^{2}+2p^{2}\right)B_{1}\left[p^{2},m_{u_{i}}^{2},m_{d_{l}}^{2}\right]\right\}\\
&&\nonumber \\
\Sigma_{WW}^{2\tilde{q}}&= & -\sum_{m,n}^{6}\sum_{i,j,k,l}^{3}\frac{3\alpha}{12\pi s_{W}^{2}}\VCKM^{k,i}\VCKM^{l,j*}R_{m,k}^{\tilde{u}}R_{m,l}^{\tilde{u}*}R_{n,j}^{\tilde{d}}R_{n,i}^{\tilde{d}*}\left\{ A_{0}\left[m_{\tilde{d}_{n}}^{2}\vphantom{m_{\tilde{d}_{l}}^{2}}\right]\right.\nonumber \\
&&+2m_{\tilde{u}_{m}}^{2}B_{0}\left[p^{2},m_{\tilde{u}_{m}}^{2},m_{\tilde{d}_{n}}^{2}\vphantom{m_{\tilde{d}_{l}}^{2}}\right]\nonumber \\
 && \left.+\left(p^{2}+m_{\tilde{u}_{m}}^{2}-m_{\tilde{d}_{n}}^{2}\right)B_{1}\left[p^{2},m_{\tilde{u}_{m}}^{2},m_{\tilde{d}_{n}}^{2}\vphantom{m_{\tilde{d}_{l}}^{2}}\right]-\frac{p^{2}}{3}+m_{\tilde{u}_{m}}^{2}+m_{\tilde{d}_{n}}^{2}\right\}\\
&&\nonumber \\
\Sigma_{WW}^{1\tilde{q}}&= & \sum_{l=1}^{6}\sum_{i=1}^{3}\frac{3}{8\pi s_{W}^{2}}\alpha A_{0}\left[m_{\tilde{u}_{l}}^{2}\vphantom{m_{\tilde{d}_{l}}^{2}}\right]R_{l,i}^{\tilde{u}}R_{l,i}^{\tilde{u}*}+\sum_{l=1}^{6}\sum_{i=1}^{3}\frac{3}{8\pi s_{W}^{2}}\alpha A_{0}\left[m_{\tilde{d}_{l}}^{2}\vphantom{m_{\tilde{d}_{l}}^{2}}\right]R_{l,i}^{\tilde{d}}R_{l,i}^{\tilde{d}*}
\end{eqnarray}

\begin{itemize}
\item Tadpoles
\end{itemize}
\begin{eqnarray}
T_{h}^{q} & = & -\sum_{i}^{3}\frac{3}{8\pi^{2}M_{W}s_{\beta}s_{W}}c_{\alpha}m_{u_{i}}^{2}eA_{0}\left[m_{u_{i}}^{2}\right]\\
T_{h}^{\tilde{q}} & = & \sum_{m}^{6}\sum_{i,j}^{3}\frac{1}{32\pi^{2}c_{W}M_{W}s_{\beta}s_{W}}eA_{0}\left[m_{\tilde{u}_{m}}^{2}\right]\nonumber \\
 &  & \times\left[\vphantom{\sum_{k=1}^{3}}\left\{ \vphantom{\sum_{k=1}^{3}}\delta_{i,j}\left(M_{W}m_{Z}s_{\alpha+\beta}s_{\beta}\left(-3+4s_{W}^{2}\right)+6c_{\alpha}c_{W}m_{u_{i}}^{2}\right)R_{m,j}^{\tilde{u}}\right.\right.\nonumber \\
 &  & \left.+3c_{W}\left(c_{\alpha}A_{i,j}^{u}+\mu^{*}s_{\alpha}\delta_{i,j}\right)m_{u_{i}}R_{m,3+j}^{\tilde{u}}\vphantom{\sum_{k=1}^{3}}\right\} R_{m,i}^{\tilde{u}*}
\nonumber \\
&&+\left\{ \vphantom{\sum_{k=1}^{3}}3c_{\alpha}c_{W}A_{j,i}^{u*}m_{u_{j}}R_{m,j}^{\tilde{u}}+3c_{W}\mu s_{\alpha}\delta_{i,j}m_{u_{i}}R_{m,j}^{\tilde{u}}\right.\nonumber \\
 &  & \left.\left.+2\delta_{i,j}\left(-2M_{W}m_{Z}s_{\alpha+\beta}s_{\beta}s_{W}^{2}+3c_{\alpha}c_{W}m_{u_{i}}^{2}\right)R_{m,3+j}^{\tilde{u}}\vphantom{\sum_{k=1}^{3}}\right\} R_{m,3+i}^{\tilde{u}*}\vphantom{\sum_{k=1}^{3}}\right]\nonumber \\
 &  & +\sum_{i}^{3}\frac{3}{8\pi^{2}M_{W}c_{\beta}s_{W}}s_{\alpha}m_{d_{i}}^{2}eA_{0}\left[m_{d_{i}}^{2}\right]-\sum_{m}^{6}\sum_{i,j}^{3}\frac{1}{32\pi^{2}c_{W}M_{W}c_{\beta}s_{W}}eA_{0}\left[m_{\tilde{d}_{m}}^{2}\right]\nonumber \\
 &  & \times\left[\vphantom{\sum_{k=1}^{3}}\left\{ \vphantom{\sum_{k=1}^{3}}\delta_{i,j}\left(M_{W}m_{Z}s_{\alpha+\beta}c_{\beta}\left(-3+2s_{W}^{2}\right)+6s_{\alpha}c_{W}m_{d_{i}}^{2}\right)R_{m,j}^{\tilde{d}}\right.\right.\nonumber \\
 &  & \left.+3c_{W}\left(s_{\alpha}A_{i,j}^{d}+\mu^{*}c_{\alpha}\delta_{i,j}\right)m_{d_{i}}R_{m,3+j}^{\tilde{d}}\vphantom{\sum_{k=1}^{3}}\right\} R_{m,i}^{\tilde{d}*}
\nonumber \\
&&+\left\{ \vphantom{\sum_{k=1}^{3}}3s_{\alpha}c_{W}A_{j,i}^{d*}m_{d_{j}}R_{m,j}^{\tilde{d}}+3c_{W}\mu c_{\alpha}\delta_{i,j}m_{d_{i}}R_{m,j}^{\tilde{d}}\right.\nonumber \\
 &  & \left.\left.-2\delta_{i,j}\left(M_{W}m_{Z}s_{\alpha+\beta}c_{\beta}s_{W}^{2}-3s_{\alpha}c_{W}m_{d_{i}}^{2}\right)R_{m,3+j}^{\tilde{d}}\vphantom{\sum_{k=1}^{3}}\right\} R_{m,3+i}^{\tilde{d}*}\vphantom{\sum_{k=1}^{3}}\right]\end{eqnarray}

The self-energies for $T_{H}$ are obtained using the replacements
of eq.(\ref{eq:cambio}) on the results of~$T_{h}$.

\end{appendix}
%%%%%%%%%%%%%%%%%%%%%%%%%%%%%%%%%%%%%%%%%%%%%%%%%%%%%%%%%%%%%%%%%%%%%%%%%%%%%%%

\newpage
% \pagebreak
% \clearpage


\begin{thebibliography}{99} 

%1
\bibitem{mssm} H.P.~Nilles, 
               {\em Phys.\ Rep.} {\bf 110} (1984) 1; \\ 
               %%CITATION = PRPLC,110,1;%%
               H.E.~Haber and G.L.~Kane, 
               {\em Phys.\ Rep.} {\bf 117} (1985) 75; \\  
               %%CITATION = PRPLC,117,75;%%
               R.~Barbieri, 
               {\em Riv.\ Nuovo Cim.} {\bf 11} (1988) 1. 
               %%CITATION = RNCIB,11,1;%%
%2
\bibitem{reviews} S.~Heinemeyer, 
  {\em Int.\ J.\ Mod.\ Phys.} {\bf A 21} 2659 (2006)
  [arXiv:hep-ph/0407244];
  %%CITATION = HEP-PH 0407244;%%
  A.~Djouadi, 
  {\em Phys.\ Rept.} {\bf 459} (2008) 1
  [arXiv:hep-ph/0503173].
  %%CITATION = HEP-PH 0503173;%%  
%3
\bibitem{ATLASsusy11} W.~Ehrenfeld, talk given at {\em SUSY11}, Fermilab, August 2011,\\
 {\tt https://indico.fnal.gov/contributionDisplay.py?sessionId=10\&\\contribId=257\&confId=3563}.
%4
\bibitem{CMSsusy11} I.~Melzer-Pellmann, talk given at {\em SUSY11}, Fermilab, August 2011,\\
 {\tt https://indico.fnal.gov/contributionDisplay.py?sessionId=10\&\\contribId=258\&confId=3563}. 


%5

\bibitem{Nakamura:2010zzi}
  K.~Nakamura {\it et al.}  [Particle Data Group],
  J.\ Phys.\ G {\bf 37}, 075021 (2010).
  %%CITATION = JPHGB,G37,075021;%%


%6
\bibitem{LEPHiggsSM} [LEP Higgs working group],
                     {\em Phys. Lett.} {\bf B 565} (2003) 61
                     [arXiv:hep-ex/0306033].
                     %%CITATION = HEP-EX 0306033;%%

%7
\bibitem{LEPHiggsMSSM} [LEP Higgs working group],
                       {\em Eur.\ Phys.\ J.} {\bf C 47} (2006) 547
                       [arXiv:hep-ex/0602042].
                       %%CITATION = HEP-EX 0602042;%%

%8
\bibitem{Tevbounds} {\tt http://tevnphwg.fnal.gov/} and references therein.

%9
\bibitem{LHCHiggsbounds} M. Vazquez Acosta, talk given at {\em SUSY11}, Fermilab, August 2011,\\
 {\tt https://indico.fnal.gov/contributionDisplay.py?sessionId=10\&\\contribId=261\&confId=3563}. 

%10
\bibitem{lhctdrs} G.~Aad et al. [The ATLAS Collaboration],
                  arXiv:0901.0512;\\
                  %%CITATION = ARXIV:0901.0512;%%
                  G.~Bayatian et al. [CMS Collaboration],
                  {\em J.\ Phys.} {\bf G 34} (2007) 995.
                  %%CITATION = JPHGB,G34,995;%%
%
%
%

%11
\bibitem{jakobs} V.~B\"uscher and K.~Jakobs,
                 {\em Int.\ J.\ Mod.\ Phys.} {\bf A 20} (2005) 2523
                 [arXiv:hep-ph/0504099].
                 %%CITATION = HEP-PH 0504099;%%
%

%12
\bibitem{tesla} J.~Aguilar-Saavedra et al.,
                TESLA TDR Part~3: 
                ``Physics at an $e^+e^-$ Linear Collider'', 
                arXiv:hep-ph/0106315,
                %%CITATION = HEP-PH 0106315;%%
                see: {\tt tesla.desy.de/tdr/};\\
                K.~Ackermann et al.,
                DESY-PROC-2004-01,
                {\em prepared for 4th ECFA / DESY Workshop on Physics
                and Detectors for a 90-GeV to 800-GeV Linear e+ e-
                Collider, Amsterdam, The Netherlands, 1-4 Apr 2003}.
%
%

%13
\bibitem{Snowmass05Higgs} S.~Heinemeyer et al.,
                          arXiv:hep-ph/0511332.
                          %%CITATION = HEP-PH 0511332;%%



%14
\bibitem{lhcilc} [LHC / ILC Study Group], G.~Weiglein et al.,
                 {\em Phys. Rept.} {\bf 426} (2006) 47
                 [arXiv:hep-ph/0410364];\\
                 %%CITATION = HEP-PH 0410364;%%
  A.~De Roeck et al.,
  {\em Eur.\ Phys.\ J.} {\bf C 66} (2010) 525
  [arXiv:0909.3240 [hep-ph]].
  %%CITATION = EPHJA,C66,525;%%
%

%15
\bibitem{mhcMSSM2L} S.~Heinemeyer, W.~Hollik, H.~Rzehak and G.~Weiglein,
                    {\em Phys. Lett.} {\bf B 652} (2007) 300
                    [arXiv:0705.0746 [hep-ph]].
                    %%CITATION = ARXIV:0705.0746;%%

%16
\bibitem{mhcMSSMlong} M.~Frank, T.~Hahn, S.~Heinemeyer, W.~Hollik,  
                      H.~Rzehak and G.~Weiglein,
                      {\em JHEP} {\bf 0702} (2007) 047
                      [arXiv:hep-ph/0611326].
                      %%CITATION = HEP-PH 0611326;%%


%17
\bibitem{mhiggsAEC} G.~Degrassi, S.~Heinemeyer, W.~Hollik,
                   P.~Slavich and G.~Weiglein,
                   {\em Eur. Phys. J.} {\bf C 28} (2003) 133
                   [arXiv:hep-ph/0212020].
                   %%CITATION = HEP-PH 0212020;%%

%18
\bibitem{mhiggsEP3l} S.~Martin,
                     {\em Phys.\ Rev.} {\bf D 75 } (2007)  055005
                     [arXiv:hep-ph/0701051].

%19
\bibitem{mhiggsFD3l} R.~Harlander, P.~Kant, L.~Mihaila and M.~Steinhauser,
                     {\em Phys.\ Rev.\ Lett.} {\bf 100} (2008) 191602
                     [{\em Phys.\ Rev.\ Lett.} {\bf 101} (2008) 039901]
                     [arXiv:0803.0672 [hep-ph]];
                     %%CITATION = PRLTA,101,039901;%%
                     {\em JHEP} {\bf 1008 } (2010) 104
                     [arXiv:1005.5709 [hep-ph]].


%20
\bibitem{mhNMFVearly} S.~Heinemeyer, W.~Hollik, F.~Merz and S.~Pe\~naranda,
                      {\em Eur.\ Phys.\ J.} {\bf C 37} (2004) 481
                      [arXiv:hep-ph/0403228].
                      %%CITATION = EPHJA,C37,481;%%

%21

\bibitem{feynhiggs} S.~Heinemeyer, W.~Hollik and G.~Weiglein,
                   {\em Comput. Phys. Commun.} {\bf 124} (2000) 76
                   [arXiv:hep-ph/9812320];
                   %%CITATION = HEP-PH 9812320;%%
                   see {\tt www.feynhiggs.de} .
%22
\bibitem{mhiggslong} S.~Heinemeyer, W.~Hollik and G.~Weiglein,
                    {\em Eur. Phys. J.} {\bf C 9} (1999) 343
                    [arXiv:hep-ph/9812472].
                    %%CITATION = HEP-PH 9812472;%%

%23
\bibitem{HdecNMFV} J.~Guasch and J.~Sola,
                   {\em Nucl. Phys.} {\bf B 562} (1999) 3, 
                   hep-ph/9906268;\\
                   %%CITATION = HEP-PH 9906268;%%
                   S.~Bejar, F.~Dilme, J.~Guasch and J.~Sola,
                   hep-ph/0402188.
                   %%CITATION = HEP-PH 0402188;%%

%24
\bibitem{HdecNMFV2} A.~Curiel, M.~Herrero and D.~Temes,
                    {\em Phys. Rev.} {\bf D 67} (2003) 075008, 
                    hep-ph/0210335;\\
                    %%CITATION = HEP-PH 0210335;%%
                    A.~Curiel, M.~Herrero, W.~Hollik, F.~Merz and 
                    S.~Pe{\~n}aranda, 
                    {\em Phys. Rev.} {\bf D 69} (2004) 075009, 
                    hep-ph/0312135.
                    %%CITATION = HEP-PH 0312135;%%

%25
\bibitem{HdecNMFV3} T.~Hahn, W.~Hollik, J.~Illana and S.~Pe\~naranda,
                    arXiv:hep-ph/0512315.
                    %%CITATION = HEP-PH/0512315;%%

    %\cite{Cao:2006xb}
    \bibitem{Cao1}
      J.~Cao, G.~Eilam, K.~i.~Hikasa and J.~M.~Yang,
      %``Experimental constraints on stop-scharm flavor mixing and implications  in
      %top-quark FCNC processes,''
      {\em Phys.\ Rev.} {\bf D 74} (2006) 031701
      [arXiv:hep-ph/0604163].
      %%CITATION = PHRVA,D74,031701;%%



\bibitem{Dittmaier:2007uw}
  S.~Dittmaier, G.~Hiller, T.~Plehn and M.~Spannowsky,
  %``Charged-Higgs Collider Signals with or without Flavor,''
  Phys.\ Rev.\  D {\bf 77} (2008) 115001
  [arXiv:0708.0940 [hep-ph]].
  %%CITATION = PHRVA,D77,115001;%%


%\cite{Cao:2007dk}
\bibitem{Cao2}
  J.~J.~Cao, G.~Eilam, M.~Frank, K.~Hikasa, G.~L.~Liu, I.~Turan and J.~M.~Yang,
  %``SUSY-induced FCNC top-quark processes at the Large Hadron Collider,''
  Phys.\ Rev.\  D {\bf 75} (2007) 075021
  [arXiv:hep-ph/0702264].
  %%CITATION = PHRVA,D75,075021;%%



%26
\bibitem{Degrassi:2007kj}
 G.~Degrassi, P.~Gambino, P.~Slavich,
 Comput.\ Phys.\ Commun.\  {\bf 179 } (2008)  759-771.
 [arXiv:0712.3265 [hep-ph]].
%27
\bibitem{Degrassi:2006eh}
 G.~Degrassi, P.~Gambino, P.~Slavich,
 Phys.\ Lett.\  {\bf B635 } (2006)  335-342.
 [arXiv:hep-ph/0601135 [hep-ph]].


%28
\bibitem{sufla}
 G.~Isidori and P.~Paradisi,
  Phys.\ Lett.\ B {\bf 639} (2006) 499
  [arXiv:hep-ph/0605012];
  %%CITATION = HEP-PH 0605012;%%
  G.~Isidori, F.~Mescia, P.~Paradisi and D.~Temes,
  Phys.\ Rev.\  D {\bf 75} (2007) 115019
  [arXiv:hep-ph/0703035], and references therein.
  %%CITATION = PHRVA,D75,115019;%% 


%29

\bibitem{Baek:2001kh}
  S.~Baek, T.~Goto, Y.~Okada and K.~i.~Okumura,
  Phys.\ Rev.\  D {\bf 64}, 095001 (2001)
  [arXiv:hep-ph/0104146].
  %%CITATION = PHRVA,D64,095001;%%  

%30
\bibitem{Foster:2005wb}
  J.~Foster, K.~i.~Okumura and L.~Roszkowski,
  JHEP {\bf 0508}, 094 (2005)
  [arXiv:hep-ph/0506146].
  %%CITATION = JHEPA,0508,094;%%


%31
\bibitem{Gabbiani:1996hi}
  F.~Gabbiani, E.~Gabrielli, A.~Masiero and L.~Silvestrini,
  Nucl.\ Phys.\  B {\bf 477}, 321 (1996)
  [arXiv:hep-ph/9604387].
  %%CITATION = NUPHA,B477,321;%%

%32
\bibitem{Becirevic:2001jj}
  D.~Becirevic {\it et al.},
  Nucl.\ Phys.\  B {\bf 634}, 105 (2002)
  [arXiv:hep-ph/0112303].
  %%CITATION = NUPHA,B634,105;%%


%33
\bibitem{Hall:1985dx}
  L.~J.~Hall, V.~A.~Kostelecky and S.~Raby,
  %``New Flavor Violations in Supergravity Models,''
  Nucl.\ Phys.\  B {\bf 267}, 415 (1986).
  %%CITATION = NUPHA,B267,415;%%


%34
\bibitem{Gunion:1986yn}
J.~Gunion and H.~Haber, 
{\em Nucl. Phys.} {\bf B 272} (1986) 1
[Erratum-ibid. {\bf B 402} (1993) 567].
%

%35
\bibitem{feynarts} J.~K\"ublbeck, M.~B\"ohm and A.~Denner, 
                   {\em Comput. Phys. Commun.} {\bf 60} (1990) 165;\\
                   %%CITATION = CPHCB,60,165;%%
                   T.~Hahn, 
                   {\em Comput. Phys. Commun.} {\bf 140} (2001) 418
                   [arXiv:hep-ph/0012260].
                   %%CITATION = HEP-PH 0012260;%%
                   The program and the user's guide 
                   are available via {\tt www.feynarts.de} .

%36
\bibitem{formcalc} T.~Hahn and M.~P\'erez-Victoria,
                   {\em Comput. Phys. Commun.} {\bf 118} (1999) 153
                   [arXiv:hep-ph/9807565].
                   %%CITATION = HEP-PH 9807565;%%


%37
\bibitem{famssm}   T.~Hahn and C.~Schappacher, 
                   {\em Comput. Phys. Commun.} {\bf 143} (2002) 54
                   [arXiv:hep-ph/0105349].
                   %%CITATION = HEP-PH 0105349;%%


%38
\bibitem{Gambino:2001ew}
  P.~Gambino and M.~Misiak,
  Nucl.\ Phys.\  B {\bf 611} (2001) 338
  [arXiv:hep-ph/0104034].
  %%CITATION = NUPHA,B611,338;%%

%39
\bibitem{AbdusSalam:2011fc}
 S.~S.~AbdusSalam {\it et al.},
 [arXiv:1109.3859 [hep-ph]].


%40
\bibitem{Porod:2011nf}
  W.~Porod and F.~Staub,
  arXiv:1104.1573 [hep-ph].
  %%CITATION = ARXIV:1104.1573;%%
%

%41
\bibitem{Allanach:2002nj}
  B.~C.~Allanach {\it et al.},
   Eur.\ Phys.\ J.\  C {\bf 25}, 113 (2002)
  [arXiv:hep-ph/0202233].
  %%CITATION = EPHJA,C25,113;%%
%

%42
\bibitem{Buchmueller:2011ki}
  O.~Buchmueller {\it et al.},
  arXiv:1106.2529 [hep-ph].
  %%CITATION = ARXIV:1106.2529;%%  
%


%43
\bibitem{Bertolini:1990if}
  S.~Bertolini, F.~Borzumati, A.~Masiero and G.~Ridolfi,
  Nucl.\ Phys.\  B {\bf 353}, 591 (1991).
  %%CITATION = NUPHA,B353,591;%%  

%44
\bibitem{Cho:1996we}
  P.~L.~Cho, M.~Misiak and D.~Wyler,
  Phys.\ Rev.\  D {\bf 54}, 3329 (1996)
  [arXiv:hep-ph/9601360].
  %%CITATION = PHRVA,D54,3329;%%

%45
\bibitem{Degrassi:2000qf}
  G.~Degrassi, P.~Gambino and G.~F.~Giudice,
  JHEP {\bf 0012}, 009 (2000)
  [arXiv:hep-ph/0009337].
  %%CITATION = JHEPA,0012,009;%%  

%53
\bibitem{Hurth:2003dk}
  T.~Hurth, E.~Lunghi and W.~Porod,
  Nucl.\ Phys.\  B {\bf 704}, 56 (2005)
  [arXiv:hep-ph/0312260].
  %%CITATION = NUPHA,B704,56;%%

%46
\bibitem{Ciuchini:1997xe}
  M.~Ciuchini, G.~Degrassi, P.~Gambino and G.~F.~Giudice,
  Nucl.\ Phys.\  B {\bf 527}, 21 (1998)
  [arXiv:hep-ph/9710335].
  %%CITATION = NUPHA,B527,21;%%
%

%47
\bibitem{Hall:1993gn}
  L.~J.~Hall, R.~Rattazzi and U.~Sarid,
  Phys.\ Rev.\  D {\bf 50}, 7048 (1994)
  [arXiv:hep-ph/9306309].
  %%CITATION = PHRVA,D50,7048;%%

%48
\bibitem{Carena:2000uj}
  M.~S.~Carena, D.~Garcia, U.~Nierste and C.~E.~M.~Wagner,
  Phys.\ Lett.\  B {\bf 499}, 141 (2001)
  [arXiv:hep-ph/0010003].
  %%CITATION = PHLTA,B499,141;%%

%49
\bibitem{Carena:1999py}
  M.~S.~Carena, D.~Garcia, U.~Nierste and C.~E.~M.~Wagner,
  Nucl.\ Phys.\  B {\bf 577}, 88 (2000)
  [arXiv:hep-ph/9912516].
  %%CITATION = NUPHA,B577,88;%% 

%50
\bibitem{Isidori:2001fv}
  G.~Isidori and A.~Retico,
  JHEP {\bf 0111}, 001 (2001)
  [arXiv:hep-ph/0110121].
  %%CITATION = JHEPA,0111,001;%%

%51
\bibitem{Buras:2002vd}
  A.~J.~Buras, P.~H.~Chankowski, J.~Rosiek and L.~Slawianowska,
  Nucl.\ Phys.\  B {\bf 659}, 3 (2003)
  [arXiv:hep-ph/0210145].
  %%CITATION = NUPHA,B659,3;%%

%52
\bibitem{Isidori:2002qe}
  G.~Isidori and A.~Retico,
  JHEP {\bf 0209}, 063 (2002)
  [arXiv:hep-ph/0208159].
  %%CITATION = JHEPA,0209,063;%%
%


%\cite{Crivellin:2009ar}
\bibitem{Crivellin1}
  A.~Crivellin and U.~Nierste,
  %``Chirally enhanced corrections to FCNC processes in the generic MSSM,''
  Phys.\ Rev.\  D {\bf 81} (2010) 095007
  [arXiv:0908.4404 [hep-ph]].
  %%CITATION = PHRVA,D81,095007;%%


%\cite{arXiv:0810.1613}
\bibitem{Crivellin2}
  A.~Crivellin and U.~Nierste,
  %``Supersymmetric renormalisation of the CKM matrix and new constraints on the squark mass matrices,''
  Phys.\ Rev.\ D\ {\bf 79} (2009) 035018
  [arXiv:0810.1613 [hep-ph]].
  %%CITATION = PHRVA,D79,035018;%%

\bibitem{Okumura:2002wa}
  K.~i.~Okumura and L.~Roszkowski,
  %``Weakened Constraints from b ---> s gamma on supersymmetry flavor mixing due to next-to-leading-order corrections,''
  Phys.\ Rev.\ Lett.\  {\bf 92} (2004) 161801
  [hep-ph/0208101].
  %%CITATION = HEP-PH/0208101;%%

\bibitem{Okumura:2003hy}
  K.~i.~Okumura and L.~Roszkowski,
  %``Large beyond leading order effects in b ---> s gamma in supersymmetry with general flavor mixing,''
  JHEP {\bf 0310} (2003) 024
  [hep-ph/0308102].
  %%CITATION = HEP-PH/0308102;%%

%60
\bibitem{Foster:2004vp}
  J.~Foster, K.~i.~Okumura and L.~Roszkowski,
  Phys.\ Lett.\  B {\bf 609}, 102 (2005)
  [arXiv:hep-ph/0410323].
  %%CITATION = PHLTA,B609,102;%%


%


%54
\bibitem{Asner:2010qj}
  D.~Asner {\it et al.}  [Heavy Flavor Averaging Group],\\
  {\tt http://www.slac.stanford.edu/xorg/hfag/}, 
  arXiv:1010.1589 [hep-ex].
  %%CITATION = ARXIV:1010.1589;%%
%

%55
\bibitem{Misiak:2009nr}
  M.~Misiak,
  Acta Phys.\ Polon.\  B {\bf 40}, 2987 (2009)
  [arXiv:0911.1651 [hep-ph]].
  %%CITATION = APPOA,B40,2987;%%

%56
\bibitem{Chankowski:2000ng}
  P.~H.~Chankowski and L.~Slawianowska,
  Phys.\ Rev.\  D {\bf 63}, 054012 (2001)
  [arXiv:hep-ph/0008046].
  %%CITATION = PHRVA,D63,054012;%%  

%57
\bibitem{Bobeth:2002ch}
  C.~Bobeth, T.~Ewerth, F.~Kruger and J.~Urban,
  Phys.\ Rev.\  D {\bf 66}, 074021 (2002)
  [arXiv:hep-ph/0204225].
  %%CITATION = PHRVA,D66,074021;%%

%58
\bibitem{Babu:1999hn}
  K.~S.~Babu and C.~F.~Kolda,
  Phys.\ Rev.\ Lett.\  {\bf 84}, 228 (2000)
  [arXiv:hep-ph/9909476].
  %%CITATION = PRLTA,84,228;%%
%

%59
\bibitem{Bobeth:2001sq}
  C.~Bobeth, T.~Ewerth, F.~Kruger and J.~Urban,
  Phys.\ Rev.\  D {\bf 64}, 074014 (2001)
  [arXiv:hep-ph/0104284].
  %%CITATION = PHRVA,D64,074014;%%

%61
\bibitem{CMSLHCb} CMS and LHCb Collaborations, \\
{\tt http://cdsweb.cern.ch/record/1374913/files/BPH-11-019-pas.pdf}.


%62
\bibitem{Buras:2009if}
  A.~J.~Buras,
  PoS E PS-HEP2009, 024 (2009)
  [arXiv:0910.1032 [hep-ph]].
  %%CITATION = POSCI,EPS-HEP2009,024;%%

%63
\bibitem{Becirevic:2001xt} 
  D.~Becirevic, V.~Gimenez, G.~Martinelli, M.~Papinutto and J.~Reyes,
 JHEP {\bf 0204}, 025 (2002)
  [arXiv:hep-lat/0110091];
%%CITATION = HEP-LAT 0110091;%%
Nucl.\ Phys.\ Proc.\ Suppl.\  {\bf 106}, 385 (2002)
[arXiv:hep-lat/0110117].
%%CITATION = HEP-LAT 0110117;%%


%64
\bibitem{Buras:1990fn}
  A.~J.~Buras, M.~Jamin and P.~H.~Weisz,
  Nucl.\ Phys.\  B {\bf 347}, 491 (1990).
  %%CITATION = NUPHA,B347,491;%%

%65
\bibitem{Golowich:2011cx}
  E.~Golowich, J.~Hewett, S.~Pakvasa, A.~A.~Petrov and G.~K.~Yeghiyan,
  Phys.\ Rev.\  D {\bf 83}, 114017 (2011)
  [arXiv:1102.0009 [hep-ph]].
  %%CITATION = PHRVA,D83,114017;%%
%

%66
\bibitem{refLunghi} Talk given by Enrico Lunghi at Lattice 2011 - July 11-16, 2011
http://tsailab.chem.pacific.edu/lat11/plenary/lunghi/lunghi-lattice2011.pdf

%67
\bibitem{hhg} J.~Gunion, H.~Haber, G.~Kane and S.~Dawson,
              {\em The Higgs Hunter's Guide}, Addison-Wesley, 1990.

%68
\bibitem{mhiggsf1lC} A.~Dabelstein,
                     {\em Nucl. Phys.} {\bf B 456} (1995) 25
                     [arXiv:hep-ph/9503443];
                     %%CITATION = HEP-PH 9503443;%%
                     {\em Z. Phys.} {\bf C 67} (1995) 495
                     [arXiv:hep-ph/9409375].
                     %%CITATION = HEP-PH 9409375;%%

%69
\bibitem{a0b0c0} A.~Denner,
  %``Techniques for calculation of electroweak radiative corrections at the one
  %loop level and results for W physics at LEP-200,''
  Fortsch.\ Phys.\  {\bf 41} (1993) 307
  [arXiv:0709.1075 [hep-ph]].
  %%CITATION = FPYKA,41,307;%%


\end{thebibliography}
\end{document}